\documentclass[binding=0.6cm]{sapthesis}

\usepackage{microtype}
\usepackage{hyperref}

\usepackage{amsmath}
\usepackage{amsfonts}
\usepackage{amssymb}
\usepackage{slashbox}

\font\msytw=msbm9 scaled\magstep1

\let\a=\alpha \let\b=\beta  \let\g=\gamma  \let\d=\delta \let\e=\varepsilon
  \let\h=\eta   \let\th=\theta \let\k=\kappa \let\l=\lambda
\let\m=\mu    \let\n=\nu    \let\x=\xi     \let\p=\pi    \let\r=\rho
\let\s=\sigma \let\t=\tau    \let\ph=\varphi\let\c=\chi
   \let\o=\omega
\let\G=\Gamma \let\D=\Delta  \let\L=\Lambda 
         
\let\O=\Omega 

\def\PPP{{\cal P}}\def\EE{{\cal E}} \def\VV{{\cal V}}
\def\CC{{\cal C}} \def\HHH{{\cal H}}\def\WW{{\cal W}}
\def\TT{{\cal T}}\def\NN{{\cal N}} \def\BBB{{\cal B}}\def\III{{\cal I}}
\def\RR{{\cal R}}\def\LL{{\cal L}}  \def\OO{{\cal O}}
\def\DD{{\cal D}}\def\AAA{{\cal A}}\def\GG{{\cal G}} \def\SS{{\cal S}}
\def\KK{{\cal K}}  

  \def\AA{{\bf A}} \def\qq{{\bf q}}
   \def\pp{{\bf p}}
 \def\xx{{\bf x}} \def\yy{{\bf y}} 

\def\kk{{\bf k}}
\def\TTT{{\bf T}}

\def\RRR{\hbox{\msytw R}} 
 \def\CCC{\hbox{\msytw C}}
 
\def\NNN{\hbox{\msytw N}} 
 \def\ZZZ{\hbox{\msytw Z}}

\def\\{\hfill\break}
\def\={:=}
\let\io=\infty
\let\0=\noindent
\def\media#1{{\langle#1\rangle}}
\let\dpr=\partial

\def\const{{\rm const}}
\def\tende#1{\,\vtop{\ialign{##\crcr\rightarrowfill\crcr\noalign{\kern-1pt
    \nointerlineskip} \hskip3.pt${\scriptstyle #1}$\hskip3.pt\crcr}}\,}
\def\otto{\,{\kern-1.truept\leftarrow\kern-5.truept\to\kern-1.truept}\,}

\def\to{\rightarrow}

\def\qed{\hfill\raise1pt\hbox{\vrule height5pt width5pt depth0pt}}
\def\Val{{\rm Val}}
\def\ul#1{{\underline#1}}
\def\lis{\overline}
\def\V#1{{\bf#1}}
\def\be{\begin{equation}}
\def\ee{\end{equation}}
\def\bea{\begin{eqnarray}}
\def\eea{\end{eqnarray}}
\def\nn{\nonumber}
\def\pref#1{(\ref{#1})}

\def\lb{\label}

\def\Tr{\mathrm{Tr}}

\newcommand{\sgn}{\text{sgn}}

\newtheorem{lemma}{Lemma}[section]
\newtheorem{theorem}{Theorem}[section]

\newtheorem{oss}{Remark}

\author{Marcello Porta}
\title{A lattice gauge theory model for graphene}
\submitdate{December 2010}
\cycle{XXIII}
\IDnumber{698454}
\advisor{Prof. Giovanni Gallavotti}
\advisor{Prof. Vieri Mastropietro}

\examdate{4 February 2011}
\examiner{Prof. Carlo Mariani (``Sapienza'', Roma)}
\examiner{Prof. Arianna Montorsi (Politecnico di Torino)}
\examiner{Prof. Olle S\"oderman (Lund University)}
\versiondate{8 December 2010}
\authoremail{marcello.porta@roma1.infn.it}

\begin{document}

\frontmatter
\maketitle

\tableofcontents

\mainmatter
\chapter{Introduction}\label{cap1}
\setcounter{equation}{0}
\renewcommand{\theequation}{\ref{cap1}.\arabic{equation}}

\section{Motivations and main results}\label{secmot}
\setcounter{equation}{0}
\renewcommand{\theequation}{\ref{secmot}.\arabic{equation}}

\subsection{Motivations}

Graphene is a newly discovered material, and its discovery \cite{N1} in 2004 by the group of A. Geim and K. Novoselov elicited an enormous interest in the physical community\footnote{Geim and Novoselov have been awarded the 2010 Nobel prize for Physics, ``for groundbreaking experiments regarding the two-dimensional material graphene''.}. Consisting in a {\it single} monoatomic layer of graphite, graphene can be seen as the first realization of a {\it two-dimensional crystal}; its stability is a remarkable and not yet completely understood issue, since, according to Mermin-Wagner theorem, two-dimensional systems are not expected to exist in nature. Before its experimental discovery, graphene was known to theorists as an interesting academical problem; the first research paper on graphene dates back to the work of Wallace in 1947, \cite{W}, where graphene was studied as the ``building brick'' of graphite. Some of the very peculiar features of graphene were brought to the attention of the physical community many years later, see \cite{S, F1, F2, Ha, V1} for instance; already at the time of these works it was understood that a similar physical compound, {\it if existing in nature}, would exibit unique physical properties.

\begin{figure}[htbp]
\centering
\includegraphics[width=0.8\textwidth]{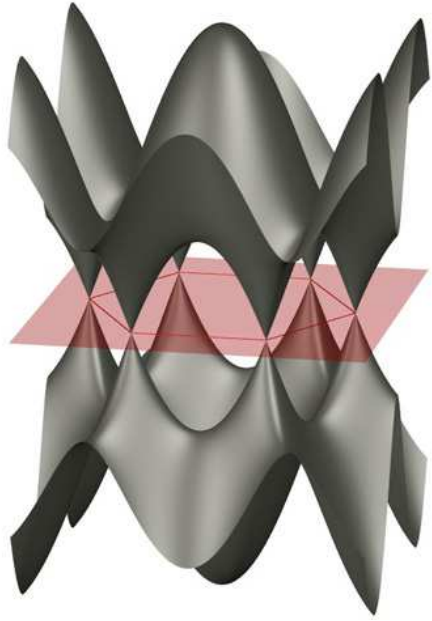} \caption{Sketch of the energy dispersion relation of graphene; the plot reports the energy $\EE(\vec k)$ as function of the momentum $\vec k$. The upper/lower manifolds correspond respectively to the conduction and valence bands; the bands meet at six points, among which only two are independent (the others lie outside the first Brillouin zone), and are called {\it Fermi points} $\vec p_{F}^{+}$, $\vec p_{F}^{-}$.} \label{figcone}
\end{figure}

\begin{figure}[htbp]
\centering
\includegraphics[width=0.5\textwidth]{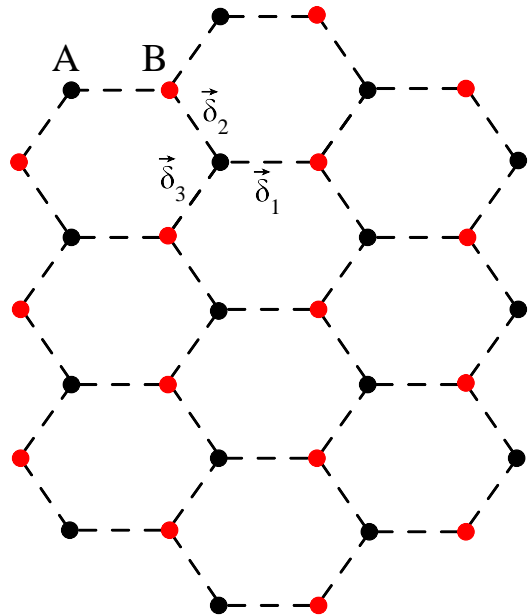} \caption{Sketch of the honeycomb lattice of graphene; it can be seen as the superposition of two shifted triangular sublattices, namely $\L_{A}$ (black dots) and $\L_{B}$ (red dots). On each site sits a carbon atom, which is $sp^2$-bonded to its nearest neighbours; the length of the carbon-carbon bonds is about $0.142$ nm.} \label{fighex}
\end{figure}

From the pioneering experimental work of Novoselov, Geim {\it et al.}, \cite{N1, N2}, graphene has shown immediatly a rich variety of interesting electronic and structural properties. For instance, the hallmark of graphene is the shape of its electronic dispersion relation, which vanishes {\it linearly} at the Fermi surface, given by only two inequivalent points at neutrality ({\it i.e.} when there is only one electron per site in average, that is at {\it half-filling}); see Fig. \ref{figcone} for a sketch of the energy dispersion relation in graphene.
As expected from the early works, \cite{W}, the conical shape of the low-energy excitations in graphene is a consequence of the {\it hexagonal} geometry of its two-dimensional lattice, see Fig. \ref{fighex}. 
Such property implies that the low energy excitations of graphene are effectively described by {\it massless Dirac fermions}, with velocity $v$ much smaller than the speed of light (experimentally, $v\sim c/300$); this fact stimulated a fruitful cross-fertilization between different areas of Physics, namely high energy physics and condensed matter. For instance, in the work of Semenoff \cite{S} graphene has been proposed as a condensed matter realization of $2+1$ dimensional Quantum Electrodynamics (QED); and since then, a lot of theoretical research in graphene has been motivated by fundamental questions in quantum field theory, {\it e.g.} the understanding of quantum anomalies \cite{S, Ha}, or more recently of electron fractionalization in two dimensional systems, \cite{HCM, JP, SeF, WeF}. 

Other very interesting physical properties of graphene are for instance the presence of a {\it minimal value} of the electric conductivity \cite{N2}, of the order of the conductance quantum $\s\sim e^2 / h$ (the exact prefactor is still debated), reached in presence of a vanishing density of charge carries; or the insensitivity to localization effects which are usually induced by disorder, a phenomenon which was already guessed by Fradkin in \cite{F1, F2}. Graphene also shows an unusual quantum Hall effect: the spacing between the Hall plateaus is $4e^2/h$ \cite{N2, ZTSK}, i.e. bigger than for the usual quantum Hall effect, and the steps occur at {\it half integer} multiples of this value. Finally, graphene is considered by many as the ideal canditate for interesting technological applications \cite{GeNo}, due to presence of {\it extremely high} electric and thermal condictivities (higher than copper and silver, respectively), and of the overall high electronic quality (the electrons have very long mean free paths); moreover, it is substantially stronger than steel and at the same time very stretchable.

Aim of this Thesis is to contribute to the understanding of the effect of the electron-electron interactions in graphene; this is a highly debated topic, also in view of the recent experimental realization of {\it suspended} graphene samples, \cite{Bo, Du, Li}, a playground in which electron-electron interactions - so far obscured by the presence of substrates - are expected to play a major role. This is confirmed by recent measurements of fractional quantum Hall energy gaps, \cite{Du, Bo, GZCBK}.
 
 We can consider two types of interactions: short range and long range interactions, corresponding respectively to {\it screened} or {\it unscreened} Coulomb interactions; in this Thesis we shall consider the latter, however let us brefly mention what it is known for the former. It is widely believed that the presence of weak short range interactions shoud not affect too much the low energy properties of electrons on the honeycomb lattice; this was argued on the basis of one-loop computations, see for instance \cite{V1} and references therein. Remarkably, this belief has been recently {\it rigorously proved} by Giuliani and Mastropietro in \cite{GM, GM2}; they considered the Hubbard model on the honeycomb lattice and proved that the free energy and the Schwinger functions are {\it analytic} functions of the coupling constant uniformly in the temperature and in the system size, provided the coupling is small enough. From the mathematical physics point of view, this is one of the two cases in which the ground state of a two dimensional interacting fermionic system can be rigorously constructed; the other one is \cite{FKT}, where two dimensional fermions with highly asymmetric Fermi surface were considered.

Regarding unscreened Coulomb interactions, a general picture of their effect on the low energy physics of graphene is still lacking; however, there is a wide consensus that they should play a nontrivial role. For instance, it has been first proposed by Gonz\'alez, Guinea and Vozmediano, \cite{V1, V2, V3} that the Fermi velocity is {\it strongly renormalized} by the interaction; in particular, it is believed that the presence of unscreened Coulomb interaction between the electrons produces an unbounded logarithmic growth of the effective Fermi velocity close to the Fermi points. This behavior has been claimed by many authors, see \cite{Mi, He1, KUC} for instance; all these arguments heavily rely on lowest order perturbation theory, and are affected by crude approximations: for instance, the presence of the lattice is always neglected and a momentum cutoff is imposed by hand. Attempt to include higher order corrections have been done in \cite{Mi0}, and the same logarithmic behavior has been found at two-loops in perturbation theory. 
 The effect of such a strong renormalization of the Fermi velocity on physical observables is still puzzling and not yet fully understood; for instance, in \cite{Mi0} it was claimed that because of this phenomenon graphene in the vacuum is an insulator, a claim which was subsequently proven to be wrong, \cite{SS, He1}. 

A quite unsatisfactory feature of all the above results is that they are {\it regularization dependent}; in fact, because of the absence of the lattice an ultraviolet regularization scheme has to be introduced. This led to an explicit dependence of physical quantities on the cutoffs, and to controversies in the literature: see \cite{Mi2, He1, He3}, where {\it three} different expressions for the optical conductivity of graphene were found using momentum \cite{Mi2, He1} or dimensional \cite{He3} regularization. And no one of these results succeeds in explaining why experimentally \cite{Na} the optical conductivity of graphene appears to be {\it universal}, that is independent of electron-electron interactions and lattice parameters.

Another interesting issue is whether interaction-dependent anomalous exponents are present in the many body correlations of graphene. The presence of anomalous scaling was first argued in \cite{V2} on the basis of lowest order perturbation theory, and considering a continuum model in presence of dimensional regularization; in fact, in \cite{V2} a log-correction to the two point correlation function was found, and this was {\it interpreted} as the lowest order contribution to an anomalous power law decay. The presence of anomalous scaling in the correlations would suggest that the system behaves as a {\it Luttinger liquid} \cite{Ha2}, a quite common feature of one dimensional systems, which has never been established in two dimensions. A proof of this would be of great physical interest, for instance in view of the role that two dimensional Luttinger liquids are expected to play in high temperature superconductivity, as argued by Anderson \cite{A}. Recently, \cite{KUC, Son}, on the basis of ``large $N$ expansions'', where $N$ is the number of fermionic ``species'', anomalous exponents in the scaling of various physical quantities were found; but still, the contact with actual graphene is not clear, since in graphene $N=4$. 

Large $N$ expansions are widely used tools in perturbative analyses of quantum field theory models; in particular, in the eighties they allowed to argue the emergence of spontaneous chiral symmetry breaking in $2+1$ dimensional QED,  \cite{Ap1, Ap2}. Recently, the same mechanism has been revisited and applied to graphene models \cite{K1, KL, GGMS, K2, KUC}, in order to predict the occurrence of a metal-insulator transition parametrized by the strength of the many body coupling; this issue is of great interest for possible technological applications. The presence of a mass gap has also been claimed on the basis of Monte Carlo calculations, see \cite{DL1, DL2}. But, again, from a theoretical point of view it is not clear how to control the error involved when choosing $N=4$ in the large $N$ expansion. Finally, another debated point is the effect of the strong renormalization of the Fermi velocity in the gap generation; it has been recently proposed in \cite{SSG} that the growth of the velocity due to the presence of unscreened Coulomb interactions depresses the phenomenon of gap opening, while in \cite{SZ} it is claimed that Coulomb interactions support the gap once it is open. 

Therefore, at present time no unambiguous prediction on the effect of electron-electron interactions in graphene can be made. The main way to overcome these difficulties is to ``start from scratch'', that is from the definiton of the model that one wish to study. In fact, the typical model which has been considered so far to describe Coulomb interactions in graphene has the following Hamiltonian:
\bea
&&\HHH = \HHH^{D} + \HHH^{C}\;,\label{ris1}\\
&&\HHH^{D} = \mbox{free Hamiltonian of $2+1$ dim. massless Dirac fermions with velocity $v$,}\nn\\
&&\HHH^{C} = \mbox{three dimensional instantaneous Coulomb interaction;}\nn
\eea
therefore, in (\ref{ris1}) the honeycomb structure of the crystalline lattice is completely lost; moreover, the coupling with the electromagnetic field is not complete, since the vector potential does not enter in $\HHH$. Usually, this omission is justified by arguing that the contribution from the vector potential should be negligible in view of the fact that the ratio $v/c$ is small; however, this argument cannot be consistent at all energy scales, since in this kind of models the interacting Fermi velocity as function of the momentum of the quasi-particles {\it blows up} at the Fermi surface. Moreover, the omission of the vector potential breaks {\it gauge invariance}, which has to be respected in order to correctly define the coupling with the electromagnetic field.

\subsection{Main results}\label{secmaires}

In this Thesis we shall introduce a {\it new model} for graphene in presence of electromagnetic interactions, which takes into account the honeycomb lattice, and where the coupling with the electromagnetic field is gauge symmetry preserving; in the language of high energy physics, the model that we shall consider is a ``lattice gauge theory model''. Then, we will study the model using rigorous Renormalization Group (RG) methods, which have been developed in the last $25$ years by the Roma school of Constructive Quantum Field Theory, see \cite{BG1, GeM, M} for extensive reviews; this approach is based on the {\it functional Renormalization Group} developed in the 1980's starting from \cite{GN1, GN2, P}. The methods have been originally devised to prove the ultraviolet stability of various quantum field theories, like $\varphi^{4}$ theory in four dimensions, see \cite{GN1, GN2, G84}; subsequently, the same ideas have been extended to develop a general and rigorous RG approach to interacting fermionic systems, \cite{BG}. Since then, these techniques have been proved very effective to study many interacting quantum physical models; in particular, they allowed to rigorously prove the Luttinger liquid behavior of various one dimensional systems, see \cite{BGPS, BoM, BM2, M5} for instance, and more recently to investigate the ground state and low energy properties of two dimensional systems \cite{BGM, BGM2, GM, GM2, GMP1}.

Our results, contained in \cite{GMP2}, can be briefly and informally summarized as follows; see Section \ref{secres} for a more extensive discussion.

\begin{itemize}
\item {\it Anomalous scaling of the two point Schwinger function.} We show that the scaling properties of the ``dressed propagator'' is dramatically changed by the interaction; in particular, interaction dependent anomalous exponents appear, as usual in Luttinger liquids. 
\item {\it Emergent Lorentz symmetry.} The effective Fermi velocity, that is the one appearing in the leading contribution to the two point Schwinger function at momenta close to the Fermi surface, tends to the speed of light.
\item {\it Enhancement of the ``excitonic'' and ``charge density wave'' response functions.} We compute various response functions, and we show that their space-time decays are governed by anomalous exponents. In particular, we find that the response functions associated to special lattice distortions, the {\it Kekul\' e} ones (see Fig. \ref{figkek0}), or to a periodic charge inequivalence on the sites of the honeycomb lattice, a {\it charge density wave}, are {\it amplified} by the interaction.
\item {\it Enhancement of Kekul\' e distortions.} We show that small Kekul\' e distortions of the honeycomb lattice are {\it dramatically enhanced} by the electron-electron interactions.
\item {\it A mechanism for spontaneous lattice distortion.} Considering the lattice distortion as a dynamical variable, we show that the energy of the system is extremized in correspondence of a Kekul\' e distortion; the amplitude of the distortion is determined by a self-consistence non-BCS equation, from which we argue that strong electromagnetic interactions favor the emergence of a spontaneous Kekul\' e distortion and the opening of a gap in the fermionic energy spectrum.
\end{itemize}

Our results are true at {\it all orders} in renormalized perturbation theory; in particular, we provide explicit (``$N!$'') bounds on the coefficient of the series of all the quantities that we compute, which are uniform in the temperature and in the system size. These bounds are not enough to prove absolute convergence of the series; however, here we shall prove that perturbation theory is {\it consistent to all orders}, in the sense of \cite{BG}.

A key role in the derivation of the above results is played by gauge invariance and lattice Ward identities (WI); in particular, following a strategy similar to the one introduced in \cite{BM} to prove Luttinger liquid behavior in one dimensional systems, the WIs allow us to prove that the Beta function of the effective charge is {\it asymptotically vanishing}, which means that the renormalized charge is close to the bare one.

\section{A honeycomb lattice gauge theory}\label{sec1}
\setcounter{equation}{0}
\renewcommand{\theequation}{\ref{sec1}.\arabic{equation}} 

Let us start by defining the model that we introduced and studied; our goal is to describe electrons hopping on the hexagonal lattice interacting through a quantized three dimensional electromagnetic field. For this purpose, the Hamiltonian we consider has the following structure, see \cite{GMP2}:
\be
\HHH_{\L} = \HHH^{hop}_{\L} + \HHH^{C}_{\L} + \HHH^{A}_{\L}\;,\label{1.1.1}
\ee
where: (i) $\HHH^{hop}_{\L}$ is the {\it gauge invariant hopping term}; (ii) $\HHH^{C}_{\L}$ contains the contribution due to the instantaneous Coulomb interaction between the electrons on the hexagonal lattice; (iii) $\HHH^{A}_{\L}$ is the energy of the free photon field. Choosing units such that $\hbar=c=1$, these three terms are given by:
\bea
\HHH^{hop}_{\L} &:=& -t\sum_{\substack{\vec x\in \L \\ i=1,2,3}}\sum_{\s = \uparrow\downarrow} a^{+}_{\vec x,\s}b^{-}_{\vec x + \vec\d_i,\s}\eu^{\iu e A_{(\vec x,i)}} + b^{+}_{\vec x + \vec\d_i,\s}a^{-}_{\vec x,\s}\eu^{-\iu e A_{(\vec x,i)}}\;,\label{1.1.1b}\\
\HHH^{C}_{\L} &:=& \frac{e^2}{2}\sum_{\vec x\in \L_{A}\cup \L_{B}}\big( n_{\vec x} - 1 \big)\varphi(\vec x - \vec y,0)\big( n_{\vec y} - 1 \big)\nn\;,\\
\HHH^{A}_{\L} &:=& \frac{1}{L\AAA_{\L}}\sum_{r=1,2}\sum_{\ul p \in \widetilde\PPP_{L}} |\ul{p}|c^{+}_{\ul{p},r}c^{-}_{\ul{p},r}\;,\nn
\eea
where:
\begin{enumerate}
\item $\L_{A} = \L$ is a periodic triangular lattice, defined as $\L = \mathbb{B}/L\mathbb{B}$, where $L\in \NNN$ and $\mathbb{B}$ is the triangular lattice with basis vectors $\vec a_{1} = \frac{1}{2}(3,\sqrt{3})$, $\vec a_{2} = \frac{1}{2}(3,-\sqrt{3})$; $\AAA_{\L}$ is the area of $\L$. We denote by $\L^{*}$ the dual of $\L$; its basis vectors are $\vec b_{1} = \frac{2\pi}{3}(1,\,\sqrt{3})$, $\vec b_{2} = \frac{2\pi}{3}(1,\,-\sqrt{3})$.
\item The vectors $\vec \d_{1} := (1,0)$, $\vec \d_{2} := \frac{1}{2}(-1, \sqrt{3})$, $\vec\d_{3} := \frac{1}{2}(-1,-\sqrt{3})$ connect each site $\vec x$ to its three nearest neighbours.
\item The operators $a^{\pm}_{\vec x,\s}$, $b^{\pm}_{\vec x + \vec\d_{j},\s}$ are real space {\it fermionic creation/annihilation operators}, acting respectively on $\L_{A}$, $\L_{B} := \L_{A} + \vec\d_{1}$ and satisfying periodic boundary conditions in $\vec x\in \L_{A}$; they satisfy the anticommutation relations
\bea
&&\{ a^{+}_{\vec x,\s},\, a^{-}_{\vec x',\s'} \} = \{ b^{+}_{\vec x + \vec\d_1,\s},\, b^{-}_{\vec x' + \vec\d_1,\s'} \}= \d_{\vec x,\vec x'}\d_{\s,\s'}\,,\nn\\
&&\{a^{\e}_{\vec x,\s},\, b^{\e'}_{\vec y,\s'}\} = \{ a^{\e}_{\vec x,\s},\, a^{\e}_{\vec y,\s'} \} = \{ b^{\e}_{\vec x + \vec\d_1,\s},\,b^{\e}_{\vec y + \vec\d_1,\s'} \}=0\;.
\eea
Setting $\ul v = (\vec v,v_3)$ with $\vec v\in \RRR^{2}$, the operators $c^{\pm}_{\ul{p},r}$ are momentum space {\it bosonic creation/annihilation operators}, defined on $\widetilde \PPP_{L} := \PPP_{L}\cup \{p_{3} = \frac{2\pi n}{L},\, n\in \ZZZ\}$ with $\PPP_{L} := \{\vec p = \frac{n_{1}}{L}\vec b_{1} + \frac{n_{2}}{L}\vec b_{2}:\,\vec n\in \ZZZ^{2}\}$, and satisfying the commutation relations
\be
[c^{-}_{\ul p,r}, c^{+}_{\ul p',r'}] = L\AAA_{\L}\d_{\ul p,\ul p'}\d_{r,r'}\;,\qquad [c^{\e}_{\ul p,r}, c^{\e}_{\ul p',r'}] = 0\;.\label{1.1.2a}
\ee
%
\item $A_{(\vec x,i)} := \int_{0}^{1} ds\,\vec\d_{i}\cdot \vec A_{(\vec x + s\vec\d_{i},0)}$, and $\ul{A}_{\ul y}$ is the {\em three} dimensional quantized electromagnetic vector potential in the Coulomb gauge and in presence of an ultraviolet and an infrared cutoff:
\bea
&&\ul{A}_{\ul y} = \frac{1}{\AAA_{\L} L}\sum_{\ul p \in \widetilde\PPP_{L}}\sqrt{\frac{\chi_{[h^{*},0]}(|\ul{p}|)}{2|\ul{p}|}}\Big(\ul{\varepsilon}_{\ul{p},r}c^{-}_{\ul{p},r}\eu^{-\iu \ul{p}\cdot \ul{y}} + \ul{\varepsilon}^{*}_{\ul{p},r}c^{+}_{\ul{p},r}\eu^{\iu \ul{p}\cdot \ul{y}} \Big)\;,\label{1.1.2}\\
&& \ul{\varepsilon}^{*}_{\ul k,r}\cdot \ul{\varepsilon}_{\ul{k},r'} = \d_{r,r'}\;,\qquad\qquad \ul{\varepsilon}_{\ul{k},r}\cdot \ul{k} =0\;;\nn
\eea
$\chi_{[h^*,0]}(t) := \chi(t) - \chi(M^{-h^*}t)$ with $h^*\in \ZZZ^{-}$, and $\chi(t)$ is a smooth compact support function equal to $1$ for $t\leq a_0$ and equal to $0$ for $t\geq a_0 M$, with $M>1$ and $a_0$  constant to be chosen below. The presence of a finite $h^*$ plays the role on an infrared cutoff, and it will be removed in the computation of physical quantities; instead, the ultraviolet cutoff will be kept fixed. Definition (\ref{1.1.2}) implies that $\vec A_{(\vec x,x_3)}$ is periodic in $\vec x$ according to the periodicity of the triangular lattice $\L_{A}$, and in $x_{3}$ along the line $[0,L]$.
\item $\varphi(\ul{y})$ is a regularized version of the Coulomb potential,
\be
\varphi(\ul y) = \frac{1}{\AAA_{\L} L}\sum_{\ul p \in \widetilde\PPP_{L}}\frac{\chi_{[h^{*},0]}(|\ul{p}|)}{|\ul{p}|^2}\eu^{-\iu \ul{p}\cdot \ul{y}}\;.\label{1.1.3}
\ee
Definition (\ref{1.1.3}) implies that $\varphi(\vec x,x_3)$ is periodic in $\vec x$ according to the periodicity of the triangular lattice $\L_{A}$, and in $x_{3}$ along the line $[0,L]$.
\item $n_{\vec x}$ is the {\em density operator}, and it is given by $\sum_{\s}a^{+}_{\vec x,\s}a^{-}_{\vec x,\s}$ or by $\sum_{\s}b^{+}_{\vec x,\s}b^{-}_{\vec x,\s}$, depending on whether $\vec x\in \L_A$ or $\vec x\in \L_B$.
\end{enumerate}

\noindent Notice that the Hamiltonian (\ref{1.1.1}) is invariant under the {\it gauge transformation}
\be
a^{\pm}_{\vec x,\s} \rightarrow a^{\pm}_{\vec x,\s}\eu^{\pm\iu e\a_{\vec x}}\;,\quad b^{\pm}_{\vec x + \vec\d_i,\s}\rightarrow b^{\pm}_{\vec x + \vec\d_i,\s}\eu^{\pm \iu e\a_{\vec x + \vec\d_i}}\;,\quad \vec A_{(\vec x,0)} \rightarrow \vec A_{(\vec x,0)} + \vec\partial \a_{\vec x}\;; \label{1.1.4}
\ee
this is a property which has to be fulfilled in order to correctly describe the interaction of matter with the electromagnetic field. As far as we know, this is the first time that a lattice gauge theory model is considered in the analysis of condensed matter systems; usually, lattice gauge theory models are introduced in quantum field theory as gauge symmetry preserving regularizations of continuum models, to be recovered in the limit in which the bond length is sent to zero. Here the bond length is kept fixed, and we shall be interested in the infrared, {\it i.e.} large distance, properties of the model. Moreover, notice also that the Hamiltonian is {\it particle-hole symmetric}, that is it is invariant under the exchange
\be
a^{+}_{\vec x,\s}\leftrightarrow a^{-}_{\vec x,\s}\;,\quad b^{+}_{\vec x + \vec\d_i,\s}\leftrightarrow -b^{-}_{\vec x + \vec\d_i,\s}\;,\quad \vec A_{(\vec x,0)} \rightarrow - \vec A_{(\vec x,0)}\;;\label{1.1.5}
\ee
this invariance implies in particular that, if we define the average density of the system to be 
\be
\r := \frac{1}{2|\L|}\frac{\Tr\{ e^{-\b\HHH_{\L}}N \}}{\Tr\{e^{-\b\HHH_{\L}}\}}\;\label{1.1.5b}
\ee
where $N := \sum_{\vec x,\s}\big( a^{+}_{\vec x,\s}a^{-}_{\vec x,\s} + b^{+}_{\vec x + \vec\d_1,\s}b^{-}_{\vec x + \vec\d_1,\s} \big)$ is the total particle number operator, then one has $\r\equiv 1$ for all $|\L|$ and $\b$. This is the so-called {\it half-filling condition}.

\section{The free theory}\label{secfree}

Before discussing the properties of the model in presence of interaction, that is for $e\neq 0$, we briefly review what it is known in the case $e=0$. In absence of interaction the fermionic and bosonic degrees of freedom are completely decoupled, and we shall be interested in the fermionic sector only. The Hamiltonian of the system is given by:
\be
\HHH^{\Psi}_{0,\L} := -t\sum_{\substack{\vec x\in \L \\ i=1,2,3}}\sum_{\s = \uparrow\downarrow} a^{+}_{\vec x,\s}b^{-}_{\vec x + \vec\d_i,\s} + b^{+}_{\vec x + \vec\d_i,\s}a^{-}_{\vec x,\s}\;;\label{free1}
\ee
the ground state of this model can be explicitly determined, in the sense that the free energy and the {\it $n$-point Schwinger functions} can be explicitly computed. In fact, being the Hamiltonian quadratic in the fermionic creation/annihilation operators, this last fact simply follows from the knowledgle of the two point Schwinger functions and from the Wick rule. Let us call $\hat \Psi^{\pm}_{\kk,\s,\r}$, $\r=1,2$, the Fourier transform of the imaginary time evolutions of $a^{\pm}_{\vec x,\s}$, $b^{\pm}_{\vec x+\vec\d_1,\s}$, namely $e^{\HHH_{\L} x_0}a^{\pm}_{\vec x,\s}e^{-\HHH_{\L} x_0}$ and $e^{\HHH_{\L} x_0}b^{\pm}_{\vec x + \vec\d_1,\s}e^{-\HHH_{\L} x_0}$; $\kk = (k_0,\vec k)\in \DD_{\b,L} := \DD_{\b}\times \DD_{L}$, where $\DD_{\b}$ is the set of fermionic Matsubara frequencies and $\DD_{L}$ is the first Brillouin zone, that is
\bea
\DD_{\b} &:=& \Big\{ k_0= \frac{2\pi}{\b}(n_0 + \frac{1}{2}): n_0\in \ZZZ \Big\}\;,\nn\\
\DD_{L} &:=& \Big\{ \vec k = \frac{n_1}{L}\vec b_{1} + \frac{n_2}{L}\vec b_{2}: \vec n \in \ZZZ^{2},\,0\leq n_{1},n_{2}\leq L-1 \Big\}\;,\label{free0}
\eea
where $\vec b_{1} = \frac{2\pi}{3}(1,\,\sqrt{3})$, $\vec b_{2} := \frac{2\pi}{3}(1,\,-\sqrt{3})$ form a basis of the dual lattice $\L^{*}$. We define the non-interacting two point Schwinger function in momentum space, or {\it free propagator}, as
\be
\big[\hat S^{\b,L}_{0}(\kk)\big]_{\r,\r'} := \frac{\media{\hat\Psi^{-}_{\kk,\s,\r}\hat\Psi^{+}_{\kk,\s,\r'}}_{\b,L}}{\b|\L|}\Big|_{e=0}\;,
\ee
where $\media{\cdot}_{\b,L}$ denotes the grand-canonical average with respect to the Hamiltonian $\HHH_{\L}$. An explicit computation shows that, see \cite{GM} or Appendix \ref{app1.1.1}:
\be
\hat S^{\b,L}_{0}(\kk) = \frac{1}{k_0^2 + t^2 |\O(\vec k)|^2}\begin{pmatrix} \iu k_0 & -t\O^{*}(\vec k) \\ -t\O(\vec k) & \iu k_0 \end{pmatrix}\;,
\ee
where $\O(\vec k) = \sum_{i=1}^{3} \eu^{\iu \vec k(\vec \d_i - \vec\d_1)}$; its modulus $|\O(\vec k)|$ is the {\it energy dispersion relation}, and it is plotted in Fig. \ref{figcone}. Interestingly, the function $\O(\vec k)$ is vanishing if and only if $\vec k = \vec p_{F}^{\pm}$, where $\vec p_{F}^{\pm}$ are the two {\it Fermi points},
\be
\vec p_{F}^{\pm} = \Big(\frac{2\pi}{3},\,\pm\frac{2\pi}{3\sqrt{3}}\Big)\;;\label{free4}
\ee
the fact that the energy dispersion relation vanishes only at two points is very unusual for two dimensional systems, where the Fermi surface typically consists of a closed curve. Close to the Fermi points $\O(\vec k' + \vec p_{F}^{\pm}) = \frac{3}{2}(\iu k'_{1} \pm k'_{2}) + O(|\vec k'|^2)$; therefore, setting $\pp_{F}^{\o} = (0,\vec p_{F}^{\o})$ and $\kk' = \kk - \pp_{F}^{\o}$, we can rewrite
\be
\hat S_{0}(\kk' + \pp_{F}^{\o}) = \frac{1}{Z}\begin{pmatrix} -\iu k_0 & -v(-\iu k'_{1} + \o k'_{2}) + r_{\o}(\vec k')\\  -v(\iu k'_{1} + \o k'_{2}) + r^{*}_{\o}(\vec k') & -\iu k_{0}   \end{pmatrix}^{-1}\;,\label{free5}
\ee
where $Z=1$ is the {\it bare wave function renormalization}, $v = \frac{3}{2}t$ is the {\it bare Fermi velocity}, and $|r_{\o}(\vec k')|\leq C|\vec k'|^{2}$ for some $C>0$. Therefore, from (\ref{free5}) we see that the free propagator is asymptotically the same of {\it massless Dirac fermions} in $2+1$ dimensions.

\section{Results and discussion}\label{secres}
\setcounter{equation}{0}
\renewcommand{\theequation}{\ref{secres}.\arabic{equation}}

Below we shall discuss more extensively our results, which have been presented in \cite{GMP2}, obtained with methods similar to those of \cite{GMP1}; as already mentioned in Section \ref{secmaires}, they concern the effect of the electromagnetic interaction on the two point Schwinger function, various response functions, the effect of lattice distortions and a possible mechanism for gap generation. A sketch of the proof together with the main ideas underlying the methods is given in Section \ref{secsk}. In what follows we will assume that the infrared limit $h^{*}\rightarrow-\infty$ has been taken.

\subsection{The two point Schwinger function}

We start by discussing the result on the two point Schwinger function; as we are going to see, the interaction dramatically modifies its scaling properties. In fact, the interacting two point Schwinger function is given by, for $\b,L\rightarrow +\infty$ and close to the singularities $\pp_{F}^{\o}$:
\bea
&&\hat S(\kk' + \pp_{F}^{\o}) =\label{res1}\\&&\frac{-1}{Z(\kk')}\begin{pmatrix} \iu k_0 & v(\kk')(-\iu k'_1 + \o k'_2) + \tilde r_{\o}(\vec k')\\ v(\kk')(\iu k'_1 + \o k'_2) +  \tilde r_{\o}^{*}(\vec k') & \iu k_0 \end{pmatrix}^{-1}\big(1 + R(\kk')\big)\nn
\eea
with $|\tilde r(\vec k')|\leq C|\vec k'|^2$, $R(\kk') = O(e^2)$ and
\bea
&&Z(\kk') \simeq |\kk'|^{-\eta}\;,\qquad v(\kk') \simeq 1 - (1 - v)|\kk'|^{\tilde\eta}\;,\\
&&\eta = \frac{e^2}{12\pi^2} + \eta^{(>2)}\;,\qquad \tilde\eta = \frac{2 e^2}{5\pi^2} + \tilde\eta^{(>2)}\;,\label{res2}
\eea
where $v := \frac{3}{2}t$, $\eta^{(>2)} = O(e^4)$, $\tilde\eta^{(>2)} = O(e^4)$; $R(\kk')$, $\eta^{(>2)}$ and $\tilde\eta^{(>2)}$ are expressed as renormalized series with finite coefficients admitting $N!$ bounds, in the sense of Theorem \ref{thm1}. The functions $Z(\kk')$ and $v(\kk')$ are called respectively the {\it wave function renormalization} and the {\it effective Fermi velocity}; as we see from (\ref{res2}), both these objects have a non-trivial dependence on the quasi-momentum $\kk'$: they scale with interaction dependent {\it anomalous exponents}.\\

This results tells us that the scaling of the two point Schwinger function is {\em dramatically modified} by the presence of the electromagnetic interaction; in particular, in absence of interaction $\hat S_{0}(\kk') \sim |\kk'|^{-1}$, see (\ref{free5}), while in the interacting case $\hat S(\kk') \sim |\kk'|^{-1 + \eta}$ with $\eta  = O(e^2)>0$, therefore
\be
\lim_{\kk'\rightarrow \V0}\frac{\hat S_{0}(\kk' + \pp_{F}^{\o})}{\hat S(\kk' + \pp_{F}^{\o})} = +\infty\;.\label{res2b}
\ee
This is due to the fact that the quasi-particle weight $Z(\kk')^{-1}$ {\em vanishes} as a power law at the singularity; this suggests that the interacting system is a {\em Luttinger liquid}, in the sense of \cite{Ha2}. If convergence of the series is proved, this would be the first rigorous proof of Luttinger liquid behavior in more than one spatial dimension. Regarding one dimensional systems, these have been widely studied in the last $15$ years, starting from the work of Benfatto, Gallavotti, Procacci and Scoppola in \cite{BGPS}, where the first rigorous proof of Luttinger liquid behavior for a non solvable one dimensional system was given; in particular, quantities of physical interest like the Schwinger functions, the anomalous exponents and the free energy were explicitly computed as convergent series in the coupling constant. A crucial ingredient in the proof of \cite{BGPS} is the {\it vanishing of the Beta function} of the effective coupling; in \cite{BGPS} this was shown using informations coming from the exact solution of the Luttinger model. Recently, an independent strategy to prove this remarkable cancellation has been proposed and developed by Benfatto and Mastropietro, \cite{BM}, which completely avoids the use of the exact solution; this new method is based on the rigorous implementation of Ward identities in the RG, and similar ideas will be adopted here.

Another remarkable feature of the two point Schwinger function is the non-trivial behavior of the effective Fermi velocity $v(\kk')$; in fact, in the limit $|\kk'|\rightarrow 0$ the effective Fermi velocity tends to $1$, which is the speed of light in our units, for any value of the bare Fermi velocity $v$.

Finally, we stress that our result {\em does not} neglect the lattice, and {\em does not} assume unphysical regularization procedures, like the dimensional one. The first attempt to investigate the effect of electromagnetic interactions in graphene is due to Gonz\` alez, Guinea and Vozmediano in \cite{V1}; the model they considered described $2+1$ dimensional Dirac fermions interacting with a $3+1$ dimensional photon field in the Feynman gauge; both the fields lived in the continuum, and dimensional regularization was used to avoid ultraviolet divergences. One-loop computations suggested that logarithmic corrections to the scaling of the two point function were present, and these corrections were {\em interpreted} as the first order contributions to the expansions of $|\kk'|^{-\eta}$, $|\kk'|^{\tilde\eta}$. For the same model, anomalous scaling has been recently established at all orders by Giuliani, Mastropietro and Porta in \cite{GMP1}, using momentum regularization (in order to mimic the presence of the lattice) instead of the dimensional one.

\begin{oss}
In general, we shall represent the grand-canonical average $\media{\cdot}_{\b,L}$ using the functional integral representation in the Feynman gauge, see Section \ref{sec2.2}; if the averaged observable is gauge invariant the result of the computation is equal to the corresponding grand-canonical average with the Hamiltonian $\HHH_{\L}$, see Appendices \ref{app1}, \ref{app2}. All the quantities that we shall compute in this Thesis, except the two point Schwinger function, will be gauge invariant; however, the two point Schwinger function appears as ``dressed propagator'' in the perturbative series of physical observables, and therefore it is interesting it its own right.
\end{oss}

\subsection{The response functions}

The same methods used to evaluate the two point Schwinger function can be used to compute other correlations; in particular, the analysis of various response functions allows us to investigate the effect of the electromagnetic interactions on possible quantum instabilities which may take place at strong coupling. Let us define:
\bea
C^{(\a)}_{i,j}(\xx - \yy) &:=& \media{\r^{(\a)}_{\xx,i};\r^{(\a)}_{\yy,j}}_{\b,L}\;,\label{res3}\\
\r^{(E_{\pm})}_{\xx,j} &:=& \sum_{\s=\uparrow\downarrow} a^{+}_{\xx,\s}b^{-}_{\xx + (0,\vec\d_j),\s}\eu^{\iu e A_{(\xx,j)}} \pm b^{+}_{\xx + (0,\vec\d_j),\s}a^{-}_{\xx,\s}\eu^{-\iu e A_{(\xx,j)}}\;,\nn\\
\r^{(CDW)}_{\xx,j} &:=& \sum_{\s=\uparrow\downarrow} a^{+}_{\xx,\s}a^{-}_{\xx,\s} - b^{+}_{\xx + (0,\vec\d_j),\s}b^{-}_{\xx + (0,\vec\d_j),\s}\;,\nn\\
\r^{(D)}_{\xx,j} &:=& \sum_{\s = \uparrow\downarrow} a^{+}_{\xx,\s}a^{-}_{\xx,\s} + b^{+}_{\xx + (0,\vec\d_j),\s} b^{-}_{\xx + (0,\vec\d_j),\s}\;,\nn
\eea
where $a^{\pm}_{\xx,\s}$, $b^{\pm}_{\yy,\s}$, $A_{(\xx,i)}$ are respectively the imaginary time evolutions under the Hamiltonian $\HHH_{\L}$ of $a^{\pm}_{\vec x,\s}$, $b^{\pm}_{\vec y,\s}$, $A_{(\vec x,i)}$; these correlations are called the {\it excitonic, charge density wave } or {\it density-density} susceptibilities, depending on whether $\a = E_{\pm}$, $\a= CDW$ or $\a = D$. The correlations $C^{(E_{\pm})}$ measure the tendency of the system to form {\it particle-hole} pairs between nearest neighbours on the honeycomb lattice, and in particular $C^{(E_{+})}$ measures the response of the system to {\it Kekul\' e lattice distortions}, see next section; $C^{(CDW)}$ measures the tendency of the system to produce charge asymmetries between the two triangular sublattices; finally, $C^{(D)}$ measure the correlations of on-site electronic densities.\\

\noindent It follows that, for $\b,L\rightarrow+\infty$ and $|\xx|\gg 1$, the $C^{\a}_{1,1}$ correlations are given by\footnote{The $C_{j,j}^{\a}(\xx)$ correlations are obtained from $C_{1,1}^{\a}(\xx)$ using that $C^{\a}_{j+1,j+1}(\xx) = C^{\a}_{j,j}(x_0,T\vec x)$.}
\be
C^{\a}_{1,1}(\xx) =  \frac{G^{\a}_{1}(\xx)}{|\xx|^{4 - \xi^{\a}_{++}}} + \cos(\vec p_{F}^{+}\cdot\vec x)\frac{G^{\a}_{2}(\xx)}{|\xx|^{4 - \xi^{\a}_{+-}}} + r^{\a}_{1,1}(\xx)\;,\label{res4}
\ee
where: (i) the functions $G^{\a}_{i}(\xx)$ are given by, if $b = 8\pi^2/27$:
\bea
&& G^{D}_{1}(\xx) = \frac{-x_0^2 + |\vec x|^2}{b |\xx|^{2}}\;,\qquad G^{D}_{2}(\xx) = \frac{-x_0^2 + x_1^2 - x_2^2}{b |\xx|^2}\;,\nn\\
&& G^{CDW}_{1}(\xx) = -\frac{1}{b}\;,\qquad G^{CDW}_{2}(\xx) = \frac{-x_0^2 - x_1^2 + x_2^2}{b |\xx|^2}\;,\nn\\
&& G^{E_+}_{1}(\xx) = \frac{-x_0^2 - x_1^2 + x_2^2}{b |\xx|^2}\;,\qquad G^{E_+}_{2}(\xx) = -\frac{1}{b}\;,\nn\\
&& G^{E_-}_{1}(\xx) = \frac{x_0^2 - x_1^2 + x_2^2}{b |\xx|^2}\;,\qquad G^{E_-}_{2}(\xx) = \frac{x_0^2 - |\vec x|^2}{b |\xx|^2}\,;\label{res5}
\eea
\\

\noindent(ii) the anomalous exponents $\xi^{\a}_{++}$, $\xi^{\a}_{+-}$ are given by renormalized series in the renormalized charge, and are equal to:
\bea
&&\xi^{\a}_{\ul\o} = 0 + \xi^{\a,(>2)}_{\ul\o}\;,\qquad \mbox{$\forall (\a,\ul\o) : (\a,\ul\o)\neq (E_{+},+-)\,,(CDW,++)$}\;,\nn\\
&&\xi^{E_{+}}_{+-} = \xi + \xi^{E_+,(>2)}_{+-}\;,\quad \xi^{CDW}_{++} = \xi + \xi^{CDW,(>2)}_{++}\;,\quad \xi = \frac{4e^2}{3\pi^2}\;,\label{res6}
\eea
where $\xi^{\a,(>2)}_{\ul\o}$ are expressed as renormalized series starting from fourth order, admitting $N!$ bounds in the sense of Theorem \ref{thm1};\\

\noindent(iii) $r^{\a}_{11}(\xx)$ contains corrections which are either bounded by $|\xx|^{-(4 + \th - \max_{\ul\o}\xi^{\a}_{\ul\o})}$ 
for $\frac{1}{2}\leq \th < 1$ or by $e^{2}|\xx|^{-(4 - \max_{\ul\o}\xi^{\a}_{\ul\o})}$.\\

The above results show that the electromagnetic interaction has the effect of {\em depressing} the space-time decay of the correlations $C^{(E_{+})}$, $C^{(CDW)}$; this suggests that the interaction {\it may favor} excitonic or charge density wave instabilities at strong coupling. The depression of the decay of the oscillating part of the $\a=E_+$ susceptibilities is associated with a particular lattice distortion, the {\it Kekul\' e} one; as we are going to see with the next result, the amplitude of Kekul\' e distortions is {\it greatly enhanced} by the interactions. The relevance of charge density wave and Kekul\' e instabilities in graphene have been first discussed in \cite{K1, KL} and \cite{HCM, JP, SeF}, respectively. As far as we know, this is the {\em first time} that these correlations are computed for graphene; regarding one dimensional systems, analogous computations have been performed in \cite{BM0, BM2}.

\subsection{Lattice distortions}

In this Section we shall discuss the effect of the interaction on some special lattice distortions, the {\it Kekul\' e ones}. First, we will report the computation of the two point Schwinger function in presence of Kekul\' e distortion; then, we will discuss a possible mechanism for {\it spontaneous} lattice distortion in presence of strong enough electron-electron interactions. Remarkably, the amplitude of the Kekul\' e distortion behaves as a {\it bare mass} for the fermion propagator; therefore, the problem of spontaneous lattice distortion is equivalent to the one of mass generation in quantum field theory.

\subsubsection{The two point function in presence of Kekul\' e distortion}

\begin{figure}[htbp]
\centering
\includegraphics[width=0.7\textwidth]{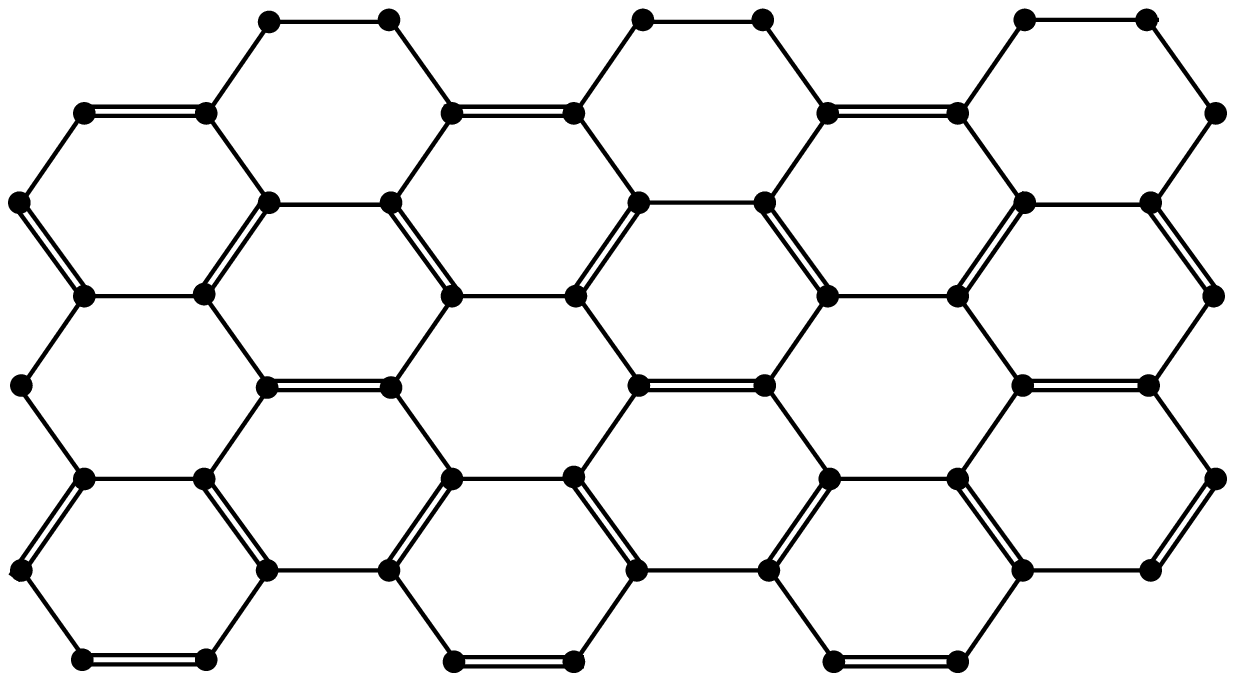} \caption{The Kekul\' e dimerization pattern for $j_0=1$. Double and single bonds are respectively ``long'' and ``short'', and correspond to the hopping parameters $t - \frac{\D_0}{3}$, $t + \frac{2}{3}\D_0$.} \label{figkek0}
\end{figure}

The presence of a small distortion can be taken into account by assuming that the hopping parameter $t$ is replaced by a lattice site and bond dependent one; namely by replacing $t$ in (\ref{1.1.1}) with $t + \D_0 \phi_{\vec x,i}$, where $\phi_{\vec x,i}$ is the distortion of the bond connecting $\vec x$ to $\vec x + \vec\d_i$, and $\D_0$ is a small parameter. This approximation has been widely used to study the Peierls instability in one dimensional systems, see \cite{SSH, LN}, for instance. In this framework, the Kekul\' e distortion is obtained by replacing $t$ in (\ref{1.1.1b}) with $t + \frac{2}{3}\D_0 \cos(\vec p_{F}^{+}(\vec\d_j - \vec\d_{j_0} - \vec x))$; this choice corresponds to the dimerization pattern represented in Fig. \ref{figkek0}.

To express the result for the two point Schwinger function is convenient to adopt a ``relativistic'' notation; let 
\bea
&&\psi^{T}_{\kk',\s} := \begin{pmatrix} \hat \Psi^{-}_{\kk' + \pp_{F}^{+},\s,1} & \hat \Psi^{-}_{\kk' + \pp_{F}^{+},\s,2} & \hat \Psi^{-}_{\kk' + \pp_{F}^{-},\s,2} & \hat \Psi^{-}_{\kk' + \pp_{F}^{-},\s,1} \end{pmatrix}\;,\nn\\
&&\lis\psi_{\kk',\s} := \begin{pmatrix} \hat \Psi^{+}_{\kk' + \pp_{F}^{-},\s,2} & \hat \Psi^{+}_{\kk' + \pp_{F}^{-},\s,1} & -\hat \Psi^{+}_{\kk' + \pp_{F}^{+},\s,1} & -\hat \Psi^{+}_{\kk' + \pp_{F}^{+},\s,2} \end{pmatrix}\;,\nn
\eea
where $\hat\Psi^{\pm}_{\kk,\s,\r}$, $\r=1,2$, are the Fourier transform of the imaginary time evolutions of $a^{\pm}_{\vec x,\s}$, $b^{\pm}_{\vec x + \vec\d_1,\s}$, and define
\be
\big[ \hat S_{\D}^{\b,L}(\kk') \big]_{jk} := \frac{\media{\psi_{\kk',\s,j}\lis\psi_{\kk',\s,k}}_{\b,L}}{\b|\L|}\;.\label{res7}
\ee
It follows that, for $\b,L\rightarrow+\infty$, and $|\kk'|$ small, the limiting propagator is given by:
\be
\hat S_{\D}(\kk') = \frac{1}{\bar Z(\kk')}\frac{1}{\iu k_0 \g_0 + \bar v(\kk')\iu k'_{i}\g_{i} + \iu \g^{(j_0)}\D(\kk')}\big(1 + \bar R(\kk')\big)\;,\label{res8}
\ee
where: (i) $\g_0,\g_1,\g_2$ are Euclidean gamma matrices (defined in (\ref{1.2.57}), (\ref{1.2.52b})), and
\be
\g^{(j_0)} := \begin{pmatrix} -\iu I \eu^{\iu \vec p_{F}^{+}(\vec\d_{j_0} - \vec\d_1)} & 0 \\ 0 & \iu I \eu^{\iu \vec p_{F}^{-}(\vec\d_{j_0} - \vec\d_1)} \end{pmatrix}\;;\label{res9}
\ee
(ii) setting
\be
\D := \D_0^{\frac{1}{1 + \eta_K}}\;,\qquad \eta_K = \frac{2e^2}{3\pi^2} + \eta_K^{(>2)}\;,\label{res10}
\ee
where $\eta_K^{(>2)}$ is given by a series in the renormalized charge starting from the fourth order and admitting $N!$ bounds, the functions $\D(\kk')$, $\bar v(\kk')$, $\bar Z(\kk')$ are given by:
\bea
&&\bar Z(\kk') \simeq \max\{|\kk'|,\D\}^{-\eta}\;,\quad \bar v(\kk') \simeq 1 - (1 - v)\max\{|\kk'|,\D\}^{\tilde\eta}\;,\nn\\&&\D(\kk') \simeq \D_0\max\{|\kk'|,\D\}^{-\eta_K}\;;\label{res11}
\eea
(iii) $\bar R(\kk') = \bar R_{1}(\kk') + \bar R_2 (\kk')$ with $\bar R_1(\kk') = O(e^2)$ and $\bar R_{2}(\kk') = O(|\kk'|)$.\\

Therefore, the presence of a Kekul\' e distortion with a non-vanishing amplitude $\D_0$ produces an {\em effective mass} $\D(\kk')$ for the fermion field, which is {\em strongly renormalized} by the interaction; in fact,
\be
\lim_{\D_0\rightarrow 0}\frac{\D}{\D_0} = +\infty\;.\label{res12a}
\ee
In this sense we say that the presence of electromagnetic interactions {\em enhances} the Kekul\' e distortion of the honeycomb lattice; as far as we know, this is the first time that the effect of electromagnetic interaction on a preexisting Kekul\' e distortion is discussed. 

\subsubsection{A mechanism for spontaneous lattice distortion}

Finally, we conclude the summary of the results of this Thesis by discussing a mechanism for spontaneous lattice distortions. We consider the lattice distortion $\phi_{\vec x,i}$ as a dynamical variable; the full Hamiltonian of the model is:
\be
\widetilde \HHH_{\L}(\{\phi\}) = \HHH_{\L}(\{\phi\}) + \KK_{\L}(\{\phi\}) := \HHH_{\L}(\{\phi\}) + \frac{\kappa}{2 g^2}\sum_{\substack{\vec x\in\L\\i=1,2,3}}\phi^{2}_{\vec x,i}\;,\label{res12}
\ee
where: $\HHH_{\L}(\{\phi\})$ is given by (\ref{1.1.1}) after the replacement $t\rightarrow t + \phi_{\vec x,i}$; the second term is the {\it elastic energy} of the distortion and $\k,g$ are respectively the {\it stiffness constant} and the {\it phonon coupling}. Let us define the {\it specific free energy} in the {\it Born-Oppenheimer} approximation as:
\bea
F^{BO}_{\b,L} &:=& -\frac{1}{\b|\L|}\log \int \Big[ \prod_{\substack{\vec x\in \L \\ i = 1,2,3}}d\phi_{\vec x,i} \Big]\,e^{-\b\KK_{\L}(\{\phi\}) - \b|\L|F_{\b,L}(\{\phi\})}\;,\\
&=:& -\frac{1}{\b|\L|}\log \int \Big[ \prod_{\substack{\vec x\in \L \\ i = 1,2,3}}d\phi_{\vec x,i} \Big]\,e^{-\b|\L| E^{BO}_{\b,L}(\{\phi\})}\;,
\eea
where $F_{\b,L}(\{\phi\})$ is the specific free energy of the model with Hamiltonian $\HHH_{\L}(\{\phi\})$ in presence of a fixed distortion. It follows that $\partial_{\phi_{\vec x,i}}E_{\b,L}^{BO}(\{\phi\}) = 0$ in correspondence of a Kekul\' e distortion 
\be
\phi^{*,(j_0)}_{\vec x,i} = \phi_0 + \frac{2}{3}\D_0\cos(\vec p_F^{+}(\vec\d_j - \vec\d_{j_0} - \vec x))\;\label{res12b}
\ee
 with $\phi_0$ and $\D_0$ self-consistently determined by the following equations:
\bea
\phi_0 &=& \frac{g^2}{\kappa}\frac{1}{|\L|}\media{\hat \r^{(E_{+})}_{\vec 0,1}}^{(1)}_{L}\;,\nn\\
\frac{\D_0}{3} &=& \frac{g^2}{\k}\frac{1}{|\L|}\media{\hat \r^{(E_{+})}_{\vec p_{F}^{+},1}}^{(1)}_{L}\;,\label{res13}
\eea
where: the symbol $\media{\cdots}_{L}^{(1)}$ denotes the limit $\b\rightarrow+\infty$ of the grand-canonical average $\media{\cdots}_{\b,L}$ computed with the Hamiltonian $\HHH_{\L}(\{\phi^{*,(1)}\})$; the operator $\hat\r^{(E_+)}_{\vec k,i}$ is the Fourier transform in $\vec x$ of $\r^{(E_+)}_{(\vec x,0),i}$ defined in (\ref{res3}). Moreover, for $g$ small enough the solution (\ref{res12b}), (\ref{res13}) corresponds to a local minimum of the energy.

The first equation gives $\phi_0 = O(g^2)$; this amounts to a small renormalization of the hopping parameter. The second equation is the most interesting one; it can be seen as a {\it non-BCS gap equation} for the fermion field. To understand it from a qualitative viewpoint, we replace it with:
\be
\frac{\D_0}{3} = \frac{8 g^2}{\k}\int\limits_{\D\leq |\kk'|\leq 1}\frac{d\kk'}{D}\,\frac{1}{Z(\kk')}\frac{\D(\kk')}{k_0^2 + v(\kk')^2|\kk'|^2}\;,\label{res14}
\ee
where $D = (2\pi)|B_{1}|$ and $B_1 = 8\pi^2/(3\sqrt{3})$ is the volume of the first Brillouin zone; Eq. (\ref{res14}) is very similar to the gap equation first derived by Mastropietro in \cite{M1}, in the context of certain Luttinger superconductors. It follows that:
\begin{itemize}
\item for small $e$, that is when $\eta_K - \eta = \frac{7e^2}{12\pi^2} + \ldots$ is negligible with respect to $1$, the equation (\ref{res14}) admits the solution
\be
\D_0 \simeq \Big( 1 - \frac{g_0^2}{g^2} \Big)^{\frac{1 + \eta_K}{1 + \eta - \eta_K}}\;,\qquad \mbox{for $|g|>g_0 = O(\sqrt{k}v)$}\;,\label{res15}
\ee
where $g_0$ is positive and essentially independent of $e$; moreover, since the exponent in (\ref{res15}) is greater than $1$, $\partial_{g}\D(g)$ is continuous at $|g|=g_0$: the transition is smoothened by the interaction.
\item For large $e$, that is when $\eta_K - \eta \rightarrow 1$, then $g_0\rightarrow 0$. In particular, if $\eta_K - \eta - 1$ exceeds $0$, then 
\be
\D_0 \simeq \Big( 1 + \frac{\kappa(\eta_K - \eta - 1)}{g^2} \Big)^{\frac{1}{1 + \eta - \eta_K}}\;,\qquad \mbox{for $|g|>0$.}
\ee
\end{itemize}

Putting these results together with those regarding the excitonic susceptibility $C^{(E_+)}$ and the two point Schwinger function in presence of Kekul\' e distortion, we argue that strong enough electron-electron interactions may favor the occurrence of a possible ``excitonic quantum instability'' at strong coupling, in the form of a spontaneous Kekul\' e distortion of the honeycomb lattice. This issue is interesting not only because of a possible interaction driven metal-insulator transition in graphene (the amplitude of the distortion produces a gap in the energy spectrum of the fermions); but also because, as proposed first by Hou, Chamon and Mudry in \cite{HCM}, and by Jackiw and Pi in \cite{JP}, the Kekul\' e distortion may be a prerequisite for {\em electron fractionalization}, that is for the existence of collective excitations with net charge equal to a {\em half} the electric charge. This would be the two dimensional analogous of the Peierls instability studied in one dimensional systems by Su, Schrieffer and Heeger in \cite{SSH} and by Jackiw and Rebbi in \cite{JR}.

\section{Sketch of the proof}\label{secsk}
\setcounter{equation}{0}
\renewcommand{\theequation}{\ref{secsk}.\arabic{equation}}

Here we shall give a sketch of the proof of the above results, trying to explain in an informal way the ideas behind our RG methods. Our framework consists in a rigorous formulation of the Wilsonian Renormalization Group, \cite{Wi, Wi2}, based on the functional RG developed in the 1980's in \cite{GN1, GN2, P}, see also \cite{G84}. In the last $25$ years these methods have been successfully applied to different fields, such as quantum field theory, condensed matter and classical statistical mechanics. See \cite{BG1, GM, M} for extensive reviews.

The underlying physical idea is very simple: we look for a sequence of {\it effective theories} that describe the system at larger and larger scales, whose parameter are determined by the integration of the degrees of freedom corresponding to smaller scales. Each theory is determined starting from the ones on smaller scales by applying a well defined procedure, the ``RG map''; the goal is to find a fixed point of this map, and to determine the scaling behavior of the corresponding many-body correlations.

First of all, we rewrite our many-body problem in terms of a quantum field theory model, in the sense that we rewrite the partition function and the generating functional of the correlations as suitable functional integrals, involving fermionic, {\it i.e.} Grassmann, and bosonic, {\it i.e.} real, Gaussian {\it fields}. The equivalence is established noting that the perturbative series are equal order by order; see Appendix \ref{app1}, where the case of the partition function is discussed. Our goal is to evaluate such functional integrals, at least in terms of perturbative series with coefficients bounded uniformly in $\b$, $L$, which may or may not converge (the problem of proving convergence will be not adressed in this Thesis). Let us consider for simplicity the partition function $\Xi_{\b,L}$; the generating functional of the correlations can be studied in a similar way, just more technically involved, and we will not discuss it here. It follows that:
\be
\Xi_{\b,L} = \int P(d\Psi)P(dA)\eu^{V(\Psi,A)}\;,\label{sk1}
\ee
where $V(\Psi,A)$ is the {\it interaction} and $P(d\Psi)$, $P(dA)$ are Gaussian fermionic and bosonic measures respectively, with propagators $\hat g(\kk)$, $\hat w(\pp)$; the interaction and the gaussian measures are defined in Section \ref{sec2.2}. The identity has to be understood in the sense that the perturbative series in $e$ of the l.h.s. and of the r.h.s. are equal order by order for any $\b$, $L$, $h^{*}$ fixed; moreover, by {\it gauge invariance}, see Appendix \ref{app2}, we are free to express the photon propagator in the Feynman gauge, which is more convenient for our purposes. The momenta $\kk = (k_0,\vec k)$, $\pp = (p_0,\vec p)$ appearing at the argument of the propagators are constrained by the fact that: (i) $\vec k$ has to belong to the first Brillouin zone; (ii) $|\vec p|$ cannot be too big, because of the bosonic ultraviolet cutoff in the definition of the model, which will be kept fixed. However, the ``Matsubara frequency'' variables $k_0$, $p_0$ are unbounded; therefore, to have a well defined problem we fix an ultraviolet cutoff on both of them, by imposing that the arguments of the propagators must satisfy $k_{0}\leq M^{K}$, $p_{0}\leq M^{K}$, with $M>1$ a scaling parameter and $K$ a positive integer, which at the end will be sent to $+\infty$.

One may naively think of evaluating (\ref{sk1}) by expanding the exponential in (\ref{sk1}) and taking averages; as it is well known, the result can be expressed in terms of a sum over Feynman graphs. However, an easy computation shows that, already at lowest order, these graphs are not bounded uniformly in $\b$, $L$, $h^{*}$:  this is so because the fermionic and bosonic propagator are {\it singular} at $\kk= \pp_{F}^{\pm}$ and $\pp = \V0$. The hope is that these infinities {\it cancel} and produce a finite result. Roughly speaking, to see these cancellations we shall write each graph as a sum of many pieces, where in every piece each propagator carries a momentum which is ``close'' to some prefixed value, labelled by some momentum scale label. The goal is to show that classes of finite pieces cancel, and produce a result which is summable over the scale labels.

To see these cancellations directly from the perturbative series would be an almost desperate task; the startegy that we shall adopt to reorganize in a convenient way the series and to keep track of all the contributions has been introduced by Gallavotti in 1984, see \cite{G84}. First of all, setting $\kk' = \kk - \pp_{F}^{\o}$, we rewrite the propagators as 
\be
\hat g(\kk) = \sum_{\o = \pm}\hat g_{\o}^{(\leq 0)}(\kk') + \hat g^{(> 0)}(\kk)\;,\quad \hat w(\pp) = \hat w^{(\leq 0)}(\pp) + \hat w^{(> 0)}(\pp)\;,\label{sk2}
\ee
where $\hat g^{(\leq 0)}_{\o}(\kk')$, $\hat w^{(\leq 0)}(\pp)$ are supported respectively on momenta close to $\pp_{F}^{\o}$ and $\V0$; correspondingly, we rewrite the fermionic and bosonic fields as
\be
\hat\Psi_{\kk,\s,\r} = \sum_{\o = \pm}\hat\Psi^{(\leq 0)}_{\kk',\s,\r,\o} + \hat\Psi^{(>0)}_{\kk',\s,\r}\;,\qquad \hat A_{\m,\pp} = \hat A^{(\leq 0)}_{\m,\pp} + A^{(>0)}_{\m,\pp}\;,\label{sk3}
\ee
where the fields labelled by $(\leq 0)$ and $(>0)$ are {\it independent} Gaussian variables, with propagators given by (\ref{sk2}). After this, we {\it integrate} the fields labelled by $(>0)$ and we get a theory which depends only on the $(\leq 0)$ fields; this is done by writing
\bea
&&\Xi_{\b,L} = \int P_{\leq 0}(d\Psi^{(\leq 0)})P_{\leq 0}(dA^{(\leq 0)})e^{\VV^{(0)}(\Psi^{(\leq 0)},A^{(\leq 0)})}\;,\label{sk4}\\
&& \VV^{(0)}(\cdot,\cdot) = \log \int P_{>0}(d\Psi^{(>0)})P_{>0}(dA^{(>0)})\Big[1 + V(\cdot + \Psi^{(>0)},\cdot + A^{(>0)}) + \ldots \Big]\nn
\eea
where the integration measures labelled by $\leq 0$, $>0$ have propagators given by (\ref{sk2}), and the dots stand for the higher order terms coming from the expansion of the exponential in (\ref{sk1}). The expectations in the second line of (\ref{sk4}) can be graphically represented in terms of {\it Gallavotti - Nicol\`o} (GN) {\it trees}, \cite{GN1, GN2, G84}, which in turn can be evaluated as sums over connected Feynman graphs, with an arbitrary number of external lines denoting the fields labelled by $(\leq 0)$; the values of these graphs, and therefore of the trees, are bounded uniformly in $K$, see Appendix \ref{app2b}. Therefore, the limit $K\rightarrow+\infty$ can be safely taken. 

So far we discussed the ``easy'' part of the work; now comes the ``hard'' one, since singularities $\kk' = \V0$, $\pp = \V0$ belong to the domains of the fields labelled by $(\leq 0)$. We rewrite:
\be
\hat\Psi^{(\leq 0)\pm}_{\kk',\s,\r,\o} = \sum_{h=-\io}^{0}\hat\Psi^{(h)\pm}_{\kk',\s,\r,\o}\;,\qquad \hat A^{(\leq 0)}_{\m,\pp} = \sum_{h=-\io}^{0}\hat A^{(h)}_{\m,\pp}\;, \label{sk5}
\ee
where $\hat\Psi^{(0)},\,\hat A^{(0)},\, \hat\Psi^{(-1)},\,\hat A^{(-1)},\,\ldots$ are independent Gaussian fields supported on momenta $|\kk'|\sim M^{h}$, $|\pp|\sim M^{h}$ ($h$ is the {\it scale label}), whose propagators will be determined inductively. We integrate the fields iteratively expanding the exponential as in (\ref{sk4}), starting from scale $0$ going down to $-\io$; each step of the integration can be organized graphically in terms of GN trees. It follows that, after the integration of the first $|h|$ scales, the partition function (\ref{sk4}) can be rewritten as:
\be
\Xi_{\b,L} = e^{-\b|\L|F_{h}}\int P_{\leq h}(d\Psi^{(\leq h)})P_{\leq h}(dA^{(\leq h)})e^{\VV^{(h)}(\sqrt{Z_{h}}\Psi^{(\leq h)},A^{(\leq h)})}\;,\label{sk6}
\ee
where:
\begin{itemize}
\item $F_{h}$ is the specific free energy on scale $h$.
\item The integration measures have propagators given by $\hat g^{(\leq h)}_{\o}(\kk')$, $\hat w^{(\leq h)}(\pp)$, both supported on momenta $\leq M^{h+1}$. For these values of the momenta, the photon propagator is essentially equal to $\hat w^{(\leq 0)}(\pp)$, while $\hat g^{(\leq h)}(\kk')$ is morally equal to (\ref{free5}) after having relaced $Z$ and $v$ with $Z_{h}$ and $v_h$, the {\it wave function renormalization} and the {\it effective Fermi velocity} on scale $h$.
\item The interaction on scale $h$ has the form
\be
\VV^{(h)}(\Psi,A) = \int \frac{d\pp}{(2\pi)^3}\Big[ e_{\m,h}\hat j^{(\leq h)}_{\m,\pp}\hat A^{(\leq h)}_{\m,\pp} - M^{h}\n_{\m,h}\hat A^{(\leq h)}_{\m,\pp}A^{(\leq h)}_{\m,-\pp}\Big] + \RR\VV^{(h)}\;,\label{sk7}
\ee
where $\hat j^{(\leq h)}_{\m,\pp}$ is a ``relativistic'' fermionic current, $e_{\m,h}$, $M^{h}\n_{\m,h}$ are respectively the {\it effective charges} and the {\it effective mass terms} for the photon field, and $\RR\VV^{(h)}$ contains irrelevant terms in the RG sense.
\end{itemize}

 The parameters $Z_{h}$, $v_{h}$, $e_{\m,h}$, $M^{h}\n_{\m,h}$ are called {\it running coupling constants}; it follows that, collecting all these numbers in a vector $\vec \a_{h}$, the iterative integration implies an ``evolution equation'' for $\vec \a_{h}$, namely:
\be
\vec\a_{h} = \vec\a_{h+1} + \vec\b_{h+1}(\vec \a_{h+1},\ldots, \vec\a_{0})\;;\label{sk8}
\ee
the vector $\vec\b_{h+1}$ in (\ref{sk8}) is called the {\it Beta function} of the theory, and it can be computed as series in the running coupling constant on scales $\geq h+1$. It turns out that the Beta function, the free energy and the correlation functions can be expressed as perturbative series in the {\it renormalized couplings} $\{e_{\m,k},\,\n_{\m,k}\}$, with coefficients depending on $\big\{ Z_{k}/Z_{k-1},\,v_{k} \big\}$; in particular, the coefficients of the series are finite if the ratio $Z_{k}/Z_{k-1}$ is close to $1$ and if $v_{k}$ is bounded away from $0$. Therefore, the problem of computing the partition function (\ref{sk6}) is reduced to the study of a finite dimensional dynamical system, evolving under (\ref{sk8}); an explicit computation shows that {\it if} $e_{\m,h} \simeq e$ and $\n_{\m,h} \simeq e^{2}$ then by truncating (\ref{sk8}) to any finite order $p>0$, choosing $e$ small enough, we get that $Z_{h}\simeq M^{-\eta e^{2}}$, $v_{h}\simeq v_{-\io} - (1 - v)M^{\tilde\eta h}$, where $\eta$, $\tilde \eta$ are $O(e^2)$ and positive, and $v_{-\io} - 1 = O(e_{0,-\io} - e_{1,-\io})$. Clearly, one would show that the remainder of the truncation is small; however to do this one has to prove the convergence of the series, which is notoriously a difficult task for bosonic field theories, and it is outside the purposes of this Thesis. However, in the language of \cite{BG}, here we prove that perturbation theory is {\em consistent to the order $p$} for any $p>0$.

But then how to check the assumptions on the renormalized couplings? Notice that now we cannot rely on the evolution equation (\ref{sk8}), because a single non-vanishing order in the components of $\vec \b_{h}$ contributing to $e_{\m,h}$, $\n_{\m,h}$ would generically produce an unbounded flow as $h\rightarrow-\io$. The informations that we need are provided by {\it Ward identities}; the Ward identities are non perturbative identities between the Schwinger functions of a field theory, which are implied by gauge invariance. To get informations on $e_{\m,h},\,\n_{\m,h}$ we will implement Ward identities at each RG step (that is at each single scale integration), following a strategy recently proposed and developed by Benfatto and Mastropietro, \cite{BM}, in the context of one dimensional Luttinger liquids.\footnote{This strategy allowed to prove the so called {\it vanishing of the Beta function} for one dimensional fermionic systems. At the formal level, this was known since the 1970's, see \cite{DLa, DM}; however, the earlier works neglected the presence of cutoffs, which necessarily break gauge invariance. Therefore, the problem of establishing whether gauge invariance and formal Ward identities were recovered in the limit of cutoff removal was not considered by the Authors of the original proof. In \cite{BM} this problem has been considered, and the Authors proved that actually the Ward identities found after the removal of the cutoff are {\it different} from the formal ones: this is the phenomenon of {\it chiral anomaly}, well-known in the context of similar models used in Relativistic Quantum Field Theory, {\it e.g.} the Schwinger model \cite{ZJ}.} The strategy can be summarized as follows.
\begin{itemize}
\item Assume inductively that our assumptions on the renormalized couplings are true on scales $>h$.
\item At the step $-h$ of the RG we introduce a new model, the {\it reference model}, which differs from the original model because of the presence of a cutoff on the bosons that suppresses all momenta smaller that $M^{h}$; the idea of putting a cutoff only on the bosonic sector of the theory is borrowed from Adler and Bardeen, \cite{AB}. This new model can be investigated using the multiscale integration described above, and it turns out that on scales $\geq h$ it has the {\it same running coupling constants} of the full model. After the scale $h$ all the bosonic fields have been integrated out, and we are left with a purely fermonic theory; but this theory is much easier to investigate with respect to the initial one, and in particular it turns out to be {\it superrenormalizable}. This implies that the running coupling constants essentially ``cease to flow'' on scale $h$.

\item The bosonic cutoff {\it does not break gauge invariance}, and Ward identities for the Schwinger functions of the reference model can be derived; by suitably choosing the external momenta of the Schwinger functions involved in the Ward identities we will be able to check our inductive assumptions on $e_{\m,h},\,\n_{\m,h}$.
\end{itemize}

Finally, another interesting consequence of Ward identities is that $e_{0,-\io} = e_{1,-\io}$ (while $e_{1,h} = e_{2,h}$ for all $h$, as it follows from the $2\pi/3$ rotational symmetry of the model); therefore, since in general $v_{-\io} = 1 + O(e_{0,-\io} - e_{1,-\io})$, this last fact implies that $v_{-\io} =1$, which is the speed of light in our units. This concludes the sketch of the strategy that allows to compute the free energy, the Schwinger functions and the correlation functions of the model. 

Now, let us conclude the Section by briefly discussing what happens in presence of Kekul\' e distortions; the effect of Kekul\' e distortions can be taken into account by replacing the hopping parameter with $t_{\vec x,i} = t + \frac{2}{3}\D_0 \cos(\vec p_{F}^{+}(\vec\d_j - \vec\d_{j_0} - \vec x))$ with $j_0 = 1,2,3$. The main difference with respect to the case $\D_0=0$ is that $\D_0$ behaves as a {\it bare mass} for the fermion propagator, and it gives rise to a new running coupling constant $\D_{h}$, which grows as $\D_0 M^{-\eta_K h}$, with $\eta_K >0$. As for the case $\D_0=0$ we perform a multiscale integration to evaluate the partition function, but here we have to distinguish two regimes: in the first regime the mass $\D_h$ is smaller than $M^{h}$, which is the size of the momentum of the single scale field $\Psi^{(h)}_{\kk',\s,\r,\o}$, and the analysis is qualitatively the same discussed before. The second regime is defined by the scale $h(\D)$ such that $\D_{h(\D)} = M^{h(\D)}$, where the fermion propagator becomes {\it massive}; the main consequence of this fact is that the running coupling constants on scales $\leq h(\D)$ remain close to their values at the threshold $h(\D)$.

\section{Summary}\label{secsum}
\setcounter{equation}{0}
\renewcommand{\theequation}{\ref{secsum}.\arabic{equation}}

In Chapter \ref{capUV} we introduce a functional integral representation for our model and, as a pedagogical example, we start to evaluate the free energy by discussing the integration of the ultraviolet degrees of freedom; the analysis will be similar, but not equal, to the one discussed in \cite{GM} for the Hubbard model on the honeycomb lattice with short range interactions. The outcome of the ultraviolet integration is an effective infrared theory, whose action has a precise structure determined by remarkable lattice symmetries. 

In Chapter \ref{sec2.4.2} we continue the evaluation of the free energy by discussing the integration of the infrared degrees of freedom; the procedure will be similar to the one perfomed in \cite{GMP1}, in the context of an effective continuum model for graphene. Here we shall define the localization and renormalization operations, which allow to safely integrate the infrared scales. In this way we get a sequence of infrared effective theories, involving fields depending on momenta closer and closer to the singularities and a number of running coupling constants; the various contributions to the effective actions will be graphically represented in terms of Gallavotti-Nicol\` o trees. This graphical representation will be crucial to derive infrared-stable bounds on the {\it effective potentials} of the effective actions. After this, we discuss the flow of the running coupling constants, and in particular we introduce the {\it reference model} on which we derive the Ward identities that we need in order to control the flow of the effective couplings.

In Chapter \ref{seccorr} we adapt the strategy developed in the previous two Chapters to the generating functionals of the Schwinger functions and of the respose functions; here shall we prove our results on the two point Schwinger function (\ref{res1}) and (\ref{res2}), and on the response functions (\ref{res4}) -- (\ref{res6}).

In Chapter \ref{caplat} we change the definition of the model in order to take into account Kekul\' e distortions. We first study the effect of the electromagnetic interaction on a preexisting Kekul\' e distortion; then, we investigate a mechanism for a possible {\it spontaneous} Kekul\' e instability. Here we will prove the result on the two point Schwinger function in presence of Kekul\' e distortion (\ref{res8}) -- (\ref{res11}), and derive and discuss the gap equation (\ref{res12b}) -- (\ref{res15}). 

In the Appendices we collect all the technical ingredients and explicit computations that are needed in our proofs; in particular, in Appendix \ref{app1} we show how the Hamiltonian model is mapped in a Quantum Field Theory, in Appendix \ref{app2} we prove the perturbative equivalence of Coulomb and Feynman gauges, and in Appendix \ref{app2d} we collect all the explicit computations of this Thesis.

\chapter{The ultraviolet integration}\label{capUV}
\setcounter{equation}{0}
\renewcommand{\theequation}{\ref{capUV}.\arabic{equation}}
\section{Introduction}\label{secintro2}
\setcounter{equation}{0}
\renewcommand{\theequation}{\ref{secintro2}.\arabic{equation}}

In this Chapter and in the next one we introduce our rigorous RG framework, by performing as a pedagogical example the computation of the specific free energy. The same methods with some modifications will be used in Chapter \ref{seccorr} to prove our results on the two point Schwinger function and on the response functions, and in Chapter \ref{caplat} to take into account the presence of lattice distortions. Here we start by discussing the integration of the ultraviolet degrees of freedom; this is the ``easiest'' part of the work, since the bosons are regularized by an ultraviolet cutoff, which will be kept fixed, and the hexagonal lattice provides a natural ultraviolet cutoff for the fermions. As an outcome of the integration we will get an effective infrared theory, where the various contributions to the effective action will be expressed as series in the bare charge $e$ with finite coefficients, the $N$-th order bounded proportionally to $(N/2)!$. In particular, as a consequence of some remarkable lattice symmetry properties of our model, the effective action will have a precise structure, which will be preserved by the subsequent infrared integration. 

The Chapter is organized in the following way: in Section \ref{sec2.2} we introduce the functional integral representation of the model; in Section \ref{sec2.4} we discuss the symmetries of our model, while in Section \ref{sec2.4.1} we describe the outcome of the integration of the ultraviolet regime; the details of the ultraviolet integration are contained in Appendix \ref{app2b}.

\section[Functional integral representation]{Functional integral representation}\label{sec2.2}
\setcounter{equation}{0}
\renewcommand{\theequation}{\ref{sec2.2}.\arabic{equation}}

Here we shall introduce the quantum field theory associated to the model defined in Section \ref{sec1}; in fact, as shown in Appendix \ref{app1}, for any $\b$, $L$, $h^{*}$ fixed the perturbative series in $e$ of the partition function of our model, which can be computed using Trotter formula and Wick theorem, turns out ot be equal order by order to the perturbative series of a suitable functional integral, involving Grassmann ({\it i.e.} fermionic) and real ({\it i.e.} bosonic) Gaussian fields. The same is true for the average of physical observables which can be expressed as series in the fermionic and bosonic creation/annihilation operators.
\subsection{Integration measures and interaction}\label{secfint}

In this Section we shall introduce the elements that define the quantum field theory in which we will map our Hamiltonian model; namely, the free fermionic and bosonic measures, and the fermion-boson interaction.

\paragraph{Fermionic gaussian measure.} Let $K\in \mathbb{N}$, and 
\be
{\cal D}^{*}_{\b,L} := {\cal D}_{\b,L}\cap \{ k_0: \chi(M^{-K}|k_0|)>0 \}\;.
\ee 
We consider the Grassmann algebra generated by the Grassmann variables $\hat\Psi^{\pm}_{\kk,\s,\r}$ with $\s = \uparrow\downarrow$, $\r=1,2$ and $\kk\in\DD_{\b,L}^{*}$, and a Grassmann integration 
\be
\int \DD\Psi := \int \prod_{\kk\in {\cal D}^{*}_{\b,L}}\prod_{\s=\uparrow\downarrow}^{\r=1,2}d\hat\Psi^{+}_{\kk,\s,\r}d\hat \Psi^{-}_{\kk,\s,\r}\;; 
\ee
we define the Fourier transform of the fermionic field as:
\be
\Psi^{\pm}_{\xx + (\r-1)(0,\vec\d_1),\s,\r} := \frac{1}{\b|\L|}\sum_{\kk\in\DD_{\b,L}}\eu^{\pm\iu \kk\xx}\hat\Psi^{\pm}_{\kk,\s,\r}\;,\qquad \xx \in [0,\b]\times \L\;,\label{1.2.0}
\ee
where $\DD_{\b,L}$ has been defined in (\ref{free0}). In the following, we shall use the notation $\int \frac{d\kk}{D} \equiv (\b|\L|)^{-1}\sum_{\kk\in\DD_{\b,L}}$, where $D := (2\pi)B_1$ and $B_1 = 8\pi^2/(3\sqrt{3})$ is the volume of the first Brillouin zone. Let us define the {\it free fermionic propagator} $\hat{g}(\kk)$ as
\bea
\hat{g}(\kk) &:=& \chi_{K}(k_0)\begin{pmatrix} -ik_0 & -t\O^{*}(\vec k)\\ -t\O(\vec k) & -ik_0 \end{pmatrix}^{-1}\;,\label{1.2.10}\\
&=:&\chi_{K}(k_0)\big[B(\kk)\big]^{-1}\;,\qquad \chi_{K}(k_0) := \chi(M^{-K}|k_0|)\;.\nn
\eea
%
Then, we introduce the Grassmann gaussian integration $P(d\Psi)$ as
\bea
P(d\Psi) &=& \Big[ \prod_{\kk\in {\cal D}^*_{\b,L}}^{\s= \uparrow\downarrow} \frac{-\b^2|\L|^2\big[ \chi_{K}(k_0) \big]^2}{k_0^2 + |v(\vec k)|^2} d\hat\Psi^{+}_{\kk,\s,1}d\hat\Psi^{-}_{\kk,\s,1}d\hat\Psi^+_{\kk,\s,2}d\hat\Psi^{-}_{\kk,\s,2}\Big]\cdot\nn\\&&\cdot\exp\Big\{ -(\b|\L|)^{-1}\sum_{\kk\in {\cal D}^{*}_{\b,L}}^{\s=\uparrow\downarrow}\hat\Psi^{+}_{\kk,\s,\cdot}\hat{g}(\kk)^{-1}\hat\Psi^{-}_{\kk,\s,\cdot} \Big\}\;;\label{1.2.11}
\eea
by the discussion of Appendix \ref{app0} it turns out that
\be
\int P(d\Psi)\hat\Psi^{-}_{\kk,\s,\r}\hat\Psi^{+}_{\kk',\s',\r'} = \b|\L|\d_{\s,\s'}\d_{\kk,\kk'}\big[ \hat{g}(\kk) \big]_{\r,\r'}\;.\label{1.2.12}
\ee
\paragraph{Bosonic gaussian measure.} Let 
\bea
\PPP_{\b,L} &:=& {\cal P}_{L}\cup \Big\{p_0 = \frac{2\pi m}{\b},\, m\in \ZZZ\Big\}\;,\nn\\ 
{\cal P}^{*}_{\b,L} &:=& {\cal P}_{\b,L}\cap \{ \pp = (p_0,\vec p): \chi_{[h^{*},0]}(|\vec p|)\chi_{K}(p_0)>0\}\;,\\
\PPP_{\b,L}^{*,+} &:=& {\cal P}_{L}\cup \{p_0 = \frac{2\pi m}{\b},\, m\in \ZZZ^{+}\}\cap \{ \pp = (p_0,\vec p): \chi_{[h^{*},0]}(|\vec p|)\chi_{K}(p_0)>0 \},\nn
\eea
where the set $\PPP_{L}$ has been defined in the lines before (\ref{1.1.2a}), and consider the complex Gaussian variables $\hat A_{\m,\pp}$ with $\m = 0,1,2$, $\pp\in \PPP_{\b,L}$ and $\hat A^{*}_{\m,\pp} = \hat A_{\m,-\pp}$; this last condition implies that the Fourier transform of the $\hat A_{\m,\pp}$ field, namely
\be
A_{\m,\xx} := \frac{1}{\b\AAA_{\L}}\sum_{\pp\in \PPP_{\b,L}}\eu^{-\iu \pp \xx}\hat A_{\m,\pp}\;,\qquad \xx\in \RRR^{3}\;\label{1.2.12b}
\ee
is real. In the following, we shall use the notation $\int \frac{d\pp}{(2\pi)^3} \equiv (\b\AAA_{\L})^{-1}\sum_{\pp\in\PPP_{\b,L}}$. If $\xi\in [0,1]$ and $n_{0} = 1$, $\vec n = \vec 0$, we define the {\em free photon propagator} in the $\xi$-gauge as
\bea
\hat w^{\xi}_{\m,\n}(\pp) &:=& \int\frac{dp_3}{(2\pi)}\,\frac{\chi_{K}(p_0)\chi_{[h^{*},0]}(|{\ul p}|)}{\pp^2 + p_3^2}\D^{\xi}_{\m,\n}(\pp,p_3)\;,\label{1.2.13}\\
\D^{\xi}_{\m,\n}(\pp,p_3) &:=&\d_{\m,\n} - \xi\frac{p_{\m}p_{\n} - p_0(p_\m n_{\n} + p_{\n} n_{\m})}{{\ul p}^2}\;,\nn\\
\eea
where $\m,\n\in\{0,1,2\}$, and the gaussian integration $P^{\xi}(dA)$ as
\bea
P^{\xi}(dA) &=& \left[\prod_{ \pp\in {\cal P}^{*,+}_{\b,L}}\Big[ \frac{1}{(\pi\b \AAA_{\L})^3 \det \hat w^{\xi}(\pp)} \Big] \prod_{\m = 0,1,2} d\hat A_{\m,\pp} d\hat A_{\m,-\pp}\right]\cdot\nn\\&&
\cdot \exp\Big\{ -(2\b \AAA_{\L})^{-1}\sum_{\pp\in {\cal P}^*_{\b,L}} \hat A_{\cdot,\pp}\big[\hat w^{\xi}(\pp)\big]^{-1} \hat A_{\cdot, -\pp} \Big\}\;;\label{1.2.14}
\eea
notice that if $\xi\in[0,1]$ then $\pp\in \PPP^{*}_{\b,L}\Rightarrow\det \hat w^{\xi}(\pp) >0$, and 
\be
\int P^{\xi}(dA) \hat A_{\m,\pp} \hat A_{\n,-\pp'} = \b \AAA_{\L} \d_{\pp,\pp'}\hat w^{\xi}_{\m,\n}(\pp)\;.\label{1.2.15}
\ee

\paragraph{Interaction.} Finally, we define the {\it interaction} $V(\Psi,A)$ as
\bea
V(\Psi,A) &=& \int_{0}^{\b} dx_0 \sum_{\substack{\vec x\in \L \\ i\in [1,3]}} \sum_{\s = \uparrow\downarrow}t\Big[\Psi^{+}_{\xx,\s,1}\Psi^{-}_{(x_0,\vec x + \vec\d_i),\s,2}\Big(\eu^{\iu eA_{(\xx,i)}} - 1\Big) + c.c.\Big]  \nn\\&&- \iu e\sum_{\vec x\in \L_{A}\cup\L_{B}} n_\xx A_{0,\xx}\;,\label{1.2.20}
\eea
where $A_{(\xx,i)} := \int_{0}^{1} ds\,\vec\d_{i}\cdot\vec A_{\xx + s(0,\vec\d_i)}$. It is useful to rewrite (\ref{1.2.20}) in momentum space; it follows that
\bea
&&V(\Psi,A) = \nn\\&& = \int\sum_{\s,j} t\Big[\hat\Psi^{+}_{\kk+\pp,\s,1}\hat\Psi^{-}_{\kk,\s,2}\hat F_{j,\pp}\eu^{-\iu\vec k(\vec \d_j - \vec\d_1)} + \hat\Psi^{+}_{\kk+\pp,\s,2}\hat\Psi^{-}_{\kk,\s,1}\hat F^{*}_{j,-\pp}\eu^{\iu (\vec k + \vec p)(\vec \d_j - \vec \d_1)}\Big] \nn\\&& \quad 
-\iu e\int \sum_{\s}\Big[\hat \Psi^{+}_{\kk+\pp,\s,1}\hat \Psi^{-}_{\kk,\s,1}\hat A_{0,\pp} + \hat \Psi^{+}_{\kk+\pp,\s,2}\hat \Psi^{-}_{\kk,\s,2}\hat A_{0,\pp}\eu^{-\iu\vec p\vec\d_1}\Big]\;\label{1.2.20b}
\eea
where $\kk+\pp$ is constrained to live in $\DD_{\b,L}$ and
\bea
\hat F_{j,\pp} &:=& \int_{0}^{\b} dx_{0}\sum_{\vec x\in \L_{A}} \eu^{\iu\pp\xx}\Big[\exp\Big\{ \iu e A_{(\xx,j)} \Big\} - 1\Big]\label{1.2.20c}\\
&=& \sum_{N\geq 1}\frac{(\iu e)^{N}}{N!} \int\frac{d\pp_1}{(2\pi)^3}\ldots\frac{d\pp_N}{(2\pi)^3}\,\tilde\d(\pp - \sum_{i}\pp_{i})\Big[\prod_{i=1}^{N}\eta_{j}(\vec p_{i})\,\vec\d_{j}\cdot \vec{\hat{A}}_{\pp_{i}}\Big]\;,\nn
\eea
with $\eta_{j}(\vec p) := \frac{1 - \eu^{-\iu \vec p\cdot\vec \d_{j}}}{\iu\vec p\cdot\vec\d_{j}}$, and $\tilde\d(\pp) := (\AAA_\L |\L|^{-1}) \d(\pp)$ with $\d(\pp) := \d(p_0)\d(\vec p)$ where
\be
\d(\vec k) := |\L|\sum_{n_{1},n_{2}\in\ZZZ}\d_{\vec k,n_{1}\vec b_{1} + n_{2}\vec b_{2}}\;,\quad \d(k_{0}) := \b\d_{k_{0},0}\;.\label{1.2.20d}
\ee
%
%

\subsection{Averages}\label{secrav}
Consider a generic function $\OO(\Psi,A)$ of the fields $\Psi,A$; we shall consider only functions $\OO(\Psi,A)$ which are {\it integrable} in the sense of \cite{BG}, {\it i.e.} that can be expressed as series in the fields $\Psi,A$, with bounded coefficients. For any positive $\b,L$, we define the average in the $\xi$-gauge of $\OO(\Psi,A)$ as
\bea
\media{\OO(\Psi,A)}^{\xi}_{\b,L} &:=& \lim_{K\rightarrow+\infty}\lim_{h^{*}\rightarrow-\infty}\frac{1}{\Xi^{[h^{*},K]}_{\b,L}}\int P(d\Psi)P^{\xi}(dA)\eu^{V(\Psi,A)}\OO(A,\Psi)\;,\nn\\
\Xi^{[h^{*},K]}_{\b,L} &:=& \int P(d\Psi)P^{\xi}(dA)\eu^{V(\Psi,A)}\;,\label{av1}
\eea
where $\Xi_{\b,L} := \lim_{K\rightarrow+\infty}\lim_{h^{*}\rightarrow-\infty}\Xi^{[h^{*},K]}_{\b,L}$ is the {\it partition function} of the model. As explained in Appendix \ref{app1}, the quantity (\ref{av1}) can be formally expressed as a perturbative series in the electric charge $e$; in particular, if $\widetilde \OO$ is the operator obtained by suitably replacing the fermionic and bosonic fields in $\OO$ with fermionic and bosonic creation/annihilation operators, it follows that
\be
\frac{\Tr\{e^{-\b \HHH_{\L}}\widetilde \OO)\}}{ \Tr\{e^{-\b\HHH_{\L}}\}} = \media{\OO(\Psi,A)}^{1}_{\b,L}\;.\label{av1b}
\ee
Moreover, as shown in Appendix \ref{app2}, if $\OO(\Psi,A)$ is {\it gauge invariant}, that is if it is left unchanged by the transformation $\Psi_{\xx,\s,\r}^{\varepsilon}\rightarrow \Psi_{\xx,\s,\r}^{\varepsilon}e^{\varepsilon \iu \a_{\xx}}$, $A_{\m,\xx}\rightarrow A_{\m,\xx} + \partial_{\m}\a_{\xx}$, then:
\be
\partial_{\xi}\media{\OO(\Psi,A)}^{\xi}_{\b, L}=0\;.\label{av2}
\ee
This means that in the functional integral (\ref{av1}) we are free to choose the gauge that we prefer; for convenience, in this Thesis we shall work in the {\it Feynman gauge}, corresponding to the choice $\xi=0$. We shall set 
\be
P(dA) := P^{\xi=0}(dA)\;,\qquad \media{\cdots}_{\b,L} := \media{\cdots}_{\b,L}^{0}\;.
\ee

\section{Symmetries}\label{sec2.4}
\setcounter{equation}{0}
\renewcommand{\theequation}{\ref{sec2.4}.\arabic{equation}}
Before discussing the multiscale integration, it is important to note that both the Gaussian integrations and the interaction $V(\Psi,A)$ are invariant under the action of suitable symmetry transformations; this invariance will be preserved by the subsequent iterative integration procedure, and will guarantee the vanishing of some running coupling constants. In the following Lemma we collect all the symmetry properties that will be needed in the following.
\begin{lemma}\label{lem2.4}
For any choice of $K$, $h^{*}$, $\b$, $L$, the fermionic Gaussian integration $P(d\Psi)$, the bosonic Gaussian integration in the Feynman gauge $P(dA)$ and the interaction $V(\Psi,A)$ are invariant under the following transformations: 
\begin{enumerate}
\item[(1)] \underline{spin exchange:} $\hat\Psi^{\varepsilon}_{\kk,\s,\r}\leftrightarrow \hat\Psi^{\varepsilon}_{\kk,-\s,\r}$;
\item[(2)] \underline{global $U(1)$:} $\hat\Psi^{\varepsilon}_{\kk,\s,\r}\rightarrow \eu^{\iu \varepsilon \a_{\s}}\hat\Psi^{\varepsilon}_{\kk,\s,\r}$, with $\a_{\s}\in \RRR$ independent of $\kk$;
\item[(3)] \underline{spin $SO(2)$:} $\begin{pmatrix} \hat\Psi^{\varepsilon}_{\kk,\uparrow,\r} \\ \hat\Psi^{\varepsilon}_{\kk,\downarrow,\r} \end{pmatrix}  \rightarrow R_{\th}\begin{pmatrix} \hat\Psi^{\varepsilon}_{\kk,\uparrow,\r} \\ \hat\Psi^{\varepsilon}_{\kk,\downarrow,\r} \end{pmatrix}$, with $R_{\th}=\begin{pmatrix} \cos\th & \sin\th \\ -\sin\th & \cos\th \end{pmatrix}$ and $\th\in \mathbb{T}$ independent of $\kk$, where $\mathbb{T}$ is the one-dimensional torus;
\item[(4)] \underline{discrete spatial rotations:} $\hat \Psi^{\varepsilon}_{\kk,\s,\r}\rightarrow
e^{\varepsilon \iu T\vec k(\vec\d_2 - \vec\d_1)(\r - 1)}\hat \Psi^{\pm}_{(k_0,T\vec k),\s,\r}$,
$(\hat A_{0,\pp},\,\vec{\hat{A}}_{\pp})\rightarrow (\hat A_{0,(p_{0},T\vec p)},\,T^{-1}\vec{\hat{A}}_{(p_0,T\vec p)})$,
where $T$ is the $2\pi/3$ rotation matrix.
\item[(5)] \underline{complex conjugation:} $\hat\Psi^{\varepsilon}_{\kk,\s,\r}\rightarrow \hat\Psi^{\varepsilon}_{-\kk,\s,\r}$,
$\hat A_{\m,\pp}\rightarrow -\hat A_{\m,-\pp}$, $c\rightarrow c^{*}$, where $c$ is generic constant appearing in $P(d\Psi)$, $P(dA)$ and/or in $V(\Psi,A)$; 
\item[(6.a)] \underline{horizontal reflections:} $\hat\Psi^{\varepsilon}_{\kk,\s,1}\leftrightarrow
\hat\Psi^{\varepsilon}_{(k_0,-k_1,k_2),\s,2}$, $\hat A_{1,\pp}\rightarrow -\eu^{\iu \vec p \vec \d_1}\hat A_{1,(p_0,-p_1,p_2)}$,
$\hat A_{\n,\pp}\rightarrow \eu^{\iu \vec p \vec \d_1}\hat A_{\n,(p_0,-p_1,p_2)}$ if $\n\neq 1$.
\item[(6.b)] \underline{vertical reflections:} $\hat\Psi^{\varepsilon}_{\kk,\s,\r}\rightarrow \hat\Psi^{\varepsilon}_{(k_0,k_1,-k_2),\s,\r}$,
$\hat A_{2,\pp}\rightarrow -\hat A_{2,(p_0,p_1,-p_2)}$, $\hat A_{\n,\pp}\rightarrow \hat A_{\n,(p_0,p_1,-p_2)}$ if $\n\neq 2$.
\item[(7)] \underline{particle-hole:} $\hat\Psi^{\varepsilon}_{\kk,\s,\r}\rightarrow \iu\hat\Psi^{-\varepsilon}_{(k_0,-\vec k),\s,\r}$,
$\vec{\hat{A}}_{\pp}\rightarrow -\vec{\hat{A}}_{(-p_0,\vec p)}$, $\hat A_{0,\pp}\rightarrow \hat A_{0,(-p_0,\vec p)}$.
\item[(8)] \underline{inversion:} $\hat\Psi^{\varepsilon}_{\kk,\s,\r}\rightarrow \iu(-1)^{\r}\hat\Psi^{\varepsilon}_{(-k_0,\vec k),\s,\r}$, $\vec{\hat{A}}_{\pp}\rightarrow \vec{\hat{A}}_{(-p_0,\vec p)}$,
$\hat A_{0,\pp}\rightarrow -\hat A_{0,(-p_0,\vec p)}$.
\end{enumerate}
\end{lemma}
\noindent{\it Proof.} For notational simplicity, here we shall consider the limit $K\rightarrow+\infty$; the case of a finite $K$ can be worked out in exactly the same way.
\vskip.2cm
\noindent The invariance of $P(d\Psi)$, $P(dA)$, $V(\Psi,A)$ under (1), (2), (3) is obvious, and so is the invariance of $P(dA)$ under (4)--(8). 
\vskip.2cm
\noindent{\it Symmetry (4).} Let us prove the invariance of
\bea
&&\sum_{\kk} \hat\Psi^{+}_{\kk,\s,\cdot}\hat{g}(\kk)^{-1}\hat\Psi^{-}_{\kk,\s,\cdot} = \nn\\
&& \quad = -\iu\sum_{\kk}\hat\Psi^{+}_{\kk,\s,1}k_{0}\hat\Psi^{-}_{\kk,\s,1} - \sum_{\kk}\hat\Psi^{+}_{\kk,\s,1}t\O^{*}(\vec k)\hat\Psi^{-}_{\kk,\s,2} - \sum_{\kk}\hat\Psi^{+}_{\kk,\s,2}t\O(\vec k)\hat\Psi^{-}_{\kk,\s,1}\nn\\&&\qquad - \iu\sum_{\kk}\hat\Psi^{+}_{\kk,\s,2}k_{0}\hat\Psi^{-}_{\kk,\s,2}\;,\label{1.2.24}
\eea
which implies the invariance of $P(d\Psi)$. The first and the fourth term in the second line of (\ref{1.2.24}) are obviously invariant, while the sum of the second and the third term is changed into
\bea
&&-t\sum_{\kk}\hat\Psi^{+}_{(k_0,T\vec k),\s,1}\O^*(\vec k)\eu^{\iu \vec k(\vec\d_3 - \vec\d_1)}\hat\Psi^{-}_{(k_0,T\vec k),\s,2} + c.c.\label{1.2.25}\\&& = -t\sum_{\kk}\hat\Psi^{+}_{\kk,\s,1}\O^*(T^{-1}\vec k)\eu^{\iu \vec k(\vec\d_1 - \vec\d_2)}\hat\Psi^{-}_{\kk,\s,2} -t\sum_{\kk} \hat\Psi^{+}_{\kk,\s,2}\O(T^{-1}\vec k)\eu^{\iu \vec k(\vec\d_2 - \vec\d_1)}\hat\Psi^{-}_{\kk,\s,1}\;.\nn
\eea
Using that $\O(T^{-1}\vec k) = \eu^{\iu \vec k(\vec\d_1 - \vec\d_2)}\O(\vec k)$, as it follows by the definition of $\O(\vec k)$, we find that the last line of (\ref{1.2.25}) is equal to the sum of the second and third term in (\ref{1.2.24}), as desired. Let us now prove the invariance of the temporal part of the interaction, that is:
\be
-\iu e\sum_{\kk,\pp}\hat \Psi^{+}_{\kk+\pp,\s,1}\hat \Psi^{-}_{\kk,\s,1}\hat A_{0,\pp} - \iu e\sum_{\kk,\pp}\hat \Psi^{+}_{\kk+\pp,\s,2}\hat \Psi^{-}_{\kk,\s,2}\hat A_{0,\pp}\eu^{-\iu\vec p\vec\d_1}\;.\label{1.2.26}
\ee
The first term in (\ref{1.2.26}) is obviously invariant under (4), while the second is changed into
\be
-\iu e\sum_{\kk,\pp}\hat \Psi^{+}_{\kk+\pp,\s,2}\hat \Psi^{-}_{\kk,\s,2}\hat A_{0,\pp}\eu^{-\iu\vec p\cdot T\vec\d_1}\eu^{\iu \vec p\cdot (\vec \d_2 - \d_1)}\;;\label{1.2.27}
\ee
using that $T\vec \d_j = \d_{j+1}$, the invariance of (\ref{1.2.26}) follows. Finally, let us prove the invariance of
\be
t\sum_{\kk,\pp}\sum_{j}\hat \Psi^{+}_{\kk+\pp,\s,1}\hat \Psi^{-}_{\kk,\s,2}\hat F_{j,\pp}\eu^{-\iu\vec k(\vec\d_j - \d_1)} + t\sum_{\kk,\pp}\sum_{j}\hat \Psi^{+}_{\kk+\pp,\s,2}\hat \Psi^{-}_{\kk,\s,1}\hat F^{*}_{j,-\pp}\eu^{\iu(\vec k+\vec p)(\vec\d_j - \vec\d_1)}\;.\label{1.2.28}
\ee
Notice that under (4) $A_{(\xx,j)}\rightarrow A_{((x_0,T\vec x),j+1)}$, which means that $\hat F_{j,\pp}\rightarrow \hat F_{j+1,(p_0,T\vec p)}$; therefore, (\ref{1.2.28}) is changed into
\bea
&&t\sum_{\kk,\pp}\sum_{j} \hat \Psi^{+}_{\kk+\pp,\s,1}\hat \Psi^{-}_{\kk,\s,2}\hat F_{j+1,\pp}\eu^{-\iu \vec k(\vec\d_{j+1} - \vec \d_2)}e^{-i\vec k(\vec\d_2 - \vec \d_1)} + \nn\\&&+ t\sum_{\kk,\pp}\sum_{j} \hat \Psi^{+}_{\kk+\pp,\s,2}\hat \Psi^{-}_{\kk,\s,1}\hat F^{*}_{j+1,-\pp}\eu^{\iu (\vec k+\vec p)(\vec\d_{j+1} - \vec \d_2)}e^{i(\vec k + \vec p)(\vec\d_2 - \vec \d_1)}\;,\label{1.2.29}
\eea
which shows the invariance of (\ref{1.2.29}) and concludes the proof of the invariance of  $V(\Psi,A)$ under (4).
\vskip.2cm
\noindent{\it Symmetry (5).} The invariance of $P(d\Psi)$ follows simply by noting that $\O(-\vec k) = \O^{*}(\vec k)$. Concerning $V(\Psi,A)$, (\ref{1.2.26}) is obviously invariant; the invariance of (\ref{1.2.28}), and therefore of $V(\Psi,A)$, follows by noting that under (5) $A_{(\xx,j)}\rightarrow -A_{(\xx,j)}$ and hence $\hat F_{j,\pp}\rightarrow \hat F_{j,-\pp}$.
\vskip.2cm
\noindent{\it Symmetry (6.a).} To check the invariance of $P(d\Psi)$ notice that under (6.a) the sum of the first and the fourth term in the second line of (\ref{1.2.24}) is obviously invariant, while the sum of the second and the third is changed into
\bea
&&-t\sum_{\kk}\hat\Psi^{+}_{(k_0,-k_1,k_2),\s,2}\O^{*}(\vec k)\hat\Psi^{-}_{(k_0,-k_1,k_2),\s,1} + c.c.\label{1.2.30}\\&& = -t\sum_{\kk}\hat\Psi^{+}_{\kk,\s,2}\O^{*}((-k_1,k_2))\hat\Psi^{-}_{\kk,\s,1} - t\sum_{\kk}\hat\Psi^{+}_{\kk,\s,1}\O((-k_1,k_2))\hat\Psi^{-}_{\kk,\s,2}\;;\nn
\eea
noting that $\O((-k_1,k_2)) = \O^{*}(\vec k)$, one sees that (\ref{1.2.30}) is equal to the sum of the second and third term in the second line of (\ref{1.2.24}), as desired. The invariance of the temporal part of the interaction is obvious. To conclude the proof of the invariance of $V(\Psi,A)$ notice that $\hat F_{j,(p_0,-p_1,p_2)}\rightarrow e^{i\vec p(\vec\d_{j'} - \vec\d_1)}\hat F^{*}_{j',-\pp}$, with $1' = 1$, $2' = 3$, $3' = 2$. In fact, setting $\tilde\xx := (x_0,-x_1,x_2)$, after (6.a) $\vec A_{\xx}\rightarrow \big(-A_{1,\tilde\xx + (0,\vec\d_1)},A_{2,\tilde\xx + (0,\vec\d_1)}\big)$ and $A_{(\xx,j)}\rightarrow -\int _{0}^{1}ds\,\vec\d_{j'}\vec A_{\tilde \xx - (0,s\vec\d_{j'} - \vec\d_1)}$; therefore, setting $\int_{\#} d\xx = \int dx_0\sum_{\vec x\in \L_{\#}}$ with $\#=A,B$,
\bea
&&\int_{A} d\xx\, \eu^{i\pp\xx}\exp\big\{ i e A_{(\xx,j)} \big\}\rightarrow \eu^{i\vec p\vec\d_1}\int_{B} d\xx\, \eu^{i\tilde\pp\xx}\exp\big\{ -i e \int_{0}^{1}ds\,\vec\d_{j'}\vec A_{\xx - (0,s\vec\d_{j'})}\big\}\nn\\
&&\quad = \eu^{i\vec p(\vec \d_1 - \vec\d_{j})}\int_{A} d\xx\, e^{i\tilde\pp\xx} \exp\big\{ -ie \int_{0}^{1}\vec\d_{j'}\vec A_{\xx - (0,(s-1)\vec\d_{j'})} \big\}\nn\\&&\quad = \eu^{i\vec p(\vec \d_1 - \vec\d_{j})}\int_{A} d\xx\, e^{i\tilde\pp\xx} \exp\big\{ -ie A_{(\xx,j')}\big\}\;,\label{1.2.31}
\eea
where in the first equality we used that $\L_{B} = \L_{A} + \vec\d_{j'}$ and that $\tilde\pp\cdot(0,\vec \d_j) = -\pp\cdot(0,\vec\d_{j'})$. Formula (\ref{1.2.31}), together with the fact that $\int d\xx\, \eu^{i\pp\xx} = \eu^{i\vec p(\vec\d_1 - \vec\d_j)}\int d\xx\, \eu^{i\tilde\pp\xx}$ implies that $\hat F_{j,\pp}\rightarrow \eu^{ i\vec p(\vec\d_{1} - \vec\d_j)}\hat F^{*}_{j',-\tilde\pp}$; therefore, (\ref{1.2.28}) changes into
\bea
&&t\sum_{\kk,\pp}\sum_{j} \hat\Psi^{+}_{\kk+\pp,\s,2}\hat\Psi^{-}_{\kk,\s,1}\eu^{ i \vec k(\vec\d_{j'} - \vec\d_1)}\eu^{ i\vec p(\vec\d_{j'} - \vec\d_1)}\hat F^{*}_{j',-\pp} +\nn\\&&+ t\sum_{\kk,\pp}\sum_{j}\hat\Psi^{+}_{\kk+\pp,\s,1}\hat\Psi^{-}_{\kk,\s,2}\eu^{- i(\vec k + \vec p)(\vec\d_{j'} - \vec\d_1)}\eu^{ i\vec p(\vec\d_{j'} - \vec\d_1)}\hat F_{j',\pp}\label{1.2.32}
\eea
which proves the invariance of $V(\Psi,A)$ under $(6.a)$.
\vskip.2cm
\noindent{\it Symmetry (6.b).} The invariance of $P(d\Psi)$ follows by noting that $\O((k_1,-k_2)) = \O(\vec k)$. The invariance of the temporal part of $V(\Psi,A)$ is obvious. To show the invariance of (\ref{1.2.28}), and therefore to conclude the proof of the invariance of $V(\Psi,A)$, notice that under (6.b) $\hat F_{j,\pp}\rightarrow \hat F_{j',\tilde \pp}$ with $\tilde\pp := (p_0,p_1,-p_2)$; then, (\ref{1.2.28}) changes into
\be
t\sum_{\kk,\pp}\sum_{j}\hat \Psi^{+}_{\kk+\pp,\s,1}\hat \Psi^{-}_{\kk,\s,2}\hat F_{j',\pp}\eu^{-\iu\vec k(\vec\d_{j'} - \d_1)} + t\sum_{\kk,\pp}\sum_{j}\hat \Psi^{+}_{\kk+\pp,\s,2}\hat \Psi^{-}_{\kk,\s,1}\hat F^{*}_{j',-\pp}\eu^{\iu(\vec k+\vec p)(\vec\d_{j'} - \vec\d_1)}\;,\label{1.2.33}
\ee
and (\ref{1.2.33}) concludes the proof of the invariance under (6.b).
\vskip.2cm
\noindent{\it Symmetry (7).} To show the invariance of $P(d\Psi)$, notice that the sum of the first and the fourth term in (\ref{1.2.24}) is left unchanged, while the sum of the second and the third term is changed into
\bea
&&+t\sum_{\kk}\hat\Psi^{-}_{(k_0,-\vec k),\s,1}\O^{*}(\vec k)\hat\Psi^{+}_{(k_0,-\vec k),\s,2} + t\sum_{\kk}\hat\Psi^{-}_{(k_0,-\vec k),\s,2}\O(\vec k)\hat\Psi^{+}_{(k_0,-\vec k),\s,1}\nn\\&& = -t\sum_{\kk}\hat\Psi^{+}_{\kk,\s,2}\O^{*}(-\vec k)\hat\Psi^{-}_{\kk,\s,1} - t\sum_{\kk}\hat\Psi^{+}_{\kk,\s,1}\O(-\vec k)\hat\Psi^{-}_{\kk,\s,2}\;;\label{1.2.34}
\eea
using that $\O(-\vec k) = \O^{*}(\vec k)$, we see that (\ref{1.2.34}) is transformed into the sum of the second and third term of (\ref{1.2.24}), as desired. Regarding the interaction, (\ref{1.2.26}) changes into, setting $\tilde \kk := (k_0,-\vec k)$:
\be
-\iu e \sum_{\kk,\pp}\hat\Psi^{+}_{\tilde\kk,\s,1}\hat\Psi^{-}_{\tilde\kk+\tilde\pp,\s,1}\hat A_{0,-\tilde\pp} - \iu e\sum_{\kk,\pp}\hat\Psi^{+}_{\tilde\kk,\s,2}\hat\Psi^{-}_{\tilde\kk+\tilde\pp,\s,2}\hat A_{0,-\tilde\pp}\eu^{-\iu \vec p\vec \d_1}\;,\label{1.2.35}
\ee
which after a change of variables gives (\ref{1.2.26}). To show the invariance of the spatial part of $V(\Psi,A)$ (\ref{1.2.28}) first notice that under (7) $F_{j,\pp}\rightarrow F^{*}_{j,\tilde\pp}$; therefore, (\ref{1.2.28}) is changed into
\bea
&&t\sum_{\kk,\pp}\sum_{j}\hat \Psi^{+}_{(k_0,-\vec k),\s,2}\hat \Psi^{-}_{(k_0 + p_0,-\vec k - \vec p),\s,1}\hat F^{*}_{j,(p_0,-\vec p)}\eu^{-\iu\vec k(\vec\d_j - \d_1)} +\nn\\&&+ t\sum_{\kk,\pp}\sum_{j}\hat \Psi^{+}_{(k_0,-\vec k),\s,1}\hat \Psi^{-}_{(k_0 + p_0, -\vec k - \vec p),\s,2}\hat F_{j,(-p_0,\vec p)}\eu^{\iu(\vec k+\vec p)(\vec\d_j - \vec\d_1)}\;,\label{1.2.36}
\eea
which is equal to (\ref{1.2.28}).
\vskip.2cm
\noindent{\it Symmetry (8).} Under (8) all the terms appearing in (\ref{1.2.24}) are separately invariant; this proves the invariance of $P(d\Psi)$; the same is true for $V(\Psi,A)$. This concludes the proof of Lemma \ref{lem2.4}.\qed

\section{The ultraviolet integration}\label{sec2.4.1}
\setcounter{equation}{0}
\renewcommand{\theequation}{\ref{sec2.4.1}.\arabic{equation}}

We start by studying the partition function $\Xi^{[h^{*},K]}_{\b,L} =: e^{-\b|\L| F^{[h^{*},K]}_{\b,L}}$. Note that in our model the fermionic momenta $\vec k$ have an intrinsic ultraviolet cut-off induced by the lattice, while the bosonic momenta $\vec p$ have an ultraviolet cut-off because of the definition of the photon field; on the contrary, the $k_0,p_0$ variables are not bounded uniformly in $K$. A preliminary step to our infrared analysis is the integration of the ultraviolet degrees of freedom corresponding to large values of $k_0,p_0$. We proceed in the following way. We decompose the free propagators $\hat{g}(\kk)$, $\hat{w}(\pp)$ into sums of two propagators supported in the regions of $k_0,p_0$ ``large'' and ``small''. The regions of $k_{0},p_0$ large and small are defined in terms of the smooth support function $\chi(t)$ introduced in Section \ref{sec1}; the constant $a_0$ entering in its definition is chosen so that the supports of $\chi\Big( \sqrt{k_0^2 + |\vec k - \vec p_{F}^{+}|^{2}} \Big)$ and $\chi\Big( \sqrt{k_0^2 + |\vec k - \vec p_{F}^{-}|^{2}} \Big)$ are disjoint (here $| \cdot |$ is the Euclidean norm over $\RRR^{2}\setminus \L^{*}$). In order for this condition to be satisfied, it is enough to choose $2 a_0 M < 4\pi /(3\sqrt{3})$; in the following, for reasons that will become clearer later (see discussion after Lemma \ref{lem2.4b}), we shall assume the slightly more restrictive condition $a_0 M < \pi/6$. We rewrite $\hat{g}(\kk)$ and $\hat w(\pp)$ as
\be
\hat{g}(\kk) = \hat{g}^{(u.v.)}(\kk) + \hat g^{(i.r.)}(\kk)\;,\qquad \hat{w}(\pp) = \hat{w}^{(u.v.)}(\pp) + \hat{w}^{(i.r.)}(\pp)\;\label{1.2.37}
\ee
where, setting $\pp_{F}^{\o} = (0,\vec p_{F}^{\o})$ with $\o=\pm$:
\bea
&&\hat{g}^{(u.v.)}(\kk) = \hat{g}(\kk) - \hat g^{(i.r.)}(\kk)\;,\qquad g^{(i.r.)}(\kk) = \sum_{\o = \pm}\chi(|\kk - \pp_{F}^{\o}|)\hat{g}(\kk)\;,\label{1.2.38}\\
&&\hat{w}^{(u.v.)}(\pp) = \hat{w}(\pp) - \hat{w}^{(i.r.)}(\pp)\;,\qquad \hat{w}^{(i.r.)}(\pp) = I\frac{\chi_{[h^{*},0]}(|\pp|)}{|\pp|}\frac{\arctan\big(\frac{a_0 M}{|\pp|}\big)}{\pi}\;,\nn
\eea
with $\hat{w}(\pp) := \hat w^{0}(\pp)$ is the photon propagator in the Feynman gauge and $I$ is the $3\times 3$ identity matrix; notice that $\arctan\big(\frac{a_0 M}{|\pp|}\big) = \frac{\pi}{2} + O(|\pp|)$. We now introduce four sets of independent Gaussian fields $\{\hat \Psi^{(u.v.)\pm}_{\kk,\s,\r}\}$, $\{ \hat \Psi^{(i.r.)\pm}_{\kk,\s,\r} \}$ and $\{\hat A^{(u.v.)}_{\m,\pp}\}$, $\{\hat A^{(i.r.)}_{\m,\pp}\}$ defined by, if $\# = u.v.,\,i.r.$:
\bea
\int P(d\Psi^{(\#)})\hat \Psi^{(\#)-}_{\kk,\s,\r}\hat\Psi^{(\#)+}_{\kk',\s',\r'} &=& \b|\L|\d_{\s,\s'}\d_{\kk,\kk'}\hat{g}^{(\#)}_{\r,\r'}(\kk)\;,\nn\\
\int P(dA^{(\#)})\hat A^{(\#)}_{\pp,\m}\hat A^{(\#)}_{-\pp',\n} &=& \b|\L|\d_{\pp,\pp'}\hat{w}^{(\#)}_{\m,\n}(\pp)\;.\label{1.2.39}
\eea
Similar to $P(d\Psi)$, $P(dA)$, the gaussian integrations $P(d\Psi^{(\#)})$, $P(dA^{(\#)})$ also admit an explicit representation analogous to (\ref{1.2.11}), (\ref{1.2.14}), with $\hat{g}(\kk)$, $\hat{w}(\pp)$ replaced by $\hat{g}^{(\#)}(\kk)$, $\hat{w}^{(\#)}(\pp)$ respectively, and the sums over $\kk,\pp$ restricted to the values in the support the corresponding propagator. The definition of Grassmann integration implies the following identity (``addition principle''):
\bea
&&\int P(d\Psi)P(dA)\eu^{V(\Psi,A)} = \label{1.2.41}\\&&= \int P(d\Psi^{(i.r.)})P(dA^{(i.r.)})\int P(d\Psi^{(u.v.)})P(dA^{(u.v.)})\eu^{V(\Psi^{(i.r.)} + \cdot, A^{(i.r.)} + \cdot)}\;,\nn
\eea
so that we can rewrite the partition function as
\bea
\Xi^{[h^{*},K]}_{\b,L} &=& \int P(d\Psi^{(i.r.)})P(dA^{(i.r.)})\exp\Big\{ \sum_{n\geq 1}\frac{1}{n!}\EE^{T}_{u.v.}\big(V(\Psi^{(i.r.)} + \cdot, A^{(i.r.)} + \cdot);n\big) \Big\}\nn\\
&\equiv& e^{-\b|\L|F_{0,K}}\int P(d\Psi^{(i.r.)})P(dA^{(i.r.)})e^{\VV_{K}(\Psi^{(i.r.)},A^{(i.r.)})}\;,\label{1.2.42}
\eea
where the {\it truncated expectation} $\EE^{T}_{u.v.}$ is defined, given any series $X(\Psi^{(u.v.)},A^{(u.v.)})$ in $A^{(u.v.)}$, $\Psi^{(u.v.)}$ depending on $A^{(i.r.)}$, $\Psi^{(i.r.)}$, as
\be
\EE^{T}_{u.v.}(X(\cdot);n) := \frac{\partial^{n}}{\partial \l^{n}}\log \int P(d\Psi^{(u.v.)})P(dA^{(u.v.)}) e^{\l X(\Psi^{(u.v.)}, A^{(u.v.)})}\Big|_{\l=0}\;,\label{1.2.43}
\ee
and $\VV_{K}$ is fixed by the condition $\VV_{K}(0,0) =0$. It can be shown, see Appendix \ref{app2b}, that $\VV_{K}$ can be written as
\bea
&&\VV_{K}(\Psi,A) =\label{1.2.44}\\&&= \sum_{\substack{n,m\geq 0 \\ n+m\geq 1}}\sum_{\ul\s,\ul\r,\ul\m} \int \Big[\prod_{j=1}^{n}\hat\Psi^{(i.r.)+}_{\kk_{2j-1},\s_{j},\r_{2j-1}}\hat\Psi^{(i.r.)-}_{\kk_{2j},\s_{j},\r_{2j}}\Big]\Big[ \prod_{i=1}^{m}\hat A^{(i.r.)}_{\m_{i},\pp_{i}}\Big]\cdot\nn\\
&&\quad\cdot W_{K,2n,m,\underline{\r},\underline{\m}}(\{\kk_{j}\},\{\pp_{i}\})\,\d\Big( \sum_{j=1}^{n}\big(\kk_{2j-1} - \kk_{2j}\big) - \sum_{i=1}^{K}\pp_{i}\Big)\;,
\eea
where $\int = \int \frac{d\pp_1}{(2\pi)^3}\ldots\frac{d\pp_m}{(2\pi)^3}\frac{d\kk_1}{D}\ldots\frac{d\kk_{2n}}{D}$, $\underline{\r} := (\r_{1},\ldots,\,\r_{2n})$, $\underline{\s} := (\s_{1},\ldots,\,\s_{n})$, $\underline{\m} := (\m_{1},\ldots,\,\m_{m})$, and the symbol $\d(\kk)$ has been defined in Eq.(\ref{1.2.20d}); the possibility of representing $\VV_{K}$ in the form (\ref{1.2.44}), with the kernels independent of the spin indeces $\s_{i}$, follows from the symmetry listed in Lemma \ref{lem2.4}. As an outcome of the discussion reported in Appendix \ref{app2b}, the constant $F_{0,K}$ and the kernels $W_{K,2n,m,\underline{\r},\underline{\m}}$ are given by formal power series in $e$,
\bea
&&F_{0,K} = \sum_{N\geq 0}F^{(N)}_{0,K}\,e^{N}\;,\nn\\
&&W_{K,2n,m,\underline{\r},\underline{\m}}(\{\kk_{j}\},\{\pp_{i}\}) = \sum_{N\geq 1}W_{K,2n,m,\underline{\r},\underline{\m}}^{(N)}(\{\kk_{j}\},\{\pp_{i}\})\,e^{N}\label{1.2.46}
\eea
with coefficients bounded uniformly in $\b,L,K,h^{*}$ independent of $e$; in particular, the following result holds.
\begin{lemma}\label{lem2.4b} There exist some positive constants $C_{n,m}$, $C$ independent of $\b$, $L$, $K$, $h^{*}$ such that
\be
\Big|F^{(N)}_{0,K}\Big| \leq  C^{N}\Big(\frac{N}{2}\Big)!e^{N}\;,\quad\max_{\{\kk_{j},\,\pp_i\}}\Big| W^{(N)}_{K,2n,m,\underline{\r},\underline{\m}}(\{\kk_{j}\},\{\pp_{i}\}) \Big| \leq C_{n,m}^{N}\Big(\frac{N}{2}\Big)!e^{N}\;;\label{1.2.47}
\ee
moreover, the following limits exist:
\bea
&&F^{(N)}_{0}:=\lim_{K\rightarrow+\infty}F_{0,K}^{(N)}\;,\label{1.2.46b}\\
&&W^{(N)}_{2n,m,\underline{\r},\underline{\m}}(\{\kk_{j}\},\{\pp_{i}\}) := \lim_{K\rightarrow +\infty}W^{(N)}_{K,2n,m,\underline{\r},\underline{\m}}(\{\kk_{j}\},\{\pp_{i}\})\;.\nn
\eea
\end{lemma}
\noindent{\it Proof.} See Appendix \ref{app2b}.\\

%
%
%
Regarding the choice of the constant $a_0$ in the definition of $\chi$, this is motivated as follows. The quadratic part in the fermionic fields of $\VV^{[h^{*},K]}(\Psi,A)$ is
\be
\sum_{\s,\r,\r'} \int \frac{d\kk}{D}\frac{d\qq}{D}\, \d(\qq) \hat\Psi^{(i.r.)+}_{\kk,\s,\r}\hat\Psi^{(i.r.)-}_{\kk+\qq,\s,\r'}W_{K,2,0,\r,\r'}(\kk,\kk+\qq)\;.\label{1.2.44b}
\ee
From the definitions (\ref{1.2.37}), (\ref{1.2.38}), the support of $\hat\Psi^{(i.r.)}$ consists of two disjoint regions around $\vec p_{F}^{+}$ and $\vec p_{F}^{-}$; because of the assumption $a_0 M < \pi/6$ if $\kk$ and $\kk+\pp$ belong to the support of $\hat\Psi^{(i.r.)}$ then $|\pp|<4\pi/3$, which implies that the only non-zero contribution in the sum over $\pp$ in (\ref{1.2.44b}) is the one corresponding to $\pp=\V0$. Analogously, the contributions to $\VV_{K}$ with $n=1, m=1$ and $n=0,m=2$ are, respectively:
\bea
&&\sum_{\substack{\s,\m \\ \r,\r'}} \int \frac{d\kk}{D}\frac{d\qq}{D}\frac{d\pp}{(2\pi)^3}\,\d(\qq)\hat\Psi^{(i.r.)+}_{\kk+\pp+\qq,\s,\r}\hat\Psi^{(i.r.)-}_{\kk,\s,\r'}\hat A^{(i.r.)}_{\m,\pp}W_{K,2,1,\r,\r',\m}(\kk,\kk+\pp+\qq,\pp)\nn\\
&&\sum_{\m,\m'} \int \frac{d\pp}{(2\pi)^3}\frac{d\qq}{(2\pi)^3}\, \d(\qq)\hat A_{\m,-\pp-\qq}^{(i.r.)}\hat A_{\m',\pp}^{(i.r.)}W_{K,0,2,\m,\m'}(\pp,-\pp-\qq)\;;\label{1.2.44c}
\eea
the only non-vanishing contribution in the sums over $\qq$ are those associated to $\qq=\V0$. Moreover, in the first line of (\ref{1.2.44c}), because of the assumption $a_0 M < \pi/6$, if $\vec k\in \BBB_{\o} = \{\vec q: |\vec q - \vec p_{F}^{\o}|\leq a_0 M\}$ and $\kk+\pp$ is in the support of $\Psi^{(i.r.)}$ then $\vec k + \vec p \in \BBB_{\o}$.
\begin{oss}
From now, with a little abuse of notation we shall only write the independent momenta at the argument of the kernels.
\end{oss}
Finally, it is important for the forthcoming discussion to note that the symmetries listed in Lemma \ref{lem2.4} imply some non-trivial invariance properties of the kernels.

\begin{lemma}\label{lem2.4c}
Let $W_{K,2,0}(\kk)$, $W_{K,2,1,\m}(\kk,\pp)$ be the matrices with entries given by $W_{K,2,0,\r,\r'}(\kk)$, $W_{K,2,1,\r,\r',\m}(\kk,\pp)$, respectively; in the limit $\b\rightarrow+\infty$, $|\L|\rightarrow+\infty$ the following properties are true:
\bea
W_{K,2,0}(\kk' + \pp_{F}^{\o}) &=& - z_{\m}\kk'_{\m}\G_{\m}^{\o} + O(\big|\kk' \big|^{2})\;,\label{1.2.49}\\
W_{K,0,2,\m,\n}(\pp) &=& \d_{\m\n}\n_{\m} + O(|\pp|^{2})\;,\label{1.2.50}\\
W_{K,2,1,\m}(\kk' + \pp_{F}^{\o},\pp) &=& \l_{\m}\G_{\m}^{\o} + O\Big(|\kk'| + |\pp|\Big)\;,\label{1.2.51}\\
W_{K,0,3,{\ul \m}}(\pp_1,\pp_2) &=& O\Big(|\pp_1| + |\pp_2|\Big)\;,\label{1.2.52}
\eea
where: (i) $z_{\m}$, $\l_{\m}$, $\n_{\m}$ are real and such that $z_{1} =z_{2}$, $\l_{1} = \l_{2}$, $\n_{1} = \n_{2}$; (ii) the matrices $\G^{\m}_{\o}$ are given by:
\be
\G^{0}_{\o} := \begin{pmatrix} -\iu & 0 \\ 0 & -\iu \end{pmatrix}\;,\quad \G^{1}_{\o} := \begin{pmatrix} 0 & \iu \\ -\iu & 0 \end{pmatrix}\;,\quad \G^{2}_{\o} := \begin{pmatrix} 0 & -\o \\ -\o & 0 \end{pmatrix}\;.\label{1.2.52b}
\ee
\end{lemma}
\noindent{\it Proof.} See Appendix \ref{app2c}.

\chapter{The infrared integration}\label{sec2.4.2}
\setcounter{equation}{0}
\renewcommand{\theequation}{\ref{sec2.4.2}.\arabic{equation}}

\section{Introduction}\label{secintro3}
\setcounter{equation}{0}
\renewcommand{\theequation}{\ref{secintro3}.\arabic{equation}}

As an outcome of the ultraviolet integration discussed in the previous Chapter and in Appendix \ref{app2b}, we obtained that the low energy physics of our model is described by an infrared effective theory. In this Chapter we analyze this theory by Renormalization Group; this is the ``hardest'' part of the work, since the remaining degrees of freedom that we are left with can be arbitrarily close to the singularities $\kk=\pp_{F}^{\o}$ and $\pp=\V0$. First of all, we will rewrite the fermion field as a sum over two independent fields with disjoint supports, each one supported around one of the two singularities; these new fields will be called {\it quasi-particle fields}, \cite{BG}. Then, we rewrite the quasi-particle fields and the photon fields as sums over independent fields, depending on momenta closer and closer to the infrared singularities; the propagators of these fields will be determined inductively. As for the ultraviolet integration, see Appendix \ref{app2b}, we will integrate scale after scale these new fields, and at each step of the integration we will be left with a new effective theory, whose parameters (wave function renormalization, effective masses, effective couplings, and effective Fermi velocity), also called {\it running coupling constants}, are determined by the integration of the higher energy degrees of freedom. Roughly speaking, the single step is performed by splitting the effective potential in the sum of two contributions, the {\em local part} and the {\em renormalized part}; the local part of the effective potential is absorbed in the redefinition, {\it i.e.} in the {\it renormalization}, of the fermionic integration measure and in the definition of the effective couplings: this step is the ``heart'' of the RG, and it corresponds to an infinite resummation of Feynman graphs. The new fermionic measure is similar to the old one, except for the fact that now the wave function renormalization and the effective Fermi velocity appearing in the fermion propagator are slightly changed with respect to the ones before the redefinition. Then, we {\it rescale} the fermionic fields in a suitable way, and we perform the single scale integration; the outcome of the integration has the same form of the effective potential from which we started, and we can iterate the process.

The outcome of the integration, that is the new effective potential, can be expressed graphically in terms of {\it Gallavotti-Nicol\` o trees}, \cite{GN1, GN2}; each tree has a value which can be computed as a sum over renormalized Feynman graphs, and it turns out that {\it if} the running coupling constants verify suitable bounds then the value of each tree is {\it finite}. To fully appreciate this fact one should compare with the naive perturbative series, which is plagued by infrared divergences.

Because of the iterative integration scheme, the running coupling constants on a given scale are determined starting from those on higher scales, and ultimately from the bare ones, exploiting a non-trivial recursion relation; in fact, the running coupling constants on a given scale can be obtained starting from those on the previous scale {\em plus} a well-defined function of all the running coupling constants on higher scales: this function is called the {\it Beta function}. The Beta function is expressed as a series in the running coupling constants, with finite coefficients; by truncating the Beta function to {\it any finite order} in the couplings and choosing the bare charge $e$ small enough {\it uniformly} in the momentum scale we will be able to check our assumptions on the running coupling constants and to close the single scale integration. Clearly, one would be able to prove that the remainder of the truncation is small; however to prove this one should be able to prove convergence of the series, which is notoriously a very hard task in bosonic theories, and it is outside the purposes of this Thesis.

However, the fact that even truncating the Beta function the flow of the running coupling constants is bounded is remarkable, and follows from the implementation of {\it Ward identities} at each integration step; this should be compared with what happens for instance in ``asymptotically slave'' theories ({\it e.g.} the $\varphi^{4}$ theory in $4$ dimensions, \cite{ZJ, G84}), where the ``wrong'' sign of the lowest order contribution to the Beta function makes the flow of the effective coupling unbounded. Here the phenomenon taking place is well known in one dimensional systems, and it is called {\it vanishing of the Beta function} \cite{BGPS, BM2, BM}. This is crucial to establish Luttinger liquid behavior; as far as we know, this is the first time that the same cancellation is found in a two dimensional system.

The Chapter is organized in the following way: in Section \ref{multi} we describe the infrared multiscale analysis; in Section \ref{sec4} we discuss the graphical representation of the effective potentials in terms of Gallavotti-Nicol\` o trees; in Section \ref{sec3a} we discuss the flow of the running coupling constants; in Section \ref{secWI} we derive the Ward Identities that allow to control the flows of the effective charge and of the effective photon mass, and finally in Section \ref{secWI} we show that the infrared fixed point of the RG procedure is described by a Lorentz invariant effective theory.
\begin{oss}
From now on we shall assume that the limits $h^{*}\rightarrow-\infty$, $K\rightarrow+\infty$ have been taken; this can be done safely since, as it is implicit from the discussion of Appendix \ref{app2b} and from the one that follows, the perturbative series that we shall get in presence of finite $h^{*}$, $K$ would converge uniformly order by order to a limit as $h^{*}\rightarrow-\infty$, $K\rightarrow+\infty$. 
%
%
\end{oss}

\section{Multiscale analysis}\label{multi}
\setcounter{equation}{0}
\renewcommand{\theequation}{\ref{multi}.\arabic{equation}}
In order to compute the partition function (\ref{1.2.42}) we will use standard functional Renormalization Group methods, \cite{G84, BG, M}. The integration of (\ref{1.2.42}) will be performed in an iterative way, moving from high to small momentum scales; the procedure will be similar to the one discussed for an effective continuum model of graphene in \cite{GMP1}. At the $n$-th step of the iteration the functional integral (\ref{1.2.42}) is rewritten as an integral involving only the momenta at a  distance proportional to $M^{-n}$ from the singularities  $\kk = \pp_{F}^{\o}$ for the fermions and $\pp = \V0$ for the bosons, and both the propagators and the interaction will be replaced by ``effective'' ones, which are {\em renormalized} by the integration of the momenta on higher scales. We will start by setting the notations of the zero-th scale, corresponding to the outcome of the ultraviolet integration; in particular, we rewrite the fermionic field as the sum of two {\it quasi-particle} fields, supported on disjoint sets centered around the two singularities $\pp_{F}^{\o}$. Then, we will discuss the single scale integration, and finally we will show that the outcome of the integration can be rewritten using a ``relativistic'' notation, that will allow us to show an emergent {\it Lorentz invariance}.
\paragraph{The zero-th scale.} As a starting point, we rewrite the propagators as
\be
\hat{g}^{(i.r.)}(\kk' + \pp_{F}^{\o}) =: \hat g^{(\leq 0)}_{\o}(\kk')\;,\qquad \hat{w}^{(i.r.)}(\pp) =: \hat{w}^{(\leq 0)}(\pp)\;,\label{3.01}
\ee
where
\bea
\hat g^{(\leq 0)}_{\o}(\kk') &:=& \chi_0(\kk')\begin{pmatrix} -\iu k_0 & -t\O^{*}(\vec k' + \vec p_{F}^{\o}) \\ - \O(\vec k' + \vec p_{F}^{\o}) & -\iu k_0 \end{pmatrix}^{-1}\;,\label{3.02}\\
&=& \frac{\chi_0(\kk')}{Z_0}\frac{1}{k_0 \G^0_{\o} + v_0 \vec k'\cdot \vec \G_{\o}}(1 + R_{0,\o}(\kk'))\;,\nn\\
\hat w^{(\leq 0)}(\pp) &:=& \frac{\chi_0(\pp)}{2|\pp|}(1 + R'(\pp))\;,\nn
\eea
with $\chi_0(\kk') := \chi(|\kk'|)$, $\| R_{0,\o}(\kk') \| \leq (\const.)|\kk'|$, $|R'(\pp)|\leq (\const.)|\pp|$, $v_0 = \frac{3}{2}t$, $Z_0=1$ and the matrices $\G^{\m}_{\o}$ have been defined in (\ref{1.2.52b}). Correspondingly, we define
\be
\hat\Psi^{(i.r.)\pm}_{\kk' + \pp_{F}^{\o},\s,\r} =: \hat\Psi^{(\leq 0),\pm}_{\kk',\s,\r,\o}\;,\qquad \hat A^{(i.r.)}_{\pp,\m} =: \hat A^{(\leq 0)}_{\pp,\m}\;;\label{3.03}
\ee
moreover, the following notations will be useful in the following:
\be
\hat\Psi^{\pm,T}_{\kk',\s,\o} := \begin{pmatrix} \hat\Psi^{\pm}_{\kk',\s,1,\o} & \hat\Psi^{\pm}_{\kk',\s,2,\o} \end{pmatrix}\;,\quad \int \frac{d\kk'}{D} := \frac{1}{\b|\L|}\sum_{\kk'}\;.\label{3.03b}
\ee
Then, we rewrite (\ref{1.2.42}) as, in the limit $K\rightarrow+\infty$, $h^{*}\rightarrow-\infty$,
\be
\Xi_{\b,L} = e^{-\b|\L| F_{0}}\int P_{\leq 0}(d\Psi^{(\leq 0)})P_{\leq 0}(dA^{(\leq 0)})e^{\VV^{(0)}(\Psi^{(\leq 0)},A^{(\leq 0)})}\;,\label{3.05}
\ee
where: (i) $\VV^{(0)}(\Psi^{(\leq 0)},A^{(\leq 0)})$ is equal to $\VV(\Psi^{(i.r.)},A^{(i.r.)})$ once that $\Psi^{(i.r.)}$ and $A^{(i.r.)}$ have been rewritten according to (\ref{3.03}) that is:
\bea
\VV^{(0)}(\Psi,A) &=& \sum_{\substack{n,m \geq 0\\ n+m\geq
1}}\sum_{\substack{\ul\s, \ul\r \\ \ul\m, \ul\o}}\int \prod_{i=1}^{n}\hat \Psi^{+}_{\kk'_{2i-1},\s_{i},
\r_{2i-1},\o_{2i-1}}\hat \Psi^{-}_{\kk'_{2i},\s_{i},\r_{2i},\o_{2i}}\prod_{i=1}^{m}\hat A_{\m_i, \pp_i}\cdot \nn\\&&
\cdot W^{(0)}_{2n,m,{\ul \r},{\ul \o},\ul\m}(\{\kk'_i\},\{\pp_j\})\d\left(\sum_{j=1}^{m}\pp_j + \sum_{i=1}^{2n}(-1)^{i}\kk'_i\right)\;,\label{3.06}
\eea
where $\ul\o = (\o_1,\ldots,\o_{2n})$, $\ul\s = (\s_1,\ldots,\s_n)$, $\ul\m = (\m_{1},\ldots, \m_{m})$, and the kernels are defined as, see (\ref{1.2.44}),
\be
W^{(0)}_{2n,m,{\ul \r}, {\ul \o},\ul\m}(\{\kk'_i\},\{\pp_j\}) := W_{2n,m,{\ul \r},\ul\m}(\{\kk'_i + \pp_{F}^{\o_i}\},\{\pp_j\})\;;
\ee
(ii) the integration measures are:
\bea
P_{\leq 0}(d\Psi^{(\leq 0)}) &:=& \NN_{0,\Psi}^{-1}\Big[ \prod_{\kk'\in \DD^{\o,*}_{\b,L}}\prod_{\s,\o,\r} d\hat\Psi^{(\leq 0)+}_{\kk',\s,\r,\o}d\hat\Psi^{(\leq 0)-}_{\kk',\s,\r,\o} \Big]\cdot\nn\\&&
\quad \cdot \exp\Big\{ -(\b|\L|)^{-1}\sum_{\o,\s}\sum_{\kk'\in \DD^{\o,*}_{\b,L}} \hat\Psi^{(\leq 0)+,T}_{\kk',\s,\o} \hat g^{(\leq 0)}_{\o}(\kk)^{-1} \hat\Psi^{(\leq 0)-}_{\kk',\s,\o}  \Big\}\;,\nn\\
P_{\leq 0}(dA^{(\leq 0)}) &:=& \NN_{0,A}^{-1}\left[\prod_{ \pp\in {\cal P}^{*,+}_{\b,L}}\prod_{\m=0,1,2} d\hat A^{(\leq 0)}_{\m,\pp} d\hat A^{(\leq 0)}_{\m,-\pp}\right]\cdot\nn\\&&\cdot
\exp\Big\{ -(2\b \AAA_{\L})^{-1}\sum_{\pp\in {\cal P}^*_{\b,L}} \hat A^{(\leq 0)}_{\cdot,\pp}\big[\hat w^{\xi}(\pp)\big]^{-1} \hat A^{(\leq 0)}_{\cdot, -\pp} \Big\}\;,\label{3.07}
\eea
where $\NN_{0,\Psi}$, $\NN_{0,A}$ are normalization factors.
\paragraph{Single scale integration and renormalization.} Setting $\chi_{h}(\kk'):=
\chi(M^{-h}|\kk'|)$, we start from the following identity:
\be \chi(|\kk'|)=\sum_{h=-\io}^0 f_h(\kk')\;,\quad\quad
f_h(\kk'):=\chi_h(\kk')-\chi_{h-1}(\kk')\;;\lb{3.1} \ee
let $\hat \Psi_{\kk',\s,\r,\o} = \sum_{h=-\io}^{0}\hat\Psi^{(h)}_{\kk',\s,\r,\o}$ and $\hat A_{\m,\pp} = \sum_{h=-\io}^{0}\hat A^{(h)}_{\m,\pp}$,
where $\{\hat\Psi^{(h)}\}$, $\{\hat A^{(h)}\}$ are independent
free fields with the same support of the functions $f_{h}$ introduced above. We evaluate the functional integral (\ref{1.2.42}) by integrating the fields in
an iterative way starting from $\hat\Psi^{(0)}$, $\hat A^{(0)}$. We want to inductively prove that after the
integration of $\hat\Psi^{(0)},\hat A^{(0)},\ldots,\hat\Psi^{(h+1)},\hat A^{(h+1)}$
we can rewrite:
\be \Xi_{\b,L} = e^{-\b|\L| F_{h}}\int
P_{\leq h}(d\Psi^{(\le h)})P_{\leq h}(dA^{(\le h)}) e^{\VV^{(h)}(\sqrt{Z_{h}}\Psi^{(\le h)},A^{(\le
h)})}\;,\lb{3.3}\ee
where: (i) $P_{\leq h}(d\Psi^{(\le h)})$ and $P_{\leq h}(d A^{(\le h)})$ have propagators
\bea \hat g_{\o}^{(\le h)}(\kk') &=& \frac{\chi_{h}(\kk')} {\tilde
Z_h(\kk')}\frac{1}{k_0 \G^0_{\o} + \tilde v_h(\kk') \vec k'\cdot \vec \G_{\o}}
\big(1 + R_{h,\o}(\kk')\big)\;,\nn\\
\hat w^{(\le h)}(\pp) &=& I\frac{\chi_{h}(\pp)}{2|\pp|}\big(1 + R'(\pp)\big)\;,\label{3.4}
\eea
with $\big\|R_{h,\o}(\kk')\big\|\leq C|\kk'|$; 
(ii) $\VV^{(h)}$ has the form
\bea \VV^{(h)}(\Psi,A) &=& \sum_{\substack{n,m \geq 0\\ n+m\geq
1}}\sum_{\substack{\ul\s, \ul\r \\ \ul\m, \ul\o}}\int \prod_{i=1}^{n}\hat\Psi^{+}_{\kk'_{2i-1},\s_{i},
\r_{2i-1},\o_{2i-1}}\hat\Psi^{-}_{\kk'_{2i},\s_{i},\r_{2i},\o_{2i}}\prod_{i=1}^{m}\hat A_{\m_i, \pp_i}\cdot\nn\\&&\cdot
W^{(h)}_{2n,m,{\ul \r},\ul\o,\ul\m}(\{\kk'_i\},\{\pp_j\})\d\left(\sum_{j=1}^{m}\pp_j
+ \sum_{i=1}^{2n}(-1)^{i}\kk'_i\right)\;,\label{3.5} \eea
and $F_{h}$, $\tilde Z_{h}(\kk')$, $\tilde v_{h}(\kk')$ and the
kernels $W^{(h)}_{2n,m,{\ul \r},\ul\m,\ul\o}$ will be defined
recursively. Formulas (\ref{3.3}) -- (\ref{3.5}) are trivially true 
for $h=0$, see (\ref{3.01}), (\ref{3.02}) and (\ref{3.06}).

In order to inductively prove (\ref{3.3}), we split $\VV^{(h)}$ as
$\LL \VV^{(h)} + \RR \VV^{(h)}$, where $\RR =
1-\LL$ and $\LL$, the {\em localization operator}, is a linear
operator on functions of the form (\ref{3.5}), defined by its action on the
kernels $
W^{(h)}_{2n,m,{\ul r},\ul\o,\ul\m}$ in the following way:
\bea &&\mathcal{L}W^{(h)}_{2,0,{\ul \r},\ul\o}(\kk') :=
W^{(h)}_{2,0,{\ul \r},\ul\o}({\bf 0}) +
k'_\a \partial_{\a}W^{(h)}_{2,0,{\ul \r},\ul\o}({\bf 0})\;,\label{3.6}\\
&&\mathcal{L}W_{2,1,{\ul \r},\ul\o,\mu}^{(h)}(\pp,\kk')
:= W_{2,1,{\ul \r},\ul\o,\mu}^{(h)}({\bf 0},{\bf 0})\;,\nn\\
&&\mathcal{L}W_{0,2,\ul\mu}^{(h)}(\pp) :=
W_{0,2,\ul\mu}^{(h)}({\bf 0}) + p_{\a}\partial_{\a}
W_{0,2,\ul\mu}^{(h)}({\bf 0})\;,\quad \mathcal{L}
W_{0,3,\ul\mu}^{(h)}(\pp_1,\pp_2) := W_{0,3,\ul\mu}^{(h)}({\bf
0},\V0)\;,\nn\eea
and $\mathcal{L}W_{P}^{(h)} := 0$ otherwise. By the
 symmetries listed in Lemma \ref{lem2.4}, which are preserved by the multiscale analysis, it turns out that, see Appendix \ref{app2c}:
\bea &&W^{(h)}_{0,1,\m}(\V0) = 0\;,\quad W_{0,3,\ul
\mu}^{(h)}({\bf 0},{\bf 0}) =0\;,\quad
W^{(h)}_{2,0,{\ul \r},\ul\o}({\bf 0})= 0\;,\nn\\
&& W^{(h)}_{0,2,\ul \mu}(\V0) =
- \delta_{\mu_{1},\mu_{2}} M^{h}\n_{\m_1,h}\;,\quad \partial_{\a}\hat
W_{0,2,\ul\mu}^{(h)}({\bf 0})=0\;\qquad\label{3.7} \eea
and, moreover, that
\be
\partial_{\m}W^{(h)}_{2,0,\ul\r,\ul\o}(\V0) = -\d_{\o_1,\o_2} z_{\m,h} \big[\G_{\o_1}^{\m}\big]_{\r_1,\r_2}\;,
\quad W^{(h)}_{2,1,\ul\r,\ul\o,\m}(\V0,\V0) = \d_{\o_1,\o_2}\l_{\m,h}\big[\G_{\o_1}^{\m}\big]_{\r_1,\r_2}\;,\label{3.8} 
\ee
with $z_{\m,h}$, $\l_{\m,h}$, $\n_{\m,h}$ real, and $z_{1,h} = z_{2,h}$, $\l_{1,h} =
\l_{2,h}$, $\n_{1,h} = \n_{2,h}$. We can {\em renormalize} $P_{\leq h}(d\psi^{(\le h)})$ by adding to the
exponent of its gaussian weight
the local part of the quadratic terms in the fermionic fields; we get that
\bea &&\int P_{\leq h}(d\Psi^{(\le h)})P_{\leq h}(d A^{(\le h)})
e^{\VV^{(h)}(\sqrt{Z_{h}}\Psi,A)} =\label{3.9}\\&&\hskip3cm= e^{-\b|\L| t_h} \int
\widetilde P_{\leq h}(d\Psi^{(\le h)}) P_{\leq h}(d A^{(\le h)})
\eu^{\widetilde \VV ^{(h)}(\sqrt{Z_{h}}\Psi,A)}\;,\nn \eea
where $t_h$ takes into account the different normalization of the
two functional integrals, $\widetilde \VV^{(h)}$ is given by
\bea
\widetilde\VV^{(h)}(\Psi,A) &=& \VV^{(h)}(\Psi,A) + \int \frac{d\kk'}{D} \sum_{\s,\o} z_{\m,h} 
k'_\m \hat\Psi^{+,T}_{\kk',\s,\o} \G^{\m}_{\o} \hat\Psi^{-}_{\kk',\s,\o}\nn\\&=:&
\VV^{(h)}(\Psi,A) - \LL_\Psi\VV^{(h)}(\Psi,A)\;,\label{3.10}\eea
and $\widetilde P_{\leq h}(d\psi^{(\le h)})$ has propagator equal to
\be \tilde g^{(\le h)}_{\o}(\kk') = \frac{\chi_{h}(\kk')} {\tilde
Z_{h-1}(\kk')}\frac{1}{k_0 \G^0_{\o} + \tilde v_{h-1}(\kk') \vec k'\cdot \vec \G_{\o}}
\big(1 + R_{h-1,\o}(\kk')\big)\;,\label{3.11} \ee
with
\bea \tilde Z_{h-1}(\kk') &=& \tilde Z_{h}(\kk') +
Z_{h}z_{0,h}\chi_{h}(\kk')\;,\label{3.12}\\\tilde Z_{h-1}(\kk')\tilde
v_{h-1}(\kk') &=& \tilde Z_{h}(\kk')\tilde v_{h}(\kk') +
Z_{h}z_{1,h}\chi_{h}(\kk')\;\nn \eea
and, setting $(1 + R_{h-1,\o}(\kk'))^{-1} =: 1 + T_{h-1,\o}(\kk')$:
\be
T_{h-1,\o}(\kk') = T_{h,\o}(\kk')\frac{\tilde Z_{h}(\kk')}{\tilde Z_{h-1}(\kk')}
\frac{k_{0}\G^{0}_{\o} + \tilde v_{h}(\kk')\vec k'\cdot \vec\G_{\o}}{k_{0}\G^{0}_{\o} + \tilde v_{h-1}(\kk')\vec k'\cdot \vec\G_{\o}}\;.\label{3.12b}
\ee
After this, defining $Z_{h-1} := \tilde Z_{h-1}(\V0)$,
we {\em rescale} the fermionic field so that
\be \widetilde \VV^{(h)}(\sqrt{Z_h}\Psi,A) = \hat
\VV^{(h)}(\sqrt{Z_{h-1}}\Psi,A)\;;\label{3.13} \ee
therefore, setting
\be
v_{h-1}:= \tilde v_{h-1}(\V0)\;,\quad e_{0,h}:=
\frac{Z_{h}}{Z_{h-1}}\l_{0,h}\;,\quad e_{1,h}v_{h-1}=e_{2,h}v_{h-1}:=
\frac{Z_{h}}{Z_{h-1}}\l_{1,h}\;,\label{3.13b}
\ee
we have that:
\be \LL \hat \VV^{(h)}(\sqrt{Z_{h-1}}\,\Psi^{(\le h)},A^{(\le h)}) = \int \frac{d\pp}{(2\pi)^3}\,
Z_{h-1} e_{\m,h} j^{(\leq h)}_{\m,\pp}\hat A^{(\leq h)}_{\m,\pp} - M^{h}\n_{\m,h}\hat A^{(\leq h)}_{\m,\pp}\hat A^{(\leq h)}_{\m,-\pp}
\label{3.14}\ee
where
\bea
j^{(\leq h)}_{0,\pp} &:=& \sum_{\s,\o}\int \frac{d\kk'}{D}\, \hat\Psi^{(\leq h)+,T}_{\kk'+\pp,\s,\o}\G^{0}_{\o}\hat\Psi^{(\leq h)-}_{\kk',\s,\o}\;,\nn\\ 
\vec j^{(\leq h)}_{\pp} &:=& v_{h-1}\sum_{\s,\o}\int \frac{d\kk'}{D}\, \hat\Psi^{(\leq h)+,T}_{\kk'+\pp,\s,\o}\vec \G_{\o}\hat\Psi^{(\leq h)-}_{\kk',\s,\o}\;.\label{3.15b}
\eea
After this rescaling, we can rewrite (\ref{3.9}) as
\bea && \int P_{\leq h}(d\Psi^{(\le h)} d A^{(\le h)})
e^{\VV^{(h)}(A,\sqrt{Z_{h}}\Psi)} = e^{\b|\L| t_h}
\int P_{\leq h-1}(d\Psi^{(\le h-1)} d A^{(\le h-1)})\cdot
\nn\\
&&\hskip1cm\cdot
\int P_{h}(d\Psi^{(h)} d A^{(h)})
e^{\hat \VV ^{(h)}(A^{(\le h-1)}+A^{(h)}
,\sqrt{Z_{h-1}}(\Psi^{(\le h-1)}+\Psi^{(h)}))}\;,\label{3.14a}
\eea
where $\Psi^{(\le h-1)},A^{(\le h-1)}$ have propagators given by (\ref{3.4})
(with $h$ replaced by $h-1$) and $\Psi^{(h)},A^{(h)}$
have propagators given by
\bea &&\frac{\hat{g}_{\o}^{(h)}(\kk')}{Z_{h-1}} = \frac{\tilde f_h(\kk')}{Z_{h-1}}
\frac{\big(1 + R_{h-1,\o}(\kk')\big)}{k_0 \G^{0}_{\o} + 
\tilde v_{h-1}(\kk')\vec k'\cdot\vec\G_{\o}}\;,\quad \tilde f_h(\kk') =
\frac{Z_{h-1}}{\tilde Z_{h-1}(\kk')}f_h(\kk')\nn\\
&&\hat{w}^{(h)}(\pp) =
\frac{f_h(\pp)}{2|\pp|}\big( 1 + R'(\pp)\big)\;.\label{3.16} \eea
At this point, we can integrate the scale $h$ and, defining
\be e^{\VV^{(h-1)}(A,\sqrt{Z_{h-1}}\Psi) - \b|\L| \tilde
F_{h}} := \int P_{h}(d\Psi^{(h)}) P_{h}(dA^{(h)})e^{\hat \VV ^{(h)}
(A +
A^{(h)} ,\sqrt{Z_{h-1}}(\Psi + \Psi^{(h)}))}\;,\label{3.17} \ee
our inductive assumption (\ref{3.3}) is reproduced at the scale
$h-1$ with $F_{h-1} := F_{h} + t_{h} + \tilde F_{h}$. Notice that
(\ref{3.17}) can be seen as a recursion relation for the effective 
potential, since from (\ref{3.10}), (\ref{3.13}) it follows that
\be
\hat \VV^{(h)}(A,\sqrt{Z_{h-1}}\psi) = 
\widetilde \VV^{(h)}(A,\sqrt{Z_h}\psi) = \VV^{(h)}(A,\sqrt{Z_h}\psi) - 
\LL_{\psi}\VV^{(h)}(A,\sqrt{Z_h}\psi)\;.\label{3.17b}
\ee
The integration in (\ref{3.17}) is performed by expanding in series
the exponential in the r.h.s. (which involves interactions of any
order in $\Psi$ and $A$, as apparent from (\ref{3.5})), and
integrating term by term with respect to the gaussian integration
$P(d\Psi^{(h)}) P(dA^{(h)})$. This procedure gives rise to an
expansion for the effective potentials $\VV^{(h)}$ (and to an
analogous expansion for the correlations) in terms of the
renormalized parameters
$\{e_{\m,k},\n_{\m,k-1},Z_{k-1},v_{k-1}\}_{h< k\le 1}$, where $e_{1} = e$, which can be
conveniently represented as a sum over {\em Feynman graphs}
according to rules that will be explained below. We will call
$\{e_{\m,k},\n_{\m,k}\}_{h< k\le 0}$ {\it effective couplings} or
{\it running coupling constants} while $\{e_{\m,k}\}_{h< k\le 0}$
are the {\it effective charges}

Note that such {\it renormalized} expansion is significantly
different from the power series expansion in the bare coupling
$e$; while the latter is plagued by {\it logarithmic
divergences}, the former is {\it order by order finite}.

By comparing (\ref{3.05}) with (\ref{3.3}),
(\ref{3.5}) and (\ref{3.14}), we see that the integration of the
fields living on momentum scales $\ge M^h$ produces an {\it
effective theory} very similar to the original one, modulo the
presence of a new propagator, involving a renormalized velocity
$v_h$ and a renormalized wave function $Z_h$, and the presence of
a modified interaction $\VV^{(h)}$.
\paragraph{Relativistic notations.} We conclude this Section by showing that the 
effective theory on a given scale $h$ can be rewritten as a ``relativistic'' theory, where
the fermions propagate with a velocity $v_{h-1}$. Let us define the 4- component Grassmann 
fields $\psi^{(\leq h)}_{\kk,\s}$, $\lis\psi^{(\leq h)}_{\kk,\s}$ as follows:
\bea
\psi^{(\leq h),T}_{\kk',\s} &:=&  \begin{pmatrix} \hat\Psi^{(\leq h)-}_{\kk',\s,1,+} & \hat\Psi^{(\leq h)-}_{\kk',\s,2,+} & \hat\Psi^{(\leq h)-}_{\kk',\s,2,-} & \hat\Psi^{(\leq h)-}_{\kk',\s,1,-} \end{pmatrix}\;,\\
\lis\psi^{(\leq h),T}_{\kk',\s} &:=& \begin{pmatrix} \hat\Psi^{(\leq h)+}_{\kk',\s,2,-} & \hat\Psi^{(\leq h)+}_{\kk',\s,1,-} & - \hat\Psi^{(\leq h)+}_{\kk',\s,1,+} & - \hat\Psi^{(\leq h)+}_{\kk',\s,2,+} \end{pmatrix}\;,\label{rel1}
\eea
and the {\it Euclidean gamma matrices} $\g_{\m}$, $\m=0,1,2$ as:
\be
\g_{0} := \begin{pmatrix} 0 & \iu \G_{0}^{-} \\ -\iu \G^{0}_{+} & 0\end{pmatrix}\;,\quad \g_{1} := \begin{pmatrix} 0 & -\iu \G_{1}^{-} \\ -\iu \G^{1}_{+} & 0 \end{pmatrix}\;,\quad \g_{2} := \begin{pmatrix} 0 & \iu \G_{2}^{-} \\ -\iu \G_{2}^{+} & 0 \end{pmatrix}\;,\label{1.2.57}
\ee
it is easy to see that $\{\g_{\m},\g_{\n}\} = -2\d_{\m,\n}$. With these notations, the fermionic integration measure can be rewritten as:
\be
P_{\leq 0}(d\Psi^{(\leq h)}) = \NN_{h,\psi}^{-1}\prod_{\kk',\s,j}d\lis\psi_{\kk',\s,j}d\psi_{\kk',\s,j}\exp\Big\{-\sum_{\s}\int\frac{d\kk'}{D}\, \lis\psi_{\kk',\s}^{(\leq h)} \hat g^{(\leq h)}(\kk')^{-1}\psi_{\kk',\s}^{(\leq h)}\Big\}\label{1.2.56b}
\ee
where $\NN_{h,\psi}$ is a normalization factor, and the fermion propagator $\hat g^{(\leq h)}(\kk')$ is given by
\be
\hat g^{(\leq h)}(\kk') := \frac{\chi_{h}(\kk')}{\tilde Z_{h}(\kk')}\frac{1}{\iu k_{0}\g_0 + \tilde v_{h}(\kk')\vec k'\cdot\vec\g}(1 + R_{h}(\kk'))\;,\label{rel2}
\ee
with $\|R_{h}(\kk')\|\leq (\const.)|\kk'|$; analogously, we can rewrite the fermionic current defined in (\ref{3.15b}) as:
\be
j^{(\leq h)}_{0,\pp} := \sum_{\s}\int \frac{d\kk'}{D}\, \iu \lis\psi_{\kk' + \pp,\s}^{(\leq h)}\g_0\psi_{\kk',\s}^{(\leq h)}\;,\quad 
\vec j^{(\leq h)}_{\pp} := v_{h-1}\sum_{\s}\int \frac{d\kk'}{D}\, \iu \lis\psi_{\kk' + \pp,\s}^{(\leq h)}\vec \g\psi_{\kk',\s}^{(\leq h)}\;.\label{rel3}
\ee
In a fully relativistic theory, that is with bare propagators given by (\ref{rel2}), (\ref{3.16}) with $h=0$ and $\tilde Z_0 = \tilde v_0 = 1$, $R_{0}(\pp) = R'(\pp)=0$, and bare interaction given by (\ref{3.15b}) with $h=0$ and $e_{\m,0} = e$, $\n_{\m,0}=\n$, Lorentz symmetry implies that $v_{h}=1$ ({\it the speed of light is not renormalized}) $e_{\m,h} = e_{0,h}$, $\n_{\m,h} = \n_{0,h}$; see \cite{GMP1}. On the contrary, the lack of Lorentz symmetry in our model, due to the presence of the lattice, has two main effects: (i) the Fermi velocity $v_{h}$ has a non trivial flow; (ii) the marginal terms in the effective potential are defined in terms of {\it two} charges, namely $e_{0,h}$ and $e_{1,h}=e_{2,h}$, which are {\it different}, in general.

\section{Tree expansion}\label{sec4}
\setcounter{equation}{0}
\renewcommand{\theequation}{\ref{sec4}.\arabic{equation}}

\begin{figure}[htbp]
\centering
\includegraphics[width=0.95\textwidth]{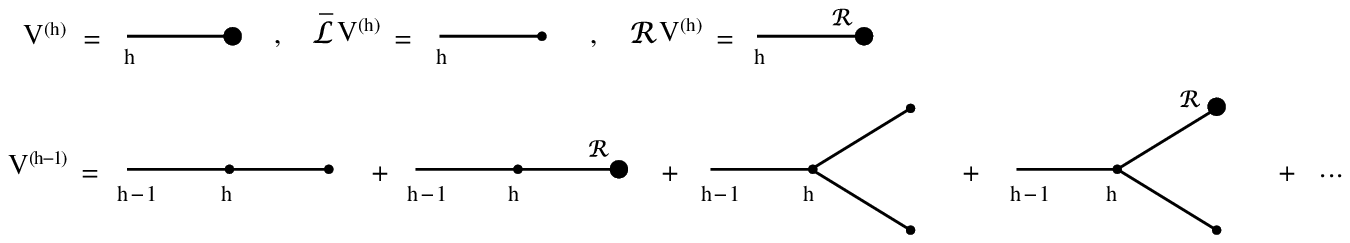} \caption{Graphical
interpretation of Eq.(\protect\ref{4.1}). The graphical equations
for $\lis\LL \VV^{(h-1)}$, $\RR \VV^{(h-1)}$ are obtained from the
equation in the second line by putting an $\lis\LL$, $\RR$ label,
respectively, over the vertices on scale $h$.} \label{fig4.2}
\end{figure}

\begin{figure}[htbp]
\centering
\includegraphics{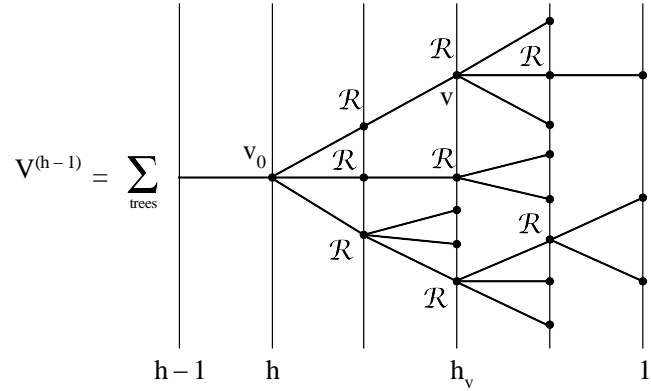} \caption{The effective potential $\VV^{(h-1)}$
can be represented as a sum over {\em Gallavotti -- Nicol\`o}
trees. The black dots will be called {\it vertices} of the tree.
All the vertices which are not endpoints except the first ({\it i.e.} the one on scale
$h$) have an $\RR$ label attached, which
 means that they correspond to the action of $\RR\EE^{T}_{h_v}$, while the
first represents $\EE^{T}_{h}$. The endpoints on scales $<1$ correspond to the first two
graph elements in Fig.\protect\ref{fig.3b}, which are associated to the two
terms in $\LL \hat \VV^{(h)}$, see (\protect\ref{3.14}); instead, the endpoints on 
scale $1$ correspond to $\LL\hat\VV^{(0)} + \RR \VV^{(0)}$.} \label{fig:3a}
\end{figure}

The iterative integration procedure described above leads to a representation of 
the effective potentials in terms of a sum over connected Feynman diagrams, 
as explained in the following.
The key formula, which we start from, is (\ref{3.17}), which can
be rewritten as
\bea &&-\b|\L|\tilde F_{h} + \VV^{(h-1)}(\sqrt{Z_{h-1}}\, \Psi^{(\le h-1)},A^{(\le
h-1)})=\nn\\&&\quad= \sum_{n\geq
1}\frac{1}{n!}\EE_{h}^{T}\left(\hat \VV^{(h)} (\sqrt{Z_{h-1}}\Psi^{(\le h)},A^{(\le
h)});n\right)\;,\label{4.1}\qquad\eea
with $\EE_h^T$ the truncated expectation on scale $h$, defined as
\be \EE_{h}^{T}(X(A^{(h)},\Psi^{(h)});n) :=
\frac{\partial^{n}}{\partial \lambda^{n}}\log \int P_h(d\Psi^{(h)})
P_h(dA^{(h)}) e^{\lambda
X(A^{(h)},\Psi^{(h)})}\Big|_{\lambda=0}\;\label{4.2}\ee
If $X$ is graphically represented as a vertex with external lines corresponding to
$A^{(h)}$ and $\Psi^{(h)}$, the truncated expectation (\ref{4.2})
can be represented as the sum over the Feynman diagrams obtained
by contracting in all possible connected ways the lines exiting
from $n$ vertices of type $X$. Every contraction corresponds to a
propagator on scale $h$, as defined in (\ref{3.16}). Since
$\hat{\VV}^{(h)}$ is related to $\VV^{(h)}$ by a rescaling and a
subtraction, see (\ref{3.10}) and (\ref{3.13}), Eq.(\ref{4.1}) can
be iterated until scale $0$, and $\VV^{(h-1)}$ can be written as a
sum over connected Feynman diagrams with lines on all possible
scales between $h$ and $0$. The iteration of (\ref{4.1}) induces a
natural hierarchical organization of the scale labels of every
Feynman diagram, which will be conveniently represented in terms
of tree diagrams. In fact, let us rewrite $\hat \VV^{(h)}$ in the
r.h.s. of (\ref{4.1}) as $\hat
\VV^{(h)}(\sqrt{Z_{h-1}}\Psi,A)=\lis\LL\VV^{(h)}
(\sqrt{Z_{h}}\Psi,A)+\RR\VV^{(h)}(\sqrt{Z_{h}}\Psi,A)$, where
$\lis\LL:=\LL-\LL_{\Psi}$, see (\ref{3.10}). Let us graphically
represent $\VV^{(h)}$, $\lis\LL\VV^{(h)}$ and $\RR \VV^{(h)}$ as
in the first line of Fig. \ref{fig4.2}, and let us represent
Eq.(\ref{4.1}) as in the second line of Fig. \ref{fig4.2}; in the
second line, the node on scale $h$ represents the action of
$\EE^T_h$.
Iterating the graphical equation in Fig. \ref{fig4.2} up to scale
0, we end up with a representation of $\VV^{(h)}$ in terms of a
sum over {\it Gallavotti-Nicol\`o} trees $\t$ \cite{G84,BG,GeM}:
\be \VV^{(h)}(\sqrt{Z_h}\,\Psi^{(\le h)},A^{(\le h)}) =
\sum_{N\ge
1}\sum_{\tau\in\mathcal{T}_{h,N}}\VV^{(h)}(\t)\;,\label{4.1a}\ee
where $\TT_{h,N}$ is the of rooted trees with {\em root} $r$ on
scale $h_{r}=h$ and $N$ endpoints, see Fig. \ref{fig:3a}.
The tree value $\VV^{(h)}(\t)$ can be evaluated in terms of a sum
over connected Feynman diagrams, defined by the following rules.

\begin{figure}[htbp]
\centering
\includegraphics[width=.6\textwidth]{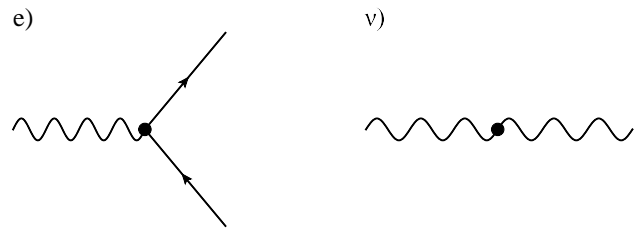}
\caption{The graph elements corresponding to the two terms in the r.h.s.
of (\protect\ref{3.14}).} \label{fig.3b}
\end{figure}

With each endpoint $v$ of $\t$ on scale $<1$ we associate a graph element of
type $e$ or $\n$, corresponding to the two terms in the r.h.s. of
(\ref{3.14}), see Fig. \ref{fig.3b}; if $v$ is on scale $1$ we also have to consider the graphs 
arising in the integration of the ultraviolet degrees of freedom, contributing to the kernels of $\RR \VV^{(0)}$, see Appendix \ref{app2b}.

%

%
%

We introduce a {\it field label} $f$ to distinguish the fields
associated to the graph elements (any field label can
be either of type $A$ or of type $\Psi$); the set of field labels
associated with the endpoint $v$ will be called $I_v$.
Analogously, if $v$ is not an endpoint, we call $I_v$ the set of
field labels associated with the endpoints following the vertex
$v$ on $\t$.

\begin{figure}[hbtp]
\centering
\includegraphics[width=.8\textwidth]{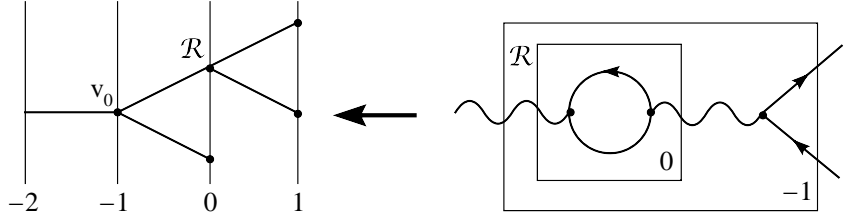}
\caption{A possible Feynman diagram contributing to $V^{(-2)}$ and
its cluster structure.}\label{fig:1e}
\end{figure}

We start by looking at the graph elements corresponding to
endpoints on scale $1$: we group them in {\it clusters}, each
cluster $G_v$ being the set of endpoints attached to the same
vertex $v$ on scale 0, to be graphically represented by a box
enclosing its elements. For any $G_v$ on scale 0 (associated to a
vertex $v$ on scale $0$ that is not an endpoint), we contract in
pairs some of the fields in $\cup_{w\in G_v}I_w$, in such a way
that after the contraction the elements of $G_v$ are connected;
each contraction produces a propagator $\hat g_{\o}^{(0)}$ or $\hat w^{(0)}$,
depending on whether the two fields are of type $\Psi$ or of type
$A$. We denote by $\III_v$ the set of contracted fields inside the
box $G_v$ and by $P_v=I_v\setminus \III_v$ the set of external
fields of $G_v$; if $v$ is not the vertex immediatly following the
root we attach a label $\RR$ over the box $G_v$, which means that
the $\RR$ operator, defined after (\ref{3.5}), acts on the value
of the graph contained in $G_v$. Next, we group together the
scale-0 clusters into scale-(-1) clusters, each scale-(-1) cluster
$G_v$ being a set on scale-0 clusters attached to the same vertex
$v$ on scale $-1$, to be graphically represented by a box
enclosing its elements, see Fig. \ref{fig:1e}.

Again, for each $v$ on scale $-1$ that is not an endpoint, if we
denote by $v_1,\ldots,v_{s_v}$ the vertices immediately following
$v$ on $\t$, we contract some of the fields of
$\cup_{i=1}^{s_v}P_{v_i}$ in pairs, in such a way that after the
contraction the boxes associated to the scale-0 clusters contained
in $G_v$ are connected; each contraction produces a propagator
$g^{(-1)}$ or $w^{(-1)}$. We denote by $\III_v$ the set of fields
in $\cup_{i=1}^{s_v}P_{v_i}$ contracted at this second step and by
$P_v=\cup_{i=1}^{s_v}P_{v_i}\setminus\III_v$ the set of fields
external to $G_v$; if $v$ is not the vertex immediatly following
the root we attach a label $\RR$ over the box $G_v$.

Now, we iterate the construction, producing a sequence of boxes
into boxes, hierarchically arranged with the same partial ordering
as the tree $\t$. Each box $G_v$ is associated to many different
Feynman (sub-)diagrams, constructed by contracting in pairs some
of the lines external to $G_{v_i}$, with $v_i$, $i=1,\ldots,s_v$,
the vertices immediately following $v$ on $\t$; the contractions
are made in such a way that the clusters
$G_{v_1},\ldots,G_{v_{s_v}}$ are connected through propagators of
scale $h_v$. We denote by $P_v^A$ and by $P_v^\Psi$ the set of
fields of type $A$ and $\Psi$, respectively, external to $G_v$.
The set of connected Feynman diagrams compatible with this
hierarchical cluster structure will be denoted by $\G(\t)$. Given
these definitions, we can write:
\bea &&\VV^{(h)}(\t)=\sum_{\GG\in\G(\t)}\int \prod_{f\in P_{v_0}^\Psi}
\frac{d\kk'_f}{D}\,\prod_{f\in P_{v_0}^A}\frac{d\pp_f}{(2\p)^3}
\Val(\GG)\;,\label{4.4}\\
&&\Val(\GG)=\Big[ \prod_{f\in P_{v_0}^A}\hat A_{\m(f),\,\pp_f}^{(\le
h)}\Big] \Big[\prod_{f\in P_{v_0}^\Psi}
\sqrt{Z_{h-1}}\,\hat \Psi_{\kk'_f,\s(f),\r(f),\o(f)}^{(\le h),\e(f)}\Big]\d(v_0)\widehat \Val(\GG)\;,\nn\\
&&\widehat \Val(\GG)=(-1)^\p\int\!\!\!\prod_{v\ {\rm not}\ {\rm e.p.}}
\Big(\frac{Z_{h_v-1}}{Z_{h_v-2}}\Big)^{\frac{|
P_v^\Psi|}2}\frac{\RR^{\a_v}}{s_{v}!} \Biggl[\Big(\prod_{\ell\in
v}g_\ell^{(h_v)}\Big) \Big(\prod_{\substack{v^*\ {\rm e.p.}\\
v^*>v,\\h_{v^*}=h_v+1}} K_{v^*}^{(h_v)}\Big)\Biggr]\nn \eea
where: $(-1)^\p$ is the sign of the permutation necessary to bring
the contracted fermionic fields next to each other; in the product
over $f\in P_v^\Psi$, $\e(f) = \pm$ depending on the specific field label $f$; 
$\d(v_0)=\d\Big(\sum_{f\in P_{v_0}^A}\pp_f-\sum_{f\in P_{v_0}^{\Psi}}
(-1)^{\e(f)}\kk'_f\Big)$; the integral in the third
line runs over the independent loop momenta; $s_v$ is the
number of vertices immediately following $v$ on $\t$; $\RR=1-\LL$
is the operator defined in (\ref{3.6}) and preceding lines);
$\a_v=0$ if $v=v_0$, and otherwise $\a_v=1$; $g_\ell^{(k)}$ is
equal to $g^{(k)}$ or to $w^{(k)}$ depending on the fermionic or
bosonic nature of the line $\ell$, and $\ell\in v$ means that
$\ell$ is contained in the box $G_v$ but not in any other smaller
box; finally, $K_{v^*}^{(k)}$ is the matrix associated to the
endpoints $v^*$ on scale $k+1$: setting 
\be
\bar e_{0,k} := e_{0,k}\,,\qquad \bar e_{j,k} := v_{k-1}e_{j,k}\,,\label{4.4b}
\ee
it is given by $\bar e_{\m,k}\G^{\m}_{\o}$ if $v^*$
is of type $e$, by $-M^k\n_{\m,k}$ if $v^*$ is
of type $\n$, or, if $v^*$ is on scale $1$ and is not of type 
$e,\n$, it is equal to one of the kernels contributing to $\RR V^{(0)}$, see Appendix \ref{app2b}. 
In (\ref{4.4}) it is understood that the operators
$\RR$ act in the order induced by the tree ordering (i.e.,
starting from the endpoints and moving toward the root); moreover,
the matrix structure of $g^{(k)}_\ell$ is neglected, for
simplicity of notations. 
\subsection{An example of Feynman graph}
\begin{figure}[hbtp]
\centering
\includegraphics[height=3.5truecm]{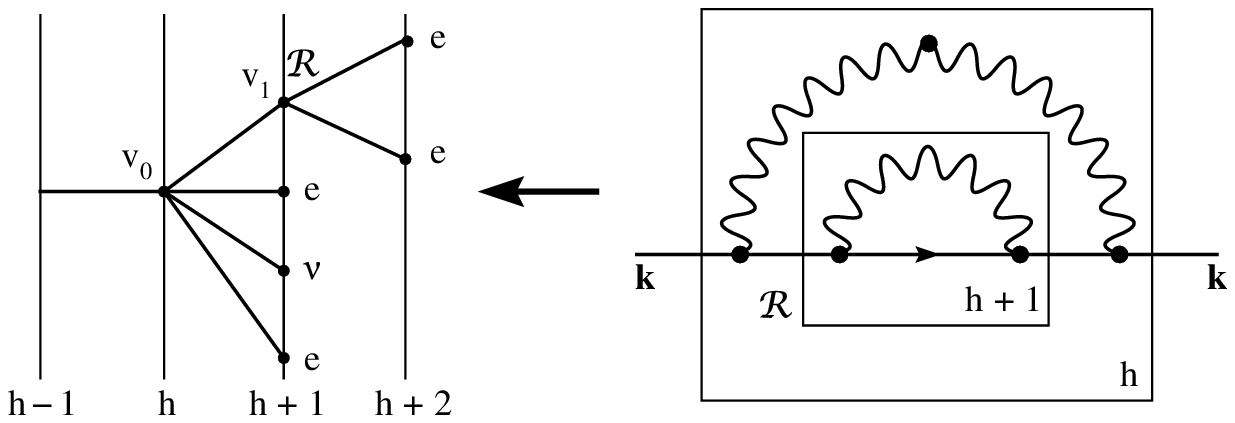}
\caption{A possible Feynman diagram contributing to $\VV^{(h-1)}$
and its cluster structure.}\label{figex}
\end{figure}

To be concrete, let us apply the rules described above in the
evaluation of a simple Feynman graph $\GG$ arising in the tree
expansion of $\VV^{(h-1)}$. Let $\GG$ be the diagram in
Fig. \ref{figex}, associated to the tree $\t$ drawn in the left
part of the figure; let us assume that the sets $P_v$ of the
external lines associated to the vertices of $\t$ are all
assigned.

We can write:
\bea &&\Val(\GG) = -\frac{1}{4!2!}\frac{Z_h}{Z_{h-1}}
\frac{Z_{h-1}}{Z_{h-2}}\bar e_{\m_1,h}^2\bar
e_{\m_2,h+1}^{2}M^{h}\n_{\m_1,h}\hat\Psi_{\kk,\s,\o}^{(\leq h-1)+}\cdot\nn\\&&\quad\cdot\Biggl\{\int
\frac{d\pp}{(2\pi)^3}\,|\hat{w}^{(h)}(\pp)|^2 \G^{\m_1}_{\o}
\hat g_{\o_1}^{(h)}(\kk+\pp)\cdot\label{4.6}\\&& \cdot\RR\Big[ \int
\frac{d\qq}{(2\pi)^3}\, \G^{\m_2}_{\o}\hat g_{\o}^{(h+1)}(\kk+\pp+\qq)\G^{\m_2}_{\o}
\hat w^{(h+1)}(\qq) \Big]
\hat g_{\o}^{(h)}(\kk+\pp)\G^{\m_1}_{\o}\Biggr\}\hat\Psi^{(\leq h)-}_{\kk,\s,\o}\;,\nn\eea
where $\RR \big[F(\kk+\pp)\big] = F(\kk+\pp)-F(\V0)-(\kk+\pp)\cdot
\nabla F(\V0)\equiv \frac12(k_\m+p_\m)(k_\n+p_\n)\dpr_\m\dpr_\n
F(\kk^*)$. Notice that the same
 Feynman graph appears in the evaluation of other trees, which are
topologically equivalent to the one represented in the left part
of Fig. \ref{figex} and that can be obtained from it by: (i)
relabeling the fields in $P_{v_1}$, $P_{v_0}$, (ii) relabeling the
endpoints of the tree, (iii) exchanging the relative positions of
the topologically different subtrees with root $v_0$. If one sums
over all these trees, the resulting value one obtains is the one
in Eq.(\ref{4.6}) times a combinatorial factor $2^2\cdot 3\cdot 4$
($2^2$ is the number of ways for choosing the fields in $P_{v_1}$
and in $P_{v_0}$; $3$ is the number of ways in which one can
associate the label $\n$ to one of the endpoints on scale $h+1$;
$4$ is the number of distinct unlabelled trees that can be
obtained by exchanging the positions of the subtrees with root
$v_0$). 
\subsection{Dimensional bounds}\label{sec4b}
We are now ready to derive
a general bound for the Feynman graphs produced by the multiscale
integration; before stating our result, we define the {\it order} of a tree as follows. Let $\widetilde\TT_{K;h,N}$ be the set of trees with $N$ endpoints 
obtained expanding into trees the effective potentials $\RR \VV^{(0)}$ attached to the endpoints of the
elements of $\bigcup_{n = 1}^{N}\TT_{h,n}$. Let us define the order of an endpoint $v$ in the following way: the order of the endpoints of type $e,\n$ is $1$; the order of an endpoint on scale $K+1$ is equal to the number of wavy lines exiting from the corresponding graph element. We define the order $O(\t)$ of a tree $\t\in \widetilde\TT_{K;h,N}$ as the sum of the orders of its endpoints. In general, trees belonging to $\widetilde\TT_{K;h,N}$ may be of arbitrary order $\geq N$, since the endpoints on scale $K+1$ correspond to vertices with an arbitrary number of wavy lines (because of the expansion of the exponential of the photon field performed before the ultraviolet integration, see Eq. (\ref{1.2.20c})).

Now, let $W^{(h);N}_{2n,m,\ul{\r},\ul\o,\ul{\m}}$ be the
contribution due to the trees of order $N$ to the kernel $W^{(h)}_{2n,m,\ul{\r},\ul\o,\ul{\m}}$ (to which in the following we shall simply refer as ``the contribution of order $N$''), that is
\bea W^{(h)}_{2n,m,\ul{\r},\ul{\o},\ul{\m}}(\{\kk'_i\},\{\pp_j\}) &=& \sum_{N= 1}^{+\io}\sum_{N'\leq N}
\sum_{\substack{\t\in \widetilde\TT_{K;h,N'} \\ O(\t) = N}}\sum^*_{\substack{\GG\in\G(\t)\\
|P^A_{v_0}| = m,\\|P^{\psi}_{v_0}| = 2n}} \widehat\Val(\GG) \nn\\&=:&\sum_{N=1}^\io W^{(h);N}_{2n,m,\ul{\r},\ul\o,\ul{\m}}(\{\kk'_i\},\{\pp_j\})\;,\label{WN}\eea
where the * on the sum indicates the constraints that: 
$\cup_{f\in P_{v_0}^A}\{\pp_f\}=\cup_{j=1}^m\{\pp_j\}$; $\cup_{f\in P_{v_0}^\Psi}\{\kk'_f\}=
\cup_{i=1}^{2n}\{\kk'_i\}$; $\cup_{f\in P_{v_0}^\Psi}\{\r(f)\}=\underline\r$;
 $\cup_{f\in P_{v_0}^\Psi}\{\o(f)\}=\underline \o$; $\cup_{f\in P_{v_0}^A}\{\m(f)\}=\underline\m$.

The $N$-th order contribution to the kernel of the effective potential 
admits the following bound. 
\begin{theorem}{\bf{($N!$ bound)}}\label{thm1} Let $\bar\e_h =
\max_{h< k\leq 1} \{|e_{\m,k}|,|\n_{\m,k-1}|\}$ be small enough. If
$Z_{k}/Z_{k-1} \le e^{C\bar \e_h^2}$ and ${C}^{-1}\le v_{k-1}\le C$,
for all $h< k\le 0$ and a suitable constant $C>0$, then
\be ||W^{(h);N}_{2n,m,\ul{\r},\ul\o,\ul{\m}}|| \le (\const.)^N\bar\e_h^N
\Big(\frac{N}2\Big)!\; M^{h(3-m-2n)}\;,\label{4.11a}\ee
where $||W^{(h);N}_{2n,m,\ul{\r},\ul\o,\ul{\m}}||:=\sup_{\{\kk'_i\},\{\pp_j\}}
|W^{(h);N}_{2n,m,\ul{\r},\ul\o,\ul{\m}}(\{\kk'_i\},\{\pp_j\})|$.
\end{theorem}
The factor $3-2n-m$ in \pref{4.11a} is referred to as
the {\it scaling dimension} of the kernel with $2n$ external
fermionic fields and $m$ external bosonic fields; according to the
usual RG therminology, kernels with positive, vanishing or
negative scaling dimensions are called {\it relevant, marginal} or
{\it irrelevant} operators, respectively. Notice that, if we tried
to expand the effective potential in terms of the bare coupling $e$, 
the $N$-th order contributions in this ``naive''
perturbation series could {\it not be bounded uniformly in the
scale $h$} as in (\ref{4.11a}), but rather by the r.h.s. of
(\ref{4.11a}) times $|h|^N$, an estimate which blows up order by
order as $h\to-\io$.
\vskip.2cm
\noindent{\it Proof.} To begin, let us assume that all the endpoints are on scales $<1$; the general case can be worked out in a completely analogous way, and it will be discussed later. Therefore, for the moment we shall consider the contribution of trees $\t\in\TT_{h,N}$. Plugging the bounds
\bea&&\big\| \hat{g}^{(h)}_{\o}(\kk') \big\| \leq \const\cdot
M^{-h}\;,\,\qquad
\int d\kk' \big\| \hat{g}^{(h)}_{\o}(\kk') \big\| \leq \const\cdot M^{2h}\;,\nn\\
&&\big\| \hat{w}^{(h)}(\pp) \big\| \leq \const\cdot M^{-h}\;,\,\qquad \int
d\kk \big\|\hat{w}^{(h)}(\pp) \big\| \leq\const\cdot M^{2h}\;,\label{4.7}
\eea
into (\ref{4.4}), we find that, if $\t\in\TT_{h,N}$ and
$\GG\in\G(\t)$,
\bea &&|\widehat\Val(\GG)|\le (\const.)^N\bar\e_h^N\cdot\label{4.8}\\
&&\hskip.5cm\cdot\prod_{v\ {\rm not}\ {\rm e.p.}} \frac{e^{\frac{C}2\bar \e_h^2
|P_v^\Psi|}}
{s_v!}M^{-3 h_v(s_v-1)}M^{2 h_v n^0_v} M^{h_v m^\n_v} \prod_{\substack{v\ {\rm
not}\ {\rm e.p.}\\ v>v_0}} M^{-z_v(h_v-h_{v'})}\;,\nn\eea
where: $n^0_v$ is the number of propagators $\ell\in v$, i.e., of
propagators $\ell$ contained in the box $G_v$ but not in any
smaller cluster; $s_v$ is the number of vertices immediately
following $v$ on $\t$; $m^\n_v$ is the number of end-points of
type $\n$ immediately following $v$ on $\t$ (i.e., contained in
$G_v$ but not in any smaller cluster); $v'$ is the vertex
immediately preceding $v$ on $\t$ and $z_v=2$ if
$|P_v^\Psi|=|P_v|=2$ or $|P_v^{A}| = |P_v| =2$, $z_v=1$ if $|P_v^\Psi|=2|P_v^A|=2$ or $|P_{v}^{\Psi}|=0$ and $|P_{v}^{A}|=3$, and
$z_v=0$ otherwise. The last product in (\ref{4.8}) is due to the
action of $\RR$ on the vertices $v>v_0$ that are not end-points.
In fact, the operator $\RR$, when acting on a kernel
$W^{(h_v)}_{2,1}(\pp,\kk')$ associated to a vertex $v$ with
$|P_v^\Psi|=2,\;|P_v^A|=1$, extracts from $W^{(h_v)}_{2,1}$ the rest
of first order in its Taylor expansion around $\pp=\kk'=\V0$: if
$|W^{(h_v)}_{2,1}(\pp,\kk')|\le C$, then $|\RR
W^{(h_v)}_{2,1}(\pp,\kk')|=\frac12|(\pp\dpr_\pp+\kk'\dpr_\kk')
W^{(h_v)}_{2,1}(\pp^*,\kk^*)|\le (\const.) M^{-h_v+h_{v'}}C$,
where $M^{-h_{v}}$ is a bound for the derivative with respect to
momenta on scale $h_v$ and $M^{h_{v'}}$ is a bound for the
external momenta $\pp$, $\kk'$; i.e., $\RR$ is dimensionally
equivalent to $M^{-(h_v-h_{v'})}$. 
The same is true if $\RR$ acts on kernels $W^{(h_{v})}_{0,3}(\pp,\qq)$. 
Similarly, if $\RR$ acts on a
terms with $|P_v|=2$, it extracts the rest of second
order in the Taylor expansion around $\kk'=\V0$, and it is
dimensionally equivalent to ${\kk'}^2\dpr_{\kk'}^2\sim
M^{-2(h_v-h_{v'})}$. As a result, we get (\ref{4.8}).

Now, let $n^e_v$ ($n^\n_v$) be the number of vertices of type $e$
(of type $\n$) following $v$ on $\t$. If we plug in (\ref{4.8})
the identities
\bea &&\sum_{v\ {\rm not}\ {\rm e.p.}} (h_v-h)(s_v-1) =
\sum_{v\ {\rm not}\ {\rm e.p.}} (h_v-h_{v'})(n^e_v+n^\n_v-1)\nn\\
&&\sum_{v\ {\rm not}\ {\rm e.p.}}(h_v - h)n^{0}_v = \sum_{v\ {\rm
not}\ {\rm e.p.}}(h_v - h_{v'})\Big( \frac{3}{2}n^{e}_v +
n^{\n}_{v} -
\frac{|P_{v}|}{2} \Big)\nn\\
&&\sum_{v\ {\rm not}\ {\rm e.p.}} (h_v - h)m^{\n}_{v} = \sum_{v\
{\rm not} \ {\rm e.p.}}(h_{v} - h_{v'})n^{\n}_{v}\label{4.9} \eea
we get the bound
\be|\widehat\Val(\GG)|\le (\const.)^N{\bar\e}_h^N\frac1{s_{v_0}!}
M^{h(3-|P_{v_0}|)}\!\!\!\prod_{ \substack{v\ {\rm not}\ {\rm e.p.}\\
v>v_0}}\frac{e^{\frac{C}2\bar \e_h^2 
|P_v^\Psi|}}{s_v!}M^{(h_v - h_{v'}) (3-|P_v|-z_v)}\;.\label{4.10}\ee
In the latter equation, $3-|P_v|$ is the {\it scaling dimension}
of the cluster $G_v$, and $3-|P_v|-z_v$ is its renormalized
scaling dimension. Notice that the renormalization operator $\RR$
has been introduced precisely to guarantee that $3-|P_v|-z_v<0$
for all $v$, by construction. This fact allows us to sum over the
scale labels $h\le h_v\le 1$, and to conclude that the
perturbative expansion is well defined at any order $N$ of the
renormalized expansion. More precisely, the fact that the
renormalized scaling dimensions are all negative implies, via a
standard argument (see, e.g., \cite{BG,GeM}), the following
bound, valid for a suitable constant $C$:
\bea &&||W^{(h);N}_{2n,m,\ul{j},\ul{\m}}|| \le (\const.)^N\bar\e_h^N 
\frac1{s_{v_0}!}M^{h(3-m-2n)}\cdot\label{A.11}\\&&
\hskip2.truecm\cdot\sum_{\t\in \TT_{h,N}}\sum_{\substack{\GG\in\G(\t)\\
|P^A_{v_0}| = m,\\|P^{\psi}_{v_0}| = 2n}}\prod_{\substack{v\ {\rm
not}\ {\rm e.p.}\\ v>v_0}}\frac{e^{\frac{C}2\bar \e_h^2
|P_v^\Psi|}}{s_v!}M^{(h_v - h_{v'})(3 - |P_v| - z_v)}\;,
\nn\eea
from which, after counting the number of Feynman graphs
contributing to the sum in (\ref{A.11}), (\ref{4.11a})
follows.

An immediate corollary of the proof leading to (\ref{4.11a}) is
that contributions from trees in $\TT_{h,N}$ with a vertex $v$ on
scale $h_v=k>h$ admit an improved bound with respect to
(\ref{4.11a}), of the form $\le (\const.)^N\bar\e_h^N (N/2)!\,
M^{h(3-|P_{v_0}|)}M^{\th(h-k)}$, for any $0<\th<1$; the factor
$M^{\th(h-k)}$ can be thought of as a dimensional gain with
respect to the ``basic'' dimensional bound in (\ref{4.11a}). This
improved bound is usually referred to as the {\it short memory
property} (i.e., long trees are exponentially suppressed); it is due 
to the fact that the renormalized scaling dimensions $D_v=3-|P_v|-z_v$
in (\ref{4.10})  are all negative, and can be obtained by taking a fraction of 
the factors $M^{(h_v-h_{v'})D_v}$ associated to the branches of the tree
$\t$ on the path connecting the vertex on scale $k$ to the one on scale $h$.

We conclude by discussing the contribution of the trees $\t\in \widetilde\TT_{K;h,N}$ with endpoints on scales $\geq 1$. In this case, the bound 
(\ref{4.10}) is replaced by the analogous bound obtained by taking into account the contribution of vertices on scales $\geq 0$ by matching the bound (\ref{4.10}) with the ultraviolet bound (\ref{C14}); the final result is that
\be|\widehat\Val(\GG)|\le (\const.)^N{\bar\e}_h^N\frac1{s_{v_0}!}
M^{h(3-|P_{v_0}|)}\!\!\!\prod_{ \substack{v\ {\rm not}\ {\rm e.p.}\\
v>v_0}}\frac{e^{\frac{a_{v} C}2\bar \e_h^2 
|P_v^\Psi|}}{s_v!}M^{(h_v - h_{v'}) D_{v}}\;.\label{4.11}
\ee
where the renormalized scaling dimensions $D_{v}$ are always negative by construction, $a_{v} = 1$ if $h_{v}\leq 0$ or $0$ otherwise.
\qed
\begin{table}[htbp]
\begin{center}
    \begin{tabular}{|l|p{11cm}|}
  \hline
    Symbol & Description \\ \hline
    $\t$ & Gallavotti -- Nicol\` o (GN) tree.\\
    $r$ & Root label of the tree.\\
    $v_0$ & First vertex of the tree, immediately following the root.\\
    $h_v$ & Scale label of the tree vertex $v$.\\
    $\TT_{h,N}$ & Set of GN trees with root on scale $h_r = h$ 
    and with $N$ endpoints.\\
    $I_v$ & Set of field labels associated with the endpoint of the tree $v$.\\
    $G_v$ & Cluster associated with the tree vertex $v$.\\
    $\III_v$ & Set of contracted fields inside the box corresponding to the cluster $G_v$.\\
    $P_v$ & Set of external fields of $G_v$.\\
    $v_{i}$ & $i$-th vertex immediately following $v$ on the tree.\\
    $s_{v}$ & Number of vertices immediately following the vertex $v$ on the tree.\\
    $P_{v}^{\#}$ & Set of fields of type $\# = A,\Psi$ external to $G_v$.\\
    $\G(\t)$ & Set of connected Feynman diagrams compatible 
    with the hierarchical cluster structure of the tree $\t$.\\
    $n^{0}_{v}$ & Number of propagators contained in $G_{v}$ but not in any smaller cluster.\\
    $m^{\n}_{v}$ & Number of end-points of type $\n$ immediately following $v$ on the tree.\\
    $v'$ & Vertex immediately preceding $v$ on the tree.\\
    $n^{\#}_{v}$ & Number of vertices of type $\# = e,\n$ following $v$ on the tree.\\
    $z_v$ & Improvement on the scaling dimension due to the renormalization.\\
    $D_{v}$ & Renormalized scaling dimension (equal to the scaling dimension if $h_{v}>0$).\\
    $a_{v}$ & Integer equal to $0,1$ depending on whether $h_{v}\leq 0$, $h_{v}>0$.\\
    \hline
    \end{tabular}
\end{center}
\caption{List of symbols introduced in Section \ref{sec4}.}
\end{table}

\section{The flow of the running coupling constants}\label{sec3a}
\setcounter{equation}{0}
\renewcommand{\theequation}{\ref{sec3a}.\arabic{equation}}

As a consequence of the interative integration scheme the effective parameters $\{e_{\m,h},\,\n_{\m,h},\,Z_{h},\,v_{h}\}$ obey to a nontrivial recursion relation; in fact, the difference of renormalized parameters on scales $h$ and $h+1$ can be expressed in terms of a well defined function of all the renormalized parameters on scales $\geq h+1$. This function is called the {\it Beta function}, and it can be written as a series in $\{e_{\m,k},\,\n_{\m,k}\}_{k\geq h+1}$ with coefficients depending on $\{ Z_{k}/Z_{k-1},\,v_{k} \}_{k\geq h+1}$, admitting $N!$ bounds in the sense of Theorem \ref{thm1}. A crucial point for the consistency of our approach is that the running coupling constants $e_{\m,h},\n_{\m,h}$ are small for all $h\le 0$, that the ratios $Z_h/Z_{h-1}$ are close to 1, and the effective Fermi velocity $v_h$ does not approach zero; even if we do not prove the convergence of the series but only $N!$ bounds, we expect that our series gives meaningful information only as long as the running coupling constants satisfy these conditions. In this Section we describe how to control their flow; we shall proceed by induction: we will first assume that $\bar\e_{h}=\max_{k\geq h+1}\{|e_{\m,k}|,|\n_{\m,k}|\}$ is small, that $Z_k/Z_{k-1}\le e^{C\bar\e_{k}^2}$ and $C^{-1}\le v_k\le C$ for all $k\geq h+1$ and a suitable constant $C>0$, and we will study the flow of $Z_h$ and $v_h$ under these assuptions. We will show that, asymptotically as $h\to-\io$, 
\bea
&&Z_{h} \sim M^{-\eta h}\;,\qquad v_{h} \sim v_{-\io} - (v_{-\io} - v)M^{\tilde\eta h}\;,\\
&&\eta = \frac{e^2}{12\pi^2} + O(e^4)\;,\qquad \tilde\eta = \frac{2 e^2}{5\pi^2} + O(e^4)\;,\qquad v_{-\io} = 1 + O\Big(1 - \frac{e_{1,-\io}}{e_{0,-\io}}\Big)\;,\nn
\eea
where $1$ is the speed of light in our units. These behaviors are true {\it at all orders}, in the sense that they are obtained by truncating the Beta functions of $Z_{h}$ and $v_{h}$ to any finite order in the running coupling constants and taking the bare charge small enough {\it uniformly} in the momentum scale $h$. Finally, as discussed in Section \ref{secWI}, to show the boundedness of the flow of the running coupling constants we shall use nonperturbative informations coming from {\it lattice Ward identities}; in particular, we will get that $e_{0,-\io} = e_{1,-\io}$, which implies that $v_{-\io}=1$.

\subsection{The Beta function}

The multiscale integration, and in particular formulas (\ref{3.7}), (\ref{3.8}),
(\ref{3.12}), (\ref{3.13b}), imply the following flow equations:
\bea
\frac{Z_{h-1}}{Z_h} &=& 1 + z_{0,h} =: 1 + \b^{z}_{h}\;,\label{3a.18}\\
v_{h-1} &=& \frac{Z_h}{Z_{h-1}}(v_{h} + z_{1,h}) =: v_{h} + \b^{v}_{h}\;,\\
\n_{\m,h} &=& -M^{-h}\, W^{(h)}_{0,2,\m,\m}(\V0)
=: M \n_{\m,h+1} + \b^{\n}_{\m,h+1}\;,\qquad\label{3a.18n}\\
e_{0,h} &=& \frac{Z_h}{Z_{h-1}}\l_{0,h} =: e_{0,h+1}+\b^{e}_{0,h+1}\;,
\label{3a.18e0}\\
e_{1,h} &=& \frac{Z_h}{Z_{h-1}}\frac{\l_{1,h}}{v_{h-1}} =:
e_{1,h+1}+\b^{e}_{1,h+1}\;,\label{3a.18e1}
\eea
and $e_{2,h}=e_{1,h}$. The {\it Beta functions} appearing in the
r.h.s. of flow equations are related, see (\ref{3.7}), (\ref{3.8}), to the kernels
$W^{(h);N}_{2n,m,\ul{\r},\ul\o,\ul{\m}}$, so that they are expressed by
series in the running coupling constants admitting the bound
(\ref{4.11a}). For the explicit expressions of the one-loop
contributions to the Beta function, see below.
\subsection{The flow of Fermi velocity}\label{secZv}
In this section we show that, under proper assumptions on the flow of the effective charges,
the effective Fermi velocity $v_h$ tend to a limit value $v_{-\io}$, which will be explicitly computed in Section \ref{LI}; remarkably, we will show that, no matter how small the bare velocity $v$ is, the effective Fermi velocity $v_{-\io}$ is equal to $1$, which is the speed of light in the units we have chosen.

Let us make the following assumption on effective charges and on the effective photon masses: 
\bea
e_{\m,h}&=&e_{\m,-\io}+O(e^3(v_{-\io}-v_h))+O(e^3M^{\th h})\;,\label{ch1}\\
\n_{\m,h} &=& O(e^2)\;,\nn
\eea
with $0<\th<1$ and $e_{\m,-\io}=e+O(e^3)$; we shall refer to the first line of (\ref{ch1}) by saying that the charge tens to a {\it line of fixed points}. The remarkable properties (\ref{ch1}) will be proven order by order in perturbation theory 
by using WIs, see Section \ref{secWI}. We start by studying the flow of the Fermi velocity; at lowest order
(see Appendix \ref{secwave}), its beta 
function reads, setting $\xi_{h} := \sqrt{v_{h}^{-2} - 1}$:
\bea&&
\b^{v,(2)}_{h} =\label{3a.19v}\\&&=\frac{ \log M}{4\p^2}
\Biggl[\frac{e^2_{0,h}v_h^{-1}}2\frac{\arctan \x_h}{\x_h}- \Big(2
e^2_{1,h}v_{h} - \frac{e^2_{0,h}v_h^{-1}}2\Big)
\frac{\x_h-\arctan\x_h}{\x_h^3}\Biggr]\;.\nn \eea
Note that if $e_{0,k}\=e_{1,k}$, then the r.h.s. of (\ref{3a.19v}) is strictly 
positive for all $\x_h>0$ and it vanishes quadratically in $\x_h$ at 
$\x_h= 0$. 
The higher order contributions to $\b_h^v$ have similar properties. 
This can be proved as follows: we observe that the
Beta function $\b^{v}_h$ is a function of the renormalized 
couplings and of the Fermi velocities on scales $\geq h$, {\it i.e.}:
\be
\b^{v}_{h} = \b^{v}_h\big(\big\{(e_{0,k},e_{1,k},e_{2,k}), (\n_{0,k}, \n_{1,k}, \n_{2,k}), v_k
\big\}_{k\ge h}\big)\;.
\label{dec1}
\ee
%
We can rewrite $\b^{v}_h$ as $\b^{v}_{h,0} + \b^{v}_{h,>0}$, where $\b^{v}_{h,>0}$ takes into account all the contributions coming from trees with at least one end-point on scale $>0$; by the short memory property (see discussion after Eq. (\ref{A.11})), the bound on $\b^{v}_{h,>0}$ admit an improvement of a factor $M^{\th h}$ with respect to the basic one. Moreover, we rewrite $\b^{v}_{h,0}$ as $\b^{v,rel}_{h,0} + \tilde\b^{v}_{h,0}$, where $\b^{v,rel}_{h,0}$ is given by $\b^{v}_{h,0}$ after replacing all the propagators with their relativistic approximations, namely those obtained setting $R_{h-1,\o}(\kk') = R'(\pp)=0$ in (\ref{3.16}); therefore, $\tilde\b^{v}_{h,0}$ admits an improvement of a factor $M^{\th h}$ in its dimensional bound. Then, we can rewrite $\b^{v,rel}_{h,0}$ as $\b^{v,rel,0}_{h,0} + \b^{v,rel,1}_{h,0} + \b^{v,rel,2}_{h,0} + \b^{v,rel,3}_{h,0} + \bar\b^{v}_{h,0}$, where $\bar\b^{v}_{h,0} = O(e^{2}M^{\th h})$ and
\bea
\b^{v,rel,0}_{h,0} &=& \b^{v,rel}_{h,0}\big(\big\{(e_{0,k},e_{0,k},e_{0,k}), (\n_{0,k}, \n_{0,k}, \n_{0,k}), 1
\big\}_{k\ge h}\big)\;,\label{dec2}\\
\b^{v,rel,1}_{h, 0} &=& \b^{v,rel}_{h,0}\big(\big\{(e_{0,k},e_{0,k},e_{0,k}), (\n_{0,k}, \n_{0,k}, \n_{0,k}), v_k
\big\}_{k\ge h}\big)-\nn\\
&& -\b^{v,rel}_{h,0}\big(\big\{(e_{0,k},e_{0,k},e_{0,k}), (\n_{0,k}, \n_{0,k}, \n_{0,k}), 1
\big\}_{k\ge h}\big)\;,\nn\\
\b^{v,rel,2}_{h, 0} &=& \b^{v,rel}_{h,0}\big(\big\{(e_{0,k},e_{0,k},e_{0,k}), (\n_{0,k}, \n_{1,k}, \n_{2,k}), v_k
\big\}_{k\ge h}\big)-\nn\\
&& - \b^{v,rel}_{h,0}\big(\big\{(e_{0,k},e_{0,k},e_{0,k}), (\n_{0,k}, \n_{0,k}, \n_{0,k}), v_k
\big\}_{k\ge h}\big)\;,\nn\\
\b^{v,rel,3}_{h, 0} &=& \b^{v,rel}_{h,0}\big(\big\{(e_{0,k},e_{1,k},e_{2,k}), (\n_{0,k}, \n_{1,k}, \n_{2,k}), v_k
\big\}_{k\ge h}\big)-\nn\\
&& -\b^{v,rel}_{h,0}\big(\big\{(e_{0,k},e_{0,k},e_{0,k}), (\n_{0,k}, \n_{1,k}, \n_{2,k}), v_k
\big\}_{k\ge h}\big)\;.\nn
\eea
By the short memory property we get:
\bea
&&\b^{v,rel,1}_{h, 0} = O\big(e_{0,h}^{2}(1 - v_{h})\big)\;,\qquad \b^{v,rel,2}_{h, 0} = O\big(e^2_{0,h}(\n_{0,h} - \n_{1,h})\big)\;,\nn\\
&&\b^{v,rel,3}_{h, 0} = O\big(e_{0,h}(e_{0,h} - e_{1,h})\big)\;.\label{dec3}
\eea
Using (\ref{3a.18n}) and a telescopic argument similar to the one leading to Eq.(\ref{dec3}), we also find that  $\n_{0,h} - \n_{1,h}$ can be written as a sum of contributions of order $e_{0,h}(e_{0,h} - e_{1,h})$ and of order $e_{0,h}^{2}(1 - v_{h})$.

Now, let us consider $\b^{v,rel,0}_{h,0}$; this is the Beta function of the continuum relativistic theory studied in \cite{GMP1}, where both photons and fermions live in the continuum and are regularized by an ultraviolet cutoff function $\chi_{0}$, but where the effective charges and the photon masses have been replaced by those of our lattice model. However, we can rewrite $e_{0,k} = \tilde e_{0,k} + e_{0,k} - \tilde e_{0,k}$, $\n_{0,k} = \tilde \n_{0,k} + \n_{0,k} - \tilde\n_{0,k}$, where $\tilde e_{0,k},\,\tilde \n_{0,k}$ are the running coupling constants of the theory of \cite{GMP1} with bare Fermi velocity $v=1$ and bare parameters $\tilde e,\,\tilde\n$ chosen so that $\tilde e_{0,h} = e_{0,h}$, $\tilde\n_{0,h} = \n_{0,h}$; it is straightforward to see that $e_{0,k} - \tilde e_{0,k}$, $\n_{0,k} - \tilde\n_{0,k}$ can be written as sums  of contributions $O(e^{3}M^{\th k})$, $O(e^{2}(1-v_{k}))$, $O(e(e_{1,k+1} - e_{0,k+1}))$. Moreover, by relativistic invariance we know that, see \cite{GMP1}, if we replace at the argument of $\b^{v,rel,0}_{h,0}$ the couplings $e_{0,h}$, $\n_{0,h}$ with $\tilde e_{0,h}$, $\tilde \n_{0,h}$ we get $0$; therefore, at the end we find that, using the short memory property:
\be
\b^{v,rel,0}_{h,0} = O(e_{0,h}^{4}M^{\th h}) + O(e^{3}(1 - v_{h})) + O(e_{0,h}^{2}(e_{0,h+1} - e_{1,h+1}))\;.\label{dec3b}
\ee
Therefore, we can write:
\be \frac{v_{h-1}}{v_h}=1 +\frac{ \log M}{4\p^2} \Biggl[\frac85
e^2(1-v_h)(1+A'_h)+\frac43e(1+B'_h)
(e_{0,h}-e_{1,h})\Biggr]\;,\label{5.11b}\ee
where: the numerical coefficients are obtained from the explicit lowest order computation 
(\ref{3a.19v}); $A'_h$ is a sum of contributions that are finite at all orders
in the effective couplings, which are either of order two or more in the 
effective charges, or vanishing at $v_k=1$; similarly, $B'_h$ is a sum of 
contributions that are finite at all orders in the effective couplings, which 
are of order two or more in the effective charges. From
\pref{5.11b} it is apparent that $v_h$ tends as $h\to-\io$ to a
limit value
\be v_{-\io}=1+\frac{5}{6e}(e_{0,-\io}-e_{1,-\io})(1+C'_{-\io}) \label{veff}\ee
with $C'_{-\io}$ a sum of contributions 
that are finite at all orders in the effective couplings, which 
are of order two or more in the effective charges. The fixed point (\ref{veff}) 
is found simply by requiring that in the limit $h\rightarrow-\infty$ 
the argument of the square brakets in (\ref{5.11b}) vanishes. 

Using Eq.(\ref{ch1}), we find that the expression in square 
brackets in the r.h.s. of (\ref{5.11b}) can be rewritten as 
$(8e^2/5)(v_{-\io}-v_h+R_h')(1+A''_h)$, where: (i) $A''_h$  
is a sum of contributions that are finite at all orders
in the effective couplings, which are either of order two or more in the 
effective charges, or vanishing at $v_k=v_{-\io}$; (ii) $R_h'$ 
is a sum of contributions that are finite at all orders
in the effective couplings, which are of order two or more in the 
effective charges and are bounded at all orders by $M^{\th h}$, for some 
$0<\th<1$. Therefore, (\ref{5.11b})
can be rewritten as
\be
v_{-\io} - v_{h-1} = (v_{-\io} - v_{h})
\Big(1 - v_{h}\frac{v_{-\io} - v_{h}+R_h'}{v_{-\io} - v_{h}}
\log M \frac{2 e^{2}}{5\pi^2}(1 + A''_{h}) \Big)\;,\label{veff2}
\ee
from which, using the fact that $R_h'=O(e^2 M^{\th h})$,  
we get that there exist two positive constants $C_1,C_2$
such that \footnote{Eq.(\ref{flowv1}) must be understood as an order by 
order inequality: if we truncate the theory at order $N$ in the bare coupling $e$, 
both sides of the inequality in Eq.(\ref{flowv1}) are verified asymptotically as $e\to 0$, for 
all $N\ge 1$.}:
\be C_1M^{h\tilde\eta}\le \frac{v_{-\io} - v_{h}}{v_{-\io}-v}\le C_2  M^{h\tilde\eta}\;,
\label{flowv1}\ee
with 
\be
\tilde\h = -\log_{M}\Big[ 1 - v_{-\io}\log M
\frac{2e^{2}}{5\pi^2}\big( 1 +A''_{-\io} \big) \Big]\;;\label{tildeh}
\ee
at lowest order, Eq.(\ref{tildeh}) gives $\tilde\h^{(2)}=2e^2/(5\p^2)$.

\subsection{The flow of the wave function renormalization}

In contrast to what happens in the case of the other running coupling constants, the flow of $Z_{h}$ {\it diverges} as $h\rightarrow-\infty$ with an interaction-dependent power law. In particular, as it will be clear from the analysis of Section \ref{schwing}, the divergence of the wave function renormalization, that is the vanishing of the {\it quasi-particle weight} $Z_{h}^{-1}$, implies the anomalous scaling of the Schwinger functions.

As for the Fermi velocity, assume that (\ref{ch1}) holds; from (\ref{3a.18}) it follows that, for two suitable positive constants $C_{1}$, $C_{2}$:
\be
C_1 M^{\h h}\le Z_h\le C_2 M^{\h h}\;,\qquad \eta = \lim_{h\rightarrow -\infty}\log_{M}\big( 1 + \b^{z}_{h} \big)\;.\label{eta} 
\ee
The lowest order contribution to $\b^{z}_{h}$ is computed in Appendix \ref{app2d}, and it is given by:
\be
\b^{z,(2)}_{h} = \frac{\log M}{4\p^2} \, (2
e^2_{1,h}-e^2_{0,h}v_h^{-2})
\frac{\x_h-\arctan\x_h}{\x_h^3}\;;\label{3a.19z}
\ee
therefore, from (\ref{eta}), (\ref{3a.19z}) we get that the lowest order contribution to $\eta$ is positive, and it is given by $\eta^{(2)} = \frac{e^{2}}{12\pi^2}$. Now, let us 
briefly comment about the relation between $Z_{h}$, $v_{h}$ and the functions 
$Z(\kk')$ and $v(\kk')$ appearing in the main result, see (\ref{res1}).
If $|\kk'|=M^h$, we define $Z(\kk'):=Z_h$ and $v(\kk'):=v_h$; for general $|\kk'|\le 1$, we let $Z(\kk')$
and $v(\kk')$ be smooth interpolations of these sequences. Of course, we can choose 
these interpolations in such a way that, if
$M^{h}\leq |\kk'|\leq M^{h+1}$, 
\bea
&&\Big| \frac{Z(\kk')}{Z_{h}} - 1\Big| \le \Big|\frac{Z_{h+1}}{Z_{h}} - 1\Big|=O(
\eta \log M)\;,\nn\\
&&\Big| \frac{v(\kk') - v_{h}}{v_{-\io}-v_h}\Big|\le 
\Big| \frac{v_{h+1} - v_{h}}{v_{-\io}-v_h}\Big|=
O(\tilde\eta \log M)\;.\label{cont1}
\eea
therefore, we can replace in the leading part of the 2-point Schwinger function, which will be computed in Section \ref{schwing}, 
the wave function renormalization $Z_{j}$ and the effective Fermi velocity $v_j$ by
 $Z(\kk')$ and $v(\kk')$, provided that the correction term $R(\kk')$ in (\ref{res1})
is defined so to take into account higher order corrections 
satisfying the bounds (\ref{cont1}).


\section{Ward identities}\label{secWI}
\setcounter{equation}{0}
\renewcommand{\theequation}{\ref{secWI}.\arabic{equation}}
To conclude the discussion on the flow of the running coupling constants, we need to show that our assumptions of the flow of the effective couplings $\{e_{\m,h},\n_{\m,h}\}$ are indeed valid. To prove this, we shall exploit the gauge invariance of the model to derive suitable {\it lattice Ward identities}; these are nonperturbative identities between Schwinger functions, which will imply the cancellations that we need in order to control the flows of $\{e_{\m,h},\n_{\m,h}\}$. Our strategy follows closely the one introduced by Benfatto and Mastropietro in the analysis on one dimensional Luttinger liquids, \cite{BMwi1, BM}.\\

\noindent{\bf Strategy.} Before getting into the technical details, let us first give an idea of the strategy. At each RG step we shall introduce a model, to be called the {\it reference model} in the following; this model is not Hamiltonian and it is defined only in terms of a functional integral. The main features of the reference model are that: (i) it is defined in terms of a finite number of Grassmann variables, living on a space-time lattice: the spatial part of the lattice is given by the physical honeycomb lattice, while the temporal one is introduced by hand, and its mesh will be sent to zero; (ii) the bosonic sector of the model is regularized by an infrared cut-off on the same momentum scale $h$ that we have to integrate in the full model. The key points of the strategy are the following ones.
\begin{itemize}
\item[(i)] The reference model can be studied using multiscale analysis and renormalization group, and the running coupling constants on scales greater than $h$ are {\it equal} to those of the full model.
\item[(ii)] After the scale $h$, that is after having integrated out all the photon fields, the reference model is reduced to an effective fermionic model which turns out to be {\it superrenormalizable}, and can be studied along the lines of \cite{GM}; this in particular implies that the running coupling constants of the reference model ``cease to flow'' on scale $h$. 
\item[(iii)] The bosonic infrared cut-off {\it does not break gauge invariance}; that is, it is possible to derive Ward identities for this model which are equal to those which can be formally written in the full model, except for the fact that the Schwinger functions involved in the identities are all evaluated in presence of a bosonic infrared cut-off on scale $h$; we stress that the presence of the cut-off on the photons does not produce corrections to the Ward identities, in contrast to what would happen with a cut-off on the fermions (see \cite{GMP1}).
\item[(iv)] Finally, the Ward identities provide relations for the bosonic two point function, and relating the fermionic two and three point functions; from this identities we get that the mass of the photon field on a given scale $h$ is bounded by a quantity of order $M^{h}e^{2}$, while the effective charge on scale $h$ is close to the bare charge up to corrections bounded by $O(e^{3})$ uniformly in $h$.
\end{itemize}
These identities hide remarkable cancellations between Feynman graphs, and it is interesting to check them at lowest order; see Appendix \ref{WIcheck}, where it is shown that at one loop in naive perturbation theory the renormalized mass of the photon field is zero and the effective charge is equal to the bare one.

\subsection{The reference model}\label{secWIa}
Here we shall define the reference model; In particular, we shall prove that the generating functional of the Schwinger functions is left invariant by local phase transformations on the fermions (the Jacobian of the transformation is equal to $1$). This is a well known property of Grassmannian functional integrals, see \cite{M} for instance. But at the same time we know that, because of the invariance of the model under local gauge transformations, which are nothing more than a local phase transformation on the fermions combined with a shift of the photon field, the argument of the functional integral transforms in a suitable way; in this way we shall get an identity between generating functionals that gives rise to an infinite set of Ward identities.
\paragraph{Definition and properties.} Consider the sets 
\bea
\tilde\L &:=& \L \cup \big\{ x_0 = \frac{\b n}{N},\,n=0,1,\ldots , N-1 \big\}\;,\nn\\ 
\tilde \DD_{\b,L} &:=& \DD_{L}\cup \Big\{ k_0 = \frac{2\pi}{\b}(n + \frac{1}{2}),\,n= 0,1,\ldots, N-1\Big\}\;,\nn\\
\tilde\PPP_{\b,L} &:=& \PPP_{L}\cup \Big\{p_0 = \frac{2\pi n}{\b},\,n=0,1,\ldots , N-1\Big\}\;;\label{refm1}
\eea
With each $\kk\in \tilde\DD_{\b,L}$, $\pp\in \tilde\PPP_{\b,L}$ we associate a fermion or boson gaussian field, respectively; with a slight abuse of notation, we shall denote these two fields with the same symbols used for the original model. The Fourier transform of the fermion field lives in the space-time lattice $\tilde\L$, and it is given by
\be
\Psi^{\pm}_{\xx + (\r- 1)(0,\vec\d_1),\s,\r} := \frac{1}{\b|\L|}\sum_{\kk\in \tilde\DD_{\b,L}}\eu^{\pm\iu \kk\xx}\hat\Psi^{\pm}_{\kk,\s,\r}\;.\label{refm1b}
\ee
The fermion and boson propagator are defined as follows, 
\bea
\hat g_{\e}(\kk) &:=& \chi^{\e}_{K}(k_0)\begin{pmatrix} -\iu k_0 & -t\O^{*}(\vec k) \\ -t\O(\vec k) & -\iu k_0 \end{pmatrix}\;,\nn\\
\hat w_{\e}^{(\geq h)}(\pp) &:=& \int\frac{d p_3}{(2\pi)}\,\frac{\chi^{\e}_{K}(p_0)\chi_{[h,0]}(|\ul p|)}{\pp^2 + p_3^2}\;,\label{refm2}
\eea
that is, they are equal to the ``usual'' ones {\it except} for the replacement of $\chi_{K}$, $\chi(|\ul p|)$ with $\chi^{\e}_{K}$, $\chi_{[h,0]}(|\ul p|)$, respectively. The function $\chi^{\e}_{K}$ is a $C^{\infty}$ function defined so that: (i) $\lim_{\e\rightarrow 0}\chi^{\e}_{K}(k_0) = \chi_{K}(k_0)$; (ii) for $|k_0|\leq M^{K}$ $\chi^{\e}_{K}(k_0) = 1$ and for $|k_0|>M$ $\chi^{\e}_{K}(k_0) \leq \e \eu^{-|k_0|M^{-K}}$.

A very important property of the reference model is that the Jacobian of local phase transformations on the fermion fields, namely
\be
\Psi^{\pm}_{\xx + (\r - 1)(0,\vec\d_1),\s,\r}\rightarrow \eu^{\pm \iu e\a_{\xx + (\r-1)(0,\vec\d_1)}}\Psi^{\pm}_{\xx + (\r - 1)(0,\vec\d_1),\s,\r}\;,\quad \a_{\xx}\in C^{1}\;,\label{refm3}
\ee
is equal to $1$. To see this, let
\bea
\DD\Psi &:=& \prod_{\kk \in \tilde \DD_{\b,L}}\prod_{\substack{\s = \uparrow\downarrow \\ \r = 1,2}} d\hat\Psi^{+}_{\kk,\s,\r}d\hat\Psi^{-}_{\kk,\s,\r}\;,\label{refm4}\\
Q_{A_- A_+}(\Psi^{-},\Psi^{+}) &:=& \prod_{i=1}^{|A_-|}\prod_{\s_i,\r_i}\Psi^{-}_{\xx_i + (\r_i -1)(0,\vec\d_1),\s_i,\r_i}\prod_{j=1}^{|A_+|}\prod_{\s'_i,\r'_i}\Psi^{+}_{\xx'_j + (\r'_i -1)(0,\vec\d_1) ,\s_i',\r_i'}\;,\nn
\eea
where the sequences $\{\xx_i\}$, $\{\xx'_j\}$ must be formed by distinct space-time points belonging to $\tilde\L$ (otherwise $Q_{A_-A_+} =0$, since the square of a Grassmann variable is zero, by definition), and consider the integral $\int D\Psi\, Q_{A_- A_+}(\Psi^{-},\Psi^{+})$; our claim is that
\be
\int \DD \Psi\, Q_{A_- A_+}(\Psi^{-},\Psi^{+}) = \int D\Psi\, Q_{A_- A_+}(\eu^{-\iu e \a}\Psi^{-},\eu^{\iu e \a}\Psi^{+})\;.\label{refm5} 
\ee
To prove (\ref{refm5}), we rewrite $Q_{A_- A_+}$ in momentum space; it follows that:
\bea
&&\int \DD \Psi\, Q_{A_- A_+}(\Psi^{-},\Psi^{+}) = \frac{1}{(\b|\L|)^{|A_-| + |A_+|}}\cdot\label{refm6}\\&&\sum_{\{\kk_i\}_{i=1}^{|A_-|}}e^{-4\iu \sum_{i}\kk_i \xx_i}\sum_{\{\kk'_{j}\}_{j=1}^{|A_+|}}e^{4\iu \sum_{j}\kk'_{j}\xx'_{j}}\int \DD\Psi\, \prod_{i}\prod_{\s_i,\r_i}\hat\Psi^{-}_{\kk_i,\s_i,\r_i}\prod_{j}\prod_{\s'_j,\r'_j}\hat \Psi^{+}_{\kk'_j,\s'_j,\r'_j}\;.\nn
\eea
Notice that the integral in the r.h.s. of (\ref{refm6}) is vanishing unless: (i) the sequences $\{\kk_{i}\}$ and $\{\kk'_{j}\}$ are respectively formed by distinct elements; (ii) $|A_{-}| = |A_{+}| = |\tilde\DD_{\b,L}|$. But the space-time points $\{\xx_i\}$, $\{\xx'_j\}$ are as many as the momenta $\{\kk_i\}$, $\{\kk'_j\}$, and by construction $|\tilde\DD_{\b,L}| = |\tilde\L|$; therefore, since by assumption the sequences $\{\xx_i\}$, $\{\xx'_j\}$ are formed by distinct space-time points, $Q_{A_- A_+}(\Psi^{-},\Psi^{+})$ must contain two Grassmann variables for each $\s,\,\r$ and space-time lattice site $\xx$, namely $\Psi^{+}_{\xx + (\r - 1)(0,\vec\d_1),\s,\r}$ and $\Psi^{-}_{\xx + (\r - 1)(0,\vec\d_1),\s,\r}$, otherwise the outcome of the integration is zero. This proves (\ref{refm5}).
\paragraph{Consequences of gauge invariance.} Let us define the generating functional of the Schwinger functions of the reference model as:
\be
\eu^{\WW_{\b,L}^{[h,K]}(J,\phi)} := \int P(d\Psi)P_{\geq h}(A)\eu^{V(A + J) + B(\Psi,\phi)}\;,\label{WI3}
\ee
where the integration measures are determined by the propagators introduced in (\ref{refm2}), and the {\it fermionic source term} is given by
\be
B(\Psi,\phi) := \sum_{\s,\r}\int\frac{d\kk}{D}\,\hat\Psi^{+}_{\kk,\s,\r}\hat\phi^{-}_{\kk,\s,\r} + \hat\phi^{+}_{\kk,\s,\r}\hat\Psi^{-}_{\kk,\s,\r}\;,\label{WI3b}
\ee
where $\hat \phi^{\pm}_{\kk,\s,\r}$ are Grassmann variables, with Fourier transform $\phi^{\pm}_{\xx + (\r - 1)(0,\vec\d_1),\s,\r} := (\b|\L|)^{-1}\sum_{\kk\in \tilde\DD_{\b,L}}\eu^{\pm \iu\kk\xx}\hat\phi^{\pm}_{\kk,\s,\r}$; to avoid cumbersome notations, we will not explicitly write the dependence on $\e$, $N$, since at the end we will be interested in the limits $\e\rightarrow 0$, $N\rightarrow +\infty$. For convenience, we rewrite the fermionic integration measure as follows:
\be
P(d\Psi) :=  \frac{1}{\NN_{\Psi}}D\Psi\,\exp\Big\{ - \big( \Psi^{+},\GG \Psi^{-} \big)\Big\}\label{WI4}\;,
\ee
with
\be
\big( \Psi^{+},\GG \Psi^{-} \big) := \frac{1}{\b|\L|}\sum_{\s,\r}\sum_{\kk\in\tilde\DD_{\b,L}}\hat\Psi^{+}_{\kk,\s,\r}\chi^{\e}_{K}(k_0)^{-1}\big[ B(\kk) \big]_{\r\r'} \hat\Psi^{-}_{\kk,\s,\r'}\;,\label{WI5}
\ee
where $B(\kk)$ has been defined in \ref{1.2.10}. Using that the local phase transformation (\ref{refm3}) has Jacobian equal to one, we can freely perform the replacement (\ref{refm1b}) in the Gaussian weight of the Grassmannian measure, in the interaction and in the source term; neglecting {\it formally} the presence of the ultraviolet cutoff we would get the relation
\be
\WW_{\b,L}^{[h,+\infty]}(J,\phi) = \WW_{\b,L}^{[h,+\infty]}(J + \partial\a, \phi \eu^{\iu e \a})\;,\label{WI5b}
\ee
which generates an infinite set of Ward identities, by deriving the l.h.s. and the r.h.s. once with respect to $\hat\a_{\pp}$ (which sets to zero the l.h.s.) and an arbitrary number of times with respect to the external fields $\hat J_{\m,\pp}$, $\hat \phi_{\kk,\s,\r}$; notice that because of the fact that $J_{\m,\xx}$ and $\a_{\xx}$ always appear in the combination $J_{\m,\xx} + \partial_{\m}\a_{\xx}$, as guaranteed by gauge invariance, the derivative w.r.t. $\hat\a_{\pp}$ is equal to $-\iu p_{\m}$ times the derivative w.r.t. $\hat J_{\m,\pp}$. However, to avoid formal expressions it is necessary to take into account the presence of the ultraviolet cutoff; the result is that for $K$ fixed the analogous of (\ref{WI5b}) involves a $K$-dependent {\it correction term}. The fact that cutoffs on the fermions produce corrections to the Ward identities is well known, \cite{BM2, BM}; if they are not vanishing in the limit of cutoff removal we say that the WIs have {\it anomalies}, and this is indeed the case in one dimension, \cite{BM}. Here, on the contrary, as we prove in Appendix \ref{app5}, the corrections are exponentially vanishing as $K\rightarrow+\infty$. For simplicity, let us consider a function $\a_\xx$ of the form $\a_{\xx} := (\b\AAA_{\L})^{-1}\sum_{\pp\in\PPP_{\b,L}}\eu^{-\iu \pp \xx}\hat\a_{\pp}$; it is straightforward to see that, after (\ref{refm1b}), in the limit $N\rightarrow+\infty$:
\bea
&&\eu^{\WW_{\b,L}^{[h,K]}(J,\phi)} = \int P(d\Psi)P_{\geq h}(dA)\eu^{V(A + \partial \a + J,\Psi) + B(\Psi, \phi \eu^{\iu e \a})}\cdot\label{WI7}\\&&\cdot e^{-\big(\Psi^{+}\eu^{\iu e \a},\GG \eu^{-\iu e \a}\Psi^{-}\big) + \big(\Psi^{+},\GG \Psi^{-}\big) + \big(\Psi^{+}\eu^{\iu e \a},\tilde \GG \eu^{-\iu e \a}\Psi^{-}\big) - \big(\Psi^{+},\tilde \GG \Psi^{-}\big) - \iu \eu (n,\partial_0 \a)}\;,\nn
\eea
where
\bea
\big( \Psi^{+},\tilde \GG \Psi^{-} \big) &:=& \sum_{\s,\r}\int\frac{d\kk}{D}\,\hat\Psi^{+}_{\kk,\s,\r}\big[ B(0,\vec k) \big]_{\r\r'} \hat\Psi^{-}_{\kk,\s,\r'}\;,\label{WI7aa}\\
(n,\partial_0 \a) &=& \sum_{\s,\r}\int \frac{d\kk}{D}\frac{d\pp}{(2\pi)^3}\,\hat\Psi^{+}_{\kk+\pp,\s,\r}\hat\Psi^{-}_{\kk,\s,\r}(-\iu p_{0})\hat\a_{\pp} \eu^{-\iu (\r-1)p_1}\;.\nn
\eea
The first two terms in the second line of (\ref{WI7}) come from the free Grassmannian measure; the remaining three are produced by the interaction. Being interested in the derivative w.r.t. $\hat \a_{\pp}$ of (\ref{WI7}), we can rewrite the exponent appearing in the second line of (\ref{WI7aa}) as follows:
\be
\iu e \Big[ \Big( \Psi^{+},\GG \a\Psi^{-} \Big) - \Big(\Psi^{+}\a, \GG \Psi^{-}\Big) - \Big( \Psi^{+},\tilde\GG \a\Psi^{-} \Big) + \Big(\Psi^{+}\a, \tilde\GG \Psi^{-}\Big) - \big(n,\partial_0 \a\big) \Big] + O(\a^{2})\label{WI01}
\ee
and it is easy to see that (repeated indeces are summed)
\bea
&&\Big( \Psi^{+},\GG \a\Psi^{-} \Big) - \Big(\Psi^{+}\a, \GG \Psi^{-}\Big) = \int \frac{d\kk}{D}\frac{d\pp}{(2\pi)^3}\,\hat\a_{\pp}\hat\Psi^{+}_{\kk+\pp,\s,\r}\cdot\label{WI02}\\&&\Big[ \eu^{-\iu p_1(\r' - 1)}\chi^{\e}_{K}(k_0 + p_0)^{-1}\big[B(\kk+\pp)\big]_{\r\r'} - \eu^{-\iu p_1(\r - 1)}\chi^{\e}_{K}(k_0)^{-1}\big[B(\kk)\big]_{\r\r'} \Big]\hat\Psi^{-}_{\kk,\s,\r'}\nn
\eea
and that
\bea
&&\Big( \Psi^{+},\tilde\GG \a\Psi^{-} \Big) - \Big(\Psi^{+}\a, \tilde\GG \Psi^{-}\Big) + (n,\partial_0 \a) = \label{WI03}\\
&&\int\frac{d\kk}{D}\frac{d\pp}{(2\pi)^3}\,\hat\a_{\pp}\hat\Psi^{+}_{\kk+\pp,\s,\r}\Big[ \eu^{-\iu p_1(\r' - 1)}\big[B(\kk+\pp)\big]_{\r\r'} - \eu^{-\iu p_1(\r - 1)}\big[B(\kk)\big]_{\r\r'} \Big]\hat\Psi^{-}_{\kk,\s,\r'}\nn
\eea
Therefore, deriving with respect to $\hat \a_{\pp}$ the {\it l.h.s.} and {\it r.h.s.} of (\ref{WI7}) and setting $\hat \a_{\pp} =0$ we get that, in the limit $\e\rightarrow 0$:
\be
0 = (\b\AAA_{\L})\frac{\partial}{\partial\hat \a_{\pp}}\WW_{\b,L}^{[h,K]}(J + \partial\a,\phi \eu^{\iu e \a})\Big|_{\hat \a_{\pp}=0} + \D^{[h,K]}_{\b,L}(\pp;J,\phi)\;,\label{WI7a}
\ee
with
\bea
&&\D^{[h,K]}_{\b,L}(\pp;J,\phi) := \int \frac{d\kk}{D}\,\media{\hat\Psi^{+}_{\kk+\pp,\s,\r}C^{K}_{\r\r'}(\kk,\pp)\hat\Psi^{-}_{\kk,\s,\r'}}_{J,\phi}\;,\label{WI9a}\\
&&-\iu e^{-1}C^{K}_{\r\r'}(\kk,\pp) :=\label{WI9}\\&& \eu^{-\iu p_{1}(\r' - 1)}\big(\chi_{K}^{-1}(k_0 + p_0) - 1\big)\big[B(\kk+\pp)\big]_{\r\r'} - \eu^{-\iu p_{1} (\r - 1)}\big(\chi_K^{-1}(k_0) - 1\big)\big[B(\kk)\big]_{\r\r'}\nn
\eea
where the subscript $J,\phi$ in (\ref{WI9a}) means that the average is taken in presence of the external fields. Formula (\ref{WI7a}) will be starting point in order to derive the Ward identities that we need to control the flows of $\{e_{\m,h},\,\n_{\m,h}\}$.

\subsection{Ward identity for the photon mass}

In this Section we shall derive the Ward identity for the photon field, which will imply that the photon field remains massless; the starting point will be formula (\ref{WI9a}). Let us define:
\bea
\SS^{[h,K]}_{0,2,\m,\n}(\pp) &:=& (\b\AAA_{\L})\frac{\partial^{2}}{\partial \hat J_{\m,\pp}\partial \hat J_{\n,-\pp}}\WW^{[h,K]}_{\b,L}(J,0)\Big|_{J=0}\;,\label{WI7b}\\
p_{\m}\D_{0,2,\m,\n}^{[h,K]}(\pp) &:=&  \frac{\partial }{\partial \hat J_{\n,-\pp}}\D_{\b,L}^{[h,K]}(J,0)\Big|_{J=0}\;,\\
\eea
where the first line is the definition of the bosonic two point Schwinger function. Using that the action of $\partial_{\hat\a_{\pp}}$ on $\WW_{\b,L}^{[h,K]}(J + \partial\a,0)$ is equal to the one of $-\iu p_{\m}\partial_{J_{\m,\pp}}$, from (\ref{WI9a}) it follows that:
\be
0 = p_{\m}\SS^{[h,K]}_{0,2,\m,\n}(\pp) + p_{\m}\D_{0,2,\m,\n}^{[h,K]}(\pp)\;.\label{WI8}
\ee
Apparently, this relation tells us that the regularized bosonic two point function is not vanishing; however, as it is proven in Appendix \ref{app5}, the correction term in (\ref{WI8}) vanishes exponentially in the ultraviolet limit $K\rightarrow+\infty$. Therefore, from (\ref{WI8}) we get:
\be
0 = p_{\m}\SS^{[h,+\infty]}_{0,2,\m,\n}(\pp)\;.\label{WI8ab}
\ee
Moreover, as we shall prove in Section \ref{schwing}, for $|\pp| \leq M^{h-1}$
\be
\SS^{[h,+\infty]}_{0,2,\m,\n}(\pp) = \d_{\m,\n}M^{h}\n_{\m} + O(M^{h}e^{2})\;;\label{WI8aba}
\ee
hence, choosing $\pp = (M^{h-1},\vec 0)$ or $\pp = (0,M^{h-1},0)$ in (\ref{WI8ab}) it follows that, as desired:
\be
\n_{\m,h} = O(e^{2})\;.\label{WI8ac}
\ee

\subsection{Ward identity for the effective charge}

Finally, in this Section we shall conclude the analysis of the flow of the running coupling constants by showing that the effective charge stays close to the bare one. Let us define:
\bea
\SS^{[h,K]}_{2,0,\r,\r'}(\kk) &:=&(\b|\L|)\frac{\partial^{2}}{\partial \hat \phi^{-}_{\kk,\s,\r'}\partial\hat\phi^{+}_{\kk,\s,\r}}\WW^{[h,K]}_{\b,L}(0,\phi)\Big|_{\phi=0}\;,\label{Wi8ad}\\
\SS^{[h,K]}_{2,1,\r,\r',\m}(\kk,\pp) &:=& (\b^2\AAA_{\L}|\L|)\frac{\partial^{3}}{\partial \hat\phi^{-}_{\kk,\s,\r'}\partial \hat\phi^{+}_{\kk+\pp,\s,\r}\partial\hat J_{\m,\pp}}\WW^{[h,K]}_{\b,L}(J,\phi)\Big|_{J=\phi =0}\;,\nn\\
p_{\m}\D_{2,1,\r,\r',\m}^{[h,K]}(\kk,\pp) &:=& (\b^2 \AAA_{\L}|\L|)\frac{\partial^{2}}{\partial \hat\phi^{-}_{\kk,\s,\r'}\partial \hat\phi^{+}_{\kk+\pp,\s,\r}}\D_{\b,L}^{[h,K]}(0,\phi)\Big|_{\phi=0}\;,\nn
\eea
where the first and second lines contain respectively the definition of the two and three point fermionic Schwinger functions. Proceeding analogously as for the photon mass, from (\ref{WI9a}) we get that, taking one derivative with respect to $\hat\a_{\pp}$ and two derivatives with respect to the fermionic external fields,
\bea
&&p_{\m}\SS^{[h,K]}_{2,1,\r,\r',\m}(\kk,\pp) = \label{WI8b}
\\&& = e\Big[\big[Q(\vec p)\big]_{\r\a}\SS^{[h,K]}_{2,0,\a,\r'}(\kk) - \SS^{[h,K]}_{2,0,\r,\a}(\kk+\pp)\big[Q(\vec p)\big]_{\a\r'}\Big] + p_{\m}\D_{2,1,\r,\r',\m}^{[h,K]}(\kk,\pp)\;,\nn
\eea
where again the last term is a correction due to the ultraviolet cutoff, which is exponentially vanishing as $K\rightarrow+\infty$, and the matrix $Q(\vec p)$ is due to the presence of the lattice and it is given by
\be
Q(\vec p) = \begin{pmatrix} 1 & 0 \\ 0 & \eu^{-\iu p_{1}} \end{pmatrix}\;;\label{WI8c}
\ee
therefore, taking the ultraviolet limit in $K\rightarrow+\infty$ in (\ref{WI8b}), we find that:
\be
p_{\m}\SS^{[h,+\infty]}_{2,1,\r,\r',\m}(\kk,\pp) = e\Big[\big[Q(\vec p)\big]_{\r\a}\SS^{[h,+\infty]}_{2,0,\a,\r'}(\kk) - \SS^{[h,+\infty]}_{2,0,\r,\a}(\kk+\pp)\big[Q(\vec p)\big]_{\a\r'}\Big]\;.\label{WI8ca}
\ee

The Ward identity (\ref{WI8ca}) will allow us to derive the informations that we need on the renormalized charges on scale $h$. By the analysis of Section \ref{schwing} it follows that, if $\kk = \kk'+\pp_F^{+}$ for $|\kk'| = M^{h}$, $|\kk' + \pp|\leq M^{h}$ and $|\pp|\ll M^{h}$ (we will be interested in the limit $\pp\rightarrow \V0$),
\bea
&&\SS^{[h,+\infty]}_{2,0}(\kk) = \frac{\hat g^{(h)}_{+}(\kk')}{Z_{h-1}}\big(1 + r_{2,0}(\kk)\big)\;,\label{WI14}\\
&&\SS^{[h,+\infty]}_{2,1,\m}(\kk,\pp) =\label{WI13}\\&&= Z_{h-1}\frac{\hat g^{(h)}_{+}(\kk'+\pp)}{Z_{h-1}}\Big( e_{0,h}p_0\G^{0}_{+} + v_{h-1}e_{1,h}\vec p\cdot\vec \G^{+} + p_{\m} r_{2,1,\m}(\kk,\pp)\Big)\frac{\hat g^{(h)}_{+}(\kk')}{Z_{h-1}}\;,\nn
\eea
with $r_{0,2,\m,\n}(\pp)$, $r_{2,1,\m}(\kk,\pp)$, $r_{2,0}(\kk)$ expressed respectively as sums of contributions $r_{0,2,\m,\n}^{N}(\pp)$, $r_{2,1,\m}^{N}(\kk,\pp)$, $r_{2,0}^{N}(\kk)$ with $N\geq 2$, admitting the $N!$-bounds:
\be
\big|r_{0,2,\m,\n}^{N}(\pp)\big| + \big|r_{2,1,\m}^{N}(\kk,\pp)\big| + \big|r_{2,0}^{N}(\kk)\big| \leq (\const.)^{N}\Big(\frac{N}{2}\Big)!\bar\e_{h}^{N}\;.\label{WI15}
\ee
Now, using that
\be
\hat g^{(h)}_{\o}(\kk') - \hat g^{(h)}_{\o}(\kk'+\pp) = \hat g^{(h)}_{\o}(\kk'+\pp)\Big[ p_{0}\G^{0}_{\o} + v_{h-1}\vec p\cdot\vec\G_{\o} \Big]\hat g^{(h)}_{\o}(\kk') + p_{\m}\hat r_{2,0,\m}(\kk,\pp)\;,\label{WI16}
\ee
with $\hat r_{2,0,\m}(\kk,\pp) = O(|\pp|M^{-3h})$\;, we get:
\bea
\SS_{2,0}^{[h,+\infty]}(\kk) - \SS_{2,0}^{[h,+\infty]}(\kk+\pp) &=& \frac{1}{Z_{h-1}}\hat g^{(h)}_{\o}(\kk'+\pp)\Big[ p_{0}\G^{0}_{\o} + v_{h-1}\vec p\cdot\vec\G_{\o} \Big]\hat g^{(h)}_{\o}(\kk') +\nn\\&&+ \frac{p_{\m}}{Z_{h-1}}\big( \hat r_{2,0,\m}(\kk,\pp) + \tilde r_{2,0,\m}(\kk,\pp) \big)\;,\label{WI17}
\eea
with $\tilde r_{2,0,\m}(\kk,\pp)$ expressed as a sum of contributions $\tilde r_{2,0,\m}^{N}(\kk,\pp)$ bounded as
\be
\big|\tilde r_{2,0,\m}^{N}(\kk,\pp)\big|\leq M^{-2h}(\const.)^{N}\Big(\frac{N}{2}\Big)!\bar\e_{h}^{N}\;;\label{WI18}
\ee
therefore, choosing $\kk' = (M^{h},\vec 0)$ and $\pp = (p,\vec 0)$, the Ward identity (\ref{WI8ca}) together with the properties (\ref{WI16})-(\ref{WI18}) implies that, taking the limit $p\rightarrow 0$:
\be
e_{0,h} = e + \G^{0}_{+}\Big(e M^{2h} \tilde r_{2,0,0} + r_{2,1,0}\Big) := e + eA_{0,h}\;,\label{WI19}
\ee
with $A_{0,h}$ a sum of contributions on order $N\geq 2$ admitting $N!$-bounds. Analogously, choosing $\kk' = (0,M^{h},0)$, $\pp=(0,p,0)$ and taking the limit $p\rightarrow 0$:
\be
e_{1,h} = e + \G^{1}_{+}\Big(e v_{h-1}M^{2h}\tilde r_{2,0,1} - \frac{1}{v_{h-1}} r_{2,1,1}\Big) := e + eA_{1,h}\;,\label{WI20}
\ee
with $A_{1,h}$ admitting bounds similar to those of $A_{0,h}$. This concludes the check of our assumptions on the effective couplings on scale $h$.
\section{Asymptotic Lorentz invariance}\label{LI}
\setcounter{equation}{0}
\renewcommand{\theequation}{\ref{LI}.\arabic{equation}}

In the previous Section we have found that, thanks to the Ward indentities implied by the gauge invariance of our model, the flow of the effective couplings can be controlled at all orders in renormalized perturbation theory. In this Section we shall see that the Ward identities have another remarkable consequence; they imply an emergent Lorentz symmetry in our model.

In fact, consider the identities (\ref{WI19}), (\ref{WI20}); in a Lorentz invariant theory we would get that $A_{0,h} = A_{1,h}$. In our case, where Lorentz symmetry is explicitly broken by the presence of the lattice, repeating an argument similar to the one leading to (\ref{5.11b}), by the short memory property it follows that:
\be
e_{0,h} - e_{1,h} = E_{1,h}(e_{0,h+1} - e_{1,h+1}) + E_{2,h}(1 - v_{h+1}) + E_{3,h}\;,\label{LI1}
\ee
where $E_{1,h}$, $E_{2,h}$ and $E_{3,h}$ can be expressed as series in the couplings on scales $\geq h+1$, with the $N$-th orders bounded proportionally to $\Big(\frac{N}{2}\Big)!$, $M^{\th h}\Big(\frac{N}{2}\Big)!$, respectively. Therefore, equation (\ref{LI1}) together with the explicit expression (\ref{veff}) for the fixed point $v_{-\io}$ of the effective Fermi velocity implies that:
\be
e_{0,-\io} - e_{1,-\io} = E_{-\io}(e_{0,-\io} - e_{1,-\io})\;,\label{LI2}
\ee
where $E_{-\io} = O(e_{-\io}^{2})$. Equation (\ref{LI2}) implies that $e_{0,-\io} = e_{1,-\io}$ at all orders in renormalized perturbation theory; plugging this result into the equation (\ref{veff}) for $v_{-\io}$ it follows that
\be
v_{-\io} =1\;,\label{LI3}
\ee
which is the speed of light in our units.

\chapter{The correlation functions}\label{seccorr}
\setcounter{equation}{0}
\renewcommand{\theequation}{\ref{seccorr}.\arabic{equation}}

\section{Introduction}\label{secintro4}
\setcounter{equation}{0}
\renewcommand{\theequation}{\ref{secintro4}.\arabic{equation}}

In this Chapter we shall adapt the Renormalization Group techniques introduced to study the free energy in Chapters \ref{capUV}, \ref{sec2.4.2} to evaluate the generating functionals of the Schwinger functions and of the response functions. The only difference with respect to the computation of the free energy discussed in the previous two Chapters is the presence of the external fields, against which we derive in order to get the desired Schwinger or response functions. In the first part of the Chapter, that is in Section \ref{schwing}, we repeat the multiscale analysis for the generating functional of the Schwinger functions in presence of an infrared cutoff on scale $h^{*}$; the crucial properties (\ref{WI8aba}), (\ref{WI14}) required to implement Ward identities at each step of the RG will be proved by keeping $h^{*}$ fixed, while our result (\ref{res1}) on the two point Schwinger function will be recovered by taking the limit $h^{*}\rightarrow-\infty$. 

Then, in the second part of the Chapter, namely in Section \ref{exc}, we will apply the multiscale analysis to compute the generating functional of the excitonic, charge density wave and density-density correlation functions, introduced in (\ref{res3}); here the analysis will be a bit more subtle, since we will have to introduce {\it new} running coupling constants associated with monomials containing the external fields, and we will have to control their flows. This will be done studying the Beta function, or alternatively by using Ward identities. Finally, as an outcome of this procedure and performing an explicit computation, we will get the announced results (\ref{res4}), (\ref{res5}).

\section{The Schwinger functions}\label{schwing}
\setcounter{equation}{0}
\renewcommand{\theequation}{\ref{schwing}.\arabic{equation}}

The multiscale integration used to compute the partition function, described in Chapter \ref{sec2.4.2}, can
be suitably modified in order to compute the generating functional $\WW^{[h^{*},+\infty]}_{\b,L}(J,\phi)$ (introduced in Section \ref{secWIa}), and in particular the two and three-point correlation functions appearing in the Ward identities discussed in Section \ref{secWIa}. From now on we will assume that the ultraviolet limit $K\rightarrow+\infty$ has been taken.

Without any loss of generality, let us choose the external fermionic fields as $\hat\phi^{\pm}_{\kk,\s,\r} = \sum_{\o=\pm}\hat\phi^{\pm}_{\kk',\s,\r,\o}$, where $\hat\phi^{\pm}_{\kk',\s,\r,\o}$ are supported in the infrared region $|\kk'|\leq a_0$; in this way, the integration of the ultraviolet degrees of freedom is exactly the same reported in Chapter \ref{capUV} and Appendix \ref{app2b}. We do so since at the end we will be interested to the scaling properties of the Schwinger functions for values of the fermionic and bosonic external momenta close to the Fermi points and to zero, respectively; and clearly, these properties are not affected by the ultraviolet regime. Therefore, after the integration of the ultraviolet degrees of freedom we are left with:
\be
\eu^{\WW^{[h^{*},+\infty]}_{\b,L}(J,\phi)} = e^{-\b|\L| F_{0}}\int P(\Psi^{(\leq 0)})P_{\geq h^{*}}(dA^{(\leq 0)})e^{\VV^{(0)}(\Psi^{(\leq 0)},A^{(\leq 0)} + J) + B(\Psi^{(\leq 0)},\phi)}\;,\label{A3.1}
\ee
where the fermionic source term is given by:
\be
B(\Psi^{(\leq 0)},\phi) := \sum_{\s,\r,\o}\int \frac{d\kk'}{D}\, \hat\Psi^{(\leq 0)+}_{\kk',\s,\r,\o}\hat\phi^{-}_{\kk',\s,\r,\o} + \hat\phi^{+}_{\kk',\s,\r,\o}\hat\Psi^{(\leq 0)-}_{\kk',\s,\r,\o}\;,\label{A3.1b}
\ee
Then, we proceed in a way analogous
to the one described in Section \ref{multi}; we iteratively integrate the
fields $\Psi^{(0)},A^{(0)}$, $\ldots$, $\Psi^{(h+1)},A^{(h+1)}$, $\ldots$,
and after the integration of the first $|h|$ infrared 
scales we are left with a functional integral similar to (\ref{3.3}), but now involving new terms
depending on $J,\phi$. Let us first consider the case $h\geq h^{*}$; the
regime $h<h^{*}$ will be discussed later.
\subsection{Multiscale analysis: case \texorpdfstring{$h\geq h^{*}$}{h>=h*}} 
We want to inductively prove that
\bea
&&\eu^{\WW^{[h^{*},+\infty]}_{\b,L}(J,\phi)} = \label{A3.2}\\&&=e^{-\b|\L| F_{h}+\SS^{(\ge h)}(\phi)}
\int P(d\Psi^{(\le h)})P_{\geq h^{*}}(dA^{(\le h)})
e^{\VV^{(h)}(\sqrt{Z_h} \Psi^{(\leq h)},A^{(\leq h)} + J)}
\cdot\nn\\
&&\quad\cdot e^{\BBB^{(h)}(\sqrt{Z_h}\Psi^{(\leq h)},\phi,A^{(\leq h)} + J)+W_{R}^{(h)}(\sqrt{Z_h}\Psi^{(\leq h)},\phi,A^{(\leq h)} + J)}\;,\nn
\eea
where: $\SS^{(\ge h)}(\phi)$ is independent of $A,\Psi,J$, 
$W_{R}^{(h)}$ contains terms explicitly depending on $A + J,\Psi$ and of order
$\geq 2$ in $\phi$, while $\BBB^{(h)}$ is given by:
\bea
&&\BBB^{(h)}(\sqrt{Z_h}\Psi,\phi,A) = \nn\\&&= \sum_{\s,\o}\int \frac{d\kk'}{D}\,
\Big[\hat \phi^{+,T}_{\kk',\s,\o}Q_{\o}^{(h+1)}(\kk')^{T}\hat \Psi^{-}_{\kk',\s,\o}  + \hat \Psi^{+,T}_{\kk',\s,\o}
 Q_{\o}^{(h+1)}(\kk')\hat \phi^{-}_{\kk',\s,\o}\Big] + \nn\\&&\quad
+\int \frac{d\kk'}{D}\, \Big[ \hat\phi^{+,T}_{\kk',\s,\o} G_{\o}^{(h+1)}(\kk')
^{T}\frac{\partial}{\partial\hat\Psi^{+}_{\kk',\s,\o}}\VV^{(h)}(\sqrt{Z_h}\Psi,A) +
\nn\\&&\quad+\frac{\partial}{\partial\hat\Psi^{-}_{\kk',\s,\o}}\VV^{(h)}(\sqrt{Z_h}\Psi,A)
G_{\o}^{(h+1)}(\kk')\hat\phi^{-}_{\kk',\s,\o} \Big]\;,\label{A3.3}\eea
where we used the notations (\ref{3.03b}) and
\bea
\frac{\partial}{\partial\hat\Psi^{-}_{\kk',\s,\o}}\VV^{(h)}(\Psi,A)\hat\phi^{-}_{\kk',\s,\o} &:=& \sum_{\r}\frac{\partial}{\partial\hat\Psi^{-}_{\kk',\s,\r,\o}}\VV^{(h)}(\Psi,A)\hat \phi^{-}_{\kk',\s,\r,\o}\;,\nn\\
\hat\phi^{+}_{\kk',\s,\o}\frac{\partial}{\partial\hat\Psi^{+}_{\kk',\s,\o}}\VV^{(h)}(\Psi,A) &:=& \sum_{\r}\hat\phi^{+}_{\kk',\s,\r,\o}\frac{\partial}{\partial\Psi^{+}_{\kk',\s,\r,\o}}\VV^{(h)}(\Psi,A)\;.\label{A3.3ba}
\eea
Moreover, the functions $Q_{\o}^{(h)}$, $G_{\o}^{(h)}$ are defined
by the following recursive relations:
\bea
&&G_{\o}^{(h)}(\kk'):= \sum_{i=h}^{0}\frac{\hat g_{\o}^{(i)}(\kk')}{Z_{i-1}}Q_{\o}^{(i)}(\kk')
\;,\nn\\
&& Q_{\o}^{(h)}(\kk'):= Q_{\o}^{(h+1)}(\kk') - Z_{h}z_{\m,h}k'_{\m}\G^{\m}_{\o}G_{\o}^{(h+1)}(\kk')\;,\label{A3.4}
\eea
with $Q_{\o}^{(1)}(\kk') \equiv 1$, $G_{\o}^{(1)}(\kk') \equiv 0$. Note that, if $\kk'$ is
in the support of $\hat{g}_{\o}^{(h)}(\kk')$,
\bea
&&G_{\o}^{(h)}(\kk') = \frac{\hat{g}_{\o}^{(h)}(\kk')}{Z_{h-1}}Q_{\o}^{(h)}(\kk') +
\frac{\hat g_{\o}^{(h+1)}(\kk')}{Z_{h}}\;,\label{A3.3b}\\
&&Q_{\o}^{(h)}(\kk') = 1 - z_{\m,h}k_\m \G^{\m}_{\o} \hat g_{\o}^{(h+1)}(\kk')\;,\eea
that is $||Q_{\o}^{(h)}(\kk') - 1||\leq (\const.)\,\bar\e_h^{2}$ and
$||\hat{g}_{\o}^{(h)}(\kk')||\leq (\const.)\,Z_{h}^{-1}M^{-h}$.

Clearly, formulas (\ref{A3.2})--(\ref{A3.4}) are true for $h=0$, with
\be 
\SS^{[h,+\infty]}(\phi) =0\,,\quad \BBB^{(0)}(\Psi,\phi,A) = B(\Psi,\phi)\,,\quad W_{R}^{(0)}=0\,.\nn
\ee
Let us now assume that (\ref{A3.2})--(\ref{A3.4}) are valid at scales
$\ge h$, and let us prove that the inductive assumption is reproduced at
scale $h-1$. We proceed as in Section \ref{multi}; first, we renormalize
the free measure by reabsorbing into $\widetilde P(d\Psi^{(\le h)})$ the
term $\exp\{\LL_\Psi\VV^{(h)}\}$, see (\ref{3.9})--(\ref{3.12}), and then we
rescale the fields as in (\ref{3.13}). Similarly, in the definition
of $\BBB^{(h)}$, Eq.(\ref{A3.3}), we rewrite $\VV^{(h)}=
\LL_\Psi\VV^{(h)}+\hat\VV^{(h)}$, combine the terms proportional to
$\LL_\Psi\VV^{(h)}$ with those proportional to $Q^{(h+1)}$,
and rewrite
\bea&& \BBB^{(h)}(\sqrt{Z_h}\Psi,\phi,A)=\hat \BBB^{(h)}(\sqrt{Z_{h-1}}\Psi,\phi,A)
:= \nn\\&&:=\int \frac{d\kk'}{D}\,
\Big[\hat\phi^{+,T}_{\kk',\s,\o} Q_{\o}^{(h)}(\kk')^{T}\hat\Psi_{\kk',\s,\o}  + \hat\Psi^{+,T}_{\kk',\s,\o}
 Q_{\o}^{(h)}(\kk')\hat\phi_{\kk,\s,\o}\Big] + \nn\\&&
\quad+\int \frac{d\kk'}{D}\, \Big[ \hat\phi^{+,T}_{\kk',\s,\o} G_{\o}^{(h+1)}(\kk')^{T}\frac{\partial}{\partial\hat\Psi^{+}_{\kk',\s,\o}}\hat\VV^{(h)}(\sqrt{Z_{h-1}}
\Psi,A) +\nn\\&&\quad+\frac{\partial}{\partial\hat\Psi^{-}_{\kk',\s,\o}}\hat\VV^{(h)}(\sqrt{Z_{h-1}}\Psi,A)
G_{\o}^{(h+1)}(\kk')\hat\phi_{\kk,\s,\o} \Big]\;,\nn\eea
with $Q_{\o}^{(h)}$ defined by (\ref{A3.4}). Finally, we rescale $W_{R}^{(h)}$,
by defining 
\be
\hat W_{R}^{(h)}(\sqrt{Z_{h-1}}\Psi,\phi,A + J):= W_{R}^{(h)}(\sqrt{Z_h}\Psi,\phi,A + J)\;,\nn
\ee
and perform the integration on scale $h$:
\bea
&&\int P(d\Psi^{(h)} dA^{(h)})e^{\hat \VV^{(h)}(\sqrt{Z_{h-1}}\Psi^{(\leq h)},A^{(\leq h)} + J) + \hat\BBB^{(h)}(\sqrt{Z_{h-1}}\Psi^{(\leq h)},\phi,A^{(\leq h)} +
 J) + \hat W_{R}^{(h)}} \nn\\
&&\equiv e^{-\b|\L|\tilde F_h + \SS^{(h-1)}(\phi)+
\VV^{(h-1)}(\sqrt{Z_{h-1}}\Psi^{(\leq h-1)},A^{(\leq h-1)} + J)}\cdot\nn\\&&\hskip3cm\cdot e^{\BBB^{(h-1)}(\sqrt{Z_{h-1}}\Psi^{(\leq h-1)},\phi,A^{(\leq h-1)} +
 J) + W_{R}^{(h-1)}}\;,\label{A3.7}
\eea
where $\SS^{(h-1)}(\phi)$ contains terms depending on $\phi$
but independent of $A^{(\le h-1)}$, $\Psi^{(\le h-1)}$ and $J$. Defining
$\SS^{(\geq h-1)}:=\SS^{(h-1)}+\SS^{(\geq h)}$, we immediately see that
the inductive assumption is reproduced on scale $h-1$.
\subsection{Multiscale analysis: case \texorpdfstring{$h<h^{*}$}{h<h*}} 
For scales smaller than $h^{*}$,
there are no more bosonic fields to be integrated out, and we are left with a
purely fermionic theory, with scaling dimensions $3-2n$, $2n$ being the
number of external fermionic legs, see Theorem \ref{thm1} and following lines.
Therefore, once that the two-legged
subdiagrams have been renormalized and step by step reabsorbed into the free
fermionic measure, we are left with a superrenormalizable theory, as in
\cite{GM}. In particular, the four fermions interaction is irrelevant,
while the wave function renormalization and the Fermi velocity are modified
by a finite amount with respect to their values at $h^{*}$; that is, if
$\bar\e_{h^{*}} = \max_{k\geq h^{*}}\{|e_{\m,k}|,|\n_{\m,k}|\}$:
\be
Z_{h} = Z_{h^{*}}(1 + O(\bar\e_{h^{*}}^{2}))\;,\qquad v_{h} = v_{h^{*}}
(1 + O(\bar\e_{h^{*}}^{2}))\;.\label{A3.17b}
\ee

\subsection{The fermionic two point function}

In this Section we will provide an explicit formula for the interacting two point Schwinger function in presence of a fixed infrared bosonic cutoff; keeping $h^{*}$ finite and choosing the external momenta on scale $h^{*}$ we will prove (\ref{WI14}), necessary to implement Ward identities in the mutliscale analysis, while taking the limit $h^{*}\rightarrow-\infty$ we will prove the announced result (\ref{res1}).

As for the partition function (see Section \ref{multi}), the kernels of the
effective potentials produced by the multiscale
integration of $\WW_{\b,L}^{[h^{*},+\infty]}(J,\phi)$ can be represented as sums
over trees, which in turn can be evaluated as sums over Feynman graphs.
Let us consider first the expansion for the $2$-point Schwinger
function. After having taken functional derivatives with respect
to $\hat\phi^{-}_{\kk',\s,\o}$, $\hat\phi^{+}_{\kk',\s,\o}$ and after having set $J=\phi=0$,
we get an expansion in terms of a new class of trees
$\t\in\TT^{(h^*)}_{\bar k,\bar h, N}$,
with $\bar k\in(-\io,-1]$ the scale of the root and $\bar h>\bar k$;
 these trees are similar to the ones described in
Chapter \ref{sec2.4.2},
up to the following differences.
\begin{enumerate}
\item There are $N+2$ end--points and two of them, called
$v_{1},v_{2}$, are special and, respectively, correspond to
\be
Q_{\o}^{(h_{v_1}-1)}(\kk')^{T}\hat\Psi^{(\le h_{v_1}-1)-}_{\kk',\s,\o}\quad\mbox{or to}\quad
\hat\Psi^{(\le h_{v_2}-1)+,T}_{\kk',\s,\o} Q_{\o}^{(h_{v_2}-1)}(\kk').\nn
\ee
\item The first vertex whose cluster contains both $v_{1}$, $v_{2}$, denoted by
$\bar v$, is on scale $\bar h$. No $\RR$ operation is associated to the
vertices on the line joining $\bar v$ to the root.
\item There are no lines external to the cluster corresponding
to the root.
\item There are no bosonic lines external to clusters on scale $h<h^*$.
\end{enumerate}
In terms of the new trees we can expand the two point Schwinger function as, taking $\kk = \kk' + \pp_{F}^{\o}$ with $|\kk'|\leq a_0$:
\bea
&&\frac{\media{\hat\Psi^{-}_{\kk,\s}\hat\Psi^{+}_{\kk,\s}}^{(h^{*})}_{\b,L}}{\b|\L|} :=  \b|\L|\frac{\partial^{2}}{\partial\hat\phi^{+}_{\kk,\s}\partial\hat\phi^{-}_{\kk,\s}}\WW^{[h_*,+\infty]}_{\b,L}(0,\phi) = \label{A3.21}\\
&&\quad =\sum_{j=h_{\kk'}}^{h_{\kk'}+1}
Q_{\o}^{(j)}(\kk')^{T}
\frac{\hat g_{\o}^{(j)}(\kk')}{Z_{j-1}}Q_{\o}^{(j)}(\kk') + \sum_{N=2}^{\infty}
\sum_{\bar h = -\infty}^{0}\sum_{\bar k = -\io}^{\bar h - 1}
\sum_{\t\in \TT^{(h^*)}_{\bar k,\bar h,N}}\SS_2(\t;\kk)\;,\nn
\eea
where $h_{\kk'}< 0$ is the integer such that $M^{h_{\kk'}}\leq |\kk'|<
M^{h_{\kk'} + 1}$, and $\SS_2(\t;\kk)$ is defined in a way similar to
$\VV^{(h)}(\t)$ in (\ref{4.4}), modulo the modifications described in
items (1)-(4) above. Using the bounds described immediately after
(\ref{A3.3b}), which are valid for $\kk'$ belonging to the support of
$\hat{g}_{\o}^{(h)}(\kk')$, and proceeding as in Section \ref{sec4}, we get bounds on
$\SS_2(\t;\kk)$, which are the analogues of Theorem \ref{thm1}:
\be
\sum_{\bar h = -\infty}^{0}\sum_{\bar k = -\io}^{\bar h - 1}
\sum_{\t\in \TT_{\bar k,\bar h,N}^{(h^*)}}||\SS_2(\t;\kk)||\leq
(\const.)^{N}\bar\e_{h^*}^{N}\big(\frac{N}2\big)!
\frac{\g^{-h_{\kk'}}}{Z_{h_{\kk'}}}\;.\label{A3.22}\ee
Being the bound (\ref{A3.22}) uniform in $h^{*}$, our result (\ref{res1}) 
on the two point Schwinger function
is obtained after taking the limit $h^{*}\rightarrow-\infty$ in (\ref{A3.21}), and taking into account that $Z_{j}$, $v_{j}$ can be replaced with their interpolations $Z(\kk')\simeq |\kk'|^{-\eta}$, $v(\kk')\simeq 1 - (1 - v)|\kk'|^{\tilde\eta}$, provided that the error term $R(\kk')$ in the main result (\ref{res1}) is defined so to take into account the corrections $O(e^2 \log M)$ generated by the replacements; see discussion after (\ref{3a.19z}).

In order to understand (\ref{A3.22}), it is enough to notice that,
as far as dimensional bounds are concerned, the vertices $v_{1}$
and $v_{2}$ play the role of two $\n$ vertices with an external
line (the $\phi$ line) and an extra
$Z_{h_{\kk'}}^{-1/2}M^{-h_{\kk'}}$ factor each. Moreover, since the
vertices on the path ${\mathcal P}_{r,\bar v}$ connecting the root
with $\bar v$ are not associated with any $\RR$ operation, we need
to multiply the value of the tree $\t\in\TT^{(h^*)}_{\bar k,\bar
h,N}$ by $M^{(1/2)(\bar h - \bar k)}M^{(1/2)(\bar k - \bar h)}$,
and to exploit the factor $M^{(1/2)(\bar k - \bar h)}$ in order to
renormalize all the clusters in ${\mathcal P}_{r,\bar v}$. Therefore,
\bea &&\sum_{\bar h = -\infty}^{0}\sum_{\bar k = -\io}^{\bar h - 1}
\sum_{\t\in \TT_{\bar k,\bar h,N}^{(h^*)}}||\SS_2(\t;\kk)||\leq
(\const.)^N\,\big(\frac{N}2\big)!\,\cdot\nn\\&&
\hskip2cm\cdot\frac{\bar\e_{h^*}^{N}}{Z_{h_{\kk'}}}
 \sum_{\bar h\leq h_{\kk'}}\sum_{\bar k\leq \bar h}
 M^{\bar k}M^{\bar h - h_{\kk'}}M^{(1/2)(\bar h - \bar k)}
 M^{-2h_{\kk'}}\qquad\label{A3.23}
\eea
where: the factor $M^{\bar k}$ is due to the fact that
graphs associated to the trees $\t\in \TT^{(h^*)}_{\bar k, \bar h, N}$
have two external lines; the factor $M^{\bar h - h_{\kk'}}$ is
given by the product of the two short memory factors associated to the
two paths connecting $\bar v$ with $v_1$ and $v_2$, respectively; the
``bad'' factor $M^{(1/2)(\bar h - \bar k)}$ is the price
to pay to renormalize the vertices in ${\mathcal P}_{r,\bar v}$;
the $Z_{h_{\kk'}}^{-1}$ and the
last $M^{-2h_{\kk'}}$ are due to the fact that
$v_{1}$, $v_{2}$ behave dimensionally as $\n$
vertices times an extra $Z_{h_{\kk'}}^{-1/2}M^{-h_{\kk'}}$
factor. Performing the summation over $\bar k$ and $\bar h$ in (\ref{A3.23}),
we get (\ref{A3.22}). Note also that, if $\kk'$ and $\kk'+\pp$ are
on scale $h_{\kk'}\simeq {h^*}$, then the derivatives of $||\SS_2(\t;\kk)||$
can be dimensionally bounded as
\be \sum_{\bar h = -\infty}^{0}\sum_{\bar k = -\io}^{\bar h - 1}
\sum_{\t\in \TT_{\bar k,\bar
h,N}^{(h^*)}}||\dpr_\kk^n\SS_2(\t;\kk)||\leq
(\const.)^{N}\bar\e_{h^*}^{N}\big(\frac{N}2\big)!
\frac{\g^{-(1+n)h_{\kk'}}}{Z_{h_{\kk'}}}\;,\label{A3.220}\ee
from which the bound on $\tilde r_{2,0,\m}^{(N)}(\kk,\pp)$ stated in
(\ref{WI18}) immediately follows. 

\subsection{The bosonic two point function}

In this Section we will prove the property (\ref{WI8aba}), which is necessary to extract informations on the effective photon mass on scale $h$ from the Ward identity (\ref{WI8}).

Taking functional derivatives with respect
to $\hat J_{\m,\pp}$, $\hat J_{\n,-\pp}$ and setting $J=\phi=0$,
we get an expansion in terms of a new class of trees
$\t\in\lis\TT^{(h^*)}_{\bar k, N}$,
with $\bar k\leq h^{*}$ the scale of the root;
 these trees are similar to those described in
Chapter \ref{sec2.4.2}, up to the following differences.
\begin{enumerate}
\item There are $N$ endpoints and three of them, called $\tilde v_{1}$, $\tilde v_{2}$, $\tilde v_{3}$, are special. The endpoints
$\tilde v_{1}$, $\tilde v_{2}$ correspond respectively to 
\bea
&&Z_{h_{\tilde v_1}-1}\,e_{\m,h_{\tilde v_1}} \hat j^{(\le h_{\tilde v_1})}_{\m,\pp} - M^{h_{\tilde v_1}} \n_{\m,h_{\tilde v_1}}\hat A_{\m,\pp}\;,\\
&&Z_{h_{\tilde v_2}-1}\,e_{\n,h_{\tilde v_2}} \hat j^{(\le h_{\tilde v_2})}_{\n,-\pp} - M^{h_{\tilde v_2}} \n_{\n,h_{\tilde v_2}}\hat A_{\n,-\pp}\;;\nn
\eea
instead, $\tilde v_{3}$ corresponds to $-M^{h_{\tilde v_3}}\n_{\m,h_{\tilde v_{3}}}\d_{\m,\n}$ (the only non-vanishing tree this endpoint can be atteched vertex is {\it trivial}, that is the vertex is attached directly to the root, and it has root scale $h^{*}$).
\item There are no lines external to the cluster corresponding to the root.
\item There are no bosonic lines external to the clusters on scales $h<h^{*}$. 
\end{enumerate}
Pick $|\pp| \leq M^{h^{*}-1}$; in terms of these trees, the bosonic $2$-point function can be written as
\be
\b\AAA_{\L}\frac{\partial^{2}}{\partial \hat J_{\m,\pp}\partial \hat J_{\n,-\pp}}\WW_{\b,L}^{[h^{*},+\infty]}(J,0) = 
-M^{h^{*}}\n_{\m,h^{*}}\d_{\m,\n} + \sum_{N\geq 2}\sum_{\bar k \leq h^{*}}\sum_{\t\in \lis\TT^{(h^*)}_{\bar k, N}}\tilde \SS_{2}(\t;\pp)\label{bos0}
\ee
and $\tilde\SS_{2}(\t,\pp)$ is defined in a way similar to $\VV^{(h)}(\t)$, 
modulo the modifications described in items (1) -- (4) above. Proceeding as in Section \ref{sec4}, 
we get that
\be
\sum_{\bar k \leq h^{*}}\sum_{\t \in \lis\TT^{(h^*)}_{\bar k, N}}
\big\| \tilde \SS_{2}(\t;\pp) \big\| \leq (\const.)^{N}\bar \e_{h^{*}}^{N}\Big(\frac{N}{2}\Big)! M^{h^{*}}\;.\label{bos2}
\ee
To understand formula (\ref{bos2}), notice first that now all the vertices are renormalized; this is a simple consequence of the fact that the $J$ field is completely equivalent to a $A$ field, because in the multiscale integration the fields $A$, $J$ always appear in the combination $A+J$. Then, from the bounds derived in Section \ref{sec4} it is easy to see that
\be
\t \in \lis\TT^{(h^*)}_{\bar k, N}\Rightarrow \big\| \tilde \SS_{2}(\t;\pp) \big\| \leq (\const.)^{N}\bar \e_{\bar k}^{N}\Big(\frac{N}{2}\Big)! M^{\bar k}\;,
\ee
and this concludes the proof of (\ref{bos2}).
\subsection{The three point function} 

Finally, let us consider the three point Schwinger function; in this Section we shall prove the property (\ref{WI13}), which is needed to derive the Ward identity for the effective charge on a given scale.

Let us pick $\kk = \kk' + \pp_{F}^{\o}$, with $|\kk'|=M^{h^*}$, $|\kk'+\pp|\le M^{h^*}$ and $|\pp|\ll M^{h^*}$,
which is the condition required in order to apply Ward
Identities in the form described in Section \ref{secWI}. 
The expansion of $3$-point function is very
similar to the one just described above for the $2$-point function. The
result can be written in the form
\bea &&(\b^2|\L|\AAA_{\L})\frac{\partial^{3}}{\partial\hat\phi^{-}_{\kk,\s}\partial\hat\phi^{+}_{\kk+\pp,\s}\partial \hat J_{\m,\pp}}\WW^{[h^{*},+\infty]}_{\b,L}(J,\phi) = \label{A3.24}\\&& \bar
e_{\m,h^{*}} [G_{\o}^{(h^{*}-1)}(\kk'+\pp)]^{T} \G^{\m}_{\o}
\hat g^{(\tilde h^{*})}_{\o}(\kk')Q_{\o}^{(h^*)}(\kk') + \sum_{\substack{N\ge 1,\\\bar h
\le h^*}}\sum_{\substack{\bar k<\bar h,\\ h_{v_3}>h^*}}
\sum_{\t\in \TT_{\bar k,\bar h,h_{v_3},
N}^{(h^*)}}\SS_3(\t;\kk,\pp)\nn \eea
where $\TT_{\bar k,\bar h,h_{v_3},N}^{(h^*)}$ is a new class of trees,
with $\bar k<0$ the scale of the root, similar to the trees in
$\TT_{\bar k,\bar h,N}^{(h^*)}$, up to the fact that they have
$N+3$ endpoints rather than $N+2$ (see item (1) in the list preceding
(\ref{A3.21})); three of them are special: $v_1$ and $v_2$
are associated to the same contributions described in item (1) above, while
$v_3$ is associated to a contribution $Z_{h_{\bar v_3}-1}\,
 e_{\m,h_{\bar v_3}} 
\hat j^{(\le h_{\bar v_3})}_{\m,\pp} - M^{h_{\bar v_3}} \n_{\m,h_{\bar v_3}}
\hat A_{\m,\pp}$, with $\bar v_3$ the vertex immediately
preceding $v_3$ on $\t$ (which the endpoint $v_3$ is attached to) and
$h_{v_3}>h^*$. The value of the tree, $\SS_3(\t;\kk,\pp)$, is defined in a way
similar to $\SS_2(\t;\kk)$, modulo the modifications described above.
$\SS_3(\t;\kk,\pp)$ admits bounds analogous to (\ref{A3.22})-(\ref{A3.23});
recalling that $|\kk'|=M^{h^*}$, $|\kk'+\pp|\le M^{h^*}$ and $|\pp|\ll M^{h^*}$,
we find:
\bea&&\sum_{\bar h = -\infty}^{h^*}\sum_{\bar k = -\io}^{\bar h - 1}
\sum_{h_{v_3}=h^*+1}^1
\sum_{\t\in \TT_{\bar k,\bar h,h_{v_3},N}^{(h^*)}}||\SS_3(\t;\kk,\pp)||\leq
(\const.)^N\,\big(\frac{N}2\big)!\,\bar\e_{h^*}^{N}\frac{1}{Z_{h^*-1}}\cdot
\nn\\
&&\hskip2cm\cdot\sum_{\substack{\bar h\leq h^*\\\bar k< \bar h\\ h_{v_3}>h^*}}
M^{(1/2)(\bar k-\bar h)}M^{\bar h - h^*}M^{(1/2)(h^*-h_{v_3})}
M^{-2h^*}\;,\label{A3.2301}
\eea
where: $M^{(1/2)(\bar k-\bar h)}$ is the short memory factor
associated to the path between the root and $\bar v$; $M^{\bar h -
h^*}$ is the product of the two short memory factors associated to
the paths connecting $\bar v$ with $v_1$ and $v_2$, respectively;
$M^{(1/2)(h^*-h_{v_3})}$ is the short memory factor associated to
a path between $h^*$ and $v_3$; $M^{-2h^*}/Z_{h^*-1}$ is the
product of two factors $M^{-h_{\kk'}}Z_{h_{\kk'}-1}^{-1/2}$ associated
to the vertices $v_1$ and $v_2$ (see the discussion following
(\ref{A3.22}) and recall that in this case $h_{\kk'}=h^*$). We remark
that in this case, contrary to the case of the 2-point function,
the fact that there is no $\RR$ operator acting on the vertices on
the path between the root and $\bar v$ does not create any
problem, since those vertices are automatically irrelevant (they
behave as vertices with at least 5 external lines, i.e., $J$,
$\phi^{-}$, $\phi^{+}$ and at least two fermionic lines) and,
therefore, $\RR=1$ on them. Note also that the vertices of type
$J\phi\psi$, which have an $\RR$ operator acting on, can only be
on scale $h^*-1$ or $h^*$ (by conservation of momentum) and,
therefore, the action of the $\RR$ operator on such vertices
automatically gives the usual dimensional gain of the form
$(\const.)\,M^{h_{v}-h_{v'}}$. Performing the summations over $\bar
k,\bar h, h_{v_3}$ in (\ref{A3.2301}), we find the analogue of
(\ref{A3.22}):
\be \sum_{\bar h = -\infty}^{h^*}\sum_{\bar k = -\io}^{\bar h - 1}
\sum_{h_{v_3}=h^*+1}^1
\sum_{\t\in \TT_{\bar k,\bar h,h_{v_3},N}^{(h^*)}}||\SS_3(\t;\kk,\pp)||\leq
(\const.)^N\,\big(\frac{N}2\big)!\,\bar\e_{h^*}^{N}\frac{M^{-2h^*}}{Z_{h^*-1}}
\;,\ee
from which the bound on $r_\m^{N}(\kk,\pp)$ stated in (\ref{WI15}) follows.
\section{Response functions}\label{exc}
\setcounter{equation}{0}
\renewcommand{\theequation}{\ref{exc}.\arabic{equation}}
The techniques introduced to evaluate the free energy and the Schwinger functions cas be used to compute the {\it response functions}, or {\it generalized susceptibilities}; physically, these objects provide further understanding on which is the effect of the electromagnetic interaction on the macroscopic properties of the model, in particular they suggest what possible physical instabilities are enhanced by the interaction, \cite{So}. As for the two pont Schwinger functions, we shall see that the response functions as well decay with interaction dependent anomalous exponents.

\noindent Let us define the imaginary time response functions $C^{(\a)}_{i,j}(\xx - \yy)$ as follows: 
\bea
C^{(\a)}_{i,j}(\xx - \yy) &:=& \media{\r^{(\a)}_{\xx,i};\r^{(\a)}_{\yy,j}}_{\b,L}\;,\label{exc01}\\
\r^{(E_{\pm})}_{\xx,j} &:=& \sum_{\s=\uparrow\downarrow} \Big( a^{+}_{\xx,\s}b^{-}_{\xx + (0,\vec\d_j),\s}\eu^{\iu e A_{(\xx,j)}} \pm b^{+}_{\xx + (0,\vec\d_j),\s}a^{-}_{\xx,\s}\eu^{-\iu e A_{(\xx,j)}} \Big)\;,\nn\\
\r^{(CDW)}_{\xx,j} &:=& \sum_{\s=\uparrow\downarrow} a^{+}_{\xx,\s}a^{-}_{\xx,\s} - b^{+}_{\xx + (0,\vec\d_j),\s}b^{-}_{\xx + (0,\vec\d_j),\s}\;,\nn\\
\r^{(D)}_{\xx,j} &:=& \sum_{\s = \uparrow\downarrow} a^{+}_{\xx,\s}a^{-}_{\xx,\s} + b^{+}_{\xx + (0,\vec\d_j),\s} b^{-}_{\xx + (0,\vec\d_j),\s}\;,\nn
\eea
Being $\r^{(E_{\pm})}$, $\r^{(CDW)}$ and $\r^{(D)}$ gauge-invariant operators, we can represent their grand-canonical correlation functions using a functional integral representation in the Feynman gauge. 
Therefore,
\bea
C^{(\a)}_{i,j}(\xx - \yy) &=& \frac{\partial^{2}}{\partial\Phi^{(\a)}_{\xx,i}\partial\Phi^{(\a)}_{\yy,j}} \log \int P(d\Psi)P(dA)\eu^{V(\Psi,A) + D(\Phi,\Psi,A)}\Big|_{\Phi=0}\;,\nn\\
&=:& \frac{\partial^{2}}{\partial\Phi^{(\a)}_{\xx,i}\partial\Phi^{(\a)}_{\yy,j}} \lis\WW_{\b,L}(\Phi)\Big|_{\Phi=0}\;,\nn\\
D(\Phi,\Psi,A) &:=& \int_{0}^{\b}dx_{0} \sum_{\vec x\in\L}\sum_{i=1,2,3} \Phi^{(\a)}_{\xx,i}\r^{(\a)}_{\xx,i}\;,\label{exc02}
\eea
where $\Phi_{\xx,i}^{(\a)} = (\b\AAA_{\L})^{-1}\sum_{\qq\in \PPP_{\b,L}} \eu^{\iu \kk\xx}\hat\Phi^{(\a)}_{\qq,i}$ (we choose $\hat\Phi^{(\a)}_{\qq,i}\in\RRR$ for simplicity), and $\r^{(\a)}$ are given by (\ref{exc01}) with $a^{\pm}$, $b^{\pm}$, $A$ replaced by fermionic and bosonic Gaussian fields.

\subsection{Renormalization group analysis}
All these generalized susceptibilities can be evaluated using multiscale analysis and renormalization group. As we are going to see, the external fields $\Phi^{(\a)}$ behave dimensionally as $A$ fields; the main difference with respect to the analyses of Chapter \ref{sec4} and Section \ref{schwing} is that now we will have to introduce {\it new} running coupling constants, associated with the relevant and marginal $\Phi$-dependent monomials. The flow of some of these new running coupling constants, namely those associated with the monomials $\Phi^{(\a)}\Psi^{+}\Psi^{-}$, will {\it diverge} as a power laws in the infrared, in a way similar to the wave function renormalization {\it but} in general with anomalous exponents different from $\eta = \frac{e^2}{12\pi^2} + \ldots$. A non-vanishing difference between these new anomalous exponents and $\eta$ will give rise to the anomalous scaling of the response functions.

%
\subsubsection{Multiscale analysis}\label{multiexc} The ultraviolet degrees of freedom can be integrated out by following a procedure completely analogous to the one described in Section \ref{sec2.4.1} and Appendix \ref{app2b}. It follows that:
\bea
&&e^{\lis\WW_{\b,L}(\Phi)} = \label{exc3}\\&&e^{-\b|\L|F_{0} + \CC^{(\geq 0)}(\Phi)}\int P(d\Psi^{(\leq 0)})P(dA^{(\leq 0)})e^{\VV^{(0)}(\Psi^{(\leq 0)},A^{(\leq 0)}) + \DD^{(0)}(\Phi,\Psi^{(\leq 0)},A^{(\leq 0)})}\;,\nn
\eea
where all the quantities appearing in (\ref{exc3}) have been defined Section \ref{multi}, except for $\CC^{(\geq 0)}(\Phi)$ and $\DD^{(0)}$, which contain $\Phi$-dependent terms. In particular,
\bea
&&\DD^{(0)}(\Phi,\Psi,A) = \sum_{\substack{q\geq 1\\ n+m\geq 1 }}\sum_{\substack{\ul\s,\ul\o,\ul\r \\ \ul\m,\ul\a,\ul j}}\int \prod_{i=1}^{n}\hat \Psi_{\kk'_{2i-1},\s_{i},
\r_{2i-1},\o_{2i-1}}^{+}\hat \Psi_{\kk'_{2i},\s_{i},\r_{2i},\o_{2i}}^{-}\prod_{i=1}^{m}\hat A_{\m_i, \pp_i}\cdot\nn\\&&\cdot
\prod_{r=1}^{q}\hat\Phi^{(\a_r)}_{\qq_r,j_{r}}W^{(0),\ul{\a},\ul{j}}_{2n,m,q,{\ul \r}, {\ul \o},\ul\m}(\{\kk'_i\},\{\pp_j\},\{\qq_r\})\d\left(\sum_{j=1}^{m}\pp_j
+ \sum_{r=1}^{q}\qq_r + \sum_{i=1}^{2n}(-1)^{i}\kk'_i\right)\;,\nn\\
&&\CC^{(\geq 0)}(\Phi) = \sum_{q\geq 1}\int \prod_{r=1}^{q}\hat\Phi^{(\a_r)}_{\qq_r,j_r}\,W^{(0),\ul \a,\ul j}_{0,0,q}(\{\qq_i\})\d\left(\sum_{i=1}^{q}\qq_{i}\right)\;,\label{exc4}
\eea
where the kernels are computed following the rules explained in Appendix \ref{app2b} and Section \ref{sec4}, by taking into account the presence of new vertices, corresponding to the source terms. It is easy to see that the external fields $\Phi^{(\a)}$ behave dimensionally as $A$ fields; therefore, the kernels proportional to the monomials $\Phi^{(\a)}A$ behave as relevant ones, while the ones proportional to $\Phi^{(\a)}\Psi^{+}\Psi^{-}$, $\Phi^{(\a)}AA$, $\Phi^{(\a)}\Phi^{(\a)}A$ are marginal. 

As for the effective potential $\VV^{(0)}$, we split the $\Phi$-dependent part of the interaction as $\LL \DD^{(0)} + \RR \DD^{(0)}$, with $\RR = 1-\LL$; the localization operator $\LL$ acts linearly on the kernels of $\DD^{(0)}$ in the following way:
\bea
\LL W^{(0),\a,j}_{0,1,1,\m}(\pp) &:=& W^{(0),\a,j}_{0,1,1,\m}(\V0) + p_{\n}\partial_{\n}W^{(0),\a,j}_{0,1,1,\m}(\V0) =: \n^{\a,j}_{\m,0} + p_{\n}\n^{\a,j}_{\m,\n,0}\;,\nn\\
\LL W^{(0),\a,j}_{2,0,1,{\ul \r},{\ul \o}}(\kk,\qq) &:=& W^{(0),\a,j}_{2,0,1,{\ul \r},{\ul \o}}(\V0,\pp_{F}^{\o_1} - \pp_{F}^{\o_2})\;,\nn\\
\LL W^{(0),\a,j}_{0,2,1,{\ul \m}}(\pp,\qq) &:=& W^{(0),\a,j}_{0,2,1,{\ul \m}}(\V0,\V0) =: \zeta^{\a,j}_{\ul{\m},0}\;,\nn\\
\LL W^{(0),{\ul\a},{\ul j}}_{0,1,2,\m}(\pp,\qq) &:=& W^{(0),{\ul\a},{\ul j}}_{0,1,2,\m}(\V0,\V0) =: \kappa^{\ul\a,\ul j}_{\m,0}\;.\label{exc5a}
\eea
and $\LL W^{(0)}_{2n,m,q} =0$ otherwise. As for the case $\Phi=0$, the number of running coupling constants will be reduced by the symmetry properties listed in the following Lemma, which will play the same role of Lemma \ref{lem2.4} in the analysis of the free energy.
\begin{lemma}\label{lemexc}
The Gaussian integrations $P(d\Psi)$, $P(dA)$, the interaction $V(\Psi,A)$ and the source term $D(\Phi,\Psi,A)$ are invariant under the transformations (1) -- (8) of Lemma \ref{lem2.4}, provided that $\hat \Phi^{(\a)}_{\qq,i}$ transforms in the following way (remember that $1'=1$, $2'=3$, $3'=2$):
\begin{itemize}
\item[(4)] $\hat \Phi^{(\a)}_{\qq,j}\rightarrow \hat \Phi^{(\a)}_{(q_0,T\vec q),j+1}$.
\item[(5)] $\hat \Phi^{(\a)}_{\qq,j}\rightarrow \hat \Phi^{(\a)}_{-\qq,j}$.
\item[(6.a)] $\hat\Phi^{(E_{\pm})}_{\qq,j}\rightarrow \pm\hat\Phi^{(E_{\pm})}_{\tilde\qq,j'}\eu^{-\iu \vec q(\vec \d_j - \vec\d_1)}$; $\hat\Phi^{(D)}_{\qq,j}\rightarrow \hat\Phi^{(D)}_{\tilde\qq,j'}\eu^{-\iu \vec q(\vec\d_j - \vec\d_1)}$; 

$\hat\Phi^{(CDW)}_{\qq,j}\rightarrow -\hat\Phi^{(CDW)}_{\tilde\qq,j'}\eu^{-\iu\vec q(\vec\d_j - \vec\d_1)}$, with $\tilde\qq = (q_0,-q_1,q_2)$.


\item[(6.b)] $\hat\Phi^{(\a)}_{\qq,j}\rightarrow \hat\Phi^{(\a)}_{\tilde\qq,j'}$, with $\tilde\qq = (q_0,q_1,-q_2)$;
\item[(7)] $\hat\Phi^{(E_{\pm})}_{\qq,j}\rightarrow \pm\hat\Phi^{(E_{\pm})}_{ - \tilde\qq,j}$; $\hat\Phi^{(CDW)}_{\qq,j}\rightarrow \hat\Phi^{(CDW)}_{- \tilde\qq,j}$; $\hat\Phi^{(D)}_{\qq,j}\rightarrow \hat\Phi^{(D)}_{ - \tilde\qq,j}$, with $\tilde\qq = (q_0,-\vec q)$.


\item[(8)] $\hat\Phi^{(E_{\pm})}_{\qq,j}\rightarrow \hat\Phi^{(E_{\pm})}_{\tilde\qq,j}$; $\hat\Phi^{(CDW)}_{\qq,j}\rightarrow -\hat\Phi^{(CDW)}_{\tilde\qq,j}$; $\hat\Phi^{(D)}_{\qq,j}\rightarrow -\hat\Phi^{(D)}_{\tilde\qq,j}$, with $\tilde\qq = (-q_0,\vec q)$.


\end{itemize}
\end{lemma}

\noindent{\it Proof.} The proof is a straightforward extension of the one of Lemma \ref{lem2.4}; we leave the details to the reader.\qed \\

\noindent By symmetry (4) it follows that:
\bea
&&W^{(0),\a,j}_{2,0,1,{\ul \r}, {\ul \o}}(\V0,\pp_{F}^{\o_1} - \pp_{F}^{\o_2}) =\label{exc5b}\\&& \exp\Big\{\iu \vec p_{F}^{\o_{1}}(\vec\d_2 - \vec\d_1)(\r_1 - 1) - \iu\vec p^{\o_2}_{F}(\vec\d_2 - \vec\d_1)(\r_2 - 1) \Big\} \cdot\nn\\&&\quad\cdot W^{(0),\a,j+1}_{2,0,1,{\ul \r}, {\ul \o}}(\V0,\pp_{F}^{\o_1} - \pp_{F}^{\o_2})\;,\nn
\eea
while by symmetries (4) -- (8) we get:
\bea
W^{(0),E_{\pm},1}_{2,0,1, {\ul \o}}(\V0,\pp_{F}^{\o_1} - \pp_{F}^{\o_2}) &=:& Z^{E_{\pm}}_{\ul{\o},0}\begin{pmatrix} 0 & 1 \\ \pm 1 & 0 \end{pmatrix} =: Z^{E_{\pm}}_{\ul{\o},0}\G^{E_{\pm}}\;,\nn\\
W^{(0),CDW,1}_{2,0,1, {\ul \o}}(\V0,\pp_{F}^{\o_1} - \pp_{F}^{\o_2}) &=:& Z^{CDW}_{\ul{\o},0}\begin{pmatrix} 1 & 0 \\ 0 & -1 \end{pmatrix} =: Z^{CDW}_{\ul{\o},0}\G^{CDW}\;,\label{exc6}\\
W^{(0),D,1}_{2,0,1, {\ul \o}}(\V0,\pp_{F}^{\o_1} - \pp_{F}^{\o_2}) &=:& Z^{D}_{\ul{\o},0}\begin{pmatrix} 1 & 0 \\ 0 & 1 \end{pmatrix} =: Z^{D}_{\ul{\o},0}\G^{D}\;,\nn
\eea
with $Z^{\a}_{\ul\o,0} = Z^{\a}_{-\ul\o,0}$, $Z^{\a}_{\ul{\o},0}\in \RRR$, and $Z^{\a}_{\ul\o,0} = 1 + O(e^{2})$. Therefore, from (\ref{exc5b}) and (\ref{exc6}) we find that
\bea
&&\sum_{\ul\r}\int \frac{d\kk'}{D}\frac{d\qq}{(2\pi)^3}\, \hat\Phi^{(\a)}_{\qq,j} \hat\Psi^{(\leq 0)+}_{\kk'+\qq,\s,\r_1,\o_1} \hat\Psi^{(\leq 0)-}_{\kk',\s,\r_2,\o_2}W^{(0),\a,j}_{2,0,1,{\ul\r},{\ul \o}}(\V0,\pp_{F}^{\o_1} - \pp_{F}^{\o_2})\nn\\
&& = \int\frac{d\kk'}{D}\frac{d\qq}{(2\pi)^3}\, \hat\Phi^{(\a)}_{\qq,j}\hat \Psi^{(\leq 0)+}_{\kk'+\qq,\s,\o_1}Z^{\a}_{\ul\o,0}\G^{(\a)}_{\ul\o,j}\hat \Psi^{(\leq 0)-}_{\kk',\s,\o_2}\;,\label{exc6b}
\eea
where the matrices $\G^{(\a)}_{\ul\o,j}$ are easily obtained from (\ref{exc5b}), (\ref{exc6}):
\bea
\G^{(E^{\pm})}_{\ul\o,j} &:=& \begin{pmatrix} 0 & \eu^{-\iu (j - 1) \vec p_{F}^{\o_2}(\vec\d_2 - \vec\d_1)} \\ \pm \eu^{\iu (j - 1) \vec p_{F}^{\o_1}(\vec\d_2 - \vec\d_1)} & 0 \end{pmatrix}\;,\label{exc6c}\\
\G^{(D)}_{\ul\o,j} &:=& \begin{pmatrix} 1 & 0 \\ 0 & \eu^{\iu (\vec\d_2 - \vec\d_1)(\vec p_{F}^{\o_2} - \vec p_{F}^{\o_1})(j-1)} \end{pmatrix}\;,\nn\\
\G^{(CDW)}_{\ul\o,j} &:=& \begin{pmatrix} 1 & 0 \\ 0 & -\eu^{\iu (\vec\d_2 - \vec\d_1)(\vec p_{F}^{\o_2} - \vec p_{F}^{\o_1})(j-1)} \end{pmatrix}\;.\nn
\eea

Repeating the analysis described in detail in Chapter \ref{sec2.4.2}, we can integrate the fields $\psi^{(0)},A^{(0)}$, $\Psi^{(-1)}, A^{(-1)}$, ..., $\Psi^{(h)}, A^{(h)}$, ...,  in an interative way; after the integration of the first $|h|$ infrared scales we get that:
\bea
&&e^{\lis\WW_{\b,L}(\Phi)} = e^{-\b|\L| F_{h} + \CC^{(\geq h)}(\Phi)}\cdot\label{exc7}\\
&&\quad \cdot\int P(d\Psi^{(\leq h)}dA^{(\leq h)})e^{\VV^{(h)}(\sqrt{Z_h}\Psi^{(\leq h)}, A^{(\leq h)}) + \DD^{(h)}(\Phi,\sqrt{Z_h}\psi^{(\leq h)},A^{(\leq h)})}\;,\nn
\eea
where all the objects appearing in (\ref{exc7}) have been defined in Section \ref{multi} except for $\CC^{(\geq h)}$, $\DD^{(h)}$, which are given by (\ref{exc4}) with $0$ replaced by $h$, and the kernels $W^{(h)}_{2n,m,q}$ will be defined inductively. We proceed as in Section \ref{multi}; first, we renormalize
the free measure by reabsorbing into $\widetilde P(d\Psi^{(\le h)})$ the term $\exp\{\LL_\Psi\VV^{(h)}\}$, see (\ref{3.9})--(\ref{3.12}), and then we rescale the fields as in (\ref{3.13}). Similarly, we rewrite
\be
\DD^{(h)}(\Phi,\sqrt{Z_h}\Psi,A) =: \hat \DD^{(h)}(\Phi,\sqrt{Z_{h-1}}\Psi,A)\;,\label{exc9}
\ee
and we split $\hat \DD^{h}$ as $\LL \hat \DD^{(h)} + \RR\hat \DD^{(h)}$, where the action of $\LL$ on the kernels of $\hat\DD^{(h)}$ is defined as follows:
\bea
\LL W^{(h),\a,j}_{0,1,1,\m}(\pp) &:=& W^{(h),\a,j}_{0,1,1,\m}(\V0) + p_{\n}\partial_{\n}W^{(h),\a,j}_{0,1,1,\m}(\V0) =: M^{h}\n^{\a,j}_{\m,h} + p_{\n}\n^{\a,j}_{\m,\n,h}\;,\nn\\
\LL W^{(h),\a,j}_{2,0,1,{\ul \r},{\ul \o}}(\kk,\qq) &:=& W^{(h),\a,j}_{2,0,1, {\ul \r}, {\ul \o}}(\V0,\pp_{F}^{\o_1} - \pp_{F}^{\o_2})\;,\nn\\
\LL W^{(h),\a,j}_{0,2,1,{\ul \m}}(\pp,\qq) &:=& W^{(h),\a,j}_{0,2,1,{\ul \m}}(\V0,\V0) =: \zeta^{\a,j}_{\ul{\m},h}\;,\nn\\
\LL W^{(h),{\ul\a},{\ul j}}_{0,1,2,\m}(\pp,\qq) &:=& W^{(h),{\ul\a},{\ul j}}_{0,1,2,\m}(\V0,\V0) =: \kappa^{\ul\a,\ul j}_{\m,h}\;.\label{exc5}
\eea
Moreover, the symmetry properties (1) -- (8) of Lemma \ref{lemexc}, which are preserved by the multiscale integration, imply the analogous of (\ref{exc6b}), that is:
\bea
&&\sum_{\ul\r}\int \frac{d\kk'}{D}\frac{d\qq}{(2\pi)^3}\, \hat\Phi^{(\a)}_{\qq,j} \hat\Psi^{(\leq h)+}_{\kk'+\qq,\s,\r_1,\o_1} \hat\Psi^{(\leq h)-}_{\kk',\s,\r_2,\o_2}W^{(h),\a,j}_{2,0,1, {\ul \o}}(\V0,\pp_{F}^{\o_1} - \pp_{F}^{\o_2})\nn\\
&& = \int\frac{d\kk'}{D}\frac{d\qq}{(2\pi)^3}\, \hat\Phi^{(\a)}_{\qq,j}\hat \Psi^{(\leq h)+}_{\kk'+\qq,\s,\o_1}Z^{\a}_{\ul\o,h}\G^{(\a)}_{\ul\o,j}\hat \Psi^{(\leq h)-}_{\kk',\s,\o_2}\;,\label{exc10}
\eea
where $Z^{\a}_{\ul\o,h} = Z^{\a}_{-\ul\o,h}$, $Z^{\a}_{\ul{\o},h}\in \RRR$, and the matrices $\G^{(\a)}_{\ul\o,j}$ are defined in (\ref{exc6c}). At this point, we are ready to integrate the scale $h$; proceeding as after (\ref{3.14a}), defining
\bea
&&e^{\VV^{(h-1)}(\sqrt{Z_{h-1}}\Psi,A) + \DD^{(h-1)}(\Phi,\sqrt{Z_{h-1}}\Psi,A) - \b|\L|\tilde F_{h} - \CC^{(h-1)}(\Phi)} := \label{exc11}\\
&& \int P_{h}(d\Psi^{(h)})P_{h}(dA^{(h)})e^{\hat \VV^{(h)}(\sqrt{Z_{h-1}}(\Psi + \psi^{(h)}), A + A^{(h)}) + \hat\DD^{(h)}(\Phi,\sqrt{Z_{h-1}}(\Psi + \Psi^{(h)}), A + A^{(h)})}\nn
\eea
our inductive assumption in reproduced at the scale $h-1$ with $\CC^{(\geq h-1)} = \CC^{(\geq h)} + \CC^{(h-1)}$. The integration is performed using the Gallavotti -- Nicol\` o tree expansion described in Section \ref{sec4}; the only difference with respect to what has been discussed in Section \ref{sec4} is that the resulting trees have new endpoints, corresponding to the various contributions to $\LL\hat\DD^{(h)}$. 

The outcome of this procedure is that the kernels of $\CC^{(\geq h-1)}$ are expressed in terms of a renormalized expansion which involves the old effective parameters plus the new ones $\{ \n^{\a,j}_{\m,k},\,\n^{\a,j}_{\m,\n,k},\,\zeta^{\a,j}_{\ul{\m},k},\,\kappa^{\ul\a,\ul j}_{\m,k},\,Z^{\a}_{\ul\o,k}\}_{k>h}$; to close our renormalization group analysis we have to control the flow of these quantities.
\subsubsection{The flow of the running coupling constants}\label{excrcc} In this Section we show that the flows of all the new effective parameters {\it except} $Z^{\a}_{\ul\o,h}$ are bounded; in particular, all the new running coupling constants will be bounded by $O(e)$ uniformly in $h$. To see this, we shall use Ward identities (for $\n^{\a,j}_{\m,h}$, $\zeta^{\a,j}_{\ul{\m},h}$, $\kappa^{\ul\a,\ul j}_{\m,h}$) and symmetry considerations (for $\n^{\a,j}_{\m,\n,h}$). Regarding $Z^{\a}_{\ul\o,h}$, these quantities in general will behave as $M^{-\eta^{\a}h}$, that is their evolutions are governed by {\it anomalous exponents}, which will be explicitly computed at lowest order; these computations will allow us to characterize the long distance behavior of the excitonic, charge density wave and density--density susceptibilities.
\paragraph{The flows of $\n^{\a,j}_{\m,h}$, $\zeta^{\a,j}_{\ul{\m},h}$, $\kappa^{\ul\a,\ul j}_{\m,h}$.} As for $\n_{\m,h}$, the ``mass'' of the photon field, the flow of these running coupling constants will be controlled implementing Ward identities at each RG step. Notice that, as for $\n_{\m,h}$, the Beta function does not provide any help here; for instance, the evolution equation of $\n^{\a,j}_{\m,h}$ is given by
\be
\n^{\a,j}_{\m,h} = M\n^{\a,j}_{\m,h+1} + \b^{\a,j}_{\m,h+1}\;,\label{exc12}
\ee
and a single non-vanishing order in $\b^{\a,j}_{\m,h+1}$ would generically produce an exponentially divergent flow for $\n^{\a,j}_{\m,h}$ in the limit $h\rightarrow-\infty$. Similar behaviors would take place for $\zeta^{\a,j}_{\ul{\m},h}$, $\kappa^{\ul\a,\ul j}_{\m,h}$. 

\begin{figure}[hbtp]
\centering
\includegraphics[width=.7\textwidth]{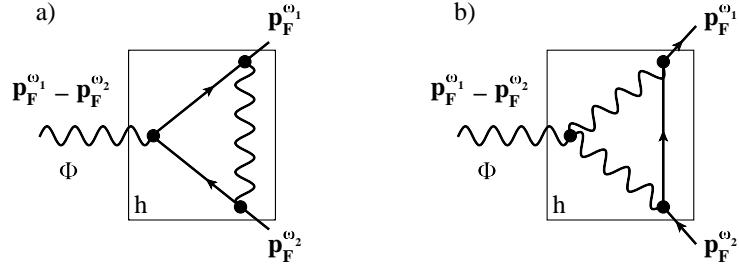}
\caption{$a)$ A graph contributing to $Z^{\a}_{\ul\o,h+1}\b^{\a}_{\ul\o,h,1}$; $b)$ a graph contributing to $\b^{\a}_{\ul\o,h,2}$. The graph $b)$ is of order $e^{2}M^{\th h}$; in fact, the same graph obtained replacing the fermionic propagator with its relativistic approximation $g^{(h)}_{rel}(\kk)$ is vanishing by parity.}\label{figexc}
\end{figure}

The Ward identities that we need are implied the gauge invariance of our theory {\it plus} the fact that the source terms appearing in the generating functional $\lis\WW_{\b,L}(\Phi)$ are separately invariant under local gauge transformations. We will proceed as for $\n_{\m,h}$, $e_{\m,h}$, see Section \ref{secWI}; we introduce a {\it reference model} which is regularized by a bosonic infrared cut-off on scale $h$, and we derive Ward identities for this new theory, exploiting the facts that: (i) the running coupling constants on scales greater than the one of the infrared cut-off are the same of the non-regularized model; (ii) the infrared cut-off does not break gauge invariance, and does not produces corrections to the WI; (iii) the new model can be studied repeating the multiscale integration described in Section \ref{secWI}, with the only difference that after the scale $h$ the theory becomes {\it superrenormalizable}, which in particular implies that the running coupling constants ``cease to flow'' on scale $h$. See Section \ref{schwing} for the discussion of the multiscale integration in presence of a bosonic infrared cut--off in the case of the generating functional of the Schwinger functions; a completely analogous analysis can be repeated in presence of $\Phi^{(\a)}$ external fields. Let:
\be
e^{\lis\WW_{\b,L}^{[h,+\infty]}(\Phi,J)} := \int P(d\Psi)P_{\geq h}(dA)\eu^{V(\Psi,A+J) + D(\Phi,\Psi,A+J)}\;,\label{exc13}
\ee
where the subscript $\geq h$ denotes the presence of an infrared bosonic cutoff on scale $h$, to be imposed as in (\ref{refm2}). We assume that the ultraviolet limit $K\rightarrow+\infty$ has been taken; the presence of a finite u.v. cutoff can be discussed as in Appendix \ref{app5}, and produces corrections that vanish exponentially in the limit of cutoff removal. The invariance of the theory under local gauge transformations implies that
\bea
0 &=& \frac{\partial^{2}}{\partial\hat\Phi^{(\a)}_{\qq,j}\partial\hat\a_{-\qq}}\lis\WW^{[h,+\infty]}_{\b,L}(\Phi,J+\partial\a)\Big|_{\Phi=J=\a=0} = \nn\\ &=& \iu q_{\n}\frac{\partial^{2}}{\partial\hat\Phi^{(\a)}_{\qq,j}\partial\hat J_{\n,-\qq}}\lis\WW^{[h,+\infty]}_{\b,L}(\Phi,J)\Big|_{\Phi=J=0}\;,\label{exc14}\\
0 &=& \frac{\partial^{3}}{\partial\hat\Phi^{(\a)}_{\qq,j} \partial\hat J_{\m_1,\pp}\partial \hat\a_{-\qq-\pp}}\lis\WW^{[h,+\infty]}_{\b,L}(\Phi, J+\partial\a)\Big|_{\Phi=J=\a=0} = \nn\\ &=& \iu(q_{\n} + p_{\n})\frac{\partial^{3}}{\partial\hat\Phi^{(\a)}_{\qq,j} \partial\hat J_{\m_1,\pp}\partial \hat J_{\n,-\qq-\pp}}\lis\WW^{[h,+\infty]}_{\b,L}(\Phi, J)\Big|_{\Phi=J=0}\;,\nn\\
0 &=& \frac{\partial^{3}}{\partial\hat\Phi^{(\a)}_{\qq,j}\partial\hat\Phi^{(\a')}_{\pp,j'}\partial\hat \a_{-\qq-\pp}}\lis\WW^{[h,+\infty]}_{\b,L}(\Phi,J+\partial\a)\Big|_{\Phi=J=\a=0} = \nn\\&=& \iu(q_{\n} + p_{\n})\frac{\partial^{3}}{\partial\hat\Phi^{(\a)}_{\qq,j}\partial\hat\Phi^{(\a')}_{\pp,j'}\partial\hat J_{\n,-\qq-\pp}}\lis\WW^{[h,+\infty]}_{\b,L}(\Phi,J)\Big|_{\Phi=J=0}\;.
\eea
At the same time we know that, repeating an analysis completely analogous to the one performed for $e_{\m,h}$ and $\n_{\m,h}$, choosing $q_{\n} = \d_{\m,\n}q$, $p_{\n} = \d_{\m,\n}p$, such that $|p|\leq M^{h-1}$, $|q|\leq M^{h-1}$, $|q+p|\leq M^{h-1}$,
\bea
\frac{\partial^{2}}{\partial\hat\Phi^{(\a)}_{\qq,j}\partial\hat J_{\m,-\qq}}\lis\WW^{[h,+\infty]}_{\b,L}(\Phi,J)\Big|_{\Phi=J=0} &=& M^{h}\big[\n^{\a,j}_{\m,h} + O(e^{2})\big]\;,\label{exc15}\nn\\
\frac{\partial^{3}}{\partial\hat\Phi^{(\a)}_{\qq,j} \partial\hat J_{\m_{1},\pp}\partial \hat J_{\m,-\qq-\pp}}\lis\WW^{[h,+\infty]}_{\b,L}(\Phi, J)\Big|_{\Phi=J=0} &=& \zeta^{\a,j}_{\ul\m,h} + O(e^2)\;,\nn\\
\frac{\partial^{3}}{\partial\hat\Phi^{(\a)}_{\qq,j}\partial\hat\Phi^{(\a')}_{\pp,j'}\partial\hat J_{\m,-\qq-\pp}}\lis\WW^{[h,+\infty]}_{\b,L}(\Phi,J)\Big|_{\Phi=J=0} &=& \kappa^{\ul\a,\ul j}_{\m,h} + O(e^2)\;;
\eea
putting together (\ref{exc15}) with the Ward identities (\ref{exc14}) we get that 
\be
\n^{\a,j}_{\m,h} = O(e^{2})\;,\qquad \zeta^{\a,j}_{\ul\m,h} = O(e^{2})\;,\qquad \kappa^{\ul\a,\ul j}_{\m,h} = O(e^{2})\;, 
\ee
as desired.
\paragraph{The flow of $\n^{\a,j}_{\m,\n,h}$.} These running coupling constants cannot be studied using Ward identities, at least not for all $\m$, $\n$; therefore, to prove the boundedness of their flows we will rely on simple symmetry considerations. The evolution equation for $\n^{\a,j}_{\m,\n,h}$ has the form:
\be
\n^{\a,j}_{\m,\n,h} = \n^{\a,j}_{\m,\n,h+1} + \b^{\a,a}_{\m,\n,h+1}\;,\label{exc16}
\ee
where, setting $\tilde\e_{h} := \max_{k>h}\{e_{\m,k},\,\n_{\n,k},\,\n^{\a,j}_{\m,k},\,\n^{\a,j}_{\m,\n,k},\,\zeta^{\a,j}_{\ul{\m},k},\,\kappa^{\ul\a,\ul j}_{\m,k}\}$ (the $\max$ is intended over all the labels), the Beta function appearing in (\ref{exc16}) is expressed as a renormalized series in the running coupling constants with the $N$-th order bounded proportionally to $(N/2)!\tilde\e_{h}^{N}$. However, this dimensional bound is too pessimistic in this case; in fact, the graphs contributing to $\b^{\a,a}_{\m,\n,h+1}$ contain an {\it even} number of fermion propagators, and one propagator (fermionic or bosonic) is derived with respect to the $\m$-th component of the momentum. Therefore, rewriting the fermion propagator as $\hat g_{\o}^{(k)}(\kk') = \hat g^{(k)}_{\o,rel}(\kk') + \tilde g_{\o}^{(k)}(\kk')$, where $\hat g^{(k)}_{\o,rel}(\kk')$ is obtained setting $R_{k-1,\o}(\kk')=0$ in (\ref{3.16}) (that is, it is the linear part of the single scale propagator) and $\|\tilde g^{(k)}(\kk)\|\leq (\const.)M^{k}\|g^{(k)}(\kk)\|$, we decompose the Beta function as
\be
\b^{\a,a}_{\m,\n,h} = \b^{\a,a,(rel)}_{\m,\n,h} + \tilde\b^{\a,a}_{\m,\n,h}\;,\label{exc17}
\ee
where:
\begin{itemize}
\item[(i)] the terms labelled by $(rel)$ are obtained by considering only trees with endpoints on scales $\leq 0$, and replacing all fermion propagators with their relativistic counterparts; 
\item[(ii)] because of the improved dimensional bound on $\tilde g_{\o}^{(k)}(\kk)$ with respect to the one on $\hat g_{\o}^{(k)}(\kk)$ and of the short memory property of the Gallavotti-Nicol\`o trees, the bounds on the terms denoted by a tilde are improved by a factor $M^{\th h}$ with respect to the basic ones.
\end{itemize}
Then, since $\hat g^{(k)}_{\o,rel}(\kk) = -\hat g^{(k)}_{\o,rel}(-\kk)$, $\hat w^{(k)}(\pp) = \hat w^{(k)}(-\pp)$, the parity of the graphs contributing to the relativistic part of the Beta function is equal to $-1$; this means that
\be
\b^{\a,a,(rel)}_{\m,\n,h+1} = 0\;,\label{exc18}
\ee 
and the equation (\ref{exc16}) can be safely iterated up to the scale $0$. This shows that $\n^{\a,j}_{\m,\n,h} = O(e)$, uniformly in $h$, as desired.
\paragraph{The flow of $Z^{\a}_{\ul\o,h}$.} Finally, we conclude the discussion of the flow of the new running coupling constants by analyzing the flow of $Z^{\a}_{\ul\o,h}$; these are marginal, and their flows are very similar to the one of the wave function renormalization $Z_{h}$, see Section \ref{secZv}. In fact, these running coupling constants evolve according to the following flow equation:
\be
Z^{\a}_{\ul\o,h} = Z^{\a}_{\ul\o,h+1}\big( 1 + \b^{\a}_{\ul\o,h+1,1} \big) + \b^{\a}_{\ul\o,h+1,2}\;,\label{exc19}
\ee
where $\b^{\a}_{\ul\o,h+1,i}$ are expressed as renormalized series admitting $N!$ bounds; in particular, $\b^{\a}_{\ul\o,h,1} = O(e^2)$, $\b^{\a}_{\ul\o,h,2} = O(e^{4})$. The graphs contributing to $\b^{\a}_{\ul\o,h,1}$ have the external field emerging from a vertex of type $\Phi^{(a)}\Psi^{+}\Psi^{-}$, while the ones appearing in the value of $\b^{\a}_{\ul\o,h,2}$ have the external field emerging from a vertex of type $\Phi^{(\a)}AA$; see Fig. \ref{figexc} for an example of two such graphs.

However, as anticipated in the caption of Fig. \ref{figexc}, the bound on $\b^{\a}_{\ul\o,h+1,2}$ admits an improvement of a factor $M^{\th h}$ with respect to the basic one. To see this, simply decompose $\b^{\a}_{\ul\o,h+1,2}$ as $\b^{\a,(rel)}_{\ul\o,h+1,2} + \tilde\b^{\a}_{\ul\o,h+1,2}$, exactly as for the flows of the other marginal terms (see discussion after (\ref{exc17})); again, $\b^{\a,(rel)}_{\ul\o,h+1,2}=0$ because it is given by a sum of graphs with an odd number of relativistic fermion propagator, while $\tilde\b^{\a}_{\ul\o,h+1,2}$ has a dimensional bound improved by $M^{\th h}$.

Hence, Eq. (\ref{exc19}) can be iterated up to the scale $0$ and it follows that
\bea
Z^{\a}_{\ul\o,h} &\simeq& M^{-h\eta^{\a}_{\ul\o}}\;,\qquad \mbox{as $h\rightarrow-\infty$}\;,\\
\eta^{\a}_{\ul\o} &:=& \lim_{h\rightarrow-\infty}\log_{M}\big(1+\b^{\a}_{\ul\o,h+1,1}\big)\;;\label{exc20}
\eea
the lowest order contribution $\eta^{\a,(2)}_{\ul\o}$ to $\eta^{\a}_{\ul\o}$ are computed in Appendix \ref{secexp}; the results are summarized in Table \ref{tabexp} (recall that by symmetry $Z_{\ul\o,h}^{\a} = Z_{-\ul\o,h}^{\a}$).
%
{
\renewcommand{\arraystretch}{1.7}
\begin{table}[htbp]
\centering
\begin{tabular}{|l|l|l|}
\hline
             \backslashbox{$\a$}{$\ul\o$}    & $(+,-)$ & $(+,+)$\\
\hline                  
 $E_{+}$ & $\frac{3 e^{2}}{4\pi^2}$ & $\frac{e^{2}}{12\pi^2}$ \\
\hline 
  $E_{-}$  & $\frac{e^{2}}{12\pi^2}$ & $\frac{e^{2}}{12\pi^2}$ \\
\hline 
  $CDW$ & $\frac{e^{2}}{12\pi^2}$ & $\frac{3 e^{2}}{4\pi^2}$ \\
\hline
  $D$ & $\frac{e^{2}}{12\pi^2}$ & $\frac{e^{2}}{12\pi^2}$ \\
 \hline
 \end{tabular}
 \caption{The lowest order contributions $\eta^{\a,(2)}_{\ul\o}$ to the anomalous exponents $\eta^{\a}_{\ul\o}$.}\label{tabexp}
 \end{table}
 }

Therefore, the conclusion is that the most divergent $Z^{\a}_{\ul\o,h}$ are those corresponding to $\a = E_{+}$, $\o_1 = - \o_2$ and $\a = CDW$, $\o_1 = \o_2$; in particular, these four running coupling constants diverge with anomalous exponents {\it bigger} than the one of $Z_h$. As we are going to see in the next Section, this implies that the spatial decays of the excitonic and charge density wave susceptibilities are {\it depressed} by the interaction.  

\subsection{Computation of the response functions}

\begin{figure}[hbtp]
\centering
\includegraphics[width=.4\textwidth]{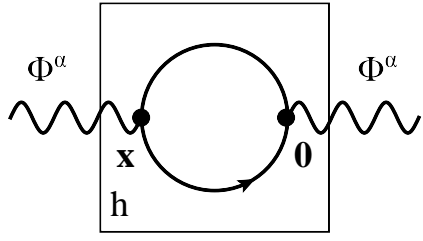}
\caption{The lowest order graph contributing to $C^{(h-1)}(\Phi)$. A sum over all the possible inner scales is understood.}\label{figexc2}
\end{figure}

Finally, we have now all the elements that we need to explicitly compute the response function; in particular, we shall prove the results (\ref{res4}), (\ref{res5}). It follows that:
\be
C^{\a}_{i,j}(\xx) := \frac{\partial^{2}}{\partial\Phi^{(\a)}_{\xx,i}\partial\Phi^{(\a)}_{\V0,j}}\lis\WW_{\b,L}(\Phi)\big|_{\Phi=0} = \frac{\partial^{2}}{\partial\Phi^{(\a)}_{\xx,i}\partial\Phi^{(\a)}_{\V0,j}}\sum_{h\leq 0}\CC^{(h)}(\Phi)\Big|_{\Phi=0}\;,\label{exc21}
\ee
where $\CC^{(h)}(\Phi)$ has been defined in (\ref{exc11}); the lowest order contribution to $\CC^{(h-1)}$ is represented in Fig. \ref{figexc2}.

Therefore, taking $i=j=1$ for simplicity, we can rewrite (\ref{exc21}) as follows:
\bea
C^{\a}_{1,1}(\xx) &=& -2\Tr \Big( \G^{\a} g^{(u.v.)}(\xx)\G^{\a}g^{(u.v.)}(- \xx) \Big)\label{exc23a}\\&&- 2\sum_{h,h'\leq 0}\,\sum_{\o_1,\o_2=\pm}\frac{(Z^{\a}_{\ul\o,h\vee h'})^{2}}{Z_{h-1}Z_{h'-1}}\Tr\Big( \G^{\a} g^{(h)}_{\o_1}(\xx)\G^{\a} g^{(h')}_{\o_2}(- \xx) \Big) \label{exc23b}\\ &&+ \sum_{h\leq 1}\O^{(h)}_{\b,L}(\xx)\;,\label{exc23c}
\eea
where the factor $2$ is due to the sum over the spins, $h\vee h' := \max\{h,h'\}$, and $\O^{(h)}_{\b,L}(\xx - \yy)$ contains higher order terms. We will get that, for $|\xx|\gg 1$:
\be
C^{\a}_{1,1}(\xx) = \frac{G^{\a}(\xx)}{|\xx|^{4 - \xi^{\a}_{++}}} + \cos(\vec p_{F}^{+}\cdot\vec x)\frac{G^{\a}_{2}(\xx)}{|\xx|^{4 - \xi^{\a}_{+-}}} + r^{\a}_{1,1}(\xx)\;,\label{exc23d}
\ee
where, for $1\geq \th > 1/2$:
\bea
&&\xi^{\a}_{\ul\o} := 2(\eta^{\a}_{\ul\o} - \eta)\;,\qquad r^{\a}_{1,1}(\xx) = r^{\a,0}_{1,1}(\xx) + \sum_{n\geq 2} r^{\a,n}_{1,1}(\xx)\;,\label{exc23e}\\
&& |r^{\a,0}_{1,1}(\xx)| \leq \frac{(\const.)}{|\xx|^{4 + \th - \max_{\ul\o}\xi^{\a}_{\ul\o}}}\;,\qquad |r^{\a,n}_{1,1}(\xx)| \leq (\const.)^{n}\Big(\frac{n}{2}\Big)!e^{n}\frac{1}{|\xx|^{4 - \max_{\ul\o}\xi^{\a}_{\ul\o}}}\;,\nn
\eea
and, if $b = 8\pi^2/27$:
\bea
&& G^{D}_{1}(\xx) = \frac{-x_0^2 + |\vec x|^2}{b|\xx|^{2}}\;,\qquad G^{D}_{2}(\xx) = \frac{-x_0^2 + x_1^2 - x_2^2}{b|\xx|^2}\;,\nn\\
&& G^{CDW}_{1}(\xx) = -\frac{1}{b}\;,\qquad G^{CDW}_{2}(\xx) = \frac{-x_0^2 - x_1^2 + x_2^2}{b |\xx|^2}\;,\nn\\
&& G^{E_+}_{1}(\xx) = \frac{-x_0^2 - x_1^2 + x_2^2}{b |\xx|^2}\;,\qquad G^{E_+}_{2}(\xx) = -\frac{1}{b}\;,\nn\\
&& G^{E_-}_{1}(\xx) = \frac{x_0^2 - x_1^2 + x_2^2}{b |\xx|^2}\;,\qquad G^{E_-}_{2}(\xx) = \frac{x_0^2 - |\vec x|^2}{b |\xx|^2}\,.\label{exc23f}
\eea
\begin{oss}
\begin{enumerate}
\item Notice that by symmetry (4) of Lemma \ref{lemexc} we know that
\be
C^{\a}_{j+1,j+1}(\xx) = C^{\a}_{j,j}((x_0, T\vec x))\;;
\ee
the correlations $C^{\a}_{i,j}(\xx)$ with $i\neq j$ can be studied in exactly the same way as $C^{\a}_{1,1}(\xx)$, and we will not discuss them here.

\item From the explicit lowest order computations of $\eta^{\a}_{\ul\o}$, $\eta$, see Table \ref{tabexp} and Appendix \ref{secexp}, we know that all $\xi^{\a}_{\ul\o}=0 + O(e^{4})$ {\it except} for $\xi^{E_+}_{+-}$, $\xi^{CDW}_{++}$, which are given by:
\be
\xi^{E_{+}}_{+-} = \frac{4e^{2}}{3\pi^2} + O(e^4)\;,\qquad \xi^{CDW}_{++} = \frac{4 e^{2}}{3\pi^2} + O(e^4)\;;\label{exc33}
\ee
therefore, we conclude that the decays of the correlations $C^{\a}_{1,1}(\xx)$ are {\it depressed} by the presence of  the electromagnetic interaction; in the spirit of \cite{So}, this suggests that the electromagnetic interaction favors instabilities of excitonic or charge density wave type.
\end{enumerate}
\end{oss}

We will study the three contributions (\ref{exc23a}), (\ref{exc23b}), (\ref{exc23c}) separately; the following bounds will play a crucial role:
\bea
&&\big\| \partial^{n_0}_{0}\partial^{n_{1}}_{1}\partial^{n_{2}}_{2} g^{(h)}_{\o}(\xx) \big\| \leq C_{N}\frac{M^{(2 +n_0 + n_1 + n_2)h}}{1 + (M^{h}|\xx|)^{N}}\;,\label{exc24}\\
&&\big\| \partial^{n_0}_{0}\partial^{n_{1}}_{1}\partial^{n_{2}}_{2} w^{(h)}(\xx) \big\| \leq C_{N}\frac{M^{(2 +n_0 + n_1 + n_2)h}}{1 + (M^{h}|\xx|)^{N}}\;,\qquad \mbox{for all $h\leq 1$ and $N\geq 0$}\;,\nn
\eea
and with $g^{(1)}_{\o}(\xx) := g^{(u.v.)}(\xx)$, $w^{(1)}(\xx) = w^{(u.v.)}(\xx)$. The proof of these bounds for $n_{i}=0$ is a consequence of the following inequalities:
\bea
\| g^{(h)}_{\o}(\xx) \| &\leq& \Big\|\int \frac{d\kk'}{D}\, \hat g^{(h)}_{\o}(\kk')\Big\| \leq C'_0 M^{2h}\;,\\
|x_{0}|^{N_{1}}|x_{1}|^{N_1}|x_{2}|^{N_{2}}\| g^{(h)}_{\o}(\xx) \| &=& \Big\| \int\frac{d\kk'}{D}\,\big[\partial^{N_{0}}_{k_{0}}\partial^{N_{1}}_{k'_{1}}\partial^{N_{2}}_{k'_{2}}\eu^{-\iu\kk'\xx}\big]\hat{g}_{\o}^{(h)}(\kk') \Big\| \nn\\
&=& \Big\| \int\frac{d\kk'}{D} \eu^{-\iu \kk'\xx}\partial^{N_{0}}_{k_{0}}\partial^{N_{1}}_{k'_{1}}\partial^{N_{2}}_{k'_{2}} \hat g^{(h)}_{\o}(\kk')\Big\| \nn\\&\leq& C'_{N} M^{(2 - N_1 - N_2 - N_3)h}\;,\label{exc25}
\eea
from which (\ref{exc24}) immediatly follows\footnote{For $N$ even use that $|\xx|^{N} = (\sum_{i=0}^{2}x_{i}^{2})^{\frac{N}{2}}$ and expand the square; for $N$ odd use that $|\xx|^{N} = |\xx||\xx|^{N-1}$, that $|\xx|\leq \sum_{i}|x_i|$ and the fact that $N-1$ is even.}. The case $n_i\neq 0$ can be derived in exactly the same way; the details are left to the reader.

\paragraph{Bound for (\ref{exc23a}).} This term can be estimated simply by using the bound (\ref{exc24}); it follows that
\be
\Big\|\Tr \Big( \G^{\a} g^{(u.v.)}(\xx)\G^{\a}g^{(u.v.)}(- \xx) \Big)\Big\|\leq \frac{\tilde C_{N}}{(1 + |\xx|^{N})^{2}}\;,\label{exc26}
\ee
for all $N\geq 0$.\\

\noindent{\bf Evaluation of (\ref{exc23b}).} This term can be evaluated explicitly, asymptotically in $|\xx|\rightarrow+\infty$ and at leading order in $e$. We will rewrite (\ref{exc23b}) as a quantity which can be explicitly evaluated, plus error terms which will be small, either because they decay faster in $|\xx|$ or because they decay with the same power but are $O(e^{2})$.
\subparagraph{Error terms.} First, we will replace the ratio appearing in (\ref{exc23b}) with a function of $|\xx|$, and we will show that the bound on the remainder is improved by a factor $O(e^{2})$; then, we will compute the leading contribution to (\ref{exc23b}) explicitly. Let us rewrite:
\bea
&&\frac{(Z^{\a}_{\ul\o,h\vee h'})^{2}}{Z_{h-1}Z_{h'-1}} =\label{exc27a}\\&&= |\xx|^{2(\eta^{\a}_{\ul\o} - \eta)}\Big[ 1 + \big( |\xx| M^{h\vee h'}\big)^{-2\eta^{\a}_{\ul\o}}\big(|\xx| M^{h}\big)^{\eta}\big( |\xx|M^{h'} \big)^{\eta}\frac{(c^{\a}_{h\vee h'})^{2}}{c^{\a}_{h}c^{\a}_{h'}} - 1\Big]\;\nn\\
&&=: |\xx|^{2(\eta^{\a}_{\ul\o} - \eta)} + |\xx|^{2(\eta^{\a}_{\ul\o} - \eta)}s_{\ul\o,h,h'}(\xx)\;,\nn
\eea
where $c^{\a}_{h} = 1 + O(e^2)$. Plugging (\ref{exc27a}), the contribution due to the error term $s_{\ul\o,h,h'}(\xx)$ can be bounded using
\begin{itemize}
\item[(i)] the inequality $r^{\a}-1\leq |\a\log r|(r^{\a} + r^{-\a})$, valid for all $\a\in\RRR$ and $r>0$;
\item[(ii)] the bound (\ref{exc26}) for the single scale propagator;
\item[(iii)] the fact\footnote{The proof of (\ref{exc28}) goes as follows. Let $h(\xx) := -[\log |\xx|]$, where $[.]$ denotes the integer part; split the sum $\sum_{h\leq 0}$ in $\sum_{h\leq h(\xx)} + \sum_{h>h(\xx)}$. The first sum can be bounded replacing the argument of the sum with $|\xx|^{\a - N}$, while the second sum is bounded replacing the argument of the sum with $(M^{h}|\xx|)^{\a}$.} that, for some $C_{N,\a}>0$ and $N>\a > 0$:
\be
\sum_{h\leq 0}\frac{(M^{h}|\xx|)^{\a}}{1 + (M^{h}|\xx|)^{N}}\leq C_{N,\a}\;.\label{exc28}
\ee
\end{itemize}
Using (i), (ii), (iii), is is easy to see that
\bea
&&|\xx|^{2(\eta^{\a}_{\ul\o} - \eta)}\sum_{h,h'\leq 0} s_{\ul\o,h,h'}(\xx)\Tr\Big( \G^{\a} g^{(h)}_{\o_1}(\xx)\G^{\a} g^{(h')}_{\o_2}(- \xx) \Big) = 
\sum_{N\geq 2}\tilde r^{\a,N}_{\ul\o}(\xx)\;,\nn\\
&&|\tilde r^{\a,N}_{\ul\o}(\xx)| \leq (\const.)^{N}\Big(\frac{N}{2}\Big)!e^{N}\frac{1}{|\xx|^{4 + 2(\eta -\eta^{\a}_{\ul\o}) }}\;.\label{exc28b}
\eea
Then, let us rewrite
\bea
\hat g^{(h)}_{\o}(\kk') &=& \hat g^{(h)}_{\o,\LL}(\kk') + \hat s_{\o,1}^{(h)}(\kk') + \hat s_{\o,2}^{(h)}(\kk')\nn\\
\hat g^{(h)}_{\o,\LL}(\kk') &:=& \frac{f_h(\kk')}{|\kk'|^{2}}(-\G^{0}_{\o}k'_0 + k'_{i}\G^{i}_{\o})\;,\label{exc27}
\eea
where: (i) $\hat s^{(h)}_{\o,1}(\kk')$ contains the error committed replacing respectively $\tilde v_{h-1}(\kk')$, $\tilde f_{h}(\kk')$ with $1$, $f_{h}(\kk')$ in $\hat g^{(h)}_{\o}(\kk')$; (ii) $\hat s_{\o,2}^{(h)}(\kk')$ is equal to $\hat g^{(h)}_{\o,\LL}(\kk') R_{h-1,\o}(\kk')$. Therefore,
\be
\big\| \hat s_{\o,1}^{(h)}(\kk') \big\| \leq (\const.)e^{2}\| \hat g^{(h)}_{\o}(\kk') \|\;,\qquad \big\| \hat s_{\o,2}^{(h)}(\kk') \big\| \leq (\const.)|\kk'| \| \hat g^{(h)}_{\o}(\kk') \|\;.\label{exc27b}
\ee
Correspondingly, we rewrite $g^{(h)}_{\o}(\xx)$ as $g^{(h)}_{\o,\LL}(\xx) + s^{(h)}_{\o,1}(\xx) + s^{(h)}_{\o,2}(\xx)$, where
\bea
g^{(h)}_{\o,\LL}(\xx) &:=& \eu^{-\iu \pp_{F}^{\o}\xx}\int \frac{d\kk'}{D}\,\eu^{-\iu\kk'\xx}\hat g^{(h)}_{\o,\LL}(\kk')\;,\label{exc29}\\
s^{(h)}_{\o,i}(\xx) &:=& \eu^{-\iu \pp_{F}^{\o}\xx}\int \frac{d\kk'}{D}\;\eu^{-\iu\kk'\xx}\hat s^{(h)}_{\o,i}(\kk')\;;
\eea
The contribution of $ s^{(h)}_{\o,1}(\xx)$ to (\ref{exc23b}) can be bounded exactly as the corresponding to $s_{h,\ul\o,h,h'}(\xx)$, and we get a bound like (\ref{exc28b}); the contribution of $s^{(h)}_{\o,2}(\xx)$ to (\ref{exc23b}) is bounded using (\ref{exc28}), (\ref{exc27b}), and it decays faster than $|\xx|^{-4 + 2(\eta -\eta^{\a}_{\ul\o})}$.
\subparagraph{Evaluation of the leading term.} Therefore, the leading contribution to (\ref{exc23b}) is
\be
C^{\a}_{1,1,\LL}(\xx) := - 2\sum_{\o_1,\o_2=\pm} |\xx|^{2(\eta^{\a}_{\ul\o} - \eta)} \Tr\Big( \G^{\a} g_{\o_1,\LL}(\xx)\G^{\a} g_{\o_2,\LL}(- \xx) \Big)\;,\label{exc30}
\ee
where $g_{\o,\LL}(\xx) := \sum_{h\leq 0}g_{\0,\LL}^{(h)}(\xx)$; an explicit computation shows that, see Appendix \ref{appgl},
\be
g_{\o,\LL}(\xx) = \eu^{-\iu\pp_{F}^{\o}\xx}\frac{-x_0 \G^{0}_{\o} + x_{i}\G^{i}_{\o}}{b' |\xx|^{3}} + o\Big( \frac{1}{1 + |\xx|^{N}} \Big)\;,\quad b' = \frac{8\pi}{3\sqrt{3}}\;,\label{exc31}
\ee
where the last term denotes corrections which vanish as $|\xx|\rightarrow+\infty$ faster than any power. Therefore, setting $\xi^{\a}_{\ul\o} = 2(\eta^{\a}_{\ul\o} - \eta)$, $b := {b'}^2/8 = 8\pi^2/27$ we get that, see Appendix \ref{appgl2}:
\bea
C^{D}_{1,1,\LL}(\xx) &=& \frac{-x_0^2 + |\vec x|^2}{b |\xx|^{6 - \xi^{D}_{++}}} + \cos(\vec p_{F}^{+}\cdot\vec x)\frac{-x_0^2 + x_1^2 - x_2^2}{b |\xx|^{6 - \xi^{D}_{+-}}}\;,\nn\\
C^{CDW}_{1,1,\LL}(\xx) &=& -\frac{1}{b |\xx|^{4- \xi^{CDW}_{++}}} + \cos(\vec p_{F}^{+}\cdot\vec x)\frac{-x_0^2 - x_1^2 + x_2^2}{b |\xx|^{6 - \xi^{CDW}_{+-}}}\;,\nn\\
C^{E_{+}}_{1,1,\LL}(\xx) &=& \frac{-x_0^2 - x_1^2 + x_2^2}{b |\xx|^{6 - \xi^{E_+}_{++}}} - \frac{\cos(\vec p_{F}^{+}\cdot\vec x)}{b |\xx|^{4 - \xi^{E_+}_{+-}}}\;,\nn\\
C^{E_{-}}_{1,1,\LL}(\xx) &=& \frac{x_0^2 - x_1^2 + x_2^2}{b |\xx|^{6 - \xi^{E_-}_{++}}} + \cos(\vec p_{F}^{+}\cdot\vec x)\frac{x_0^2 - |\vec x|^2}{b |\xx|^{6 - \xi^{E_-}_{+-}}}\;.\label{exc32}
\eea
from which (\ref{exc23f}) follows.
\paragraph{Bound for \ref{exc23c}.} Finally, we have to show that all the other Feynman graphs give a smaller contribution. This can be done repeating an analysis similar to the one of \cite{BM2}, up to a few differences that will be discussed here. We can have three situations:
\begin{itemize}
\item[(i)] all the external fields emerge from vertices of type $\Phi^{(\a)}\Psi^{+}_{\o_1}\Psi^{-}_{\o_2}$ and the graph has more than one inner loop;
\item[(ii)] at least one external field emerges from a vertex of type $\Phi^{(\a)}AA$;
\item[(iii)] at least one external field emerges from a vertex of type $\Phi^{(\a)}A$.
\end{itemize}

The vertices of type $\Phi^{(\a)}\Phi^{(\a)}A$ do not give any contribution, since the corresponding running coupling constants $\k^{\ul\a,1,1}_{\m,h}$ are vanishing by symmetries (1) -- (8) of Lemma \ref{lemexc}.

The case (i) can be worked out using that the minimal loop connecting the points $\xx$ and $\yy$ is formed by at least two single scale propagators, which satisfy the bound (\ref{exc24}); then, we can apply the analysis of \cite{BM2}, that will be not reproduced here. The $N$-th order contribution due to these graphs is bounded by $(\const.)^{N}(N/2)! e^{N}|\xx|^{-4 + \xi^{\a}_{\ul\o}}$, $N\geq 2$.

Similarly, using that the running coupling constants associated to vertices of type $\Phi^{(\a)}AA$ are $O(e^{2})$, we get that in case (ii) the $N$-th order contribution is bounded by $(\const.)^{N}(N/2)! e^{N}|\xx|^{-4 + (1/2)\max_{\ul\o}\xi^{\a}_{\ul\o}}$.

\begin{figure}[hbtp]
\centering
\includegraphics[width=.4\textwidth]{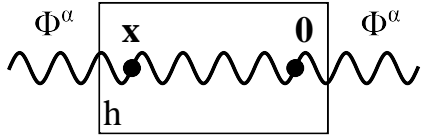}
\caption{In momentum space, the vertices of the graph correspond to $M^{h}\n_{\m,h}^{\a,1}$ or to $p_{\n}\n_{\m,\n,h}^{\a,1}$.}\label{figexc3}
\end{figure}

Let us consider case (iii); recall that the vertices of type $\Phi^{(\a)} A$ correspond to $M^{h}\n_{\m,h}^{\a,1}$ or to $p_{\n}\n_{\m,\n,h}^{\a,1}$, and both the running coupling constants are $O(e)$. Consider the simplest of these graphs, namely the one depicted in Fig. \ref{figexc3}.
Assume that both the vertices correspond to $p_{\n}\n_{\m,\n,h}^{\a,1}$; the value of the graph in Fig. \ref{figexc3} is
\be
\sum_{\n,\n',\m} \n_{\m,\n,h}^{\a,1}\n_{\m,\n',h}^{\a,1}\int \frac{d\pp}{(2\pi)^3}\,\eu^{-\iu\pp\xx} \hat{w}^{(h)}(\pp) p_{\n}p_{\n'} = -\sum_{\n,\n',\m} \n_{\m,\n,h}^{\a,1}\n_{\m,\n',h}^{\a,1} \frac{\partial^{2}}{\partial x_{\n}\partial x_{\n'}}w^{(h)}(\xx)\;.\label{exc34}
\ee
Using the bounds (\ref{exc24}), (\ref{exc28}) we get that
\be
\Big\|\sum_{h\leq 1}\n_{\m,\n,h}^{\a,1}\n_{\m,\n',h}^{\a,1}\frac{\partial^{2}}{\partial x_{\n}\partial x_{\n'}}w^{(h)}(\xx)\Big\| \leq \frac{(\const.)e^{2}}{|\xx|^{4}}\;,\label{exc35}
\ee
and the same result holds if one vertex is replaced by $M^{h}\n_{\m,h}^{\a,1}$; higher order graphs can be bounded repeating the analysis of \cite{BM2}, and the result is that the generic graph of order $N$ of type (ii) is bounded by $(\const.)^{N}e^{N}|\xx|^{-4 + (1/2)\max_{\ul\o}\xi^{\a}_{\ul\o}}$.

\chapter{Lattice distortions}\label{caplat}
\setcounter{equation}{0}
\renewcommand{\theequation}{\ref{caplat}.\arabic{equation}}

\section{Introduction}\label{secintro5}
\setcounter{equation}{0}
\renewcommand{\theequation}{\ref{secintro5}.\arabic{equation}}

In this Chapter we will discuss the effects of lattice distortions on the model that we introduced and studied in the previous Chapters. In general \cite{SSH, LN}, small lattice distortions can be taken into account by considering a position  and bond dependent hopping parameter, of the form $t_{\vec x,i} := t + \phi_{\vec x,i}$, where $\phi_{\vec x,i}$ is proportional to the distortion of the bond connecting the site $\vec x$ to its nearest neighbour $\vec x + \vec\d_i$. Again, we shall perform a RG analysis to characterize the low energy behavior of the model; as we are going to see, some special lattice distortions, the {\it Kekul\' e} ones, will be dramatically enhanced by the electromagnetic interaction. Interestingly, the amplitude of this distortion behaves as a {\it bare mass} for the fermion progagator. The presence of a non-vanishing initial distortion will give rise to a new running coupling constant with respect to the case studied in Chapter \ref{sec2.4.2}, namely the {\it effective fermionic mass} $\D_h$. We shall distinguish two regimes: in the first regime the effective mass will grow from its bare value $\D_0$ to a dressed value $\D\gg \D_0$; in this regime the fermionic field will behave essentially as a massless field, and the analysis will be qualitatively the same as in the case $\D_0=0$. After the scale $\tilde h$ such that $\D_{\tilde h} = \D$, the fermion propagator will become {\it massive}, since the typical momentum scale will be smaller than the mass $\D$; as a consequence, the flow of the running coupling constants will remain close to the values reached at the scale $\tilde h$.

After this, we will investigate the possibility of a {\it spontaneous deformation} of the honeycomb lattice; in particular, we will write a non-BCS self-consistence equation of the amplitude of the distortion, that is for the mass of the fermion field. This equation is similar to the one found by Mastropietro in the context of Luttinger superconductors, see \cite{M1}. The equation will be discussed from a qualitative point of view; as we shall see, already for small coupling the solution will be influenced in a non-trivial way by the electromagnetic interaction. For strong coupling, we will argue that the phenomenon of mass generation is {\it favored} by the interaction. 

The most direct physical interest of our study is connected with the possibility of a metal-insulator transition in graphene, an (yet open) issue on which great efforts, both theoretical and experimental, have been and are still devoted. However, there is another physical motivation for the analysis of Kekul\' e distortions; in fact, it has been recently proposed, \cite{HCM, JP, CHJMPS}, that these kind of lattice deformations may be at the basis of a possible {\it electron fractionalization} in the low energy excitations of graphene, such as the one taking place in one dimensional systems like polyacetylene, \cite{SSH, LN}.

The Chapter is organized in the following way: in Section \ref{seclat1} we slightly change the definition of the model, in order to take into account the effect of small lattice distortions; in Section \ref{secUVlat} we describe the integration of the ultraviolet degrees of freedom; in Section \ref{secIRlat} we discuss the integration of the infrared degrees of freedom, and we prove our result (\ref{res9}) on the two point Schwinger function; finally, in Section \ref{secgap} we discuss the issue of a spontaneous lattice distortion, and we derive a variational equation for the gap.

\section{The model in presence of Kekul\' e distortion}\label{seclat1}
\setcounter{equation}{0}
\renewcommand{\theequation}{\ref{seclat1}.\arabic{equation}}

\begin{figure}[htbp]
\centering
\includegraphics[width=0.7\textwidth]{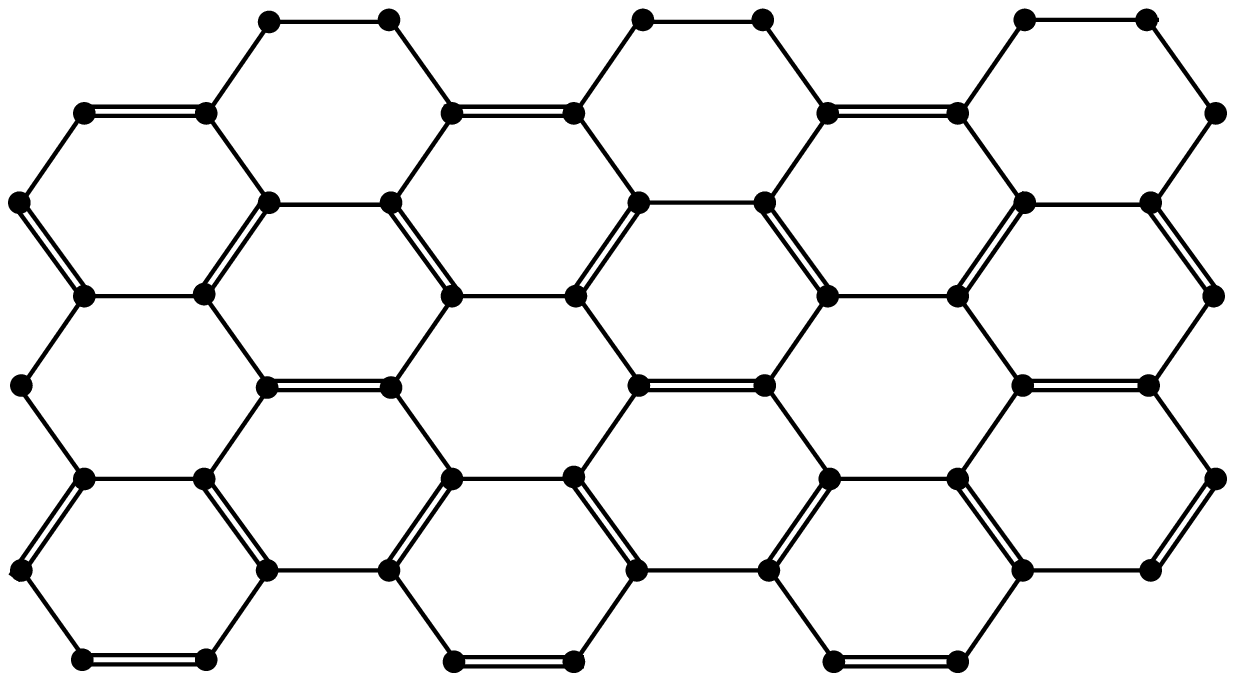} \caption{The Kekul\' e dimerization pattern for $j_0=1$. Double and single bonds are respectively ``long'' and ``short'', and correspond to the hopping parameters $t - \frac{\D_0}{3}$, $t + \frac{2}{3}\D_0$.} \label{figkek}
\end{figure}

In this Section we change the definition of the model so that the presence of a small lattice distortions may be taken into account. The Hamiltonian of the model in presence of a {\it fixed} lattice distortion is given by:
\be
\HHH_{\L}(\{\phi\}) := \HHH_{\L} + \HHH_{\L}^{e-ph}(\{\phi\})\;,\label{lat1}
\ee
where $\HHH_{\L}$ is the Hamiltonian defined in (\ref{1.1.1}) while $\HHH_{\L}^{e-ph}(\{\phi\})$ describes the {\it electron-phonon interaction} and it is given by:
\be
\HHH_{\L}^{e-ph}(\{\phi\}) := \sum_{\substack{\vec x\in \L \\ i=1,2,3}}\sum_{\s = \uparrow\downarrow}\phi_{\vec x,i}\Big( a^{+}_{\vec x,\s}b^{-}_{\vec x + \vec\d_i,\s}\eu^{\iu e A_{(\vec x,i)}} + b^{+}_{\vec x + \vec\d_i,\s}a^{-}_{\vec x,\s}\eu^{-\iu e A_{(\vec x,i)}}\Big)\;,\label{lat2}
\ee
where $\phi_{\vec x,i}$ is a real valued function corresponding to the distortion of the bond which connect $\vec x$ to $\vec x + \vec\d_i$. As for the case $\phi_{\vec x,i}=0$, this model can be studied using a functional integral representation; for any positive $\b,L$, the average of a generic physical observable $\OO(A,\Psi)$ is defined as
\bea
\media{\OO(A,\Psi)}_{\b,L}^{\phi} &:=&\frac{1}{\Xi^{\phi}_{\b,L}}\int P(d\Psi)P(dA)\eu^{V(\Psi,A) + v_{\phi}(A,\Psi)}\OO(A,\Psi)\;,\label{lat3}\\
\Xi_{\b,L}^{\phi} &:=& \int P(d\Psi)P(dA)\eu^{V(\Psi,A) + v_{\phi}(A,\Psi)}\;,\nn
\eea
where all the objects appearing in (\ref{lat3}) have been defined in Chapter \ref{cap1}, except for $v_{\phi}(A,\Psi)$ which is given by:
\bea
&&v_{\phi}(A,\Psi) := \label{lat4}\\&&\int_{0}^{\b} dx_0 \sum_{\substack{\vec x\in \L \\ i= 1,2,3}} \phi_{\vec x,i}\Big( \Psi_{\xx,\s,1}^{+}\Psi^{-}_{(x_0,\vec x + \vec\d_i),\s,2}\eu^{\iu e A_{(\xx,i)}} + \Psi^{+}_{(x_0,\vec x + \vec\d_i),\s,2}\Psi^{-}_{\xx,\s,1}\eu^{-\iu e A_{(\xx,i)}} \Big)\;.\nn 
\eea
We shall perform a specific choice for $\phi_{\vec x,i}$; namely, we shall choose $\phi_{\vec x,i}$ corresponding to the so-called {\it Kekul\' e distortion} of the honeycomb lattice. This is equivalent to the choice
\be
\phi_{\vec x,i} = \phi^{(j_{0})}_{\vec x,i} := \frac{2}{3}\D_{0}\cos(\vec p_{F}^{+}(\vec\d_{j} - \vec\d_{j_0} - \vec x))\;,\label{lat4b}
\ee
where $\D_{0}\in \RRR$ has to be understood as a ``small'' parameter. The Kekul\' e distortion corresponds to a particular dimerization pattern on the hexagonal lattice; see Fig. \ref{figkek}.

\section{Renormalization Group analysis}\label{RGlat}
\setcounter{equation}{0}
\renewcommand{\theequation}{\ref{RGlat}.\arabic{equation}}
In this Section we shall describe the RG analysis that will allow us to characterize the ground state of the model defined in Section \ref{seclat1}. For simplicity, we start by discussing the evaluation of the free energy; as already seen in the previous Chapters for the case $\D_0=0$, the analysis of the correlation functions will be a straightforward extension of the procedure that we are going to discuss.

\subsection{The ultraviolet integration}\label{secUVlat}

Proceeding in a way completely analogous to Section \ref{sec2.4.1}, it follows that after the integration of the ultraviolet degrees of freedom the partition function of the model can be rewritten as (from now on the wide tilde will recall the $\D_0$-dependence):
\be
\widetilde \Xi_{\b,L} = e^{-\b|\L| \widetilde F_{0}}\int P_{\leq 0}(d\Psi^{(\leq 0)})P_{\leq 0}(dA^{(\leq 0)})\eu^{\widetilde V^{(0)}(\Psi^{(\leq 0)},A^{(\leq 0)})}\;,\label{lat5}
\ee
where $\widetilde F_{0}$ and $\widetilde V^{(0)}$ are given by (repeated indeces are summed):
\bea
\widetilde V^{(0)}(\Psi,A) &:=& \sum_{n+m\geq 1}\sum_{\substack{\ul\s,\ul\r \\ \ul\o,\ul\m}}\int \prod_{i=1}^{n}\hat \Psi_{\kk'_{2i-1},\s_{i},
\r_{2i-1},\o_{2i-1}}\hat \Psi_{\kk'_{2i},\s_{i},\r_{2i},\o_{2i}}\prod_{i=1}^{m}\hat A_{\m_i, \pp_i}\cdot\nn\\\quad &&\cdot
\widetilde W^{(j_0)}_{2n,m,{\ul \r},{\ul \o},\ul\m}(\{\kk'_i\},\{\pp_j\})\d\left(\sum_{j=1}^{m}\pp_j
 + \sum_{i=1}^{2n}(-1)^{i}\kk'_i\right)\;,\label{lat6}\\
\widetilde F_{0} &:=& (\b|\L|)^{-1}\widetilde W^{(j_0)}_{0,0}\;,
\eea
and the kernels $\widetilde W_{2n,m}^{(j_0)}$ can be computed in terms of Feynman graphs following the rules described in Appendix \ref{app2b} and Section \ref{multi}, with the only difference that now new vertices proportional to $\D_0$ appear; these vertices have two fermionic lines and an arbitrary number $\geq 0$ of photon lines. Clearly, if we set $\D_0=0$ the kernels appearing in (\ref{lat6}) reduce exactly to the ones of Section \ref{sec2.4.1}, and in particular verify the properties stated Lemma \ref{lem2.4c}, which are implied by the symmetries listed in Lemma \ref{lem2.4}; for $\D_0\neq 0$ the role of Lemma \ref{lem2.4} is played by the following one.

\begin{lemma}\label{lemlat}
For $\phi_{\vec x,i}$ given by (\ref{lat4b}), both the Gaussian integrations $P(d\Psi)$, $P(dA)$ and the interactions $V(\Psi,A)$, $v_{\phi}(A,\Psi)$, defined in (\ref{1.2.11}), (\ref{1.2.14}), (\ref{1.2.20}), (\ref{lat4}) respectively, are invariant under the transformations (1) -- (8) of Lemma \ref{lem2.4} provided that:
\begin{itemize}
\item under (4) $j_0\rightarrow j_0 + 1$, with $3 + 1 = 1$;
\item under (6.b) $j_0 \rightarrow j_0'$, with $1' = 1$, $2' = 3$, $3' = 2$.
\end{itemize}
\end{lemma}
\noindent{\it Proof.} The proof is a trivial extension of the one of Lemma \ref{lem2.4}; the details are left to the reader.\qed\\

For $\D_0\neq 0$ the above symmetry properties no longer exclude the presence of a mass in fermion propagator; however, as proven in Appendix \ref{applat}, the following identities are true:
\bea
\widetilde W_{2,0,\r,\r,\ul{\o}}^{(j_0)}(\V0) &=& \widetilde W_{2,0,\ul{\r},\o,\o}^{(j_0)}(\V0) = 0\;,\nn\\
\widetilde W_{2,0,\r,3-\r,\o,-\o}^{(1)}(\V0) &:=& \D_0\d_0\;,\qquad \d_{0}\in \RRR\;,\nn\\
\widetilde W_{2,0,\r,3-\r,\o,-\o}^{(j_0)}(\V0) &=& \eu^{-\iu \vec p_{F}^{\o}(\vec\d_{j_0} - \vec\d_1)}\D_0\d_{0}\;.\label{lat15}
\eea
where $\d_0 = \sum_{N\geq 2}\d_0^{(N)}e^{N}\in \RRR$ with $\d_0^{(N)}$ bounded proportionally to $(N/2)!$, uniformly in $\D_0$. It is convenient to rewrite the partition function (\ref{lat5}) using the more compact relativistic notations introduced at the end of Section \ref{multi}; it follows that:
\be
\widetilde \Xi_{\b,L} = e^{-\b|\L|\widetilde F_{0}}\int P_{\leq 0}(d\psi^{(\leq 0)})P_{\leq 0}(A^{(\leq 0)})\eu^{\widetilde \VV^{(0)}(\psi^{(\leq 0)},A^{(\leq 0)})}\;,\label{lat6b}
\ee
where the interaction $\widetilde \VV^{(0)}(\psi,A)$ is given by:
\bea
&&\widetilde \VV^{(0)}(\psi,A) = -\int \frac{d\kk'}{D}\, \lis\psi_{\kk',\s}\iu \g_{\m} z_{\m,0}\psi_{\kk',\s} - \int \frac{d\kk'}{D}\, \lis\psi_{\kk',\s} \iu\g^{(j_0)} \D_0 \d_0 \psi_{\kk',\s} + \nn\\&& - \int \frac{d\pp}{(2\pi)^{3}}\, \hat A_{\m,\pp}\hat A_{\m,-\pp}\tilde \n_{\m,0} + \int \frac{d\kk'}{D}\frac{d\pp}{(2\pi)^3}\, \iu \l_{\m,0}\lis\psi_{\kk'+\pp,\s}\g_{\m}\psi_{\kk',\s}\hat A_{\m,\pp} + \nn\\
&&  + \RR \widetilde \VV^{(0)}(\psi, A)\;,\label{lat6c}
\eea
where:
\begin{itemize}
\item[(i)] the first terms takes into account the contribution of $\partial_{\m}\widetilde W_{2,\ul{\r},\ul\o}^{(j_0)}(\V0)\big|_{\D_0=0}$.
\item[(ii)] The second term is a {\it mass term} for the fermion field; the $4\times 4$ matrix $\g^{(j_0)}$ is defined as:
\be
\g^{(j_0)} := \begin{pmatrix} -\iu I \eu^{\iu \vec p_{F}^{+}(\vec\d_{j_0} - \vec\d_1)} & 0 \\ 0 & \iu I \eu^{\iu \vec p_{F}^{-}(\vec\d_{j_0} - \vec\d_1)} \end{pmatrix}\;.\label{lat6d}
\ee
\item[(iii)] The third term takes into account the contributions of $\widetilde W_{0,2,\ul{\m}}^{(j_0)}(\V0)$ (which verify the same symmetry properties of their $\D_0=0$ counterparts, see Appendix \ref{applat}).
\item[(iv)] The fourth term takes into account the contributions of $\widetilde W_{2,1,\ul{\r},\ul{\o},\m}^{(j_0)}(\V0)\big|_{\D_0=0}$.
\item[(v)] The fifth term collects all the higher powers in the fields, and the Taylor remainder of the expansions of the kernels in the momenta and in $\D_0$.
\end{itemize}

\subsection{The infrared integration}\label{secIRlat}

We proceed as in the $\D_0=0$ case; we evaluate the partition function by integrating the fields in an iterative way, starting from $\psi^{(0)}, A^{(0)}$. For $h\geq \tilde h$, where $\tilde h = \tilde h(\D_0)<0$ is a suitable scale label that will be defined in the following, the multiscale integration will proceed in a way analogous to the case $\D_0=0$; after the scale $h^{*}$ the fermion propagator will become {\it massive}.
\paragraph{Massless regime.} We want to inductively prove that after the integration of the fields on scales $\geq h+1> \tilde h$ the original functional integral can be rewritten as:
\be
\widetilde\Xi_{\b,L} = e^{-\b|\L|\widetilde F_{h}}\int P_{\leq h}(d\psi^{(\leq h)})P_{\leq h}(dA^{(\leq h)})\eu^{\widetilde\VV^{(h)}(\sqrt{Z_h}\psi^{(\leq h)},A^{(\leq h)})}\;,\label{lat6e}
\ee
where $P_{\leq h}(dA^{(\leq h)})$ has the same propagator of the $\D_0=0$ case, while the fermionic propagator is
\be
\hat g^{(\leq h)}(\kk') := \frac{\chi_h(\kk')}{\tilde Z_{h}(\kk')}\frac{\iu \g_0 k_0 + \iu \tilde v_h(\kk')\vec k'\cdot \g - \iu \g^{(j_0)*}\tilde\D_{h}(\kk')}{k_0^2 + \tilde v_h(\kk')^{2}|\vec k'|^{2} + \tilde \D_{h}(\kk')^{2}}\big(1 + R_{h}(\kk')\big)\;;\label{lat6f}
\ee
finally, the effective interaction on scale $h$ can be written as
\bea
\widetilde \VV^{(h)}(\psi,A) &:=& \sum_{n+m\geq 1}\sum_{\ul\s,\ul\a,\ul\m}\int \prod_{i=1}^{n} \lis\psi_{\kk'_{2i-1},\s_{i},
a_{2i-1}}\psi_{\kk'_{2i},\s_{i},a_{2i}}\prod_{i=1}^{m}\hat A_{\m_i, \pp_i}\cdot\nn\\\quad &&\cdot
\widetilde W^{(h),(j_0)}_{2n,m,{\ul a},\ul\m}(\{\kk'_i\},\{\pp_j\})\d\left(\sum_{j=1}^{m}\pp_j
 + \sum_{i=1}^{2n}(-1)^{i}\kk'_i\right)\;,\label{lat6g}
\eea
and the kernels $\widetilde W^{(h),(j_0)}_{2n,m}$, together with the free energy $\widetilde F_h$, will be defined recursively. In order to inductively prove formula (\ref{lat6e}), we split the kernels appearing into $\widetilde \VV^{(h)}$ as sums of local plus irrelevant parts; this procedure will be slightly different with respect to the one discussed in the case $\D_0=0$. Let us define the operators $\LL_i$, $i=0,1$, in the following way.\\

\noindent If $n=1,m=0$:
\bea
\LL_{0}\widetilde W_{2,0,\ul{a}}^{(h),(j_0)}(\kk') &:=& \widetilde W_{2,0,\ul{a}}^{(h),(j_0)}(\V0)\;,\nn\\
\LL_{1}\widetilde W_{2,0,\ul{a}}^{(h),(j_0)}(\kk') &:=& k'_{\m}\partial_{\m}\widetilde W_{2,0,\ul{a}}^{(h),(j_0)}(\V0)\;;\label{lat7}
\eea
if $n=0,m=2$:
\bea
\LL_{0}\widetilde W_{0,2,\ul{\m}}^{(h),(j_0)}(\pp) &:=& \widetilde W_{0,2,\ul{\m}}^{(h),(j_0)}(\V0)\;,\nn\\
\LL_{1}\widetilde W_{0,2,\ul{\m}}^{(h),(j_0)}(\pp) &:=& p_{\a}\partial_{\a}\widetilde W_{0,2,\ul{\m}}^{(h),(j_0)}(\V0)\;;\label{lat8}
\eea
if $n=2,m=1$:
\be
\LL_0 \widetilde W_{2,1,\ul{a},\m}^{(h),(j_0)}(\pp) := \widetilde W_{2,1,\ul{a},\m}^{(h),(j_0)}(\V0)\;;\label{lat9}
\ee
if $n=0,m=3$:
\be
\LL_0 \widetilde W_{0,3,\ul\m}^{(h),(j_0)}(\pp_1,\pp_2) := \widetilde W_{0,3,\ul\m}^{(h),(j_0)}(\V0,\V0)\;;\label{lat10}
\ee
then, given a generic kernel $\widetilde W_{2n,m,{\ul a},\ul\m}^{(h),(j_0)}$ we define the operators $\PPP_{i}$, $i=0,1$, in the following way:
\bea
&&\PPP_{0}\widetilde W^{(h),(j_0)}_{2n,m,{\ul a},\ul\m}(\{\kk'_i\},\{\pp_j\}) := \widetilde W^{(h),(j_0)}_{2n,m,{\ul a},\ul\m}(\{\kk'_i\},\{\pp_j\})\big|_{\D_0=0}\;,\label{lat11}\\
&&\PPP_{1}\widetilde W^{(h),(j_0)}_{2n,m,{\ul a},\ul\m}(\{\kk'_i\},\{\pp_j\}) :=\nn\\&&:= \sum_{k\geq h,\kk}\tilde\D_{k}(\kk')\frac{\partial}{\partial \tilde \D_{k}(\kk')}\widetilde W^{(h),(j_0)}_{2n,m,{\ul a},\ul\m}(\{\kk'_i\},\{\pp_j\})\big|_{\tilde\D_0(\kk') = ...  =\tilde\D_{h}(\kk') = 0}\;.\nn
\eea
Finally, given $\LL_{j}$, $\PPP_{j}$, $j=0,1$ as above, we can rewrite the effective potential on scale $h$ as $\LL \VV^{(h)} + \RR \VV^{(h)}$, where the {\it localization operator} $\LL$ is defined as follows.\\

\noindent If $n=1,m=0$:
\be
\LL \widetilde W_{2,0,\ul{a}}^{(h),(j_0)}(\kk') := \LL_{0}(\PPP_0 + \PPP_1)\widetilde W_{2,0,\ul{a}}^{(h),(j_0)}(\kk') + \LL_{1}\PPP_0 \widetilde W_{2,0,\ul{a}}^{(h),(j_0)}(\kk')\;;\label{lat11b}
\ee
if $n=0,m=2$:
\be
\LL \widetilde W_{0,2,\ul{\m}}^{(h),(j_0)}(\pp) := \LL_0 (\PPP_0 + \PPP_1)\widetilde W_{0,2,\ul{\m}}^{(h),(j_0)}(\pp) + \LL_{1}\PPP_0 \widetilde W_{0,2,\ul{\m}}^{(h),(j_0)}(\pp)\;;\label{lat12}
\ee
if $n=2,m=1$:
\be
\LL \widetilde W_{2,1,\ul{a},\m}^{(h),(j_0)}(\pp) := \LL_0 \PPP_0 \widetilde W_{2,1,\ul{a},\m}^{(h),(j_0)}(\pp)\;;\label{lat13}
\ee
if $n=0,m=3$:
\be
\LL \widetilde W_{0,3,\ul\m}^{(h),(j_0)}(\pp_1,\pp_2) := \LL_0 \PPP_0 \widetilde W_{0,3,\ul\m}^{(h),(j_0)}(\pp_1,\pp_2)\;.\label{lat14}
\ee
The combinations of the operators $\LL_{i}$ and $\PPP_0$ give rise to terms equal to those discussed in the case $\D_0=0$; therefore, it remains to discuss the combination of the operators $\LL_{i}$ and $\PPP_{1}$. Proceeding in a way completely analogous to the one followed in Appendix \ref{applat} to prove (\ref{lat6c}), the symmetry properties (1) -- (8) of Lemma \ref{lemlat}, which are preserved by the multiscale integration, imply that:
\bea
&&\lis\psi_{\kk',\s}\LL_0 \PPP_1 \widetilde W_{2,0}^{(h),(j_0)}(\kk')\psi_{\kk',\s} = -\iu \D_{h}\d_{h} \lis\psi_{\kk',\s}\g^{(j_0)}\psi_{\kk',\s}\;,\quad \d_h = \d_h^{*}\;.\label{lat14b}\\
&&\LL_0 \PPP_1 \widetilde W_{0,2,\ul{\m}}^{(h),(j_0)}(\pp) = -\d_{\m_1,\m_2}M^{h}\tilde \n_{\m_1,h}\;,\quad \tilde \n_{1,h} = \tilde \n_{2,h}\;,\quad \tilde \n_{i} = \tilde \n_{i}^{*}\;.\nn
\eea
As we are going to see, the remainders of the Taylor expansions in $\tilde \D_{k}(\kk')$ necessarily appearing in the definition of the $\RR = 1 - \LL$ operation, see (\ref{lat11b}) -- (\ref{lat14}), admit {\it gains} in their dimensional bounds with respect to the basic ones, and for this reason will give rise to irrelevant kernels. 

Hence, one can repeat exactly the same steps performed in Section \ref{multi} to describe the integration of the partition function of the $\D_0=0$ case; the only difference being the presence of a new running coupling function $\tilde \D_{h}(\kk')$, which evolves according to the following flow equation:
\be
\tilde Z_{h-1}(\kk')\tilde \D_{h-1}(\kk') = \tilde Z_{h}(\kk')\tilde \D_{h}(\kk') + Z_{h}\D_{h}\d_{h}\chi_{h}(\kk')\;.\label{lat16}
\ee
Setting $\D_{h-1} := \tilde\D_{h-1}(\V0)$, it follows that: 
\be
\frac{\D_{h-1}}{\D_{h}} = \frac{Z_{h}}{Z_{h-1}}\big( 1 + \b^{\D}_{h} \big)\;,\label{lat17}
\ee
where $\b^{\D}_{h} = \b^{\D,(2)}_{h} + O(e^{2}M^{\th h}) + O(e^{4})$, and (see Appendix \ref{secDelta}):
\bea
\b^{\D,(2)}_{h} &=& \big(e_{0,h}^{2} + 2v_{h-1}^{2}e_{1,h}^{2}\big)\frac{\arctan\xi_{h}}{4\pi^2}\frac{1 + \xi_h^2}{\xi_h}\log M \nn\\&=& \frac{3 e^{2}}{4\pi^2}\log M + O(e^{4}\log M)\;;\label{lat18}
\eea
therefore, the flow equation (\ref{lat17}) and the explicit computation (\ref{lat18}) together with the result (\ref{3a.19z}) for the Beta function of $Z_{h}$ imply that
\be
\D_{h}\simeq \D_{0}M^{-h\eta^{K}}\;,\qquad \eta^{K} = \frac{2e^{2}}{3\pi^2} + O(e^4)\;.\label{lat19}
\ee
As already mentioned at the beginning of this section, the multiscale integration goes on until $h=\tilde h$, where the scale $\tilde h$ is defined as the scale on which the fermionic propagator becomes massive, that is when:
\be
\D_{\tilde h} = M^{\tilde h} \Rightarrow \tilde h \simeq \frac{1}{1 + \eta^{K}}\log_{M} \D_0\;,\label{lat20}   
\ee
which implies that $\D_{\tilde h} \simeq \D_0^{\frac{1}{1 + \eta^{K}}}$. It remains to show that the actions of $1 - \PPP_{1}$ on a relevant kernel and of $1 - \PPP_0$ on a marginal one produce {\it irrelevant} kernels; in particular, we shall show that the operator $1 - \PPP_i$ acting on a kernel on scale $h>\tilde h$ is dimensionally equivalent to $M^{(\tilde h - h)(i+1)}$. In fact, it is straightforward to see that the action of $1-\PPP_{i}$ is dimensionally equivalent to $\big(\D_{h}/ M^{h}\big)^{i+1}$; and
\be
\Big(\frac{\D_{h}}{ M^{h}}\Big)^{i+1} = \Big(\frac{\D_{h}}{\D_{\tilde h}}\Big)^{i+1}M^{(\tilde h - h)(i+1)} \simeq M^{(\tilde h - h)(1 + \eta_{K})(1+i)}\;,\label{lat21}
\ee
where we used the definition (\ref{lat20}) of $\tilde h$.
\paragraph{Massive regime.} At the scale $\tilde h$ the fermion propagator has become massive; however, the photon propagator is singular in the infrared, therefore we still need the RG to analyze this regime. We could integrate the residual fermionic degrees of freedom in a single step and study the effective bosonic theory by multiscale analysis; however, in this way we would get estimates not uniform in the bare fermion mass. For this reason we will proceed in a way analougous to the massless regime, {\em i.e.} we will ``slice'' the fermionic and bosonic degrees of freedom simultaneously. 

The main difference with respect to the previous regime is that now we {\it do not expand anymore} the kernels in powers of the fermion mass; in fact, the splitting performed in the massless regime would be not convenient now, because, as we are going to see, the momentum flowing on a scale $\leq \tilde h$ will be smaller than the corresponding fermion mass (and therefore $\D_{h}/\g^{h}$ will be greater than $1$, see (\ref{lat21})). As before, we will localize the kernels corresponding to monomials of order $\leq 3$ in the fields; the main difference in the single scale integration with respect to the previous regime is that here we do not rescale the fermionic fields. In fact, because of the improved dimensional bound on the fermion propagator the flows of the running coupling constants corresponding to monomials $k_{\m}\lis\psi\psi$ converge to finite values. The same is true for the mass term $\D_{h}$, which remain close to its value on scale $\tilde h$, and for the charges $e_{\m,h}$. Finally, the photon mass is controlled using the same Ward identity exploited in the case $\D_0=0$, and the local parts of the remaining kernels vanish by symmetry. 

After the integration of the first $|\tilde h|$ infrared scales the partition function can be rewritten as
\be
\widetilde\Xi_{\b,L} = e^{-\b|\L|\widetilde F_{\tilde h}}\int P_{\leq \tilde h}(d\psi^{(\leq \tilde h)}) P_{\leq \tilde h}(dA^{(\leq \tilde h)})\eu^{\widetilde \VV^{(\tilde h)}(\sqrt{Z_{\tilde h}}\psi^{(\leq \tilde h)},A^{(\leq \tilde h)})}\;,\label{bos1}
\ee
where the effective potential $\widetilde \VV^{(\tilde h)}$ is given by (\ref{lat6g}) with $h$ replaced by $\tilde h$. We split the effective potential as $\LL \widetilde \VV^{(\tilde h)} + \RR\widetilde \VV^{(\tilde h)}$, where $\RR = 1-\LL$ and the action of $\LL$ on the kernels of $\widetilde\VV^{(\tilde h)}$ is defined as follows:
\bea
\LL \widetilde W^{(\tilde h),(j_0)}_{2,0,\ul{a}}(\kk') &=& \widetilde W^{(\tilde h),(j_0)}_{2,0,\ul{a}}(\V0) + k'_{\a}\partial_{\a}\widetilde W^{(\tilde h),(j_0)}_{2,0,\ul{a}}(\V0)\nn\\
\LL \widetilde W^{(\tilde h),(j_0)}_{2,1,\ul{a},\m}(\kk',\pp) &=& \LL \widetilde W^{(\tilde h),(j_0)}_{2,1,\ul{a},\m}(\V0,\V0)\;,\nn\\
\LL \widetilde W^{(\tilde h),(j_0)}_{0,2,{\ul \m}}(\pp) &=& \widetilde W^{(\tilde h),(j_0)}_{0,2,\ul{\m}}(\V0) + p_{\a}\partial_{\a}\widetilde W^{(\tilde h),(j_0)}_{0,2,\ul{\m}}(\V0)\;,\nn\\
\LL \widetilde W^{(\tilde h),(j_0)}_{0,3,\ul{\m}}(\pp) &=& \widetilde W^{(\tilde h),(j_0)}_{0,3,\ul{\m}}(\V0)\;.\label{bos3}
\eea
By the symmetry properties of Lemma \ref{lemlat} it follows that:
\bea
&& \lis\psi_{\kk',\s}\widetilde W^{(\tilde h),(j_0)}_{2,0}(\V0)\psi_{\kk',\s} = -\iu \D_{\tilde h}\d_{\tilde h}\lis\psi_{\kk',\s}\g^{(j_0)}\psi_{\kk',\s}\;,\quad \d_{\tilde h}\in \RRR\;,\label{bos4}\\
&&\widetilde W^{(\tilde h),(j_0)}_{0,2,\m_{1},\m_{2}}(\V0) = \d_{\m_1,\m_{2}} \widetilde W^{(\tilde h),(j_0)}_{0,2,\m_1,\m_1}(\V0) =: - M^{\tilde h}\tilde \n_{\m_{1},\tilde h}\;,\label{bos4a}\\
&&\tilde\n_{1,\tilde h} = \tilde\n_{2,\tilde h}\;,\qquad \tilde\n_{1,\tilde h} \in \RRR\;,\qquad \label{bos4b}\\
&&\partial_{\mu}\widetilde W^{(\tilde h)}_{0,2,\ul{\m}}=0\;,\qquad \widetilde W^{(\tilde h),(j_0)}_{0,3,\ul{\m}}(\V0)=0\;,\label{bos4c}
\eea
where: (i) (\ref{bos4}) is proved in Appendix \ref{applat1}; (ii) (\ref{bos4a}), (\ref{bos4b}) are proved in Appendix \ref{applat2}; the proof of the second identity in the third line of (\ref{bos4c}) is completely analogous to the one of the case $\D_0=0$, which has been discussed in Appendix \ref{secD4} (the only symmetry transformations involved are (7) and (8), and these do not involve the label $j_0$); the proof of the first one is less trivial, and it goes as follows. 

First, consider the case $j_0=1$; in this case the proof is completely analogous to the one reported in Appendix \ref{secD4} for the $\D_0=0$ case, since the symmetry transformations involved do not change the label $j_0=1$. Let us now consider the case $j_0=2$; setting $\widetilde T_{00} := 1$, $\widetilde T_{0,i} = \widetilde T_{i,0} := 0$, $\widetilde T_{ij} := T_{ij}$ where $T$ is the $2\pi/3$ rotation matrix, by symmetry (4) it follows that:
\be
\widetilde T \Big[ \big(\widetilde T\partial_{\pp}\big)_{\a} \widetilde W^{(\tilde h),(2)}_{0,2}(\V0) \Big]\widetilde T^{-1} = \partial_{\a} \widetilde W^{(\tilde h),(1)}_{0,2} =0\;,\label{bos5}
\ee
which gives
\bea
\partial_{0} \widetilde W^{(\tilde h),(2)}_{0,2,\ul{\m}}(\V0)&=&0\;,\nn\\
\Big( -\frac{1}{2}\partial_{1} - \frac{\sqrt{3}}{2}\partial_{2} \Big)\widetilde W^{(\tilde h),(2)}_{0,2,\ul{\m}}(\V0)&=&0\;,\nn\\
\Big( \frac{\sqrt{3}}{2}\partial_{1} - \frac{1}{2}\partial_{2} \Big)\widetilde W^{(\tilde h),(2)}_{0,2,\ul{\m}}(\V0)&=&0\;,\label{bos6}
\eea
and the last two equalities imply that $\partial_{i}\widetilde W^{(\tilde h),(2)}_{0,2,\ul{\m}}(\V0)=0$, for $i=1,2$. The same argument can be repeated for $j_0=3$, and this concludes the proof of (\ref{bos4}). 

Setting
\be
z^{(j_0)}_{\m,\ul a ,\tilde h} := \partial_{\m}\widetilde W^{(\tilde h),(j_0)}_{2,0,\ul{a}}(\V0)\;,\qquad e^{(j_0)}_{\m, \ul a, \tilde h} := \widetilde W^{(\tilde h),(j_0)}_{2,1,\ul{a},\m}(\V0,\V0)\;,\label{bos6a}
\ee
the partition function can be rewritten as:
\be
\widetilde \Xi_{\b,L} = \int \widetilde P_{\leq \tilde h}(d\psi^{(\leq \tilde h)})P_{\leq \tilde h}(dA^{(\leq \tilde h)})\eu^{\widetilde\VV^{(\tilde h)}(\sqrt{Z_{\tilde h}}\psi^{(\leq \tilde h)},A^{(\leq \tilde h)})}\;,\label{bos6b}
\ee
where: (i) the propagator of $\widetilde P_{\leq \tilde h}(d\psi^{(\leq \tilde h)})$ is obtained by adding to the exponential of the gaussian weight of $P_{\leq \tilde h}(d\psi^{(\leq \tilde h)})$ the mass term of (\ref{bos4}), {\it i.e.} it is given by\footnote{Notice that now $\D_{\tilde h-1} := \tilde \D_{\tilde h-1}(\V0) = \D_{\tilde h}(1 + \b^{\D}_{\tilde h})$ with $\b^{\D}_{\tilde h} := \d_{\tilde h}$; compare with (\ref{lat17}).}
\bea
&&\hat g^{(\leq \tilde h)}(\kk') := \frac{\chi_{\tilde h}(\kk')}{\tilde Z_{\tilde h}(\kk')}\frac{\iu \g_0 k_0 + \iu \tilde v_{\tilde h}(\kk')\vec k'\cdot \g - \iu \g^{(j_0)*}\tilde\D_{\tilde h-1}(\kk')}{k_0^2 + \tilde v_{\tilde h}(\kk')^{2}|\vec k|^{2} + \tilde \D_{\tilde h-1}(\kk')^{2}}\big(1 + R_{\tilde h-1}(\kk')\big)\;,\nn\\
&&\tilde Z_{\tilde h}(\kk')\tilde \D_{\tilde h-1}(\kk') := \tilde Z_{\tilde h}(\kk')\tilde \D_{\tilde h}(\kk') + Z_{\tilde h} \D_{\tilde h}\D_{\tilde h}\;;\label{bos6c}
\eea
(ii) the interaction $\widetilde \VV^{(\tilde h)}(\psi,A)$ is
\bea
&&\widetilde\VV^{(\tilde h)}(\psi,A) := \int \frac{d\pp}{(2\pi)^3}\, \Big[e_{\m,\tilde h} j^{(\leq \tilde h)}_{\m,\pp}\hat A_{\m,\pp} - M^{\tilde h}\tilde \n_{\m,\tilde h}\hat A_{\m,\pp}\hat A_{\m,-\pp}\Big] + \nn\\&&
+ \int \frac{d\kk'}{D}\,k'_{\m}\lis\psi_{\kk',a_1,\s}z^{(j_0)}_{\m,\ul a ,\tilde h}\psi_{\kk',a_2,\s} + \int \frac{d\kk'}{D}\frac{d\pp}{(2\pi)^3}\,\hat A_{\m,\pp}\lis\psi_{\kk' + \pp,a_1,\s}e^{(j_0)}_{\m,\ul a,\tilde h}\psi_{\kk',a_2,\s}\nn\\&& + \RR \widetilde \VV^{(\tilde h)}(\psi,A)\;.\label{bos6d}
\eea
Let us write, as usual, $\psi^{(\leq \tilde h)} = \psi^{(<\tilde h)} + \psi^{(\tilde h)}$, $A^{(\leq \tilde h)} = A^{(<\tilde h)} + A^{(\tilde h)}$, where $A^{(\tilde h)}$, $\psi^{(\tilde h)}$ have propagators given respectively by (\ref{3.16}) with $h$ replaced by $\tilde h$ and 
\bea
\frac{\hat g^{(\tilde h)}(\kk)}{Z_{\tilde h}} &:=& \frac{\tilde f_{\tilde h}(\kk)}{Z_{\tilde h}}\frac{\iu \g_0 k_0 + \iu \tilde v_{\tilde h}(\kk')\vec k\cdot \vec \g - \iu \g^{(j_0)*}\tilde \D_{\tilde h-1}(\kk')}{k_0^2 + \tilde v_{\tilde h}(\kk')^{2}|\vec k|^{2} + \tilde \D_{\tilde h-1}(\kk')^{2}}\big(1 + R_{\tilde h-1}(\kk')\big)\;,\nn\\
\tilde f_{\tilde h}(\kk) &=& \frac{Z_{\tilde h}}{Z_{\tilde h}(\kk')}f_{\tilde h}(\kk)\;;\label{bos6f}
\eea
integrating the fields on scale $\tilde h$ along the lines of Section \ref{sec4} we recover our starting formula (\ref{bos1}) with $\tilde h$ replaced by $\tilde h-1$. The procedure can be iterated, and to prove the infrared stability of the renormalized expansion we have to check the boundedness of the flow of the running coupling constants. First of all, the flow of $\tilde \n_{\m,h}$ can be controlled using again the Ward identity (\ref{WI8}). Regarding the remaining running coupling constants, assume inductively that $\D_{k}\sim \D_{\tilde h}$ for all $k>h$; the flows of $\D_{h}$, $z^{(j_0)}_{\m,\ul a ,h}$, $e^{(j_0)}_{\m,\ul a,h}$ can be controlled exploiting the fact that the fermion propagator on a given scale $k> h$ is bounded by $\sim M^{\tilde h}$, that is its dimensional bound admits a {\it gain} $M^{k - \tilde h}$ with respect to the corresponding dimensional bound in the massless case. In fact, consider the Feynman graph expansions of the Beta functions of $\D_{h}$, $z^{(j_0)}_{\m,\ul a ,h}$, $e^{(j_0)}_{\m,\ul a,h}$; there are two possibilities: either there is at least one (massive) fermion propagator on scale $k$ such that $h<k\leq \tilde h$, or the dimensional bound of the graph is depressed by a factor $M^{\th(h - \tilde h)}$, because of the short memory property. Therefore, in both cases the bounds admit an improvement of a factor $M^{\th(h - \tilde h)}$ with respect to the usual dimensional bounds, and for this reason the Beta functions of the remaining running coupling constants are {\it summable}. This means that for $h\leq \tilde h$
\be
\D_{h} \sim \D_{\tilde h}\;,\qquad z^{(j_0)}_{\m,\ul a ,h} \sim e^{2}\;,\qquad e^{(j_0)}_{\m,\ul a,h}\sim e^{2}\;.
\ee
This concludes the proof of the boundedness of the flow, and our discussion on the multiscale analysis for the free energy of the model in presence of Kekul\' e distortion. 

\subsection{The two point Schwinger function}

We conclude this Section by discussing the evaluation of the two point Schwinger function; in particular, we will prove our main result (\ref{res9}). To procedure is completely analogous to the one followed in the case $\D_0=0$; therefore, we shall only discuss the differences with respect to the case $\D_0=0$, without repeating the whole argument. We integrate the fields scale after scale, and the main difference with respect to the case $\D_0=0$ is that formula (\ref{A3.3}) is replaced by
\bea
&&\widetilde\BBB^{(h)}(\sqrt{Z_h}\psi,\phi,A) = \nn\\
&& = \sum_{\s}\int \frac{d\kk'}{D}\,\lis\phi_{\kk',\s}\widetilde Q^{(h+1)}(\kk')^{T}\psi_{\kk',\s} + \lis\psi_{\kk',\s}\widetilde Q^{(h+1)}(\kk')\phi_{\kk',\s} + \nn\\&&
\quad + \int \frac{d\kk'}{D}\,\lis\phi_{\kk',\s}\widetilde G^{(h+1)}(\kk')^{T}\frac{\partial}{\partial \lis\psi_{\kk',\s}}\widetilde\VV^{(h)}(\sqrt{Z_{h}}\psi,A) + \nn\\
&&\quad + \frac{\partial}{\partial \psi_{\kk',\s}}\widetilde\VV^{(h)}(\sqrt{Z_h}\psi,A)\widetilde G^{(h+1)}(\kk')\phi_{\kk',\s}\;,
\eea
where the matrix $\widetilde G^{(h)}(\kk')$ is defined as
\be
\widetilde G(\kk') := \sum_{i=h}^{0}\frac{g^{(i)}(\kk')}{Z_{i-1}}\widetilde Q^{(i)}(\kk')\;,
\ee
while the matrix $\widetilde Q^{(h)}(\kk')$ is given by (repeated indeces are summed)
\bea
&&\widetilde Q^{(h)}(\kk') = \widetilde Q^{(h+1)}(\kk') - \iu Z_{h}\big[ z_{\m,h}\g_{\m}k'_{\m} + \D_{h}\d_{h}\g^{(j_0)}\Big]\widetilde G^{(h+1)}(\kk')\;\;\;\;\;\mbox{($h\geq h^{*}$)}\;,\nn\\
&&\widetilde Q^{(h)}_{a_1,a_2}(\kk') = \widetilde Q^{(h+1)}_{a_1,a_2}(\kk') - \iu Z_{h^{*}}\Big[ \iu k'_{\m} z_{\m,a_1,a,h}^{(j_0)} + \D_{h}\d_{h} \g^{(j_0)}_{a_1,a} \Big]\widetilde G^{(h+1)}_{a,a_2}(\kk')\;\;\;\mbox{($h< h^{*}$)}\nn
\eea
The integration of the single scale and the check of the inductive assumption is done following the same strategy discussed in the case $\D_0=0$, taking into account the modifications of the multiscale analysis introduced above in the evaluation of the free energy for the case $\D_0\neq 0$; in particular, after the fermion propagator has become massive we stop to rescale the fields, and we collect all the new terms proportional to $k_{\m}\lis\psi\psi$, $A_{\m}\lis\psi\psi$ into new running coupling constants $z^{(j_0)}_{\m,\ul a, h}$, $e_{\m,\ul a, h}^{(j_0)}$ which behave dimensionally as marginal ones, but whose flows are bounded because of the improved dimensional bound on the fermion propagator.

The remaining part of the discussion is the same as for the case $\D_0=0$, and we refer the reader to Section \ref{schwing}; the two point Schwinger function is obtained deriving the generating functional with respect to the external fields $\phi_{\kk,\s}$, $\lis\phi_{\kk,\s}$. In particular, notice that the leading contribution to the two point function is given by the dressed propagator (\ref{lat6f}) or (\ref{bos6c}), depending on whether the external quasi-momentum $\kk'$ is such that $|\kk'|\geq \D$, $|\kk'|<\D$; the former propagator is essentially equal to the one of the case $\D_0=0$, while the latter is different because the running coupling constants appearing there are equal to their values reached on the mass scale $h^{*}$. This concludes the proof of our result (\ref{res8}).

\section{Gap generation}\label{secgap}
\setcounter{equation}{0}
\renewcommand{\theequation}{\ref{secgap}.\arabic{equation}}

So far, we have seen that the Kekul\'  e distortion of the honeycomb lattice is strongly renormalized by the electromagnetic interaction; in particular, the ratio of the dressed and bare ``Kekul\' e masses'' $\D/\D_0$ blows up as $\D_0\rightarrow 0$. The natural question is whether it is possible that, under suitable conditions, a {\em spontaneous} distortion (corresponding to a gap generation in the fermionic energy spectrum) may emerge. To understand this, we shall consider the hopping parameter as a dynamical variable depending on the lattice site and on the bond, and the resulting model will be studied in the {\it Born -- Oppenheimer approximation}.

\subsection{Gap equation}\label{secgapeq}

The Hamiltonian of the model in presence of a {\it dynamical} lattice distortion is
\bea
\widetilde \HHH_{\L}(\{\phi\}) &:=& \HHH_{\L}(\{\phi\}) + \KK_{\L}(\{\phi\})\;,\nn\\
\KK_{\L}(\{\phi\}) &:=& \frac{\kappa}{2g^{2}}\sum_{\substack{\vec x\in \L \\ i=1,2,3}}\phi_{\vec x,j}^{2}\;,\label{gap1}
\eea
where $\KK_{\L}(\{\phi\})$ is the elastic energy of the distortion $\phi\in \RRR$, $\kappa\in \RRR^{+}$ is the {\it stiffness} constant and $g\in \RRR$ is the (classical) {\it phonon coupling}. The free energy of the model in the Born -- Oppenheimer approximation is given by:
\bea
e^{-\b|\L| F_{\b,L}^{BO}} &:=& \int \big[\prod_{\substack{\vec x\in\L\\i=1,2,3}}d\phi_{\vec x,i}\big] e^{-\b\KK_{\L}(\{\phi\}) - \b|\L| F_{\b,L}(\{\phi\})}\nn\\ &=:&  \int \big[\prod_{\substack{\vec x\in\L\\i=1,2,3}}d\phi_{\vec x,i}\big] e^{-\b|\L| E_{\b,L}(\{\phi\})}\;,\nn\\
 F_{\b,L}(\{\phi\}) &:=& -\frac{1}{\b|\L|}\log \Tr\, e^{-\b\HHH_{\L}(\{\phi\})}\;;\label{gap2}
\eea
the total ``energy'' $E_{\b,L}(\{\phi\})$ is extremized by values of the distortion $\phi_{\vec x,i}^{*}$ satisfying the condition $\partial_{\phi_{\vec x,i}}E_{\b,L}(\{\phi\})=0$, which is equivalent to:
\bea
\phi_{\vec x,i}^{*} &=& \frac{g^{2}}{\kappa}\media{\r^{(E_{+})}_{\vec x,i}}_{\b, L}^{\phi^{*}}\;,\label{gap3}\\
\r^{(E_{+})}_{\vec x,i} &=& \sum_{\s=\uparrow\downarrow} \Big( a^{+}_{\vec x,\s}b^{-}_{\vec x + \vec\d_i,\s}\eu^{\iu e A_{(\vec x,i)}} + b^{+}_{\vec x + \vec\d_i,\s}a^{-}_{\vec x,\s}\eu^{-\iu e A_{(\vec x,i)}} \Big)\;.
\eea
Now, {\it if} the equation (\ref{gap3}) admits a non-trivial solution, and {\it if} this solution corresponds to a global minimum of the energy $E_{\b,L}(\{\phi\})$ uniformly in $\b,|\L|$, then from (\ref{gap2}):
\be
\lim_{\substack{\b\rightarrow+\infty \\ |\L|\rightarrow+\infty}}F_{\b,L}^{BO} = \lim_{\substack{\b\rightarrow+\infty \\ |\L|\rightarrow+\infty}} E_{\b,L}(\{\phi^{*}\})\;,\label{gap3b}
\ee
that is the integral over all the possible distortions in (\ref{gap2}) is dominated by a {\it single} configuration $\phi^{*}$, which therefore corresponds to the lattice distortion in the ground state of the system. 

Clearly, the problem of solving (\ref{gap3}) is very hard, and we will not be able to solve it in full generality. However, we will show that the equation (\ref{gap3}) admits as a solution a configuration $\phi^{*}$ which corresponds to a particular Kekul\' e distortion; and at least for small $g$ this solution corresponds to a {\it local} minimum of $E_{\b,L}(\{\phi\})$. This last fact follows because
\bea
\frac{\partial^{2}}{\partial \phi_{\vec x,i} \partial\phi_{\vec y, j}}\KK_{\L}(\{\phi^{*}\}) &=& \frac{\k}{g^{2}}\d_{\vec x,\vec y}\d_{i,j}\;,\nn\\
\frac{\partial^{2}}{\partial \phi_{\vec x,i} \partial\phi_{\vec y, j}} F_{\b,L}(\{\phi^{*}\}) &=& \media{\r^{(E_{+})}_{\vec x,i};\r^{(E_{+})}_{\vec y,j}}_{\b,L}^{\phi^{*}}\;,\label{gap3c}
\eea
and for $\max_{\vec x,i}|\phi^{*}_{\vec x,i}|$ small enough the $\r^{(E_{+})} - \r^{(E_{+})}$ correlations are bounded objects. The problem of showing the uniqueness of the solution and its correspondence to a global minimum of the total energy is beyond the purposes of this thesis.

\subsection{A solution to the gap equation}\label{secgapsol}

As a first step, it is convenient to rewrite the r.h.s. of (\ref{gap3}) in the functional integral representation. The observable $\r^{(E_{+})}_{\vec x,i}$ is manifestly gauge invariant; therefore, we can rewrite its statistical average as follows:
\bea
\media{\r^{(E_{+})}_{\vec x,i}}^{\phi^{*}}_{\b,L} &=& \partial_{\Phi_{(0,\vec x),i}} \log \int P(d\Psi)P(dA)\eu^{V(\psi,A) + v_{\phi^{*}}(\Psi,A) + (\Phi,\r^{(E_{+})})}\Big|_{\Phi=0}\;,\label{gap4}\\
(\Phi,\r^{(E_{+})}) &:=& \sum_{\s,i}\int d\xx\, \Phi_{\xx,i}\Big( \Psi^{+}_{\xx,1,\s}\Psi^{-}_{\xx + (0,\vec \d_i),2,\s}\eu^{\iu e A_{(\xx,i)}} + \Big)\nn
\eea
where repeated indeces are summed, and as usual $\int d\xx = \int_{0}^{\b}dx_0\sum_{\vec x\in\L}$.

We perform the following special choice for $\phi^{*}_{\vec x,i}$, corresponding to a Kekul\' e distortion of the honeycomb lattice:
\be
\phi^{*}_{\vec x,i} = \phi^{*,(j_0)}_{\vec x,i} := \phi_0 + \frac{2}{3}\D_{0}\cos(\vec p_{F}^{+}(\vec\d_{j} - \vec\d_{j_0} - \vec x))\;,\label{gap5}
\ee
where the parameters $\phi_{0}$, $\D_{0}$ will be determined starting from (\ref{gap3}); as we are going to see $\phi_{0}$ will correspond to a renormalization of the bare Fermi velocity, while, as it follows from the discussion in Section \ref{seclat1}, $\D_0$, the amplitude of the Kekul\' e distortion, behaves as a bare mass for the fermion propagator. Let us denote by $\media{\cdots}_{\beta,\L}^{(j_{0})}$ the statistical average in presence of a fixed distortion $\phi^{*,(j_0)}$, and call $\hat\r_{\vec k,i}^{(E_{+})}$ the Fourier transform of $\r^{(E_{+})}_{\vec x,i}$, see below; the following result is true. 

\begin{lemma}\label{lemgap}
For any $\b, L$, choosing $\phi^{*}_{\vec x,i}$ of the form (\ref{gap5}) with $\phi_0\in \RRR$, $\D_0\in \RRR$, the gap equation (\ref{gap3}) is equivalent to the following two equations:
\bea
\phi_0 &=& \frac{g^{2}}{\k}\frac{1}{|\L|} \media{\hat \r^{(E_{+})}_{\vec 0,1}}^{(1)}_{\b,L}\;,\\
\frac{\D_0}{3} &=& \frac{g^{2}}{\k}\frac{1}{|\L|} \media{\hat\r^{(E_{+})}_{\vec p_{F}^{+},1}}^{(1)}_{\b,L}\;.\label{gap5b} 
\eea
\end{lemma}
\noindent{\it Proof.} It follows that\footnote{To prove (\ref{gap6}), note that all the sites equivalent to $\vec x$, that is those with the same configuration of distorted nearest neighbours of $\vec x$, can be obtained starting from $\vec x$ and moving of $2\vec a_{1} - \vec a_{2}$, $2\vec a_{2} - \vec a_{1}$; therefore, $\media{\r_{\vec x,i}^{(E_{+})}}^{(j_0)}_{\b,L}$ must be invariant under $\vec x\rightarrow \vec x + 2\vec a_{1} - \vec a_{2}$, $\vec x\rightarrow \vec x + 2\vec a_{2} - \vec a_{1}$. This implies that in $\media{\hat \r_{\vec k,i}^{(E_{+})}}^{(j_0)}_{\b,L}$ only the Fourier modes verifying the following constraints can appear:
\be
\frac{3}{2}k_1 + \frac{3\sqrt{3}}{2}k_2 = 2\pi n_1\;,\qquad \frac{3}{2}k_1 - \frac{3\sqrt{3}}{2}k_2 = 2\pi n_2\;,\qquad n_{1},n_{2}\in \ZZZ\;;\label{per1}
\ee
from (\ref{per1}) we get $k_1 = \frac{2\pi}{3}(n_1 + n_2)$, $k_{2} = \frac{2\pi}{3\sqrt{3}}(n_1 - n_2)$, and the fact that $\vec k \in \DD_{L}$ implies $0\leq n_1+n_2 \leq 1$, $-2\leq n_1 - n_2\leq 2$. Therefore, the allowed integers are $\vec n = (1,0),\,(0,1),\,(-1,1),\,(1,-1)$: the first two choices correspond to $\vec p_{F}^{+}$, $\vec p_{F}^{-}$; the last two correspond to $\vec p_{F}^{-} - \vec p_{F}^{+} = \vec p_{F}^{+}$, $\vec p_{F}^{+} - \vec p_{F}^{-} = \vec p_{F}^{-}$ (these identities have to be understood modulo vectors of the reciprocal lattice). This concludes the proof of (\ref{gap6}).}
\bea
\media{\r_{\vec x,i}^{(E_{+})}}^{(j_0)}_{\b,L} &=:& \frac{1}{|\L|}\sum_{\vec k\in \DD_{L}}\eu^{-\iu \vec k\cdot \vec x}\media{\hat \r_{\vec k,i}^{(E_{+})}}_{\b,L}^{(j_0)}\label{gap6}\\
&=& \frac{1}{|\L|}\media{\hat \r_{\vec 0,i}^{(E_{+})}}_{\b,L}^{(j_0)} +\frac{1}{|\L|}\eu^{-\iu \vec p_{F}^{+}\vec x} \media{\hat \r_{\vec p_{F}^{+},i}^{(E_{+})}}_{\b,L}^{(j_0)} +  \frac{1}{|\L|}\eu^{-\iu \vec p_{F}^{-}\cdot \vec x} \media{\hat \r_{\vec p_{F}^{-},i}^{(E_{+})}}_{\b,L}^{(j_0)}\;,\nn
\eea
where the last equality is implied by the fact that $\media{\r_{\vec x,i}^{(E_{+})}}_{\b,L}^{(j_0)}$ reflects the periodicity of the distorted lattice; see Fig. \ref{figkek}. Therefore, from (\ref{gap5}) and (\ref{gap6}) we get
\bea
\phi_{0} &=& \frac{g^2}{\k}\frac{1}{|\L|} \media{\hat \r_{\vec 0,j}^{(E_{+})}}_{\b,L}^{(j_0)} \;,\label{gap7}\\
\frac{\D_0}{3} &=& \frac{g^2}{\k}\frac{1}{|\L|} \media{\hat\r_{\vec p_{F}^{\o},j}^{(E_{+})}}_{\b,L}^{(j_0)} \eu^{-\iu \vec p_{F}^{\o}(\vec\d_j - \vec\d_{j_0})}\;;\label{gap8}
\eea
from (\ref{gap7}) we see that a necessary condition for (\ref{gap5}) to be a solution of the gap equation is that the r.h.s. of (\ref{gap7}) and (\ref{gap8}) are independent of $j,j_0$ and $\o,j,j_0$, respectively. To prove this, we shall use that the generating functional appearing in (\ref{gap4}) is invariant under the symmetry transformations of Lemma \ref{lemlat}, {\it provided} the external field appearing in (\ref{gap4}) transforms as discussed in Section \ref{exc}.

Let us first consider the first equation in (\ref{gap7}). The invariance under (4) and (6.a) imply respectively that
\bea
&&\media{\hat\r^{(E_{+})}_{\vec 0,1}}^{(j_0 - 1)}_{\b,L} = \media{\hat\r^{(E_{+})}_{\vec 0,2}}^{(j_0)}_{\b,L} = \media{\hat \r^{(E_{+})}_{\vec 0,3}}^{(j_0 + 1)}_{\b,L}\;,\\
&&\media{\hat\r^{(E_{+})}_{\vec 0,2}}^{(j_0)}_{\b,L} = \media{\hat \r^{(E_{+})}_{\vec 0,3}}^{(j_0)}_{\b,L}\;,\label{gap9}
\eea
therefore $\media{\hat\r^{(E_{+})}_{\vec 0,j}}^{(j_0)}_{\b,L}$ is independent of $j$ and $j_0$; moreover, from symmetry (5) we get that
\be
\media{\hat\r^{(E_{+})}_{\vec 0,j}}^{(j_0)}_{\b,L}\in\RRR\;.\label{gap10}
\ee
Consider now the second line of (\ref{gap8}). First of all, let us prove that the r.h.s. is independent of $j,j_0$; to see this, notice that because of symmetry (4) the problem is equivalent to show that
\be
\media{\hat \r^{(E_{+})}_{\vec p_{F}^{\o},1}}^{(j_0 - j + 1)}_{\b,L}\eu^{\iu \vec p_{F}^{\o}(\vec\d_{j_0 - j + 1} - \vec\d_1)}\qquad \mbox{is independent of $j,j_0$}\;,\label{gap11}
\ee 
where we used that $\vec p_{F}^{\o}\vec\d_{i} = (T^{-1}\vec p_{F}^{\o})\vec\d_{i} = \vec p_{F}^{\o}\vec\d_{i+1}$, with $T$ the $2\pi/3$ rotation matrix. The claim (\ref{gap11}) can be checked explicitly using again symmetries (4) and (6.a); in fact the combination (4) - (6.a) - (4) gives:
\bea
&& \media{\hat\r^{(E_{+})}_{\vec p_{F}^{\o},1}}^{(2)}_{\b,L} = \media{\hat \r^{(E_{+})}_{\vec p_{F}^{\o},2}}^{(3)}_{\b,L} = \eu^{-\iu \vec p_{F}^{\o}(\vec\d_{2} - \vec\d_{1})}\media{\hat \r^{(E_{+})}_{\vec p_{F}^{\o},3}}^{(3)}_{\b,L} = \eu^{-\iu \vec p_{F}^{\o}(\vec\d_{2} - \vec\d_{1})}\media{\hat \r^{(E_{+})}_{\vec p_{F}^{\o},1}}^{(1)}_{\b,L}\;,\nn\\
&& \media{\hat\r^{(E_{+})}_{\vec p_{F}^{\o},1}}^{(3)}_{\b,L} = \media{\hat \r^{(E_{+})}_{\vec p_{F}^{\o},3}}^{(2)}_{\b,L} = \eu^{\iu \vec p_{F}^{\o}(\vec\d_{2} - \vec\d_{1})}\media{\hat \r^{(E_{+})}_{\vec p_{F}^{\o},2}}^{(2)}_{\b,L} = \eu^{-\iu \vec p_{F}^{\o}(\vec\d_{3} - \vec\d_{1})}\media{\hat \r^{(E_{+})}_{\vec p_{F}^{\o},1}}^{(1)}_{\b,L} \;,\nn
\eea
where in the last equality we used again the invariance of $\vec p_{F}^{\o}$ under $T$. Therefore, the above chain of identities proves that
\be
\media{\hat \r^{(E_{+})}_{\vec p_{F}^{\o},j}}^{(j_0)}_{\b,L}\eu^{-\iu \vec p_{F}^{\o}(\vec\d_j - \vec\d_{j_0})} = \media{\hat \r^{(E_{+})}_{\vec p_{F}^{\o},1}}^{(j_0 - j + 1)}_{\b,L}\eu^{\iu \vec p_{F}^{\o}(\vec\d_{j_0 - j + 1} - \vec\d_1)} = \media{\hat \r^{(E_{+})}_{\vec p_{F}^{\o},1}}^{(1)}\;;\label{gap12}
\ee
finally, from (6.b) and (5) we get:
\be
\media{\hat \r^{(E_{+})}_{\vec p_{F}^{\o},1}}^{(1)}_{\b,L} = \media{\hat \r^{(E_{+})}_{\vec p_{F}^{-\o},1}}^{(1)}_{\b,L}\;,\qquad \media{\hat \r^{(E_{+})}_{\vec p_{F}^{\o},1}}^{(1)}_{\b,L} \in \RRR\;,\label{gap13}
\ee
and this concludes the proof of Lemma \ref{lemgap}.\qed\\

To conclude, we have to show that the equations (\ref{gap7}), (\ref{gap8}) admit a non-trivial solution; in particular, we will be interested in understanding the effect on the electromagnetic interaction on the solution. First of all, we rewrite the r.h.s. of (\ref{gap7}), (\ref{gap8}) as (we set $\media{\cdots} \equiv \media{\cdots}_{\b,L}^{(1)}$ for notational simplicity):
\bea
&&\media{\hat \r^{(E_{+})}_{\vec 0,1}} = \sum_{\substack{\vec x\in\L \\ \s= \uparrow\downarrow}}\Big[\media{a^{+}_{\vec x,\s}b^{-}_{\vec x + \vec\d_1,\s}} + \media{b^{+}_{\vec x + \vec\d_1,\s}a^{-}_{\vec x,\s}}\Big] + \label{gap13a}\\&& + \sum_{\substack{\vec x\in\L \\ \s= \uparrow\downarrow}}\Big[\media{a^{+}_{\vec x,\s}b^{-}_{\vec x + \vec\d_1,\s}\Big( \eu^{\iu e A_{(\vec x,1)}} - 1\Big)} + \media{b^{+}_{\vec x + \vec\d_1,\s}a^{-}_{\vec x,\s}\Big( \eu^{-\iu e A_{(\vec x,1)}} - 1\Big)}\Big]\;,\nn\\
&&\media{\hat \r^{(E_{+})}_{\vec p_{F}^{+},1}}  = \sum_{\substack{ \vec x\in\L \\ \s = \uparrow\downarrow}}\eu^{-\iu \vec p_{F}^{+}\vec x}\Big[\media{a^{+}_{\vec x,\s}b^{-}_{\vec x + \vec\d_1,\s}} + \media{b^{+}_{\vec x + \vec\d_1,\s}a^{-}_{\vec x,\s}}\Big] + \label{gap13b}\\&& + \sum_{\substack{\vec x\in\L \\ \s= \uparrow\downarrow}}\eu^{-\iu \vec p_{F}^{+}\vec x}\Big[\media{a^{+}_{\vec x,\s}b^{-}_{\vec x + \vec\d_1,\s}\Big( \eu^{\iu e A_{(\vec x,1)}} - 1\Big)} + \media{b^{+}_{\vec x + \vec\d_1,\s}a^{-}_{\vec x,\s}\Big( \eu^{-\iu e A_{(\vec x,1)}} - 1\Big)}\Big]\;;\nn
\eea
the terms appearing in (\ref{gap13a}), (\ref{gap13b}) involving the exponential of the photon field can be studied again using multiscale analysis and RG, and it follows that they give rise to $O(e^{2})$ corrections (we omit the details); therefore, using the functional integral representation in the Feynman gauge, we can rewrite (\ref{gap13a}) and (\ref{gap13b}) as:
\bea
\frac{1}{|\L|}\media{\hat \r^{(E_{+})}_{\vec 0,1}} &=& -2\int \frac{d \kk}{D}\, \Big[ \media{\hat \Psi^{-}_{\kk,\uparrow,2}\hat \Psi^{+}_{\kk,\uparrow,1}} + \media{\hat \Psi^{-}_{\kk,\uparrow,1}\hat \Psi^{+}_{\kk,\uparrow,2}} \Big] + O(e^{2})\;,\label{gap13c}\\
\frac{1}{|\L|}\media{\hat \r^{(E_{+})}_{\vec p_{F}^{+},1}} &=& -2\int \frac{d\kk}{D}\, \Big[ \media{\hat \Psi^{-}_{\kk + \pp_{F}^{+},\uparrow,2}\hat \Psi^{+}_{\kk+\pp_{F}^{-},\uparrow,1}} + \media{\hat \Psi^{-}_{\kk+\pp_{F}^{+},\uparrow,1}\hat \Psi^{+}_{\kk+\pp_{F}^{-},\uparrow,2}} \Big] + \nn\\&&+ O(e^{2})\nn
\eea
Notice that in the integrals appearing in (\ref{gap13c}) the main contribution come from $\D\leq |\kk|$; in fact, for $|\kk|<\D$ the two point function is bounded by $(\const.)\D^{-1}$, and therefore the integral over $|\kk|\leq \D$ is $O(\D^{2})$. The gap equations (\ref{gap7}), (\ref{gap8}), together with (\ref{gap13c}) and our result on the two point function in presence of distortion, can be in principle studied numerically, and quantitative predictions can be made; here however we only want to get a qualitative picture of the behavior of the solution as a function on the interaction. For this purpose, we can approximate the equations for $\phi_0$, $\D_0$ as follows:
\bea
\phi_{0} &\simeq& \frac{4 g^{2}}{\k}\sum_{\o}\int\limits_{\D\leq |\kk'|\leq 1}\frac{d\kk'}{D}\,\frac{1}{Z(\kk' + \pp_{F}^{\o})}\frac{v(\kk' + \pp_{F}^{\o})\Re\, \O(\vec k' + \vec p_{F}^{\o})}{k_0^2 + v(\kk' + \pp_{F}^{\o})^2|\kk'|^2}\nn\\
\frac{\D_0}{3} &\simeq& \frac{4 g^2}{\k}\sum_{\o}\int\limits_{\D\leq |\kk'|\leq 1}\frac{d\kk'}{D}\,\frac{1}{Z(\kk' + \pp_{F}^{\o})}\frac{\D(\kk' + \pp_{F}^{\o})}{k_0^2 + v(\kk' + \pp_{F}^{\o})^2|\kk'|^2}\label{gap14}
\eea
where
\be
\D(\kk' + \pp_{F}^{\o}) = \D_0|\kk'|^{-\eta_K}\;,\quad Z(\kk' + \pp_{F}^{\o}) = |\kk'|^{-\eta}\;,\quad v(\kk') = 1 - (1 - v)|\kk'|^{\tilde\eta}\;;\label{gap15}
\ee
the gap equation (\ref{gap14}) is similar to the one found first in \cite{M1}, see also \cite{COKE}, in the context of Luttinger superconductors. From (\ref{gap14}) we get that $\phi_0 = O(g^2)$; the discussion for $\D_0$ is less trivial, and it goes as follows.

\paragraph{``Small'' electron-electron interactions.} Let us consider the case of ``small'' electron-electron interactions. Notice that in order that the effective Fermi velocity flows from $v$, the bare value, to $2v$ the momentum $\r$ has to be equal to:
\be
\r = \Big( \frac{1 - 2v}{1-v} \Big)^{\frac{1}{\tilde\eta}}\Rightarrow \r \simeq e^{-\frac{v}{\tilde\eta}} \qquad \mbox{for $\tilde\eta \ll v$;}\label{gap15b}
\ee
therefore, for $\tilde\eta\ll v$ the Fermi velocity is for all practical purposes equal to $v$, and the integral in (\ref{gap14}) is equivalent to
\be
\int_{\D}^{1} d\r \int_{1}^{-1} d\cos\th \frac{\D_0 \r^{\eta - \eta_K}}{\cos^{2}\th + v^2(1 - \cos^{2}\th)} = \frac{\arctan \xi}{\xi}\frac{2\D_0}{v^2}\frac{\big(1 - \D^{\eta - \eta_K + 1}\big)}{\eta - \eta_K + 1}\;,\label{gap16}
\ee
where $\xi = \sqrt{v^{-2} - 1}$. Plugging this result into (\ref{gap14}) we get that
\be
\D_0 \simeq \Big(1 - \frac{g_0^2}{g^2}\Big)^{\frac{1 + \eta_{K}}{1 + \eta - \eta_{K}}}\;,\qquad \mbox{for $|g|>g_0 = O(\sqrt{\k}v)$}\;,\label{gap17}
\ee
and in particular the critical coupling $g_0>0$ is essentially independent of $e$. However, still the electromagnetic interactions play a non-trivial role. In fact, since $\eta - \eta_{K} = -\frac{7 e^{2}}{12\pi^2} + O(e^{4})$, the exponent appearing in (\ref{gap17}) is greater than $1$; this means that close to $g_0$ (\ref{gap17}) behaves as $\D_0(g)\sim (g^{2} - g_0^{2})^{1 + \a }$ with $\a = \frac{15e^2}{12\pi^2} + O(e^4)>0$, that is $\partial_{g}\D_0(g)$ is {\it continuous} at $|g|=g_0$: the transition is smoothened by the interaction.

\paragraph{``Strong'' electron-electron interactions.} The discussion in this case will be more speculative. From the analyses of Sections \ref{exc} and \ref{RGlat}, one may reasonably guess that electron-electron interactions tend to favor excitonic instabilities, and in particular a Kekul\'e distortion of the honeycomb lattice: as we have seen in Section \ref{exc}, the excitonic response function $C^{E_+}$ is amplified by the interaction, while, as discussed in Section \ref{RGlat}, a preexisting Kekul\' e distortion is greatly enhanced. This belief is corroborated by the fact that the integrand appearing in the second equation in (\ref{gap14}) scales as $|\kk|^{\eta - \eta_K - 2}$, with $\eta - \eta_K = -\frac{7e^2}{12\pi^2} + O(e^4)$; therefore, {\it if} $\eta - \eta_{K}\rightarrow -1$, a situation which corresponds to a strong coupling regime, the integral in (\ref{gap14}) is bounded from below by a quantity of order $|\log\D|$. This means that in this strong coupling limit the gap equation (\ref{gap14}) admits a solution {\it for any} $|g|>0$; in particular, if $\eta_K - \eta - 1$ exceeds $0$ then
\be
\D_0 \simeq \Big( 1 + \frac{\kappa(\eta_K - \eta - 1)}{g^2} \Big)^{\frac{1}{1 + \eta - \eta_K}}\;,\qquad \mbox{for $|g|>0$.}
\ee
Hence, it is reasonable to expect the critical value of the phonon coupling $g_0$ to be generically lowered by the electromagnetic interaction.

\chapter{Conclusions and perspectives}\label{capcon}
\setcounter{equation}{0}
\renewcommand{\theequation}{\ref{capcon}.\arabic{equation}}

In this Thesis we investigated the effect of electromagnetic interaction on electrons on the honeycomb lattice at half-filling ($1$ electron per site in average), with the physical motivation of understanding the role of electron-electron interactions in the low energy physics of graphene. In particular, we introduced a new model for graphene, where the the coupling with a quantized electromagnetic field is defined so to respect gauge invariance; therefore, our model can be seen as a ``lattice gauge theory model'' for graphene. Compared to the existing literature, the model introduced here {\it does not} neglect the lattice (no Dirac approximation is involved) and {\it does not} neglect the presence of the vector potential, which is necessary in order to guarantee that the full theory is invariant under local gauge transformations. 

Using rigorous Renormalization Group methods and lattice Ward identities we succeeded in characterizing, at all orders in renormalized perturbation theory, many of its ground state and low energy properties, \cite{GMP2}; in particular, we investigated: (i) the effect of the interaction on the two point Schwinger function (the {\it dressed propagator}); (ii) the effect of the interaction on some response functions; (iii) the effect of special lattice distortions, the Kekul\' e ones; (iv) a mechanism for spontaneous lattice distortion. 

Regarding the two point Schwinger function, its scaling properties are deeply modified by the presense of the interaction; in fact, interaction-dependent {\it anomalous exponents} appear. This means that the interacting system behaves as a Luttinger liquid: as far as we know, this is the first time that Luttinger liquid behavior is found in a two dimensional system. Moreover, there is an {\it emergent Lorentz symmetry}, due to the fact that the effective Fermi velocity (that is, the one appearing in the two point Schwinger function) tends to the speed of light at the Fermi surface. After this, we studied the effect of the electromagnetic interaction on the excitonic, charge density wave and density-density response functions (or susceptibilites); we have found that, as for the two point Schwinger function, these response functions scale with interaction dependent anomalous exponents. In particular, some response functions are {\it amplified} by the interaction, namely those corresponding to Kekul\'e or charge density wave pairings; the former is associated to a Kekul\' e distortion of the honeycomb lattice, see Fig. \ref{figkek}, the second to a periodic alternation of excess/deficit of electrons on the A/B sites of the lattice.

To get a deeper understanding of the effect of lattice distortions we changed the definition of the model in order to take into account a preexisting Kekul\' e distortion; interestingly, the amplitude of the bonds deformation behaves as a bare mass for the fermion propagator, which is {\it strongly renormalized} by the electromagnetic interaction. This is interpreted by saying that Kekul\' e distortions are enhanced by the electron-electron interaction. Finally, motivated by the results of the response functions and on the mass renormalization, we investigated the possibility of {\it spontaneous} lattice distortions. We treated the lattice distortion as a classical dynamical variable (a classical {\it phonon field}), in the Born-Oppenheimer approximation; by using a variational argument we have found that the total energy is extremized in correspondence of a suitable Kekul\' e distortion, whose amplitude is determined by a well-defined self-consistence equation. We studied this gap equation from a qualitative viewpoint; in general, one sees that the solution of the equation as a function of the phonon coupling $g$ becomes non-trivial for $g$ greater than some critical value $g_0$. For weak electromagnetic interactions we have found that the transition from a gapless to a gapped phase is ``smoothened'' (the interaction removes the discontinuity at $g_0$ of the first derivative of the gap with respect to $g$), while $g_0$ is essentially interaction-independent; however, from the same equation we got evidence that strong electromagnetic interactions favor the emergence of a Kekul\' e instability by lowering the value of $g_0$.\\

Regarding the perspectives of our work, these are many both from a physical and a mathematical point of view. For instance, from the physical point of view, it would be very interesting to understand the issue of {\it universality} of the electric conductivity in graphene; in fact, it has been experimentally observed with very high precision that the optical conductivity is equal to $4e^{2}/h$: remarkably, this value is not affected by higher order corrections and material parameters, such as the Fermi velocity. The explaination of this phenomenon is very challenging from a theoretical point of view, and some attempts have already been made in the literature; however, all the existing arguments are strongly based on the logarithmic divergence of the Fermi velocity, which is, in our opinion, an unphysical feature of the models considered so far. To show that the growth of the Fermi velocity does not play any role in this remarkable cancellation it would be very interesting to see first if it can be recovered in a case in which the effective Fermi velocity is {\it weakly} renormalized; for instance, one can consider the two dimensional Hubbard model on the honeycomb lattice, whose ground state has been rigorously constructed in \cite{GM}. In that case, the effective Fermi velocity turns out to be an analytic function of the coupling, close to the bare one.

Another interesting physical problem is related to the insensitivity of graphene to localization effects usually induced by disorder; to understand this issue, it would be very interesting to add some randomness to the model defined here, and see if our RG analysis can be adapted to take into account the disorder. The effect of disorder on Dirac fermions has been investigated well before the experimental discovery of graphene. For instance, in a pioneering work, \cite{F1, F2}, Fradkin considered a model of noninteracting Dirac fermions in presence of disorder; by a replica analysis it was shown that $2$ is the lower critical dimension for a localization/delocalization transition. In particular, considering $N$ different species of fermions, in the limit $N\rightarrow+\infty$ it was shown that in two dimensions the mean free path of the electrons is always finite in presence of disorder, and that it grows exponentially as the intensity of the disorder goes to zero. Moreover, in the same formalism a universal nonvanishing minimal value of the conductivity was predicted. It would be very interesting to see if a similar analysis can be reproduced using our rigorous RG framework in the case of finite $N$, and eventually taking into account the effect of electron--electron interactions.

From the mathematical physics point of view, the model that we introduced here opens the way to many interesting problems; the most interesting one is the proof of convergence of the perturbative series. In fact, our present results are perturbative, in the sense that all the quantities that we compute are given by series in the running coupling constants with finite coefficients, where the $N$-th order contribution is bounded proportionally to $(N/2)!$; the bound is a consequence of the factorial growth of the Feynman graphs at a given order, and it is clearly not enough to prove absolute convergence. In purely fermionic theories the solution of this combinatorial problem is well-known, \cite{Le, GK1, BGPS}, and it is based on the so-called {\it Gram bounds}; basically, one exploits the $-1$ arising in the anticommutation of the fermionic fields to show that the $N$-th order contribution to the perturbative series reconstructs the determinant of an $N\times N$ matrix, which is bounded proportionally to $(\const.)^{N}$. Here the problem is that we have {\it both} fermions and bosons, and the Gram trick applies only to the fermions; to control the proliferation of graphs obtained exchanging bosonic lines we need some other argument. We expect this to be a hard but doable task; for instance, one could try to adapt the cluster expansion ideas of \cite{GK2, GK3}, or perhaps the recent ``loop vertex expansion'' of \cite{GMS, RW}. If one succeeds in this, it would be the first rigorous construction of a two-dimensional Luttinger liquid.

\appendix
\chapter{Grassmann integration}\label{app0}
\setcounter{equation}{0}
\renewcommand{\theequation}{\ref{app0}.\arabic{equation}}
In this subsection we recall some basic concepts of the Grassmann calculus.

Let us consider a finite dimensional {\it Grassmann algebra}, which is a set of anticommuting {\it Grassmann variables} $\{\Psi^{+}_{\a}\,\Psi^{-}_{\a}\}_{\a\in A}$, for some finite set $A$. This means that:
\be
\{\Psi^{\e}_{\a}\,,\Psi^{\e'}_{\a'}\} := \Psi^{\e}_{\a}\Psi^{\e'}_{\a'} + \Psi^{\e'}_{\a'}\Psi^{\e}_{\a} = 0\;,\qquad \forall \a,\a'\in A\;,\quad \forall \e,\e' = \pm\;;\label{1.2.1}
\ee
in particular $\Psi^{\e}_{\a}\Psi^{\e}_{\a} =0$ $\forall \a\in A$ and $\forall \e=\pm$.

Let us introduce another set of Grassmann variables $\{d\Psi^+_{\a}\,,d\Psi^{-}_{\a}\}_{\a\in A}$, anticommuting with $\Psi^{+}_{\a},\Psi^{-}_{\a}$, and a liner operation, the {\it Grassmann integration}, defined by
\be
\int \Psi^{\e}_{\a}d\Psi^{\e}_{\a} =1\;,\qquad \int d\Psi^{\e}_{\a}=0\;,\qquad \a\in A\;,\quad \e=\pm\;.\label{1.2.2}
\ee
If $F(\Psi)$ is a polynomial in $\Psi^{+}_{\a}$, $\Psi^{-}_{\a}$, $\a\in A$, the operation
\be
\int \prod_{\a\in A} d\Psi^{+}_{\a}d\Psi^{-}_{\a} F(\Psi)\;,\label{1.2.3}
\ee
is simply defined by iteratively applying (\ref{1.2.2}) and taking into account the anticommutation rules (\ref{1.2.1}). It is easy to check that for all $\a\in A$ and $C\in \CCC$
\be
\frac{\int d\Psi^+_{\a} d\Psi^-_{\a}e^{-\Psi^+_{\a}C\Psi^{-}_{\a}}\Psi^{-}_{\a}\Psi^{+}_{\a}}{\int d\Psi^+_{\a}d\Psi^{-}_{\s}e^{-\Psi^{+}_{\a}C\Psi^{-}_{\a}}} = C^{-1}\;;\label{1.2.4}
\ee
in fact $e^{-\Psi^{+}_{\a}C\Psi^{-}_{\a}} = 1 - \Psi^{+}_{\a}C\Psi^{-}_{\a}$ and by (\ref{1.2.2})
\be
\int d\Psi^+_{\a} d\Psi^-_{\a}e^{-\Psi^+_{\a}C\Psi^{-}_{\a}} = C\;,\label{1.2.5}
\ee
while
\be
\int d\Psi^+_{\a} d\Psi^-_{\a}e^{-\Psi^+_{\a}C\Psi^{-}_{\a}}\Psi^{-}_{\a}\Psi^{+}_{\s} = 1\;.\label{1.2.6}
\ee
More generally, if $K$ is an $|A|\times |A|$ invertible complex matrix, the above formulas are generalized in the following way:
\be
\frac{\int \prod_{\a\in A} d\Psi^+_{\a}d\Psi^{-}_{\a} e^{-\sum_{i.j\in A}\Psi^+_{i}M_{ij}\Psi^{-}_{j}}\Psi^{-}_{\a'}\Psi^{+}_{\b'}}{\int \prod_{\a\in A}d\Psi^+_{\a}d\Psi^{-}_{\a} e^{-\sum_{i,j\in A}\Psi^+_{i}M_{ij}\Psi^{-}_{j}}} = \Big[M^{-1}\Big]_{\a'\b'}\;,\label{1.2.7}
\ee
Again, (\ref{1.2.7}) can be easily verified by using (\ref{1.2.2}) and the anticommutation rules (\ref{1.2.1}); in particular it follows that
\be
\int \prod_{\a\in A} d\Psi^+_{\a}d\Psi^{-}_{\a} e^{-\sum_{i.j\in A}\Psi^+_{i}M_{ij}\Psi^{-}_{j}} = \det M\;,\label{1.2.8}
\ee
while
\be
\int \prod_{\a\in A}d\Psi^+_{\a}d\Psi^{-}_{\a} e^{-\sum_{i,j\in A}\Psi^+_{i}M_{ij}\Psi^{-}_{j}}\Psi^{-}_{\a'}\Psi^{+}_{\b'} = M'_{\a'\b'}\;,\label{1.2.9}
\ee
where $M'_{\a'\b'}$ is the minor complementary to the entry $M_{\a'\b'}$ of $K$.

\chapter{Naive perturbation theory}\label{app1}
\setcounter{equation}{0}
\renewcommand{\theequation}{\ref{app1}.\arabic{equation}}
In this Appendix we will derive a ``naive'' perturbative series in the electric charge $e$ for the grand-canonical averages computed with the Hamiltonian $\HHH_{\L}$ defined in (\ref{1.1.1}); in particular, we shall show that the Hamiltonian model  introduced in Section \ref{sec1} and the quantum field theory model defined in Section \ref{sec2.2} share the same perturbative expansion. In this sense, we say that the quantum field theory model introduced in Section \ref{sec2.2} is the {\it functional integral representation} of the Hamiltonian model introduced in Section \ref{sec1}.

The Appendix is organized in the following way: in Section \ref{app1.1} we derive a perturbative expansion for the grand-canonical averages computed with $\HHH_{\L}$; in Sections \ref{app1.1.1}, \ref{app1.1.2} we compute the fermion and boson propagators, respectively; finally, in Section \ref{sec1.2} we show how to represent the perturbative series introduced in Section \ref{app1.1} in terms of Feynman graphs, and we show that the same perturbative series is generated by the quantum field theory model introduced in Section \ref{sec2.2}.

\section{Trotter product formula}\label{app1.1}

In this Section we will derive a perturbative expansion for the grand-canonical averages of physical observables computed with the Hamiltonian $\HHH_{\L}$; for pedagogical purposes, we start by discussing the case of the partition function, namely:
\be
\Xi_{\b,L} = \Tr\big\{ e^{-\b \HHH_{\L}} \big\}\;.\label{A02}
\ee
The averages of physical observables can be studied in a completely analogous way. It is convenient to define the {\it unperturbed imaginary time evolutions} of the operators $A_{i,(\vec x,0)}$, with $i=1,2$, and of $a^{\pm}_{\vec x,\s}$, $b^{\pm}_{\vec x+\vec \d_1,\s}$ as follows ($\xx = (x_0,\vec x)$):
\bea
&&\tilde a^{\pm}_{\xx,\s} := \eu^{\HHH_{0,\L} x_0}a^{\pm}_{\vec x,\s}\eu^{-\HHH_{0,\L} x_0}\;,\quad \tilde b^{\pm}_{\xx + (0,\vec\d_1),\s} := \eu^{\HHH_{0,\L} x_0}b^{\pm}_{\vec x + \vec\d_1,\s}\eu^{-\HHH_{0,\L} x_0}\;,\label{A01}\\
&& \tilde A_{i,\xx} := \eu^{\HHH_{0,\L} x_0} A_{i,(\vec x,0)}\eu^{-\HHH_{0,\L} x_0}\;, \quad \tilde A_{(\xx,j)} := \sum_{i=1,2}\int_{0}^{1}ds\,[\vec\d_{j}]_{i}\,\tilde A_{i,(x_0,\vec x + s\vec\d_j)}\;;\nn
\eea
moreover, we define the unperturbed imaginary time evolution of $V_{\L}$ as:
\bea
&&\tilde V_{\L}(x_0) := e^{\HHH_{0,\L} x_0}V_{\L}e^{-\HHH_{0,\L}x_0} \label{A02b}\\
&&= -t \sum_{\substack{\vec x\in \L \\ i =1,2,3}}\sum_{\s = \uparrow\downarrow}\sum_{m\geq 1} \Big( \tilde a^{+}_{\xx,\s}\tilde b^{-}_{\xx + (0,\vec\d_i),\s}\frac{(\iu e)^m}{m!}\tilde A^{m}_{(\xx,i)} + \tilde b^{+}_{\xx + (0,\vec\d_i)}\tilde a^{-}_{\xx,\s}\frac{(-\iu e)^{m}}{m!}\tilde A^{m}_{(\xx,i)} \Big) + \nn\\
&&\quad + \frac{e^2}{2}\sum_{\vec x,\vec y\in \L_A \cup \L_{B}}(\tilde n_{\xx} - 1)\varphi(\vec x - \vec y,0)\d_{x_0,y_0}(\tilde n_{\yy} - 1)\;,\nn
\eea
where $\tilde n_{\xx}$ is the time evolution of the density operator. Our starting point will be the {\it Trotter product formula}, which consists in the following identity:
\be
e^{-t \HHH_{\L}} = \lim_{n\rightarrow +\infty} \Big[e^{-\frac{t}{n} \HHH_{0,\L}}\Big( 1 - \frac{t V_{\L}}{n} \Big)\Big]^{n}\;;\label{A4}
\ee
in fact, using (\ref{A4}) it is straightforward to see that the partition function (\ref{A02}) can be rewritten as follows:
\be
\frac{\Tr\big\{ e^{-\b \HHH_{\L}} \big\}}{\Tr\{ e^{-\b \HHH_{0,\L}}\}} = 1 + \frac{1}{\Tr\{ e^{-\b \HHH_{0,\L}}\}}\sum_{n\geq 1} \pm  \int \Tr\Big\{ e^{-\b\HHH_{0,\L}} \tilde V_{\L}(t_1)\tilde V_{\L}(t_{2})\cdots\tilde V_{\L}(t_{n})\Big\}\;,\label{A5}
\ee
where the integral is over all the $t_{i}$ variables with $1\leq i\leq n$, under the constraint that they decrease in their index $i$, and the sign $\pm$ is $+$ if the number of $\tilde V$ is even and $-$ otherwise. 

Being $\HHH_{0,\L}$ quadratic in the fermionic and bosonic creation/annihilation operators, the Wick rule holds for evaluating the ratios in the r.h.s. of (\ref{A5}); in particular, setting
\be
\tilde \Psi^{\pm}_{\xx,\s,1} := \tilde a^{\pm}_{\xx,\s}\;,\qquad \tilde\Psi^{\pm}_{\xx + (0,\vec\d_1),\s,2} := \tilde b^{\pm}_{\xx + (0,\vec\d_1),\s}\;,
\ee
it is possible to express the various terms in (\ref{A5}) a suitable integrals of products of expressions like, see Section \ref{sec1.2}:
\bea
\Big[g^{+}_{\s,\s'}(\xx-\yy)\Big]_{\r,\r'} &:=& \frac{\Tr\big\{e^{-\b \HHH_{0,\L}} \tilde\Psi^{+}_{\xx,\s,\r}\tilde \Psi^{-}_{\yy,\s',r'}\big\}}{\Tr\{ e^{-\b\HHH_{0,\L}}\}}\;,\label{A6a}\\
\Big[g^{-}_{\s,\s'}(\xx-\yy)\Big]_{\r,\r'} &:=& \frac{\Tr\big\{e^{-\b \HHH_{0,\L}} \tilde\Psi^{-}_{\xx,\s,\r}\tilde \Psi^{+}_{\yy,\s',\r'}\big\}}{\Tr\{ e^{-\b\HHH_{0,\L}}\}}\;,\label{A6b}\\
w_{i,i'}(\xx - \yy) &:=& \frac{\Tr\big\{ e^{-\b\HHH_{0,\L}} \tilde A_{i,\xx}\tilde A_{i',\yy} \big\}}{\Tr\big\{ e^{-\b\HHH_{0,\L}}\big\}}\;,\label{A6c}
\eea
where $x_0 - y_0 >0$, and as we are going to see in subsection \ref{app1.1.2}, $w_{i,j}(\xx - \yy) = w_{i,j}(\yy-\xx)$; $g_+$ and $g_-$ can be combined in a a single function:
\be
g_{\s,\s'}(\xx) := \left\{ \begin{array}{cc} g^{+}_{\s,\s'}(\xx) & \mbox{if $x_0 >0$} \\ -g^{-}_{\s,\s'}(-\xx) & \mbox{if $x_0\leq 0$} \end{array} \right.\label{A7}
\ee
The functions (\ref{A6c}), (\ref{A7}) are called the {\it bosonic} and {\it fermionic propagators}, respectively; they can be explicitly evaluated, and the computations are reported in Sections \ref{app1.1.1}, \ref{app1.1.2}.

\subsection{Fermion propagator}\label{app1.1.1}

In this subsection we shall compute explicitly the fermionic propagator (\ref{A7}). The free fermionic Hamiltonian in momentum space is given by:
\be
\HHH^{\Psi}_{0,\L} = -\frac{1}{|\L|}\sum_{\vec k\in \DD_{L}}\sum_{\s = \uparrow\downarrow}t\Big(\O^{*}(\vec k)\hat a^{+}_{\vec k,\s}\hat b^{-}_{\vec k,\s} + \O(\vec k)\hat b^{+}_{\vec k,\s}\hat a^{-}_{\vec k,\s}\Big)\;;\label{A7.1}
\ee
the Hamiltonian can be diagonalized by introducing the fermionic operators
\be
\hat\a^{\#}_{\vec k,\s} := \frac{\hat a^{\#}_{\vec k,\s}}{\sqrt{2}} + \frac{\O^{*}(\vec k)}{\sqrt{2}|\O(\vec k)|}\hat b^{\#}_{\vec k,\s}\;,\qquad \hat \b^{\#}_{\vec k,\s} := \frac{\hat a^{\#}_{\vec k,\s}}{\sqrt{2}} - \frac{\O^{*}(\vec k)}{\sqrt{2}|\O(\vec k)|}\hat b^{\#}_{\vec k,\s}\;,\label{A7.2}
\ee
with $\# = \pm$, in terms of which we can rewrite:
\be
\HHH^{\Psi}_{0,\L} = \frac{1}{|\L|}\sum_{\vec k\in \DD_{L}}\sum_{\s = \uparrow\downarrow}t\Big( -|\O(\vec k)|\hat \a^{+}_{\vec k,\s}\hat \a^{-}_{\vec k,\s} + |\O(\vec k)|\hat \b^{+}_{\vec k,\s}\hat \b^{-}_{\vec k,\s} \Big)\;.\label{A7.3}
\ee
Now, for $\vec x\in \L$ we define 
\be
\a^{\pm}_{\vec x,\s} := \frac{1}{|\L|}\sum_{\vec k\in \DD_{L}}e^{\pm \iu \vec k\cdot\vec x}\hat \a^{\pm}_{\vec k,\s}\;,\quad \b^{\pm}_{\vec x,\s} := \frac{1}{|\L|}\sum_{\vec k\in \DD_{L}}e^{\pm \iu \vec k\cdot\vec x}\hat\b^{\pm}_{\vec k,\s}\;;\label{A7.4}
\ee
moreover, we set $\a^{\pm}_{\xx,\s} := e^{\HHH^{\Psi}_{0,\L}x_0} \a^{\pm}_{\vec x,\s}e^{-\HHH^{\Psi}_{0,\L}x_0} $ and $\b^{\pm}_{\xx,\s} = e^{\HHH^{\Psi}_{0,\L}}\b^{\pm}_{\vec x,\s}e^{-\HHH^{\Psi}_{0,\L}x_0}$. A straightforward computation shows that, if $-\b<x_0-y_0\leq \b$, and $1(\mbox{ ``argument''})$ is equal to $1$ if ``argument'' is true otherwise is equal to $0$:
\bea
&&\media{\TTT\{\a^{-}_{\xx,\s}\a^{+}_{\yy,\s'}\}}_{\b,L} = \label{A7.5}\\ &&= \frac{\d_{\s,\s'}}{|\L|}\sum_{\vec k\in \DD_{L}}\eu^{-\iu\vec k (\vec x - \vec y)}\Big[ 1(x_0 > y_0)\frac{e^{(x_0 - y_0)|\O(\vec k)|}}{1 + e^{\b|\O(\vec k)|}} - 1(x_0 \leq  y_0)\frac{e^{(x_0 - y_0 + \b)|\O(\vec k)|}}{1 + e^{\b|\O(\vec k)|}}\Big]\;,\nn\\
&&\media{\TTT\{\b^{-}_{\xx,\s}\b^{+}_{\yy,\s'}\}}_{\b,L} = \label{A7.6}\\ &&= \frac{\d_{\s,\s'}}{|\L|}\sum_{\vec k\in \DD_{L}}\eu^{-\iu\vec k (\vec x - \vec y)}\Big[ 1(x_0 > y_0)\frac{e^{-(x_0 - y_0)|\O(\vec k)|}}{1 + e^{-\b|\O(\vec k)|}} - 1(x_0 \leq  y_0)\frac{e^{-(x_0 - y_0 + \b)|\O(\vec k)|}}{1 + e^{-\b|\O(\vec k)|}}\Big]\nn
\eea
and $\media{\TTT\{\a^{-}_{\xx,\s}\b^{+}_{\yy,\s'}\}}_{\b,L} = \media{\TTT\{ \b^{-}_{\xx,\s}\a^{+}_{\yy,\s'} \}}_{\b,L} = 0$. A priori Eq. (\ref{A7.5}) and Eq. (\ref{A7.6}) are defined only for $-\b<x_0 - y_0\leq \b$, but we can extend them periodically over the whole real axis; the periodic extension of the propagator is continuous on the time variable for $x_0 - y_0\neq \b\mathbb{Z}$, and it has jump discontinuities at the points $x_{0}-y_0\in \b\mathbb{Z}$. Note that at the points $x_{0} - y_{0} = \b n$ the difference between the right and left limits is equal to $(-1)^{n}\d_{\vec x,\vec y}$, which means that the propagator is discontinuous only at $\xx - \yy = \b\mathbb{Z}\times \vec 0$. For $\xx - \yy \notin \b\mathbb{Z}\times \vec 0$ we can write
\bea
\media{\TTT\{\a^{-}_{\xx,\s}\a^{+}_{\yy,\s'}\}}_{\b,L} &=& \frac{\d_{\s,\s'}}{\b|\L|}\sum_{\kk\in\DD_{\b,L}}\eu^{-\iu\kk(\xx - \yy)}\frac{1}{-\iu k_0 - |\O(\vec k)|}\;,\label{A7.7}\\
\media{\TTT\{\b^{-}_{\xx,\s}\b^{+}_{\yy,\s'}\}}_{\b,L} &=& \frac{\d_{\s,\s'}}{\b|\L |}\sum_{\kk\in\DD_{\b,L}}\eu^{-\iu\kk(\xx - \yy)}\frac{1}{-\iu k_0 + |\O(\vec k)|}\;.\label{A7.8}
\eea
In fact, consider Eq. (\ref{A7.7}); if $x_0 - y_0\neq 0$, by Cauchy Theorem it follows that
\be
\frac{1}{\b}\sum_{\substack{k_0 = \frac{2\pi}{\b}\big(n + \frac{1}{2}\big), \\ n\in \ZZZ}}\frac{\eu^{-\iu k_0(x_0 - y_0)}}{-\iu k_0 - |\O(\vec k)|} = \int_{\CC}\frac{dz}{(2\pi)}\,\frac{\eu^{-\iu z(x_0 - y_0)}}{-\iu z - |\O(\vec k)|}\frac{1}{\eu^{-\iu z\b \sgn(x_0 - y_0)} + 1}\;,\label{A7.9}
\ee
where the integral on the {\it r.h.s.} is taken counterclockwise along the path $\CC = \CC_+ \cup \CC_-$, with $\CC_{\pm} := \Big\{ z\in \CCC: \Im z = \pm |\O(\vec k)|/2 \Big\}$. Assume that $x_{0} - y_0 >0$; the integral over $\CC_{-}$ is vanishing (the integration path can be shifted to $\Im z = -\io$, and the integrand is exponentially vanishing) while the value of the integral over $\CC_{+}$ is determined by the residue of the pole in $z = \iu |\O(\vec k)|$. It follows that:
\be
\int_{\CC_+}\frac{dz}{(2\pi)}\,\frac{\eu^{-\iu z(x_0 - y_0)}}{-\iu z - |\O(\vec k)|}\frac{1}{\eu^{-\iu z\b} + 1} = \frac{e^{|\O(\vec k)|(x_0 - y_0)}}{1 + e^{\b|\O(\vec k)|}}\;,\label{A7.10}
\ee
which is precisely the first term appearing in the {\it r.h.s.} of Eq. (\ref{A7.5}). The same trick can be used to evaluate the other sums in $k_0$ appearing in Eqq. (\ref{A7.5}), (\ref{A7.6}), and (\ref{A7.7}), (\ref{A7.8}) follow. If $x_0 - y_0 = \b n$ and $\vec x - \vec y \neq \vec 0$, the {\it r.h.s.} of (\ref{A7.7}) is equal to
\bea
&&\frac{1}{2}\Big( \lim_{x_{0} - y_0 \rightarrow (\b n)^+}\media{\TTT\{ \a^{-}_{\xx,\s}\a^{+}_{\yy,\s'}\}}_{\b,L} + \lim_{x_0 - y_0 \rightarrow (\b n)^{-}}\media{\TTT\{ \a^{-}_{\xx,\s}\a^{+}_{\yy,\s'}\}}_{\b,L} \Big) \nn\\
&& =: \media{\TTT\{ \a^{-}_{\xx,\s}\a^{+}_{\yy,\s'} \}}_{\b,L}\Big|_{x_0 - y_0 = \b n}\;;\label{A7.11} 
\eea
the same holds for $\media{\TTT\{ \b^{-}_{\xx,\s}\b^{+}_{\yy,\s'} \}}_{\b,L}$. If we now reexpress $\a^{\pm}_{\xx,\s}$ and $\b^{\pm}_{\xx,\s}$ in terms of $\tilde a^{\pm}_{\xx,\s}$ and $\tilde b^{\pm}_{\xx + (0,\vec\d_1),\s}$, using (\ref{A7.2}) we get:
\bea
S_{0}(\xx - \yy) := \media{\TTT\{\tilde\Psi^{-}_{\xx,\s}\tilde\Psi^{+}_{\yy,\s'}\}}_{\b,L} &=& \frac{\d_{\s,\s'}}{\b |\L|}\sum_{\kk\in \DD_{\b,L}}\eu^{-\iu \kk (\xx - \yy)} \hat{g}(\kk)\label{A7.12}\\ &:=& \frac{\d_{\s,\s'}}{\b |\L|}\sum_{\kk\in \DD_{\b,L}}\frac{\eu^{-\iu \kk (\xx - \yy)}}{k_0^2 + |\O(\vec k)|^2}\begin{pmatrix} \iu k_0 & -\O^*(\vec k) \\ -\O(\vec k) & \iu k_0 \end{pmatrix}\nn
\eea
It is important to stress that (\ref{A7.12}) agrees with (\ref{A7.5}), (\ref{A7.6}) as long as $\xx - \yy \neq \b\mathbb{Z} \times \vec 0$; in fact, if $\xx = \yy$ we have
\be
\frac{1}{\b|\L|} \sum_{\kk\in \DD_{\b,L}}\eu^{-\iu \kk (\xx - \yy)} \hat{g}(\kk) = \frac{1}{\b |\L|}\sum_{\kk\in \DD_{\b,L}}\frac{1}{k_0^2 + |\O(\vec k)|^2}\begin{pmatrix} 0 & -\O^*(\vec k) \\ -\O(\vec k) & 0 \end{pmatrix}\;,\label{A7.13}
\ee
while from (\ref{A7.5}), (\ref{A7.6}) we see that
\bea
S_{0}(0^{-},\vec 0)_{1,1} = S_{0}(0^-,\vec 0)_{2,2} &=& -\frac{1}{2}\Big( \media{\a^{+}_{\vec x,\s}\a^{-}_{\vec x,\s}}_{\b,L} + \media{\b^{+}_{\vec x,\s}\b^{-}_{\vec x,\s}}_{\b,L} \Big)\label{A7.14}\\
&=& -\frac{1}{2|\L|}\sum_{\vec k\in\DD_{L}}\Big( \frac{e^{\b|\O(\vec k)|}}{1 + e^{\b|\O(\vec k)|}} + \frac{e^{-\b|\O(\vec k)|}}{1 + e^{-\b |\O(\vec k)|}} \Big) = -\frac{1}{2}\;.\nn
\eea

\subsection{Photon propagator}\label{app1.1.2}

In this subsection we shall compute explicitly the photon propagator (\ref{A6b}). The free bosonic Hamiltonian is:
\be
\HHH^{A}_{0} := \frac{1}{L\AAA_{\L}}\sum_{{\ul p}\in \widetilde\PPP_{L}}\sum_{r=1,2} |\ul{p}|\hat c^{+}_{r,\ul{p}}\hat c^{-}_{r,\ul{p}}\;,\label{A7.15}  
\ee
and the quantized vector potential is:
\be
{\ul A}_{\ul x} := \frac{1}{L\AAA_{\L}}\sum_{{\ul p}\in\widetilde\PPP_{L}}\sum_{r=1,2} \eu^{-\iu {\ul p}\cdot{\ul x}} \sqrt{\frac{\chi_{[h^{*},0]}(|{\ul p}|)}{2|{\ul p}|}}\Big[ {\ul \varepsilon}_{{\ul p},r}\hat c^{-}_{{\ul p},r} + {\ul \varepsilon}^{*}_{-{\ul p},r}\hat c^{+}_{-{\ul p},r} \Big]\;.\label{A7.16}
\ee
Let
\be
\D_{ij}({\ul p}) := \sum_{r=1,2} \big(\ul{\varepsilon}_{\ul p,r}\big)_{i}\big(\ul{\varepsilon}^{*}_{\ul p,r}\big)_{j} = \d_{ij} - \frac{p_{i} p_{j}}{{\ul p}^2}\;,\qquad i,j = 1,2,3\;,\label{A7.17}
\ee
where the last equality can be read as the completeness relation of the orthonormal basis $\Big\{\ul\e_{\ul p,1},\,\ul\e_{\ul p,1}^{*},\,\ul\e_{\ul p,2},\,\ul\e_{\ul p,2}^{*},\,\ul p/|\ul p|\Big\}$ of $\RRR^{3}\otimes\CCC$; a straightforward computation shows that:
\bea
&&\media{\TTT\{ \tilde A_{\xx,i}\tilde A_{\yy,j}\}}_{\b,L} =\label{A7.18}\\&&= \frac{1}{L\AAA_{\L}}\sum_{ {\ul p}\in\widetilde\PPP_{L}}\eu^{-\iu {\ul p}{(\ul x - \ul y)}} \frac{\chi_{[h^{*},0]}(|{\ul p}|)}{2|{\ul p}|}\D_{ij}({\ul p})\left[ \frac{e^{-|x_0 - y_0||{\ul p}|}}{1 - e^{-\b|{\ul p}|}} + \frac{e^{(|x_0 - y_0| - \b)|{\ul p}|}}{1 - e^{-\b|{\ul p}|}} \right]\;.\nn
\eea
To derive (\ref{A7.18}) it is useful to recall that, because of the commutation relation (\ref{1.1.2a}) defining the creation/annihilation operators:
\be
c^{+}_{\ul p,r}\mid n_{\ul p,r} \rangle = \sqrt{L\AAA_{\L}}\sqrt{n_{\ul p,r}+1}\mid n_{\ul p,r} + 1 \rangle\;,\quad c^{-}_{\ul p,r}\mid n_{\ul p,r} \rangle = \sqrt{L\AAA_{\L}}\sqrt{n_{\ul p,r}}\mid n_{\ul p,r} - 1 \rangle\;,
\ee
where the volume factors ensure the correct normalization of the states. Notice that this time the propagator is continuous in $\xx - \yy = \b\mathbb{Z}\times \vec 0$. Moreover, if  $-\b\leq x_0 - y_0\leq \b$ we can write
\be
\media{\TTT\{ \tilde A_{\xx,i}\tilde A_{\yy,j}\}}_{\b,L} = \frac{1}{\b L\AAA_{\L}}\sum_{ \pp\in\PPP_{\b,L}}\sum_{\substack{ p_3 = \frac{2\pi n}{L} \\ n\in \ZZZ}} \eu^{-\iu \pp(\xx - \yy)}\frac{\chi_{[h^{*},0]}(|{\ul p}|)}{p_0^2 + \ul p^2}\D_{ij}({\ul p})\;; \label{A7.19}
\ee
formula (\ref{A7.19}) can be extended periodically over the whole real axis. To get (\ref{A7.19}) we use that, by Cauchy Theorem:
\be
\frac{1}{\b}\sum_{\substack{ p_0 = \frac{2\pi n}{\b},\\n\in \ZZZ}} e^{- \iu p_0 (x_0 - y_0)} \frac{1}{p_0^2 + {\ul p}^2} = \int_{\CC'}\frac{d z}{2\pi}\,e^{-i z (x_0 - y_0)}\frac{1}{z^2 + {\ul p}^2}\frac{1}{1 - \eu^{-\iu z\b \sgn(x_0 - y_0)}}\;,\label{A7.20}
\ee
where the integral in the {\it r.h.s.} of (\ref{A7.20}) is taken counterclockwise along the path $\CC^{'} = \CC'_{+}\cup \CC'_{-}$ with $\CC'_{\pm} := \Big\{ z\in \CCC: \Im z = \pm |\ul p|/2 \Big\}$. The integrals along $\CC'_{\pm}$ are determined by the residues of the poles in $z = \pm \iu |{\ul p}|$; it is easy to see that
\be
\int_{\CC'}\frac{d z}{2\pi}\,e^{-i z (x_0 - y_0)}\frac{1}{z^2 + {\ul p}^2}\frac{1}{1 - \eu^{-\iu z\b \sgn(x_0 - y_0)}} = \frac{1}{2|{\ul p}|}\left[ \frac{e^{-|x_0 - y_0| |{\ul p}|}}{1 - e^{-\b |{\ul p}|}} + \frac{e^{(|x_0 - y_0| - \b)|{\ul p}|}}{1 - e^{-\b |{\ul p}|}} \right]\;,\label{A7.21}
\ee
which proves (\ref{A7.19}).

\section{Feynman rules}\label{sec1.2}
\setcounter{equation}{0}
\renewcommand{\theequation}{\ref{sec1.2}.\arabic{equation}}

\begin{figure}[htbp]
\centering
\includegraphics[width=0.8\textwidth]{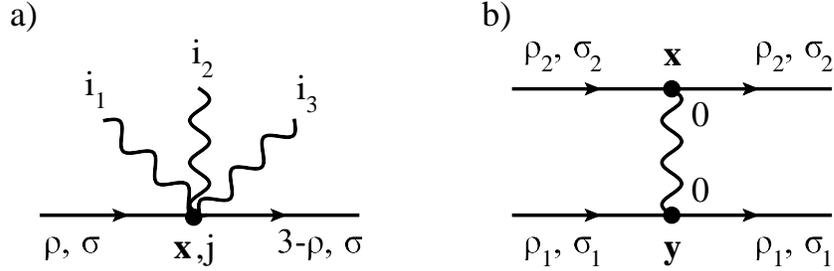}
\caption{We can have: $a)$ vertices with two external fermionic lines and an arbitrary number $m\geq 1$ of external bosonic lines (in this figure $m=3$); or $b)$ vertices with four external fermionic lines, joined in pairs connected by a wavy line labelled by $0$ at its extrema (representing the Coulomb potential $\varphi(\vec x - \vec y,0)$).} \label{figA1}
\end{figure}

The various terms contributing to (\ref{A5}) can be graphically represented in terms of {\em Feynman graphs}; here we shall briefly discuss the rules that one has to follow in order to draw and evaluate such graphs. The vertices of the graphs can be of two types, see Fig. \ref{figA1}: $a)$ they have two external fermionic lines and an arbitrary number $\geq 1$ of external bosonic (wavy) lines; $b)$ they have four external fermionic lines.
%

%
%
The value of the graph element in item {\it a)} in Fig. {\ref{figA1}} is, depending on the choice of the label $\r$:
\bea
&&(\r = 1)\quad \tilde b^{+}_{\xx + (0,\vec\d_j),\s}\Big[\prod_{r=1}^{3}-\iu e \int_{0}^{1}ds\,[\vec\d_{j}]_{i_{r}}\,\tilde A_{i_r,(x_0,\vec x + s\vec\d_j)}\Big]\tilde a^{-}_{\xx,\s}\;,\label{A9}\\
&&(\r = 2)\quad \tilde a^{+}_{\xx,\s}\Big[\prod_{r=1}^{3}\iu e\int_{0}^{1} ds\,[\vec\d_{j}]_{i_r}\,\tilde A_{i_r,(x_0,\vec x + s\vec\d_j)}\Big]\tilde b^{-}_{\xx + (0,\vec\d_j),\s}\;;\label{A10}
\eea
the value of the graph element in item {\it b)} in Fig. {\ref{figA1}} is:
\be
-\frac{e^2}{2} \tilde \Psi^{+}_{\yy,\s_1,\r_1}\tilde \Psi^{-}_{\yy,\s_1,\r_1}\varphi(\vec x - \vec y,0)\d_{y_0,x_0} \tilde \Psi^{+}_{\xx,\s_2,\r_2}\tilde \Psi^{-}_{\xx,\s_2,\r_2}\;.\label{A12}
\ee
Notice that in (\ref{A12}) we have neglected the factor $-1/2$ which originally appears in the definition of the model, see (\ref{1.1.1}); this can be justified by slightly changing the definition of the fermion propagator, see Remark \ref{osstad} below. 

Then one considers all possible Feynman graphs, that is all possible ways of joining together lines in pairs so that:
\begin{itemize}
\item no unpaired line is left over, and
\item only lines (wavy or solid) of the same type can be paired, and 
\item in the case of fermionic lines, the orientations of the paired lines have to be consistent.
\end{itemize}
To each graph we assign the sign $(-1)^{\pi}$ of the permutation $\pi$ necessary to bring next to each other the pairs of fermionic operators which, in the given graph, are paired (one says also {\it contracted}), with the $\Psi^{-}$ to the left of the associated $\Psi^{+}$.

A solid line connecting two points labelled by $\xx_1$, $\xx_2$ with an arrow pointing from the point labelled by $\xx_1$ to the point labelled by $\xx_2$ obtained contracting half-lines labelled by $\s_1,\r_1$ and $\s_2,\r_2$ corresponds to the fermionic propagator $\big[g_{\s_2,\s_1}(\xx_2 - \xx_1)\big]_{\r_2,\r_1}$ given by (\ref{A7.12}); a wavy line connecting two points labelled by $\xx_1,\,j_1$ and $\xx_2,\,j_2$ obtained contracting half-lines labelled by $i_1$ and $i_2$ corresponds to
\be
\bar w^{j_2,j_1}_{i_2,i_1}(\xx_{2} - \xx_{1}) := \int_{0}^{1}ds_{1}\int_{0}^{1}ds_{2}\,[\vec\d_{j_{1}}\otimes\vec\d_{j_2}]_{i_{1},i_{2}}w_{i_{2},i_{1}}(x_{2,0} -x_{1,0},\vec x_{2} - \vec x_{1} + s_{2}\vec\d_{j_2} - s_{1}\vec\d_{j_1})\;.\label{A13}
\ee 
To each graph we assign a {\it value}, which is the integral over all the space-time labels $\{\xx_{i}\}$ of the product of:
\begin{itemize}
\item the sign factor, times 
\item the product of functions $g(\xx_2 - \xx_1)$, $\bar w(\xx_2 - \yy_1)$ for every solid or wavy line connecting a point labelled by $\xx_1$ to a point labelled by $\xx_2$, times 
\item a factor $\pm\iu e$ for every vertex with two incident fermionic half-lines and a positive number of emerging wavy lines, where the sign is $+$ if the fermionic exiting line is of type $1$ and it is $-$ otherwise, times
\item a factor $-\frac{e^2}{2}\varphi(\vec x_2 - \vec x_1,0)\d_{t_{1},t_{2}}$ for every wavy line labelled by $0$ at its extrema.
\end{itemize}
\begin{oss}\label{osstad}
Notice that the fermionic propagator that we used to define the Feynman rules, given by Eq. (\ref{A7.12}), is {\em not} equal to $\media{\TTT\{\tilde\Psi^{-}_{\xx}\tilde \Psi^{+}_{\yy}\}}_{\b,L}$ if $\xx-\yy = \b\mathbb{Z}\times \vec 0$, see discussion at the end of subsection \ref{app1.1.1}; however, this discrepancy is absorbed by the fact that we have not considered the factor $-\frac{1}{2}$ (which is precisely the value of the fermionic ``tadpole'' graph, see Eq. (\ref{A7.14})) in the definition of the value of the graph element $b)$ in Fig. \ref{figA1}, see (\ref{A12}) and (\ref{1.1.1}).
\end{oss}

Given the above rule, it is straightforward to check that the coefficients of the perturbative series of (\ref{A5}) are {\it equal} to those generated by the functional integral
\be
\int P(d\Psi) P(d\vec A)\eu^{\widetilde V(\Psi,A)}\;,\label{A14}
\ee
where: (i) the interaction is:
\bea
&&\widetilde V(\Psi,A) := \int d\xx\,\sum_{\s,j} t\Big[ \Psi^{+}_{\xx,\s,1}\Psi^{-}_{\xx + (0,\vec\d_j),\s,2}\Big( \eu^{\iu e A_{(\xx,j)}} - 1\Big) + c.c. \Big] + V_{C}(\Psi)\;,\nn\\
&&V_{C}(\Psi) := -\frac{e^2}{2}\frac{1}{\b\AAA_{\L}}\sum_{\pp\in \PPP_{\b,L}^{*}}\sum_{\r,\r' = 1,2} \hat\varphi(\vec p)\hat n_{\r}(\pp)\hat n_{\r'}(-\pp)\eu^{\iu (\r' - \r)\vec p\vec\d_1}\;,\label{A15}\\
&&\hat \varphi(\vec p) := \frac{1}{L}\sum_{p_3 = \frac{2\pi n}{L}} \frac{\chi_{[h^{*},0]}(|\ul p|)}{|\ul p|^2}\;,\quad \hat n_{\r}(\pp) := \frac{1}{\b|\L|}\sum_{\kk\in \DD^{*}_{\b,L}}\sum_{\s} \hat \Psi^{+}_{\kk + \pp,\s,\r}\hat\Psi^{-}_{\kk,\s,\r}\;;\nn
\eea
(ii) the fermionic integration measure has been defined in (\ref{1.2.11}); (iii) the bosonic interaction measure is given by ($i,j = 1,2$):
\bea 
P(d\vec A) &:=& \prod_{\pp\in \PPP^{*,+}_{\b,L}}^{r=1,2} \Big[\frac{1}{(\pi \b \AAA_{\L})^2 \det \{\hat w^{1}_{i,j}(\pp)\}}d\hat A_{r,\pp}d\hat A_{r,-\pp}\Big]\cdot\nn\\
&&\cdot \exp\Big\{ -(2\b\AAA_{\L})^{-1}\sum_{\pp\in \PPP^{*}_{\b,L}}\hat A_{i,\pp}[\hat w^{1}(\pp)^{-1}]_{i,j}\hat A_{j,-\pp} \Big\}\;.\label{A15b}
\eea
Finally, the interaction $V(\Psi,A)$ is recovered using a {\it Hubbard-Stratonovich transformation}, namely the fact that:
\be
\eu^{V_{C}(\Psi)} = \int P(dA_{0})\exp\Big\{-(\b\AAA_{\L})^{-1} \sum_{\pp\in \PPP^{*}_{\b,L}}\sum_{\r=1,2} \iu e \hat A_{0,\pp}\hat n_{\r}(\pp)\eu^{-\iu (\r-1)\vec p\vec\d_1} \Big\}\;,
\ee
where the integration measure is defined as:
\bea
P(dA_{0}) &:=& \prod_{\pp\in \PPP^{*,+}_{\b,L}}\Big[\frac{1}{(\pi \b \AAA_{\L}) \hat\varphi(\vec p)} d\hat A_{0,\pp}d\hat A_{0,-\pp}\Big]\cdot\nn\\
&&\cdot \exp\Big\{ -(2\b\AAA_{\L})^{-1}\sum_{\pp\in \PPP^{*}_{\b,L}}\hat A_{0,\pp}\hat\varphi(\vec p)^{-1}\hat A_{0,-\pp} \Big\}\;.
\eea
Taking the product of this measure with (\ref{A15b}) we get the full interacting bosonic measure (\ref{1.2.14}) with $\xi=1$; this concludes the derivation of the functional integral representation for the Hamiltonian model. An analogous argument can be repeated for the generaing functional of the correlations, but we will not belabor the details here.

%
%
%
%
%
%

\chapter[Equivalence of the gauges]{Equivalence of the gauges}\label{app2}
\setcounter{equation}{0}
\renewcommand{\theequation}{\ref{app2}.\arabic{equation}}
In this Appendix we show that the perturbative series of gauge invariant quantities are equal in Coulomb and Feynman gauge; this fact allows us to rewrite the perturbative series obtained with the Trotter product formula, see Appendix \ref{app1}, using the functional integral representation in the Feynman gauge. To be concrete, we will perform the check on the generating functional of the generalized susceptibilities introduced in Section \ref{exc}; the same argument can be repeated for the expectation value of any gauge invariant operator which can be expressed as series in the fermionic and bosonic fields.

Pick $0 < -h,\b,L <+ \infty$, and define
\be
\lis\WW_{\b,L}^{[h,+\infty],\xi}(\Phi,J) := \log \int P(d\Psi)P_{\geq h}^{\xi}(dA)\eu^{V(\Psi, A+J) + D(\Phi,\Psi,A+J)}\;,\label{B01}
\ee
where: 
\begin{itemize}
\item[(i)] The fermion field lives on a space-time lattice: the spatial part of the lattice is given by the physical honeycomb lattice, while the temporal part is introduced by discretizing the time variables with a mesh $1/N$, as discussed in Section \ref{secWI}; each finite order of the perturbative series of (\ref{B01}) converges uniformly to a limit as $N\rightarrow+\infty$.
\item[(ii)] $P_{\geq h}^{\xi}$ is the bosonic gaussian measure in presence of an infrared cutoff on scale $M^{h}$ and in the $\xi$ gauge; the infrared cutoff is imposed as in (\ref{1.2.13}).
\item[(iii)] The source term $D(\Phi,\Psi,A+J)$ has been defined in (\ref{exc02}).
\end{itemize}

The grand-canonical susceptibilities can be obtained by taking derivatives with respect to the $\Phi^{\a}$ external fields; our purpose is to show that, order by order in perturbation theory, for any finite $h$, $\b$ and $L$, in the limit $N\rightarrow +\infty$, the result is independent of $\xi$. More precisely, denoting by $\lis\WW_{\b,L}^{[h,+\infty],\xi,m}(\Phi,J)$ the $m$-th order contribution to $\lis\WW_{\b,L}^{[h,+\infty],\xi}$, we shall prove that
\be
\partial_{\xi}\lis\WW_{\b,L}^{[h,+\infty],\xi,m}(\Phi,J) =0\;.\label{B02}
\ee 
The proof is easy, and it is based on the following Ward identity:
\be
p_{\m}\frac{\partial}{\partial \hat J_{\m,\pp}}\lis\WW_{\b,L}^{[h,+\infty],\xi}(\Phi,J) =0\;;\label{B03}
\ee
formula (\ref{B03}) can be proved along the lines of Section \ref{secWI}, see (\ref{WI5b}), (\ref{WI7a}).
%
%
%
%
%
%
%
Let us now show how (\ref{B03}) implies (\ref{B02}). First of all, we represent (\ref{B01}) in the following way:
\bea
\lis\WW_{\b,L}^{[h,+\infty],\xi}(\Phi,J) &=& \sum_{m\geq 0}\lis\WW_{\b,L}^{[h,+\infty],\xi,m}(\Phi,J)\;,\nn\\
\lis\WW_{\b,L}^{[h,+\infty],\xi,m}(\Phi,J) &=& \sum_{\GG\in \G^{m}}\Val_{\xi}(\GG)\;,\label{B07}
\eea
where $\G^{m}$ is the set of connected Feynman graphs of order $m$ generated by (\ref{B01}) (which have no $\Psi,A$ external lines), and $\Val_{\xi}(\GG)$ is the value of the graph $\GG$. Notice that for any fixed $\b$, $L$ and $h$ all the coefficients of the perturbative series of the functional integral (\ref{B01}) are {\it finite}. The photon propagator appearing in these graphs is:
\bea
\hat w^{(\geq h),\xi}_{\m,\n}(\pp) &=& \int \frac{d p_3}{(2\pi)}\,\frac{\chi_{[h,0]}(|\ul p|)}{\pp^2 + p_3^2}\D^{\xi}_{\m,\n}(\pp,p_3)\;,\nn\\
\D^{\xi}_{\m,\n}(\pp,p_3) &=& \d_{\m,\n} - \xi\frac{p_\m p_\n - p_0(p_\m n_\n + p_\n n_\m)}{\ul p^2}\;,\qquad n_{0} = 1\;, \vec n = \vec 0\;;\label{B08}\nn
\eea
therefore, it is easy to see that:
\bea
&&\partial_{\xi}\lis\WW_{\b,L}^{[h,+\infty],\xi,m}(\Phi,J) = \label{B09}\\&& -\int \frac{\chi_{[h,0]}(|\ul p|)}{\pp^2 + p_3^2}\frac{p_\m p_\n - p_0(p_\m n_\n + p_\n n_\m)}{\ul p^2}\Big[ \G_{\m,\n}^{m}(\pp) + \sum_{m_1 + m_2 = m}\G_{\m}^{m_1}(\pp)\G_{\n}^{m_2}(-\pp) \Big]\;,\nn
\eea
where $\int \equiv \int \frac{d^{4}p}{(2\pi)^4}$ and 
\begin{itemize}
\item[(i)] $\G_{\m,\n}^{m}(\pp)$ is the sum of all the possible Feynman graphs of order $m$ with two bosonic external lines labelled by $\m$, $\n$ and $\pp$ as external momentum;
\item[(ii)] $\G_{\m}^{m_{i}}(\pp)$ is the sum of all the possible Feynman graphs of order $m_i$ with one bosonic external line and $\pp$ as external momentum.
\end{itemize}
Now, both $\G^{m}_{\m,\n}(\pp)$ and $\G^{m}_{\m}(\pp)$ can be represented as functional derivatives with respect to $\hat J_{\m,\pp},\, \hat J_{\n,-\pp}$ and $\hat J_{\m,\pp}$, respectively, of $\lis\WW_{\b,L}^{[h,+\infty],\xi,m}(\Phi,J)$; and the identity (\ref{B03}) tells us that if the derivative with respect the field $\hat J_{\m,\pp}$ is contracted with $p_{\m}$ then the results is zero. But we see that the $\m,\n$ labels in (\ref{B09}) are always contracted with at least one among $p_{\m}$, $p_{\n}$: therefore, (\ref{B09}) is equal to zero. This concludes the proof of (\ref{B02}).

\chapter{The ultraviolet multiscale analysis}\label{app2b}
\begin{figure}[htbp]
\centering
\includegraphics[width=0.5\textwidth]{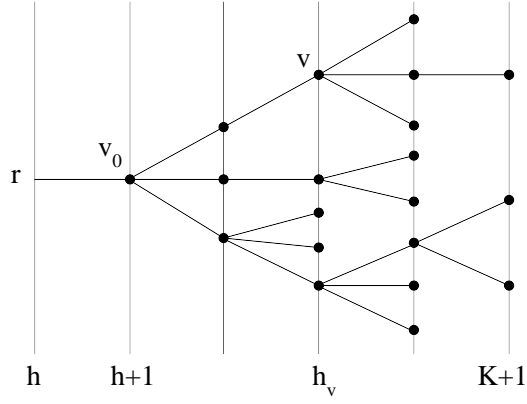}
\caption{A tree $\t\in \TT_{K;h,n}$ with its scale labels.} \label{figC1}
\end{figure}

\setcounter{equation}{0}
\renewcommand{\theequation}{\ref{app2b}.\arabic{equation}}
%
%
%
In this Appendix we prove Lemma \ref{lem2.4b}; the proof is based on a multiscale analysis similar, but much simpler, to the one discussed in Chapter \ref{sec2.4.2}. First of all, we rewrite the propagators $\hat{g}^{(u.v.)}(\kk)$, $\hat{w}^{(u.v.)}(\pp)$ as
\be
\hat{g}^{(u.v.)}(\kk) = \sum_{h=1}^{K}\hat{g}^{(h)}(\kk)\;,\qquad w^{(u.v.)}(\kk) = \sum_{h=1}^{K}\hat{w}^{(h)}(\pp)\;,\label{C1}
\ee
where
\be
\hat{g}^{(h)}(\kk) := H_{h}(k_0)\hat{g}^{(u.v.)}(\kk)\;,\qquad \hat{w}^{(h)}(\pp) := H_{h}(p_0)\hat{w}^{(u.v.)}(\pp)\;,\label{C2}
\ee
with $H_{1}(k_0) := \chi(M^{-1}|k_0|)$, $H_{h}(k_{0}) := \chi(M^{-h}|k_0|) - \chi(M^{-h+1}|k_0|)$ for $1<h<K$, $H_{K}(k_0) = \chi(M^{-K-1}|k_0|) - \chi(M^{-K+1}|k_0|)$; note that in the supports of $\hat{g}^{(u.v.)}(\kk)$, $\hat{w}^{(u.v.)}(\pp)$ it follows that $\sum_{h = 1}^{K}H_{h}(k_{0}) = \sum_{h=1}^{K}H_{h}(p_0) =1$. Moreover note that, defining the norm of a $A\times A$ matrix $B$ as $\|B\| := \max_{j}\sum_{1\leq i\leq A}|B_{i,j}|$, for some positive $C_{0},C_{1}$:
\bea
&&\max_{\kk}\big\| \hat{g}^{(h)}(\kk) \big\| \leq C_{0}M^{-h}\;,\qquad \max_{\pp}\big\| \hat{w}^{(h)}(\pp) \big\| \leq C_{0}M^{-2h}\;,\label{C3}\\
&&\frac{1}{\b|\L|}\sum_{\kk\in \DD_{\b,L}}\,\big\| \hat{g}^{(h)}(\kk) \big\| \leq C_1\;,\qquad \frac{1}{\b\AAA_{\L}}\sum_{\pp\in \PPP_{\b,L}}\big\| \hat{w}^{(h)}(\pp) \big\|\leq C_{1}M^{-h}\;.\nn
\eea
The first and the third of (\ref{C3}) are obvious; they simply follow from the definitions and from the support properties of the single scale propagators. To prove the second and the fourth of (\ref{C3}) simply note that $\hat{w}^{(h)}(\pp)$ can be rewritten as
\be
\hat{w}^{(h)}(\pp) = I H_{h}(p_0)\int_{-\a_0 M}^{a_0 M}\frac{dp_3}{(2\pi)}\,\Big( \frac{\chi(M^{-K}|p_0|)\chi(|\ul p|) - \chi(\pp)}{\pp^2 + p_3^2} \Big)\;,\label{C3b}
\ee
and the integrand is bounded proportionally to $M^{-2h}$ in the support of $H_{h}(p_0)$. Our goal is to compute
\bea
&&e^{-\b|\L|F_{0} + \VV(\Psi^{(i.r.)},A^{(i.r.)})} = \nn\\&&=\lim_{K\rightarrow+\infty} \int P(d\Psi^{[1,K]})P(dA^{[1,K]})\eu^{V(\Psi^{(i.r.)} + \Psi^{[1,K]},A^{(i.r.)} + A^{[1,K]})}\;,\label{C4}
\eea
where $P(d\Psi^{[1,K]})$, $P(dA^{[1,K]})$ are respectively the fermionic and bosonic Gaussian integrations associated with the propagators (\ref{C1}). We perform the integration in an iterative fashion, analogous (but much simplier) to the procedure described for the infrared integration. We shall inductively prove that
\bea
&&e^{-\b|\L|F_{0,K} + \VV_{K}(\Psi^{(i.r.)},A^{(i.r.)})} = \nn\\&& = e^{-\b|\L| F_{h}}\int P(d\Psi^{[1,h]})P(dA^{[1,h]}) e^{\VV_{K}^{(h)}(\Psi^{(i.r.)} + \Psi^{[1,h]}, A^{(i.r.)} + A^{[1,h]})}\;,\label{C5}
\eea
where $P(d\Psi^{[1,h]})$, $P(dA^{[1,h]})$ are the Gaussian integrations associated with the propagators $\sum_{k=1}^{h}\hat{g}^{(h)}(\kk)$, $\sum_{k=1}^{h}\hat{w}^{(h)}(\pp)$ and
\bea
&&\VV_{K}^{(h)}(\Psi^{[1,h]},A^{[1,h]}) =\label{C6}\\&&= \sum_{\substack{n,m\geq 0 \\ n+m\geq 1}}\sum_{\ul\s, \ul\r, \ul\m}\int \Big[\prod_{j=1}^{n}\hat\Psi^{[1,h]+}_{\kk_{2j-1},\s_{j},\r_{2j-1}}\hat\Psi^{[1,h]-}_{\kk_{2j},\s_{j},\r_{2j}}\Big]\Big[ \prod_{i=1}^{m}\hat A^{[1,h]}_{\m_{i},\pp_{i}}\Big]\cdot\nn\\
&&\quad\cdot W^{(h)}_{K,2n,m,\underline{\r},\underline{\m}}(\{\kk_{j}\},\{\pp_{i}\})\,\d\Big( \sum_{j=1}^{n}\big(\kk_{2j-1} - \kk_{2j}\big) + \sum_{i=1}^{m}\pp_{i}\Big)\;,
\eea
and where the constant $F_{h}$ and the kernels $W^{(h)}_{K,2n,m,\underline{\r},\underline{\m}}$ admit bounds analogous (see below) to the ones stated in Lemma \ref{lem2.4b}. In order to inductively prove (\ref{C5}), (\ref{C6}) we rewrite
\bea
&&\int P(d\Psi^{[1,h]}dA^{[1,h]})e^{\VV_{K}^{(h)}(\Psi^{(i.r.)} + \Psi^{[1,h]},A^{(i.r.)} + A^{[1,h]})} = \int P(d\Psi^{[1,h-1]})P(dA^{[1,h-1]})\nn\\
&&\times \int P(d\Psi^{(h)})P(dA^{(h)})e^{\VV_{K}^{(h)}(\Psi^{(i.r.)} + \Psi^{[1,h-1]} + \Psi^{(h)}, A^{(i.r.)} + A^{[1,h-1]} + A^{(h)})}\;,\label{C7}
\eea
where $P(d\Psi^{(h)})$, $P(dA^{(h)})$ are the Gaussian integrations with propagators $\hat{g}^{(h)}(\kk')$, $\hat{w}^{(h)}(\pp)$, respectively. After the integration of $\Psi^{(h)}$, $A^{(h)}$ we define
\bea
&&e^{\VV^{(h-1)}_{K}(\Psi^{(i.r.)} + \Psi^{[1,h-1]}, A^{(i.r.)} + A^{[1,h-1]}) - \b|\L|\tilde F_{h}} = \nn\\&& = \int P(d\Psi^{(h)})P(dA^{(h)}) e^{\VV_{K}^{(h)}(\Psi^{(i.r.)} + \Psi^{[1,h-1]} + \Psi^{(h)}, A^{(i.r.)} + A^{[1,h-1]} + A^{(h)})}\;;\label{C8}
\eea
we have that
\be
-\b|\L|\tilde F_{h} + \VV^{(h-1)}_{K}(\Psi,A) = \sum_{n\geq 1}\frac{1}{n!}\EE^{T}_{h}\big(\VV^{(h)}_{K}(\Psi + \Psi^{(h)}, A + A^{(h)});n\big)\;,\label{C9}
\ee
where
\bea
&&\EE^{T}_{h}\big(\VV^{(h)}_{K}(\Psi + \Psi^{(h)}, A + A^{(h)});n\big) = \nn\\&& = \frac{\partial^{n}}{\partial \l^{n}}\log \int P(d\Psi^{(h)})P(dA^{(h)}) e^{\l\VV_{K}^{(h)}(\Psi + \Psi^{(h)}, A + A^{(h)})}\Big|_{\l = 0}\;.\label{C10}
\eea
As described for the infrared integration, the iterative action of $\EE^{T}_{h_{i}}$ can be conveniently represented in terms of Gallavotti-Nicol\` o trees $\t\in \TT_{K;h,n}$, where $\TT_{K;h,n}$ is the set of labelled trees, completely analogous to the set $\TT_{h,n}$, unless for the following modifications (see figure \ref{figC1}): (i) a tree $\t\in \TT_{K;h,n}$ has vertices $v$ associated with scale labels $h+1\leq h_{v}\leq K+1$, while the root $r$ has scale $h$; (ii) with each end-point $v$ we associate $V(\Psi^{(i.r.)} + \Psi^{[1,K]},A^{(i.r.)} + A^{[1,K]})$.
In terms of these trees, the effective potential $\VV_{K}^{(h)}$, $0\leq h\leq K$ (with $\VV^{(0)}_{K}$ identified with $V$), can be written as
\be
-\b|\L|\tilde F_{h+1} + \VV^{(h)}_{K}(\Psi^{[1,h]},A^{[1,K]}) = \sum_{n=1}^{\infty}\sum_{\t\in \TT_{K;h,n}}\VV^{(h)}(\t;\Psi^{[1,h]},A^{[1,h]})\;,\label{C11} 
\ee
where if $v_0$ is the first vertex of $\t$ and $\t_1,\ldots\,,\t_{s}$ ($s=s_{v_0}$) are the subtrees of $\t$ with root $v_0$, $\VV^{(h)}(\t;\Psi^{[1,h]},A^{[1,h]})$ is defined recursively as follows: 
\begin{itemize}
\item if $s>1$, then
\bea
&&\VV^{(h)}(\t;\Psi^{[1,h]},A^{[1,h]}) = \label{C12}\\&&=\frac{1}{s!}\EE^{T}_{h+1}\Big( \bar\VV^{(h+1)}(\t_1;\Psi^{[1,h]}, A^{[1,h]});\ldots; \bar\VV^{(h+1)}(\t_s;\Psi^{[1,h]}, A^{[1,h]}) \Big)\;,\nn
\eea
where $\bar\VV^{(h+1)}(\t_{i};\Psi^{[1,h+1]},A^{[1,h+1]})$ is equal to $\VV^{(h+1)}(\t_{i};\Psi^{[1,h+1]},A^{[1,h+1]})$ if the subtree $\t_i$ contains more than one end-point, or if it contains one end-point but it is not a trivial subtree; it is equal to $V(\Psi^{[1,h]},A^{[1,h]})$ if $\t_{i}$ is a trivial subtree; 
\item if $s=1$ and $\t_1$ is not a trivial subtree, then $\VV^{(h)}(\t;\Psi^{[1,h]},A^{[1,h]})$ is equal to $\EE^{T}_{h+1}\big(\VV^{(h+1)}(\t_{1};\Psi^{[1,h+1]},A^{[1,h+1]})\big)$; otherwise, if $\t_1$ is a trivial subtree it corresponds to $\EE^{T}_{h+1}\Big( V(\Psi^{[1,h+1]},A^{[1,h+1]}) - V(\Psi^{[1,h]},A^{[1,h]}) \Big)$ .
\end{itemize}
Repeating step by step the discussion made for the infrared integration and using analogous notations it follows that:
\bea
&&\VV^{(h)}(\t;\Psi^{[1,h]},A^{[1,h]}) = \sum_{n = 1}^{\infty}\sum_{\GG\in \G(\t)} \int \Val(\GG)\;,\nn\\
&&\Val(\GG) = \Big[\prod_{f\in P^{A}_{v_0}}A_{\m(f),\pp(f)}^{[1,h]}\Big]\Big[\prod_{f\in P^{\psi}_{v_0}}\widetilde \Psi^{[1,h]}_{\kk(f),\s(f),\r(f)}\Big]\d(v_0)\widehat\Val(\GG)\;,\nn\\
&&\widehat\Val(\GG) = (-1)^{\p}\int \prod_{v\; {\rm not}\; {\rm e.p.}}\frac{1}{s_v!}\cdot\Big[\prod_{\ell\in v}g^{(h_v)}_{\ell}\Big]\Big[\prod_{\substack{v^{*}\,e.p. \\ v^{*}>v \\ h_{v^{*}} = h_{v} + 1}}K^{(h_{v})}_{v^{*}}\Big]\;;\label{C13}
\eea
the explicit form of the vertex functions $K^{(h_{v})}_{v^{*}}$ can be derived starting from Eq. (\ref{1.2.20c}). Note that if $K^{(h_{v})}_{v^{*}}$ corresponds to a vertex with $N$ external wavy lines then its value is bounded as $(\const.)^{N}N!^{-1}e^{N}$. Using the bounds (\ref{C3}) and proceeding as for the infrared bounds it follows that if $\t \in \TT_{K;h,N}$ and $\GG\in \G(\t)$:
\bea
\big|\widehat\Val(\GG)\big| &\leq& (\const.)^{N}e^{N}\prod_{v\;{\rm not}\;{\rm e.p.}}\frac{1}{s_v !}M^{-h_{v}(s_{v}-1 + n^{A}_{v})} \nn\\&=& (\const.)^{N}e^{N}\frac{1}{s_{v_0}!}M^{-h_{v_0}D_{v_0}}\prod_{\substack{v>v_0\\v\;{\rm not}\;{\rm e.p.}}}\frac{1}{s_v !} M^{-(h_{v} - h_{v'})D_{v}}\label{C14}
\eea
where $D_{v} := n_{v} - 1 + \sum_{\bar v\geq v}n^{A}_{\bar v}$. Therefore, the scaling dimensions are always negative, {\it except} for the case $n_{v} = 1$, $n^{A}_{\bar v} =0$ for $\bar v\geq v$, which corresponds to the fermionic tadpole graph; however, in this case the dimensional bound can be improved, noting that $\big[\hat{g}^{(h)}(\kk)\big]_{\r,\r}$ is odd in $k_0$ and that $\big[\hat{g}^{(h)}(\kk)\big]_{\r,3-\r}\sim M^{-2h}$, which means that:
\be
\frac{1}{\b}\sum_{\substack{k_{0} = \frac{2\pi n}{\b} \\ n\in \ZZZ}} \big[\hat{g}^{(h)}(\kk)\big]_{\r,\r} = 0\;,\qquad \frac{1}{\b|\L|}\Big\|\sum_{\kk} \hat{g}^{(h)}(\kk)\Big\|\leq (\const.)M^{-h}\;\label{C15}
\ee
Therefore, with any loss of generality we can assume that in (\ref{C14}) $n_{v}>1$. Now, let $W^{(h),N}_{K,2n,m,\underline{\r},\underline{\m}}$ be the $N$-th order contribution to the kernel $W^{(h),N}_{K,2n,m,\underline{\r},\underline{\m}}$, that is
\be
W^{(h),N}_{K,2n,m,\underline{\r},\underline{\m}}(\{\kk_{i}\},\{\pp_{j}\}) = \sum_{\t\in \TT_{K;h,N}}\sum^{*}_{\substack{\GG\in \G(\t) \\ |P_{v_0}^{A}| = m \\ |P_{v_0}^{\psi}| = 2n}}\widehat\Val(\GG)\;,\label{C16}
\ee
where the $*$ on the sum denotes the following constraints: $\cup_{f\in P^{A}_{v_0}}\pp(f) = \cup_{i=1}^{m}\pp_{i}$, $\cup_{f\in P^{\psi}_{v_0}}\kk(f) = \cup_{i=1}^{2n}\kk_{i}$; $\cup_{f\in P^{A}_{v_0}}\m(f) = \cup_{i=1}^{m}\m_{i}$; $\cup_{f\in P^{\psi}_{v_0}}\r(f) = \cup_{i=1}^{2n}\r_{i}$; the bound (\ref{C14}) implies that
\bea
&&\sup_{\kk_i, \pp_j}\big\| W^{(h),N}_{K,2n,m,\underline{\r},\underline{\m}}(\{\kk_i\},\{\pp_j\}) \big\| \leq \nn\\&&\leq (\const.)^{N}e^{N}\sum_{\t\in \TT_{K;h,N}}\sum^{*}_{\substack{\GG\in \G(\t) \\ |P_{v_0}^{A}| = m \\ |P_{v_0}^{\psi}| = 2n}}\frac{1}{s_v !}M^{-h D_{v_0}}\prod_{\substack{v>v_0\\v\;{\rm not}\;{\rm e.p.}}}\frac{1}{s_v !} M^{-(h_{v} - h_{v'})D_{v}}\nn
\eea
which, after counting the number of Feynman graphs, gives the second bound in (\ref{1.2.47}). The bound on the $N$-th order contribution to $F_{h}$ can be proved in the same way, and this concludes the proof of (\ref{1.2.47}).

To conclude the proof of Lemma \ref{lem2.4b}, note that the sequences $W^{(0),N}_{K,2n,m,\underline{\r},\underline{\m}}$, $F^{N}_{0,K}$ are Cauchy in $K$; in fact, the quantities $W^{(0),N}_{K,2n,m,\underline{\r},\underline{\m}} - W^{(0),N}_{K',2n,m,\underline{\r},\underline{\m}}$, $F^{N}_{0,K} - F^{N}_{0,K'}$ with $K'>K$ can be written as a sums over trees with at least one endpoint on scale $K<h^{*}\leq K'+1$. Therefore, since the scaling dimensions are all negative, by the short memory property of the GN trees it follows that the bounds on $W^{(0),N}_{K,2n,m,\underline{\r},\underline{\m}} - W^{(0),N}_{K',2n,m,\underline{\r},\underline{\m}}$, $F^{N}_{0,K} - F^{N}_{0,K'}$ are improved by a factor $M^{-K/2}$ with respect to the basic bounds (\ref{1.2.47}); taking the limit $K\rightarrow+\infty$ (\ref{1.2.46b}) follows. This conclude the proof of Lemma \ref{lem2.4b}.
\chapter{Proof of Lemma \ref{lem2.4c}}\label{app2c}
\setcounter{equation}{0}
\renewcommand{\theequation}{\ref{app2c}.\arabic{equation}}

In this Appendix we shall prove Lemma \ref{lem2.4c}; to simplify the notations, we shall drop the dependence on the infrared and ultraviolet cutoffs $h^{*},\,K$ in the kernels (whose presence does not affect the validity of the symmetries). Moreover, we shall be concerned only with the case $\b \rightarrow +\infty$, $|\L|\rightarrow +\infty$. The proof is based on the fact that $P(d\Psi^{(u.v.)}dA^{(u.v.)})$ and $P(d\Psi^{(i.r.)}dA^{(i.r.)})$ are separately invariant under the symmetries listed in Lemma \ref{lem2.4}; therefore, $\VV^{[h^{*},K]}(\Psi,A)$, see (\ref{1.2.44}) is also invariant under the same symmetries.
\section{Proof of (\ref{1.2.49})}\label{secD1}
\setcounter{equation}{0}
\renewcommand{\theequation}{\ref{secD1}.\arabic{equation}}
Let us rewrite $W_{2,0,\r,\r'}(\kk' + \pp_{F}^{\o})$ as
\be
W_{2,0,\r,\r'}(\kk'+\pp_{F}^{\o}) = W_{2,0,\r,\r'}(\pp_{F}^{\o}) + \kk'_{\a}\partial_{\a}W_{2,0,\r,\r'}(\pp_{F}^{\o}) + O(|\kk'|^2)\;.\label{D1.1}
\ee
\paragraph{Constant part.} If $\r = \r'$, by symmetry (8) it follows that
\be
W_{2,0,\r,\r}(\kk) = -W_{2,0,\r,\r}(-k_0,\vec k) \Rightarrow W_{2,0,\r,\r}(\pp_{F}^{\o}) =0\;;\label{D1.2}
\ee
if $\r\neq\r'$ by symmetry (4) we get, using that $T\vec p_{F}^{\o} = \vec p_{F}^{\o} + \o(\vec b_{2} - 2\vec b_{1})$ (which means that $\vec p_{F}^{\o}$ is left invariant by $T$),
\be
W_{2,0,\r,\r'}(\kk) = \eu^{\iu (\r - \r') \vec k(\vec \d_{2} - \vec\d_1)}W_{2,0,\r,\r'}(k_0,T^{-1}\vec k)\Rightarrow W_{2,0,\r,\r'}(\pp_{F}^{\o}) =0\;,\label{D1.3}
\ee
because $1 - e^{\pm \iu \vec p_{F}^{\o}(\vec\d_{2} - \vec\d_1)} \neq 0$.
\paragraph{Linear part in $p_0$.} By simmetry (6.b) it follows that
\be
\partial_{0}W_{2,0}(\kk) = \partial_{0}W_{2,0}(k_0,k_1,-k_2) \Rightarrow \partial_{0}W_{2,0}(\pp_{F}^{+}) = \partial_{0}W_{2,0}(\pp_{F}^{-})\;.\label{D1.4}
\ee
By simmetry (6.a) it follows that
\be
\partial_{0}W_{2,0,1,1}(\kk) = \partial_0 W_{2,0,2,2}(k_0,-k_1,k_2) \Rightarrow \partial_0 W_{2,0,1,1}(\pp_{F}^{\o}) = \partial_{0}W_{2,0,2,2}(\pp_{F}^{\o})\;,\label{D1.5}
\ee
where we used that $\vec p_{F}^{\o} + \vec p_{F}^{-\o} = \vec b_{1} + \vec b_{2}$, while if $\r\neq\r'$ by symmetry (4) we get that (see (\ref{D1.3}))
\be
0 = (1 - \eu^{\iu (\r - \r') \vec k(\vec \d_{2} - \vec\d_1)})\partial_{0}W_{2,0,\r,\r'}(\pp_{F}^{\o})\;\label{D1.6}
\ee 
 which means that $\partial_{0}W_{2,0,\r,\r'}(\pp_{F}^{\o}) =0$. Finally, from (5) we find, using (\ref{D1.4})
 \be
 \partial_{0}W_{2,0}(\kk) = -\partial_{0}W_{2,0}(-\kk)^{*}\Rightarrow \partial_0 W_{2,0,1,1}(\pp_{F}^{\o}) = -\partial_{0}W_{2,0,1,1}(\pp_{F}^{\o})^{*}\;.\label{D1.7}
 \ee
All these properties imply that $\partial_{0}W_{2,0}(\pp_{F}^{\o}) = -\iu z_{0} I$ with $z_0$ real.
\paragraph{Linear part in $p_1$.} By symmetry (6.b) it follows that
\be
\partial_{1}W_{2,0}(\kk) =  \partial_{1}W_{2,0}(k_0,k_1,-k_2) \Rightarrow \partial_{1}W_{2,0}(\pp_{F}^{+}) = \partial_{1}W_{2,0}(\pp_{F}^{-})\;.\label{D1.8}
\ee
By symmetry (8) it follows that
\be
\partial_{1}W_{2,0,\r,\r}(\kk) = -\partial_{1}W_{2,0,\r,\r}(-k_0,\vec k)\Rightarrow \partial_{1}W_{2,0,\r,\r}(\pp_{F}^{\o}) =0\;,\label{D1.9}
\ee
while from (6.a) we get
\be
\partial_{1}W_{2,0,1,2}(\kk) = -\partial_{1}W_{2,0,2,1}(k_0,-k_1,k_2) \Rightarrow \partial_{1}W_{2,0,1,2}(\pp_{F}^{\o}) = -\partial_{1}W_{2,0,2,1}(\pp_{F}^{\o})\;.\label{D1.10}
\ee
Finally, from (5) we find, using (\ref{D1.8})
\be
\partial_{1}W_{2,0}(\kk) = -\partial_{1}W_{2,0}(-\kk)^{*}\Rightarrow \partial_1 W_{2,0,1,2}(\pp_{F}^{\o}) = -\partial_{1}W_{2,0,1,2}(\pp_{F}^{\o})^{*}\;.\label{D1.11} 
\ee
All these properties imply that $\partial_{1}W_{2,0}(\pp_{F}^{\o}) = -z_{1}\s_{2}$ with $z_{1}$ real and
\be
\s_{2} = \begin{pmatrix} 0 & -\iu \\ \iu & 0 \end{pmatrix}\;.\label{D1.12}
\ee
\paragraph{Linear part in $p_2$.} By symmetry (6.b) it follows that
\be
\partial_{2}W_{2,0}(\kk) = -\partial_{2}W_{2,0}(k_0,k_1,-k_2) \Rightarrow \partial_{2}W_{2,0}(\pp_{F}^{+}) = -\partial_{2}W_{2,0}(\pp_{F}^{-})\;.\label{D1.13}
\ee
By symmetry (8) it follows that,
\be
\partial_{2}W_{2,0,\r,\r}(\kk) = -\partial_{2}W_{2,0,\r,\r}(-k_0,\vec k)\Rightarrow \partial_{2}W_{2,0,\r,\r}(\pp_{F}^{\o}) =0\;,\label{D1.14}
\ee
while from (6.a) we get
\be
\partial_{2}W_{2,0,1,2}(\kk) = \partial_{2}W_{2,0,2,1}(k_0,-k_1,k_2) \Rightarrow \partial_{2}W_{2,0,1,2}(\pp_{F}^{\o}) = \partial_{2}W_{2,0,2,1}(\pp_{F}^{\o})\;.\label{D1.15}
\ee
Finally, from (5) we find, using (\ref{D1.13})
\be
\partial_{2}W_{2,0}(\kk) = -\partial_{2}W_{2,0}(-\kk)^{*}\Rightarrow \partial_{2}W_{2,0,1,2}(\pp_{F}^{\o}) = \partial_{2}W_{2,0,1,2}(\pp_{F}^{\o})^{*}\;.\label{D1.16}
\ee
All these properties imply that $\partial_{2}W_{2,0}(\pp_{F}^{\o}) = -z_{2}\o\s_{1}$ with $z_{2}$ real and
\be
\s_{1} = \begin{pmatrix} 0 & 1 \\ 1 & 0 \end{pmatrix}\;.\label{D1.17}
\ee
Finally, the invariance under (4) implies that, using that $\vec p_{F}^{+}$ is left invariant by $T$:
\be
\begin{pmatrix} -\iu z_{1} \\ -z_2 \end{pmatrix} = \eu^{\iu \vec p_{F}^{+}(\vec\d_2 - \vec\d_1)}T\begin{pmatrix} -\iu z_1 \\ -z_2 \end{pmatrix}\;,\label{D3.16}
\ee
which means that $z_{1} = z_{2}$. This concludes the proof of (\ref{1.2.49}).
\section{Proof of (\ref{1.2.50})}\label{secD2}
\setcounter{equation}{0}
\renewcommand{\theequation}{\ref{secD2}.\arabic{equation}}
Let us rewrite $W_{0,2,\m,\n}(\pp)$ as
\be
W_{0,2,\m,\n}(\pp) = W_{0,2,\m,\n}(\V0) + \pp_{\a}\partial_{\a}W_{0,2,\m,\n}(\V0) + O(|\pp|^2)\;.\label{D2.1}
\ee
\paragraph{Constant part.} By symmetry (7), (6.a), (6.b) it follows that if $\m\neq\n$ then $W_{0,2,\m,\n}(\V0) = 0$; the invariance under (4) implies that
\be
\begin{pmatrix} W_{0,2,1,1}(\V0) & 0 \\ 0 & W_{0,2,2,2}(\V0) \end{pmatrix} = 
T\begin{pmatrix} W_{0,2,1,1}(\V0) & 0 \\ 0 & W_{0,2,2,2}(\V0) \end{pmatrix}T^{-1}\;,\label{D2.2}
\ee
which gives $W_{0,2,1,1}(\V0) = W_{0,2,2,2}(\V0)$. The reality condition $W_{0,2}(\V0)^{*} = W_{0,2}(\V0)$ is implied by (5). 

\paragraph{Linear part in $p_0$.} If $\m,\n\neq 0$ or $\m=\n=0$ symmetry (7) implies that
\be
\partial_{0} W_{0,2,\m,\n}(\pp) = \partial_{0}W_{0,2,\m,\n}(-p_0,\vec p) \Rightarrow \partial_{0} W_{0,2,\m,\n}(\V0)=0\;;\label{D2.3}
\ee
if $(\m,\n) = (1,0),(0,1)$ by symmetry (6.a) it follows that
\be
\partial_{0} W_{0,2,\m,\n}(\pp) = -\partial_{0} W_{0,2,\m,\n}(p_0,-p_1,p_2) \Rightarrow -\partial_0 W_{0,2,\m,\n}(\V0) =0\;;\label{D2.4}
\ee 
if $(\m,\n)=(2,0),(0,2)$ the same is true by symmetry (6.b).
\paragraph{Linear part in $p_1$.} If $\m,\n\neq 1$ or $\m=\n=1$ symmetry (6.a) implies that
\be
\partial_{1} W_{0,2,\m,\n}(\pp) = \partial_{1} W_{0,2,\m,\n}(p_0,-p_1,p_2) \Rightarrow \partial_{1} W_{0,2,\m,\n}(\V0) =0\;;\label{D2.5}
\ee
if $(\m,\n) = (0,1),(1,0)$ symmetry (7) implies that
\be
\partial_{1} W_{0,2,\m,\n}(\pp) = -\partial_{1} W_{0,2,\m,\n}(-p_0,\vec p) \Rightarrow \partial_{1} W_{0,2,\m,\n}(\V0) =0\;;\label{D2.6}
\ee
if $(\m,\n) = (1,2),(2,1)$ the same is true by simmetry (6.b).
\paragraph{Linear part in $p_2$.} If $\m,\n\neq 2$ of $\m=\n = 2$ symmetry (6.b) implies that
\be
\partial_{2} W_{0,2,\m,\n}(\pp) = \partial_{2} W_{0,2,\m,\n}(p_0,p_1,-p_2) \Rightarrow \partial_{2} W_{0,2,\m,\n}(\V0) = 0\;;\label{D2.7}
\ee
if $(\m,\n) = (0,2),(2,0)$ symmetry (7) implies that
\be
\partial_{2}W_{0,2,\m,\n}(\pp) = -\partial_{2}W_{0,2,\m,\n}(-p_0,\vec p) \Rightarrow \partial_{2}W_{0,2,\m,\n}(\V0) =0\;;\label{D2.7b}
\ee
if $(\m,\n) = (1,2), (2,1)$ symmetry (6.a) implies that
\be
\partial_{2}W_{0,2,\m,\n}(\pp) = -e^{\pm \iu \vec p\vec\d_1}\partial_{2}W_{0,2,\m,\n}(p_0,-p_1,p_2) \Rightarrow \partial_{2}W_{0,2,\m,\n}(\V0)=0\;.
\ee
This concludes the proof of (\ref{1.2.50}).
\section{Proof of (\ref{1.2.51})}\label{secD3}
\setcounter{equation}{0}
\renewcommand{\theequation}{\ref{secD3}.\arabic{equation}}
\paragraph{Case $\m=0$.} If $\r\neq \r'$ by symmetry (8) it follows that
\be
W_{2,1,0,\r,\r'}(\kk,\pp) = - W_{2,1,0,\r,\r'}((-k_0,\vec k), (-p_0,\vec p))\Rightarrow W_{2,1,0,\r,\r'}(\pp_F^{\o},\V0) =0\;;\label{D3.2}
\ee
if $\r=\r'$ by (6.a) it follows that
\bea
 W_{2,1,0,1,1}(\kk,\pp) &=& \eu^{-\iu \vec p\vec\d_1}W_{2,1,0,2,2}((k_0,-k_1,k_2),(p_0,-p_1,p_2))\nn\\&\Rightarrow& W_{2,1,0,1,1}(\pp_{F}^{\o},\V0) = W_{2,1,0,2,2}(\pp_{F}^{\o},\V0)\;,\label{D3.3}
\eea
where we used that $\vec p_{F}^{\o} + \vec p_{F}^{-\o} = \vec b_{1} + \vec b_{2}$. The invariance under (6.b) implies that
\bea
W_{2,1,0}(\kk,\pp) &=& W_{2,1,0}((k_{0},k_1,-k_2),(p_0,p_1,-p_2))\nn\\&\Rightarrow& W_{2,1,0}(\pp_{F}^{\o},\V0) = W_{2,1,0}(\pp_{F}^{-\o},\V0)\;,\label{D3.4}
\eea
and, finally, from symmetry (5) we get that, using (\ref{D3.4}):
\be
W_{2,1,0}(\kk,\pp) = - W_{2,1,0}(-\kk,-\pp)^{*} \Rightarrow W_{2,1,0}(\pp_{F}^{\o},\V0) = -W_{2,1,0}(\pp_{F}^{\o},\V0)^{*}\;,\label{D3.5}
\ee
which means that $W_{2,1,0}(\pp_{F}^{\o},\V0) = -\iu \l_{0}I$ with $\l_{0}$ real. 
\paragraph{Case $\m=1$.} The invariance under (6.b) implies that
\bea
W_{2,1,1}(\kk,\pp) &=& W_{2,1,1}((k_0,k_1,-k_2),(p_0,p_1,-p_2))\nn\\ &\Rightarrow& W_{2,1,1}(\pp_{F}^{\o},\V0) = W_{2,1,1}(\pp_{F}^{-\o},\V0)\;.\label{D3.6}
\eea
Symmetry (8) implies that,
\bea
W_{2,1,1,\r,\r}(\kk,\pp) &=& - W_{2,1,1,\r,\r}((-k_0,\vec k),(-p_{0},\vec p)) \nn\\&\Rightarrow& W_{2,1,1,\r,\r}(\pp_{F}^{\o},\V0) =0\;,\label{D3.7}
\eea
while symmetry (6.a) implies that, using (\ref{D3.6}),
\bea
W_{2,1,1,1,2}(\kk,\pp) &=& -\eu^{-\iu \vec p\vec\d_1}W_{2,1,1,2,1}((k_0,-k_1,k_2),(p_0,-p_1,p_2)) \nn\\
&\Rightarrow& W_{2,1,1,1,2}(\pp_{F}^{\o},\V0) = - W_{2,1,1,2,1}(\pp_{F}^{\o},\V0)\;.\label{D3.8}
\eea
Finally, symmetry (5) implies that, using (\ref{D3.6})
\be
W_{2,1,1}(\kk,\pp) = -W_{2,1,1}(-\kk,-\pp)^{*} \Rightarrow W_{2,1,1}(\pp_{F}^{\o},\V0) = -W_{2,1,1}(\pp_{F}^{\o},\V0)^{*}\;,\label{D3.9}
\ee
which means that $W_{2,1,1}(\pp_{F}^{\o},\V0) = -\l_{1}\s_{2}$ with $\l_{1}$ real and $\s_2$ given by (\ref{D1.12}).
\paragraph{Case $\m=2$.} By symmetry (6.b) it follows that
\bea
W_{2,1,2}(\kk,\pp) &=& -W_{2,1,2}((k_0,k_1,-k_2),(p_0,p_1,-p_2))\nn\\ &\Rightarrow& W_{2,1,2}(\pp_{F}^{\o},\V0) = -W_{2,1,2}(\pp_{F}^{-\o},\V0)\,.\label{D3.11}
\eea
The invariance under (8) implies that
\bea
W_{2,1,2,\r,\r}(\kk,\pp) &=& - W_{2,1,2,\r,\r}((-k_0,\vec k),(-p_{0},\vec p)) \nn\\&\Rightarrow& W_{2,1,2,\r,\r}(\pp_{F}^{\o},\V0) =0\;,\label{D3.12}
\eea
while symmetry (6.a) implies that
\bea
W_{2,1,2,1,2}(\kk,\pp) &=& W_{2,1,2,2,1}((k_0,-k_1,k_2),(p_0,-p_1,p_2)) \nn\\
&\Rightarrow& W_{2,1,2,1,2}(\pp_{F}^{\o},\V0) = W_{2,1,2,2,1}(\pp_{F}^{\o},\V0)\;.\label{D3.13}
\eea
Symmetry (5) implies that, using (\ref{D3.11}),
\be
W_{2,1,2}(\kk,\pp) = -W_{2,1,2}(-\kk,-\pp)^{*} \Rightarrow W_{2,1,2}(\pp_{F}^{\o},\V0) = W_{2,1,2}(\pp_{F}^{\o},\V0)^{*}\;,\label{D3.14}
\ee
which means that $W_{2,1,2}(\pp_{F}^{\o},\V0) = -\l_{2}\o\s_{1}$ with $\l_{2}$ real and $\s_1$ given by (\ref{D1.17}). Finally, the invariance under (4) implies that, using that $\vec p_{F}^{+}$ is left invariant by $T$:
\be
\begin{pmatrix} -\iu \l_{1} \\ -\l_2 \end{pmatrix} = \eu^{\iu \vec p_{F}^{+}(\vec\d_2 - \vec\d_1)}T\begin{pmatrix} -\iu \l_1 \\ -\l_2 \end{pmatrix}\;,\label{D3.14b}
\ee
which means that $\l_{1} = \l_{2}$. This concludes the proof of (\ref{1.2.51}).

\section{Proof of (\ref{1.2.52})}\label{secD4}
\setcounter{equation}{0}
\renewcommand{\theequation}{\ref{secD4}.\arabic{equation}}
If ${\ul \m} = (\m_1,\m_2,\m_3)$ has an odd number of labels $\m_{i} =0$ then by (8) it follows that
\be
W_{0,3,{\ul \m}}(\pp,\pp') = -W_{0,3,{\ul \m}}((-p_{0},\vec p),(-p'_{0},\vec p))\Rightarrow W_{0,3,{\ul \m}}(\V0,\V0) =0\;;\label{D4.1}
\ee
the same is true if the number of labels $\m_{i}=0$ is even by (7). This concludes the proof of (\ref{1.2.52}) and of Lemma \ref{lem2.4c}.
%
\chapter{Lowest order computations}\label{app2d}
\setcounter{equation}{0}
\renewcommand{\theequation}{\ref{app2d}.\arabic{equation}}

In this Appendix we reproduce the details of all the lowest order computations performed in this Thesis.
\section{The Beta function of the wave function renormalization}\label{secwave}
\setcounter{equation}{0}
\renewcommand{\theequation}{\ref{secwave}.\arabic{equation}}

\begin{figure}[htbp]
\centering
\includegraphics[width=0.5\textwidth]{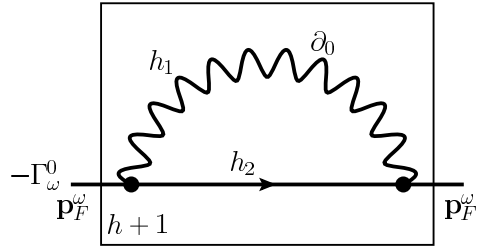}
\caption{The lowest order contribution to $\b^{z}_{h}$. The wavy line labelled by a ``$\partial_0$'' symbol corresponds to $\partial_0 \hat w^{(h_1)}(\pp)$; a sum over all the possible scales $h_{1},h_{2}\geq h+1$ is understood.}\label{figZ}
\end{figure}

We start by performing the computation of the lowest order contribution in the running coupling constants to the Beta function $\b^{z,(2)}_{h}$ of the wave function renormalization; by definition, see (\ref{3.8}) and (\ref{3a.18}),
$\b_h^{z}=z_{0,h}= \G^{0}_{\o}\partial_{0}W^{(h)}_{2,0,\o,\o}(\V0)$. Following the rules discussed in Section \ref{multi}, its lowest order contribution $\b^{z,(2)}_{h}$ is graphically represented by the Feynman graph depicted in Fig. \ref{figZ}.

Therefore, we find (repeated $\m$ labels are summed):
\bea 
\b_h^{z,(2)} &=& - \G^{0}_{\o}\bar e_{\m,h+1}^2\int\frac{d\pp}{(2\p)^3}\, \big[\partial_0 \hat w^{(h+1)}(\pp)\big]\G^{\m}_{\o}\hat g_{\o}^{(h+1)}(\pp)\G^{\m}_\o + \label{A1.1}\\
&& - \G^{0}_{\o} \bar e_{\m,h+2}^2\Big(\frac{Z_{h+1}}{Z_h}\Big)^2\int\frac{d\pp}{(2\p)^3}\, \big[\partial_{0}\hat w^{(h+2)}(\pp)\big]\G^{\m}_{\o} \hat g_{\o}^{(h+1)}(\pp)\G^{\m}_\o + \nn\\
&& -\G^{0}_{\o}\bar e_{\m,h+2}^2\frac{Z_{h+1}}{Z_h}\int\frac{d\pp}{(2\p)^3}\,\big[\partial_{0}\hat w^{(h+1)}(\pp)\big] \G^{\m}_{\o}\hat g_{\o}^{(h+2)}(\pp)\G^{\m}_{\o}\;.\nn
\eea
Using inductively the Beta function equations for $v_{h+1}$, $Z_{h+1}$, $e_{\m,h+1}$ we find that, setting $F_{h+1} := f_{h+1}^2 + 2f_{h+1}f_{h+2}$,
\bea
&&\b^{z,(2)}_{h} = -\bar e_{\m,h}^{2}\int\frac{d\pp}{(2\pi)^3}\,\frac{F_{h+1}(\pp)}{2|\pp|^3}\frac{p_0^2}{p_0^2 + v_h^2|\vec p|^2}\G^{0}_{\o}\G^{\m}_{\o}\G^{0}_{\o}\G^{\m}_{\o} + \label{A1.2}\\
&& \bar e_{\m,h}^2 \int\frac{d\pp}{(2\pi)^3}\,\big[f'_{h+1}(f_{h+1} + f_{h+2}) + f'_{h+2}f_{h+1}\big]\frac{p_0^2 \G^{0}_{\o}\G^{\m}_{\o}\G^{0}_{\o}\G^{\m}_{\o}}{2|\pp|^2 (p_0^2 + v_h^2 |\vec p|^2)} + \nn\\
&& + O(e^{4}) + O(e^{2}M^{\th h})\;;\nn
\eea
passing to radial coordinates $\pp = p(\cos\th,\sin\th\cos\ph,\sin\th\sin\ph)$, using that 
\be
\int dp\, (f_{h+1}' f_{h+1}+f_{h+1}' f_{h+2}+f_{h+2}' f_{h+1})=0\;,\quad \bar e_{\m,h}^2\G^{0}_{\o}\G^{\m}_{\o}\G^{0}_{\o}\G^{\m}_{\o} = e_{0,h}^{2} - 2v_{h-1}^2 e_{1,h}^{2}\nn
\ee
we get:
\bea 
&&\b_{h}^{z,(2)}=(2v_h^2e_{1,h}^2-e_{0,h}^2)\frac1{8\p^2} \int_0^\io dp\, \frac{F_{h+1}(\pp)}{p}\cdot\Big[\int_{-1}^1 d\cos\th\, \frac{\cos^2\th}{\cos^2\th+v^2_h\sin^2\th}\Big] + \nn\\
&&\quad + O(e^{2}M^{\th h}) + O(e^{4})\;.\label{A1.3}
\eea
The integral over the radial coordinate $p$
can be computed by using  the definition (\ref{3.1}):
\bea  &&\int_0^\io \frac{dp}{p}(f_{h+1}^2+2f_{h+1}f_{h+2}) = \label{A1.4}\\&&
=\int_0^\io \frac{dp}{p}[2(\c(p)-\c(Mp))-(\c^2(p)-\c^2(Mp))]
= \int_1^{M}\frac{dp}p=\log M\;;\nn\eea
finally, an explicit evaluation of the integral over $d\cos\th$ leads to (\ref{3a.19z}).

\section{The Beta function of the Fermi velocity}\label{secvel}
\setcounter{equation}{0}
\renewcommand{\theequation}{\ref{secvel}.\arabic{equation}}

By definition, see formulas (\ref{3.8}) and (\ref{3a.18}), $\b_h^{v,(2)} := z_{1,h}^{(2)}-v_hz_{0,h}^{(2)}$, with $z_{1,h}= \G^{1}_{\o}\partial_{1}W^{(h)}_{2,0,\o,\o}(\V0)$. At second order,
proceeding as in the derivation of (\ref{A1.2}), we find:
\be
z^{(2)}_{1,h} = \bar e^{2}_{\m,h}\int\frac{d\pp}{(2\pi)^3}\,\frac{F_{h+1}(\pp)}{2|\pp|^3}\frac{-v_h p_1^2}{p_0^2 + v_h^2 |\vec p|^2}\G^{1}_{\o}\G^{\m}_{\o}\G^{1}_{\o}\G^{\m}_{\o} + O(e^{4}) + O(e^{2}M^{\th h})\;,\label{A1.5}
\ee
(the term involving the derivatives $f'_{h+1}$, $f'_{h+2}$ of the support functions is zero, as before); using that
\be
\bar e_{\m,h}^2 \G^{1}_{\o}\G^{\m}_{\o}\G^{1}_{\o}\G^{\m}_{\o} = -e_{0,h}^2\;,\nn
\ee
passing to radial coordinates, we get that:
\be
z^{(2)}_{1,h} = e_{0,h}^2v_h\frac1{16\p^2} \Big[\int_0^\io \frac{dp}{p}(f_{h+1}^2+2f_{h+1}f_{h+2})\Big] \cdot\Big[\int_{-1}^1 d\cos\th\, \frac{\sin^2\th}{\cos^2\th+v^2_h\sin^2\th}\Big]\;.\nn
\ee
The integral over the radial coordinate can be evaluated as in (\ref{A1.4}), and gives $\log M$; therefore, an explicit evaluation of the integral over $d\cos\th$ leads to
\be z_{1,h}^{(2)} = e_{0,h}^2v_h^{-1}\frac{\log M}{8\p^2}
\Big(\frac{\arctan\x_h}{\x_h}-\frac{\x_h-\arctan\x_h}{\x_h^3}\Big)\;,
\label{A1.6}\ee
which, combined with $\b_h^{v,(2)}=z_{1,h}^{(2)}-v_hz_{0,h}^{(2)}$, proves (\ref{3a.19z}).

\section{The Beta function of the Kekul\' e mass}\label{secDelta}
\setcounter{equation}{0}
\renewcommand{\theequation}{\ref{secDelta}.\arabic{equation}}

By definition, $\D_{h}\d_{h} = \iu \g^{(j_{0})*}\LL_{0}\PPP_{1}\widetilde W^{(h)}_{2,0}(\kk)$, that is, setting $F_{h+1}(\pp) := f_{h+1}^{2}(\pp) + 2f_{h+1}(\pp)f_{h+2}(\pp)$:
\bea
\d_{h}^{(2)} &=& \iu\g^{(j_0)*}\bar e_{\m,h}^{2}\int \frac{d\pp}{(2\pi)^3}\,\frac{F_{h+1}(\pp)}{2|\pp|}\frac{-\iu}{p_0^2 + v_{h}^{2}|\vec p|^{2}}\g_{\m}\g^{(j_0)*}\g_{\m}\nn\\ 
&=& (e_{0,h}^{2} + 2v_{h}^{2}e_{1,h}^{2})\int \frac{d\pp}{(2\pi)^{3}}\,\frac{F_{h+1}(\pp)}{2|\pp|}\frac{1}{p_{0}^{2} + v_{h}^{2}|\vec p|^{2}}\label{Delta1}\\
&=& (e_{0,h}^{2} + 2v_{h}^{2}e_{1,h}^{2})\frac{1}{8\pi^2}\Big[ \int_{0}^{+\infty} d p\, \frac{F_{h+1}(p)}{p} \Big]\cdot \Big[ \int_{-1}^{1}d \cos\th \frac{1}{\cos^{2}\th + v_{h}^{2}\sin^{2}\th} \Big]\nn
\eea
where in the second equality we used that $\g_\m \g^{(j_{0})*} = -\g^{(j_{0})}\g_{\m}$, and that $\g^{(j_0)*}\g^{(j_0)} = I$. The integral over the $p$ coordinated can be computed as in (\ref{A1.4}), and an explicit evaluation of the integral over $d\cos \th$ leads to (\ref{lat18}).

\section[Anomalous exponents of the susceptibilities]{Anomalous exponents of the generalized susceptibilities}\label{secexp}
\setcounter{equation}{0}
\renewcommand{\theequation}{\ref{secexp}.\arabic{equation}}

\begin{figure}[hbtp]
\centering
\includegraphics[width=.4\textwidth]{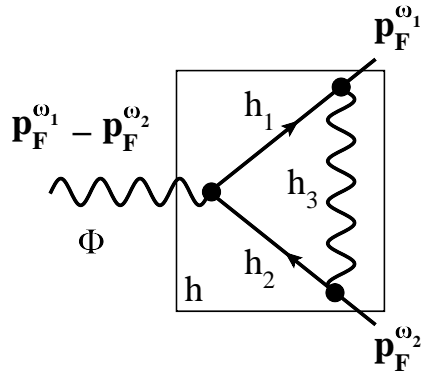}
\caption{The lowest order contribution to $\b^{\a}_{\ul\o,h,1}$; a sum over the inner scales $h_i\geq h$ is understood.}\label{figexp}
\end{figure}

In this Appendix we shall compute the lowest order contribution to the Beta functions $\b^{\a}_{\ul\o,h,1}$ of the running coupling constants $Z^{\a}_{\ul\o,h}$, whose flow has been discussed in Section \ref{excrcc}; in particular, these computations will allow us to compute the lowest order contributions to the anomalous exponents of the generalized susceptibilities. Basically, we will compute the Feynman graph depicted in Fig. \ref{figexp} for all possible values of the inner scale labels, $\a$, $\o_1$ and $\o_2$, and with the external field $\hat\Phi^{(\a)}_{\qq,1}$; $Z^{\a}_{\ul\o,h}$ does not depend on the bond label $a$, therefore we set $a=1$ for simplicity.

\paragraph{Preliminaries.} Before working out explicitly the computations, it is useful to recall a list of notations and identities that will be repeatedly used in the following. The main difference with respect to the previous lowest order computations is that here we will not use the compact relativistic notations.
\begin{itemize}
%
\item[(i)] The vertex on scale $h$ from which emerges the wavy line associated to the external field is represented as:
\be
\sum_{\s}\sum_{\o_1,\o_2}Z^{\a}_{\ul{\o},h}\int \frac{d\kk'}{D}\frac{d\pp}{(2\pi)^3}\,\hat\Phi^{(\a)}_{\qq,1}\hat\Psi^{(\leq h)+,T}_{\kk'+\qq,\s,\o_1}\G^{\a}\hat \Psi^{(\leq h)-}_{\kk,\s,\o_2}\;,\label{exp1}
\ee
where the symbol $\hat\Psi^{(\leq h)\pm}_{\kk',\s,\o}$ has been introduced in (\ref{3.03b}), and the $\G^{\a}$ matrices have been defined in (\ref{exc6}).
\item[(ii)] The $A\Psi^{+}\Psi^{-}$ vertex is represented as:
\be
\sum_{\s}\sum_{\o} \bar e_{\m,h}\int \frac{d\kk'}{D}\frac{d\pp}{(2\pi)^3}\,\hat\Phi^{(\a)}_{\qq,1}\hat\Psi^{(\leq h)+,T}_{\kk'+\qq,\s,\o}\G^{\m}_{\o}\hat \Psi^{(\leq h)-}_{\kk,\s,\o}\;,\label{exp2}
\ee 
where $\bar e_{0,h} = e_{0,h}$ and $\bar e_{1,h} = \bar e_{2,h} = v_{h-1}e_{1,h}$, and the matrices $\G^{\m}_{\o}$ are given by
\be
\G^{0}_{\o} = \begin{pmatrix} -\iu & 0 \\ 0 & -\iu \end{pmatrix}\;,\quad \G^{1}_{\o} = \begin{pmatrix} 0 & \iu \\ -\iu & 0 \end{pmatrix}\;,\quad \G^{2}_{\o} = \begin{pmatrix} 0 & -\o \\ -\o & 0 \end{pmatrix}\;.\label{exp3}
\ee
\item[(iii)] The fermion propagator on scale $h$ for $\kk = \kk' + \pp_{F}^{\o}$ and $|\kk'|$ small is given by
\be
\hat g^{(h)}_{\o}(\kk') = \frac{\tilde f_{h}(\kk')}{{k'_0}^2 + \tilde v(\kk')^2|\vec k'|^2}\Big( -\G^{0}_{\o}k'_0 + \tilde v_{h}(\kk')k'_{i}\G^{i}_{\o} \Big)\big(1 + O(\kk)\big)\;.\label{exp4}
\ee
\item[(iv)] The following identities will be useful:
\bea
&&\G_{\o_1}^{0}\G^{E_{-}}\G_{\o_2}^{0} = -\G_{\o_1}^{1}\G^{E_-}\G_{\o_2}^{1} = \o_1\o_2 \G_{\o_1}^{2}\G^{E_-}\G_{\o_2}^{2}= -\G^{E_{-}}\;,\label{exp5}\\
&&\G_{\o_1}^{0}\G^{E_{+}}\G_{\o_2}^{0} = \G_{\o_1}^{1}\G^{E_+}\G_{\o_2}^{1} = -\o_1\o_2 \G_{\o_1}^{2}\G^{E_+}\G_{\o_2} = -\G^{E_{+}}\;,\nn\\
&&\G_{\o_1}^{0}\G^{CDW}\G_{\o_2}^{0} = \G_{\o_1}^{1}\G^{CDW}\G_{\o_2}^{1} = \o_1\o_2  \G_{\o_1}^{2}\G^{CDW}\G_{\o_2}^{2} = - \G^{CDW}\;.\nn\\
&&\G_{\o_1}^{0}\G^{D}\G_{\o_2}^{0} = -\G_{\o_1}^{1}\G^{D}\G_{\o_2}^{1} = -\o_1\o_2\G_{\o_1}^{2}\G^{D}\G_{\o_2}^{2} = -\G^{D}\;,\nn
\eea
\item[(v)] Setting $F_{h}(|\pp|) := f_{h}(\pp)^{3} + 3 f_{h}(\pp)^{2}f_{h+1}(\pp) + 3f_{h}(\pp)f_{h+1}(\pp)^{2}$, it follows that
\be
\int_{0}^{+\infty} d\r\,\frac{F_{h}(\r)}{\r} = \log M\;.\label{exp5b}
\ee
\end{itemize}

The first part of the calculation is the same for all $\a$, $\ul\o$; in fact, we can write the value of the graph depicted in Fig. \ref{figexp} as follows, neglecting corrections $O(e^{4})$ and $O(e^{2}M^{h})$ (repeated $\a$ indeces are not summed):
\bea
&&\G^{\a}\b^{\a,(2)}_{\ul\o,h,1} = \nn\\
&&\bar e_{\m,h}^2\int \frac{d\pp}{(2\pi)^3}\,\G^{\m}_{\o_1}g^{(h_1)}_{\o_1}(\pp)\G^{\a}g^{(h_2)}_{\o_2}(\pp)\G^{\m}_{\o_2}w^{(h_{3})}(\pp) = \int \frac{d\pp}{(2\pi)^3}\,\frac{F_{h}(|\pp|)}{(p_0^2 + v_h^2 |\vec p|^2)^2}\frac{\bar e_{\m,h}^{2}}{2|\pp|}\cdot\nn\\&&\cdot \G^{\m}_{\o_1}\Big[ p_0^2 \G^{0}_{\o_1}\G^{\a}\G^{0}_{\o_2} + v_h^2 p_1^2 \G^{1}_{\o_1}\G^{\a}\G^{1}_{\o_2} + v_h^2 p_2^2 \G^{2}_{\o_1}\G^{\a}\G^{2}_{\o_2} \Big]\G^{\m}_{\o_2} + O(e^{4}) + O(e^{2} M^{h})\nn\\
&& = \frac{e^{2}}{12\pi^2}\log M \G^{\m}_{\o_1}\G^{\n}_{\o_1}\G^{\a}\G^{\n}_{\o_2}\G^{\m}_{\o_2} + O(e^{4}) + O(e^{2}M^{h})\;.\label{exp6}
\eea
Using (\ref{exp5}) we get that:
\bea
&& \G^{\m}_{+}\G^{\n}_{+}\G^{E_-}\G^{\n}_{-}\G^{\m}_{-} = \G^{E_-}\;,\qquad \G^{\m}_{+}\G^{\n}_{+}\G^{E_-}\G^{\n}_{+}\G^{\m}_{+} = \G^{E_-}\;,\nn\\
&& \G^{\m}_{+}\G^{\n}_{+}\G^{E_+}\G^{\n}_{-}\G^{\m}_{-} = 9 \G^{E_+}\;,\qquad \G^{\m}_{+}\G^{\n}_{+}\G^{E_+}\G^{\n}_{+}\G^{\m}_{+} = \G^{E_+}\;,\nn\\
&& \G^{\m}_{+}\G^{\n}_{+}\G^{CDW}\G^{\n}_{-}\G^{\m}_{-} = \G^{CDW}\;,\qquad \G^{\m}_{+}\G^{\n}_{+}\G^{CDW}\G^{\n}_{+}\G^{\m}_{+} = 9 \G^{CDW}\;,\nn\\
&& \G^{\m}_{+}\G^{\n}_{+}\G^{D}\G^{\n}_{-}\G^{\m}_{-} = \G^{D}\;,\qquad \G^{\m}_{+}\G^{\n}_{+}\G^{D}\G^{\n}_{+}\G^{\m}_{+} = \G^{D}\;;\label{exp7}
\eea
plugging these identities into (\ref{exp6}) we get the lowest order contribution $\b^{\a,(2)}_{\ul\o,h,1}$ to $\b^{\a}_{\ul\o,h,1}$; the results listed in Table \ref{tabexp} follow from $\eta^{\a}_{\ul\o} := \lim_{h\rightarrow-\infty}\log_{M}\big( 1 + \b^{\a}_{\ul\o,h,1} \big)$.
\section{Lowest order check of the Ward identities}\label{WIcheck}
\setcounter{equation}{0}
\renewcommand{\theequation}{\ref{WIcheck}.\arabic{equation}}

In this Section we shall check at lowest order in non-renormalized perturbation theory the validity of the Ward identities that allowed us to prove the infrared stability of the flows of the effective charge and of the photon mass.

\subsection{Ward identity for the photon mass}\label{phm}

We start with the check of the Ward identity for the photon mass; as we have discussed in Section \ref{secWI}, the gauge invariance of the theory implies that the effective photon mass is vanishing, {\it i.e.} that the electromagnetic interaction is {\it unscreened}.

\begin{figure}[htbp]
\centering
\includegraphics[width=0.6\textwidth]{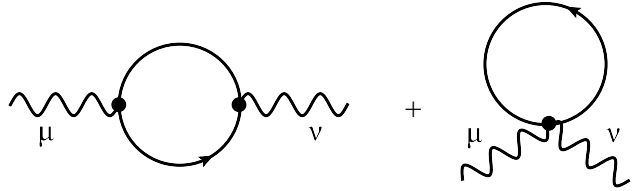}
\caption{The lowest order contribution to the renormalized photon mass in naive perturbation theory. The graphs are all computed at vanishing external momenta, and for $\m=0$ or $\n=0$ the second graph is absent. The sum of the graphs is exactly vanishing.}\label{figphm}
\end{figure}

At lowest order, the graphs contributing to the dressed photon mass are depicted in Fig. \ref{figphm}; for $\m=0$ or $\n=0$ the second graph is absent. As we are going to show here, the sum of the two graphs is {\it exactly vanishing}.

These graphs are obtained by expanding the interaction $V(\Psi,A)$ to second order in $A$; it follows that, see Eq. (\ref{1.2.20b}), assuming without any loss of generality that the field $\hat A_{\m,\pp}$ is supported inside the first Brillouin zone ($\pp=0$ in the graphs represented in Fig. \ref{figphm}):
\bea
&&V(\Psi,A) = \bar e_{\m} \int \frac{d\kk}{D}\frac{d\pp}{(2\pi)^3}\,\hat\Psi^{+}_{\kk + \pp,\s}\G^{\m}(\vec k,\vec p)\hat\Psi^{-}_{\kk,\s}\hat A_{\m,\pp} + \nn\\
&& + e^2 v \int \hat\Psi^{+}_{\kk+\pp_1+\pp_2,\s}\G^{\m_1,\m_2}(\vec k,\vec p_1,\vec p_2)\hat\Psi^{-}_{\kk,\s}\hat A_{\m_1,\pp_1}\hat A_{\m_2,\pp_2} + O(e^3)\;,\label{phm1}
\eea
where $\bar e_{0} = e$, $\bar e_{1} = \bar e_{2} =  ve$ with $v = \frac{3}{2}t$, $\vec e_{1} := (1,0)$, $\vec e_{2} = (0,1)$, and
\bea
&&\G^{0}(\vec k,\vec p) := \begin{pmatrix} -\iu & 0 \\ 0 & -\iu \eu^{-\iu \vec p\cdot \vec\d_1} \end{pmatrix}\;,\label{phm2}\\
&&\G^{i}(\vec k,\vec p) := \frac{2\iu}{3}\sum_{j=1,2,3}(\vec \d_{j}\cdot\vec e_{i})\eta_j(\vec p)\begin{pmatrix} 0 & \eu^{-\iu \vec k(\vec\d_j
 - \vec\d_1)} \\ -\eu^{\iu(\vec k + \vec{p})(\vec\d_j - \vec\d_1)} & 0 \end{pmatrix}\;.\nn\\
 && \G^{\m_1,\m_2}(\vec k,\vec p_1,\vec p_2) := \label{phm2b}\\
 && := \frac{-\prod_i (\d_{\m_i,0} - 1)}{3}\sum_{j }\big[\prod_{i=1}^{2}\eta_{j}(\vec p_i)(\vec\d_j\cdot \vec e_{\m_i})\big]\begin{pmatrix} 0 & \eu^{-\iu \vec k(\vec\d_j - \vec\d_1)} \\ \eu^{\iu (\vec k + \vec p_1 + \vec p_2)(\vec \d_j - \vec\d_1)}  & 0\end{pmatrix}\nn
\eea
Notice that $\G^{\m}(\vec p_{F}^{\o},\vec 0) = \G^{\m}_{\o}$, where $\G^{\m}_{\o}$ are the matrices introduced in Eq. (\ref{1.2.49}). Before starting, it is useful to recall that the fermion propagator is given by
\be
\hat{g}(\kk) = \frac{1}{k_0^2 + v^2|\O(\vec k)|^2}\begin{pmatrix} \iu k_0 & -v\O^{*}(\vec k) \\ -v\O(\vec k) & \iu k_0 \end{pmatrix}\;,
\ee 
with $\O(\vec k) = \frac{2}{3}\sum_{j=1,2,3}\eu^{\iu \vec k(\vec\d_j - \vec\d_1)}$. With these notations, the value of the sum of the two graphs in Fig. (\ref{figphm}) is:
\bea
&&-\bar e_{\m}\bar e_{\n}\int \frac{d\kk}{D}\,\Tr\Big\{ \hat{g}(\kk)\G^{\m}(\vec k,\vec 0)\hat{g}(\kk)\G^{\n}(\vec k,\vec 0) \Big\} + \label{phm3}\\
&& -2 e^{2}v \int \frac{d\kk}{D}\,\Tr\Big\{\hat{g}(\kk)\G^{\m,\n}(\vec k,\vec 0,\vec 0)\Big\}\;.\nn
\eea\\
We shall consider only the cases $\m=0$, $\m=2$; the case $\m=1$ can be obtained from the case $\m=2$ using the symmetry (4) of Lemma \ref{lem2.4}. 

\paragraph{Case $\m=0$.} Let us first consider the case $\m=0$; using that 
\be
\hat{g}(\kk)\G^{0}(\vec k,\vec 0)\hat{g}(\kk) = -\partial_{0}\hat{g}(\kk)\;,\label{phm3a}
\ee
 we get that (\ref{phm3}) is equal to:
\be
e\bar e_{\n} \int \frac{d\kk}{D}\, \Tr\Big\{ \partial_{0}\hat{g}(\kk) \G^{\n}(\vec k,\vec 0) \Big\} =0\;.\label{phm4}
\ee 
%
%
%
%

\paragraph{Cases $\m=2$, $\n\neq 0$.} Since
\be
\hat{g}(\kk)v\G^{2}(\vec k,\vec 0)\hat{g}(\kk) = -\partial_{2}\hat{g}(\kk)\;,\label{phm6a}
\ee
we can rewrite the integral in the first line of (\ref{phm3}) as:
\be
-ve^{2}\int\frac{d\kk}{D}\,\Tr\Big\{ \hat{g}(\kk)\partial_{2}\G^{\n}(\vec k,\vec 0) \Big\}\;;\label{phm6}
\ee
using the identity
\be
\G^{\n,2}(\vec k,\vec 0, \vec 0) = -\frac{1}{2}\partial_2 \G^{\n}(\vec k,\vec 0)\;,\label{phm6b}
\ee
we see that the second line of (\ref{phm3}) exacly cancels the first line.\\

The case $\m=\n=1$ can be obtained from $\m=\n=2$ by using symmetry (4) of Lemma \ref{lem2.4}; it follows that even in this case the sum of the two graphs is vanishing. All the other cases can be obtained using the cyclicity of the trace; this concludes the lowest order check of the Ward identity for the photon mass.

\subsection{Ward identity for the effective charge} 

In this Section we shall check at lowest order in non-renormalized perturbation theory the Ward identity for the effective charge; this identity allowed us to prove at all orders in renormalized perturbation theory that the effective charge is close to the bare one.

Graphically, the lowest order contribution to the renormalized charge in naive perturbation theory is given by Fig. \ref{figvert}:
\begin{figure}[htbp]
\centering
\includegraphics[width=0.7\textwidth]{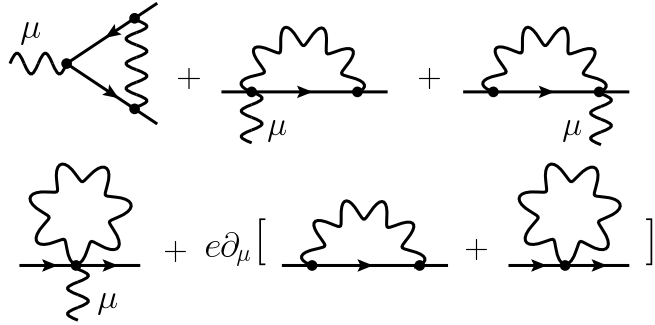}
\caption{The lowest order contribution to the renormalized charge. To avoid ambiguities, the photon propagator is regularized in the infrared by a cutoff function. The sum of the graphs is exactly vanishing, and the cancellation does not depend on the presence of the cutoff.}\label{figvert}
\end{figure}

notice that the first and fifth graph are {\it logarithmically divergent}; therefore, to avoid ambiguities we fix an infrared cutoff on the photon propagator, for instance by replacing $\hat{w}(\pp)$ with $\hat{w}^{(\geq h)}(\pp)$ defined in (\ref{refm2}) with $\varepsilon = 0$. As we are going to show, the sum of these six graphs is exactly vanishing; therefore, the dressed charge is equal to the bare one at lowest order. Remarkably, this cancellation does not depend on the presence of the infrared cutoff on the photons; this fact has been exploited in Section \ref{secWI} to derive a Ward identity for the effective charge on a given scale $h$. We shall only consider the cases $\m=0$ and $\m=2$; the renormalization of the charge corresponding to $\m=1$ is equal to the $\m=2$ one by symmetry (4) of Lemma \ref{lem2.4}. 

The graphs in Fig. \ref{figvert} can be obtained by expanding the interaction $V(\Psi,A)$ to third order in $A$; it follows that, calling $V_2(\Psi,A)$ the expansion to second order derived in Eq. (\ref{phm1}):
\bea
&&V(\Psi,A) = V_2(\Psi,A) + \label{vert1}\\&& e^3 v\int \hat\Psi^{+}_{\kk+\sum_i \pp_i,\s}\G^{\m_1,\m_2,\m_3}(\vec k,\vec p_1,\vec p_2,\vec p_3)\hat\Psi^{-}_{\kk,\s}\hat A_{\m_1,\pp_1}\hat A_{\m_2,\pp_2}\hat A_{\m_3,\pp_3} + O(e^4)\;,\nn
\eea
with
\bea
&&\G^{\m_1,\m_2,\m_3}(\vec k,\vec p_1,\vec p_2,\vec p_3) := \label{vert1b}\\&& :=  \frac{\iu\prod_i (\d_{\m_i,0} - 1)}{9}\sum_{j}\big[\prod_{i=1}^{3}\eta_{j}(\vec p_i)(\vec\d_j\cdot \vec e_{\m_i})\big]\begin{pmatrix} 0 & \eu^{-\iu\vec k(\vec\d_j - \vec\d_1)} \\ -\eu^{\iu(\vec k + \sum_i \vec p_i)(\vec\d_j - \vec\d_1)} & 0 \end{pmatrix}\;.\nn
\eea

\paragraph{Case $\m=0$.} In this case only the first and fifth graphs in (\ref{figvert}) are present; and using the identity (\ref{phm3a}) we get that their sum is exactly vanishing.

\paragraph{Case $\m=2$.} Here the situation is more complicated, because of the presence of the second, third, fourth and sixth graph in Fig. \ref{figvert}, and because the $\partial_{2}$ derivative can act on the vertex functions. However:
\begin{itemize}
\item[(i)] the first graph is cancelled by part of the fifth, because of the identity (\ref{phm6});
\item[(ii)] the contribution due to the action of the $\partial_2$ derivative on the vertices of the fourth graph are exactly cancelled by the second and third graphs; this is so because of the identity
\be
\G^{\m,2}(\vec k,\vec p,\vec 0) = -\frac{1}{2}\frac{\partial}{\partial k_2}\G^{\m}(\vec k,\vec p)\;,\label{vert2}
\ee
where the factor $\frac{1}{2}$ is compensated by a factor $2$ which counts in how many ways we can choose the external wavy line in the second graph;

\item[(iii)] the sum of the fouth and sixth graphs in Fig. \ref{figvert} is exactly zero, because of the identity
\be
\G^{\m_1,\m_2,2}(\vec k,\vec p,-\vec p,\vec 0) = -\frac{1}{3}\frac{\partial}{\partial k_2}\G^{\m_1,\m_2}(\vec k,\vec p,-\vec p)\;,\label{vert3}
\ee
\end{itemize}
where the factor $\frac{1}{3}$ is compensated by a factor $3$ which counts in how many ways we can choose the external wavy line in the third graph.\\

The case $\m=1$ can be obtained from the case $\m=2$ by symmetry (4) of Lemma \ref{lem2.4}; it follows that even in this case the sum of the graphs in Fig. \ref{figvert} is exactly vanishing. This concludes our lowest order check of the Ward identities.

\section{The relativistic propagator in real space}\label{appgl}
\setcounter{equation}{0}
\renewcommand{\theequation}{\ref{appgl}.\arabic{equation}}

In this Appendix we reproduce the details of the computation leading to (\ref{exc31}). Let us rewrite $g^{(h)}_{\o,\LL}(\xx)$ as
\be
g^{(h)}_{\o,\LL}(\xx) = \eu^{-\iu\pp_{F}^{\o}\xx}\int \frac{d\kk'}{D} \eu^{-\iu\kk'\xx}\frac{f_h(\kk')}{|\kk'|^2}\Big( -\G^{0}_{\o}k_0 + k'_i \G^{i}_{\o} \Big)\;.\label{gl1} 
\ee
Notice that
\bea
\int \frac{d\kk'}{D} \eu^{-\iu\kk'\xx}\frac{f_h(\kk')}{|\kk'|^2} k'_{\a}\eu^{-\iu \kk'\xx} &=& i\frac{\partial}{\partial x_{\a}}\int \frac{d\kk'}{D}\,\frac{f_{h}(\kk')}{|\kk'|^2}\eu^{-\iu \kk'\xx}\;,\nn\\
&=& i\frac{\partial}{\partial x_{\a}}\int \frac{d\r}{B_1}\sin\th d\th f_{h}(\r) \eu^{-\iu \r |\xx|\cos\th}\;,\label{gl2}
\eea
where to get the second line we have chosen $\th$ to be the angle of $\kk'$ with respect to $\xx$. Then, performing the integral over $d\cos\th$ and the derivative with respect to $x_{\a}$ we get that
\be
\int \frac{d\kk'}{D} \eu^{-\iu\kk'\xx}\frac{f_h(\kk')}{|\kk'|^2} k'_{\a}\eu^{-\iu \kk'\xx} = \frac{2}{B_1}\frac{x_\a}{|\xx|^3}\Big[ \int dt\, \frac{\sin t}{t} f_{h}\Big( \frac{t}{|\xx|} \Big) + \frac{1}{|\xx|}\int dt\,\sin t\, f'_{h}\Big( \frac{t}{|\xx|} \Big) \Big]\;.\label{gl3}
\ee
Plugging this result in (\ref{gl1}) and using that $\sum_{h\leq 0}f_h(\cdot) = \chi(\cdot)$, we get:
\bea
g_{\o,\LL}(\xx) &=& \sum_{h\leq 0}g^{(h)}_{\o,\LL}(\xx)\label{gl4}\\
&=& \eu^{-\iu \pp_{F}^{\o}\xx} \frac{-x_0 \G^{0}_{\o} + x_{i}\G^{i}_{\o}}{(B_1/2) |\xx|^3} \Big[\int dt\,\frac{\sin t}{t} \chi\Big(\frac{t}{|\xx|}\Big) + \frac{1}{|\xx|}\int dt\,\sin t\, \chi'\Big(\frac{t}{|\xx|}\Big) \Big]\;;\nn\\
\eea
finally, using integration by parts and Riemann - Lebesgue Lemma, we know that for any $N>0$
\be
\lim_{|\xx|\rightarrow+\infty}|\xx|^{N}\int \frac{\sin t}{t}\Big[ \chi\Big(\frac{t}{|\xx|}\Big) - 1 \Big] = 0\;,\quad \lim_{|\xx|\rightarrow+\infty} |\xx|^{N}\int dt\,\sin t\, \chi'\Big(\frac{t}{|\xx|}\Big) =0\;,\label{gl5}
\ee
that is, we can rewrite $g_{\o,\LL}(\xx)$ as:
\bea
g_{\o,\LL}(\xx) &=& \eu^{-\iu \pp_{F}^{\o}\xx}\frac{-x_0 \G^{0}_{\o} + x_{i}\G^{i}_{\o}}{(B_1/2)|\xx|^3} \int dt\,\frac{\sin t}{t} + o\Big( \frac{1}{1 + |\xx|^{N}} \Big)\;,\label{gl6}\\
&=& \eu^{-\iu \pp_{F}^{\o}\xx}\frac{-x_0 \G^{0}_{\o} + x_{i}\G^{i}_{\o}}{b' |\xx|^{3}} + o\Big( \frac{1}{1 + |\xx|^{N}} \Big)\;,\quad b' := \frac{B_1}{\pi} = \frac{8\pi}{3\sqrt{3}}\;.\nn
\eea

\section{Proof of (\protect\ref{exc32})}\label{appgl2}
\setcounter{equation}{0}
\renewcommand{\theequation}{\ref{appgl}.\arabic{equation}}

In this Appendix we shall compute the trace
\be
\Tr\Big\{ \G^{\a} g_{\o_1,\LL}(\xx)\G^{\a} g_{\o_2,\LL}(-\xx) \Big\}\;,\label{gl21}
\ee
for $\a = E_{\pm}, CDW, D$. First of all, using (\ref{exc31}), we rewrite (\ref{gl21}) as:
\be
\frac{\eu^{-\iu \xx (\pp_{F}^{\o_1} - \pp_{F}^{\o_2})}}{{b'}^2 |\xx|^6}\Tr\Big\{ \G^{\a}\big( -x_0 \G^{0}_{\o_1} + x_i \G^{i}_{\o_1} \big)\G^{\a} (x_0 \G^{0}_{\o_2} - x_i \G^{i}_{\o_2}) \Big\}\;;\label{gl22}
\ee
then, we use that:
\bea
\Tr\Big\{ \G^{D}\G^{0}_{\o_1}\G^{D}\G^{0}_{\o_2}\Big\} &=& - \Tr\Big\{ \G^{D}\G^{1}_{\o_1}\G^{D}\G^{1}_{\o_2} \Big\} \nn\\&=& - \o_1 \o_2 \Tr\Big\{ \G^{D}\G^{2}_{\o_1}\G^{D}\G^{2}_{\o_2} \Big\} = -2\;,\label{gl23}\\
\Tr\Big\{ \G^{CDW}\G^{0}_{\o_1}\G^{CDW}\G^{0}_{\o_2}\Big\} &=& \Tr\Big\{ \G^{CDW}\G^{1}_{\o_1}\G^{CDW}\G^{1}_{\o_2}\Big\}\nn\\&=& \o_1\o_2\Tr\Big\{ \G^{CDW}\G^{2}_{\o_1}\G^{CDW}\G^{2}_{\o_2}\Big\} = -2\;,\\
\Tr\Big\{ \G^{E_{\pm}}\G^{0}_{\o_1}\G^{E_{\pm}}\G^{0}_{\o_2}\Big\} &=& \pm \Tr\Big\{ \G^{E_{\pm}}\G^{1}_{\o_1}\G^{E_{\pm}}\G^{1}_{\o_2}\Big\} \nn\\&=& \mp \o_1\o_2 \Tr\Big\{ \G^{E_{\pm}}\G^{2}_{\o_1}\G^{E_{\pm}}\G^{2}_{\o_2}\Big\} = \mp 2\;,\label{gl24}
\eea
and that $\Tr\Big\{ \G^{\a}\G^{\m}_{\o_1}\G^{\a}\G^{\n}_{\o_2}\Big\} = 0$ for $\m\neq \n$, $\forall \a$. From (\ref{exc30}), (\ref{gl22}) -- (\ref{gl24}) it is easy to see that the results claimed in (\ref{exc32}) follow.


%
%

\chapter[Corrections to the Ward Identities]{Corrections to the Ward Identities}\label{app5}
\setcounter{equation}{0}
\renewcommand{\theequation}{\ref{app5}.\arabic{equation}}

\begin{figure}[htbp]
\centering
\includegraphics[width=0.25\textwidth]{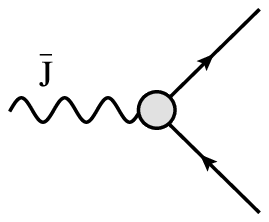}
\caption{The vertex corresponding to $\bar J_{\pp} \hat\Psi^{+}_{\kk+\pp,\s} C^{K}(\kk,\pp)\hat\Psi^{-}_{\kk,\s}$.}\label{figcoor}
\end{figure}

In this Appendix we will show that the corrections to the Ward identities for the photon mass (\ref{WI8}) and for the effective charge (\ref{WI8b}) are exponentially vanishing in the ultraviolet limit $K\rightarrow+\infty$. The corrections $p_{\m}\D^{[h,K]}_{0,2,\m,\n}(\pp)$, $p_{\m}\D^{[h,K]}_{2,1,i,j,\m}(\kk,\pp)$ can be obtained as functional derivatives with respect to the external fields $\bar J_{\pp}$, $\hat J_{\n,-\pp}$ or $\bar J_{\pp}$, $\hat \phi_{\kk+\pp,\s,i}^{+}$, $\hat \phi^{-}_{\kk,\s,j}$ of:
\be
\WW_{\b,L}^{[h,K]}(\bar J,J,\phi) := \log \int P(d\Psi)P_{\geq h}(dA)\eu^{V(\Psi,A + J) + B(\Psi,\phi) + \CC(\Psi,\bar J)}\;,\label{corr1}
\ee
where
\be
\CC(\Psi,\bar J) := \int \frac{d\kk}{D}\frac{d\pp}{(2\pi)^{3}}\,\bar J_{\pp} \hat\Psi^{+}_{\kk+\pp,\s,\r} C^{K}_{\r\r'}(\kk,\pp)\hat\Psi^{-}_{\kk,\s,\r'}\;,\label{corr2}
\ee
and $C^{K}$ given by:
\bea
&&-\iu e^{-1}C^{K}_{\r\r'}(\kk,\pp) :=\label{corr3}\\&& \eu^{-\iu p_{1}(\r' - 1)}\big(\chi_{K}^{-1}(k_0 + p_0) - 1\big)\big[B(\kk+\pp)\big]_{\r\r'} - \eu^{-\iu p_{1} (\r - 1)}\big(\chi_K^{-1}(k_0) - 1\big)\big[B(\kk)\big]_{\r\r'}\nn
\eea
As for the generating functional of the correlation functions, the functional integral (\ref{corr1}) can be studied using multiscale analysis and renormalization group; the only difference is the presence of the term $\CC(\Psi,\bar J)$. A crucial role in the analysis is playied by the properties of the function $C^{K}(\kk,\pp)$; in fact, it is straightforward to check that 
\be
\hat g(\kk+\pp)C^{K}(\kk,\pp)\hat{g}(\kk)\;,\label{corr4}
\ee
is non-vanishing only if $|k_0 + p_0|\geq M^{K}$, and/or $|k_0|\geq M^{K}$. Moreover, when it is non-vanishing (\ref{corr4}) it is dimensionally bounded as $(\const.)|\pp|M^{-2K}$. Let us choose the external fermionic field as in Section \ref{schwing}; proceeding as in Appendix \ref{app2b}, after the integration of the ultraviolet degrees of freedom we are left with:
\bea
&&\eu^{\WW_{\b,L}^{[h,K]}(\bar J,J,\phi)} = e^{-\b|\L| F_{0} + \CC^{(\geq 0)}(\phi,\bar J)}\cdot\label{corr5}\\&&\cdot\int P(\Psi^{(\leq 0)})P_{\geq h}(dA^{(\leq 0)})e^{\VV^{(0)}(\Psi^{(\leq 0)},A^{(\leq 0)} + J) + B(\Psi^{(\leq 0)},\phi) + \CC^{(0)}(\Psi^{(\leq 0)},\phi,\bar J,A^{(\leq 0)}+J)}\;,\nn
\eea
where $\CC^{(\geq 0)}$, $\CC^{(0)}$ contain all the $\bar J$-dependent terms, and $\CC^{(0)}(0,\phi,\bar J,0) := 0$. The kernels of $\CC^{(\geq 0)}$, $\CC^{(0)}$ admit a graphical representation in terms of Gallavotti-Nicol\` o trees, following the rules discussed in Section \ref{sec4b} and Appendix \ref{app2b}, with the only difference that new endpoints appear, corresponding to $\bar J_{\pp} \hat\Psi^{+}_{\kk+\pp,\s,\r} C^{K}_{\r\r'}(\kk,\pp)\hat\Psi^{-}_{\kk,\s,\r'}$; as usual, the trees can be expanded in terms of Feynman graphs, with the new endpoints corresponding to the vertex represented in Fig. \ref{figcoor}. 

Because of the properties of (\ref{corr4}), at least one of the solid lines emerging from the new vertex has to be contracted on scale $K$; therefore, using the general dimensional bound (\ref{C14}) and the short memory property of the GN trees, the kernels of $\CC^{(\geq 0)}$ are bounded proportionally to $(\const.)|\pp|M^{-\th K}\prod_{i=0}^{P_{\phi}}M^{-h_{\kk_i}}$, where $P_{\phi}$ is the number of $\phi$ external lines and $\{\kk_i\}$ are their momenta, while those in $\CC^{(0)}$ admit an improvement of a factor $|\pp|M^{-\th K}$ with respect to their basic dimensional bounds. The bound on the kernels of $\CC^{(\geq 0)}$ is derived from (\ref{C14}) taking into account that the tadpole graph associated to the new vertex depicted in Fig. \ref{figcoor} is exactly vanishing.

The single scale integration is performed as in Section \ref{schwing}, with the only difference that the $\bar J$-dependent terms independent of $\Phi^{(\leq -1)}$, $A^{(\leq -1)} + J$ are taken into account in the definition of $\CC^{(\geq -1)}(\phi,\bar J) := \CC^{(\geq 0)}(\phi,\bar J) + \tilde \CC^{(0)}(\phi,\bar J)$, where $\tilde \CC^{(0)}$ collect the new $\bar J$-dependent kernels independent of $\Psi^{(\leq -1)}$, $A^{(\leq -1)} + J$ produced in the integration of the scale $0$, while all the other $\bar J$-dependent ones are contained into $\CC^{(-1)}(\Psi^{(\leq -1)},\phi,\bar J,A^{(\leq -1)}+J)$, and so on. After the integration of the first $|k|\leq |h|$ scales we are left with:
\bea
&&\eu^{\WW^{[h,+\infty]}_{\b,L}(\bar J,J,\phi)} = \label{corr6}\\&&=e^{-\b|\L| F_{k}+\SS^{(\ge k)}(\phi) + \CC^{(\geq k)}(\phi,\bar J)}
\int P(d\Psi^{(\le k)})P_{\geq h}(dA^{(\le k)})
e^{\VV^{(k)}(\sqrt{Z_k} \Psi^{(\leq k)},A^{(\leq k)} + J)}
\cdot\nn\\
&&\cdot e^{\BBB^{(h)}(\sqrt{Z_k}\Psi^{(\leq k)},\phi,A^{(\leq k)} + J)+W_{R}^{(k)}(\sqrt{Z_k}\Psi^{(\leq k)},\phi,A^{(\leq k)} + J) + \CC^{(k)}(\Psi^{(\leq k)},\phi,\bar J,A^{(\leq k)}+J)}\;,\nn
\eea
where all the $\bar J$-dependent terms absorbed in $\CC^{(\geq k)} = \CC^{(\geq 0)} + \sum_{w = k+1}^{0}\tilde \CC^{(w)}$ and $\CC^{(k)}$, with $\CC^{(k)}(0,\phi,\bar J,0):=0$, while all the other objects appearing in (\ref{corr6}) have been defined in Section \ref{schwing}. Again, because of the short memory property, the dimensional bounds on the kernels of $\tilde\CC^{(w)}$ and of $\CC^{(k)}$ admit respectively an improvement of a factor $|\pp|M^{\th(-K + w)}$, $|\pp|M^{\th (-K +k)}$, with respect to the basic ones derived in Section \ref{sec4b}. Finally, the integration of the scales $< h$ proceeds in a way completely analogous to the one sketched in Section \ref{schwing}. The conclusion is that the kernels of $\tilde\CC^{(w)}$ are exponentially vanishing for $K\rightarrow+\infty$, uniformly in $w$; this implies that 
\be
\lim_{K\rightarrow+\infty}\D^{[h,K]}_{0,2,\m,\n}(\kk,\pp) = \lim_{K\rightarrow+\infty}\D^{[h,K]}_{2,1,\m}(\kk,\pp) =0\;,
\ee
as desired.

\chapter[Proof of \ref{lat6c}]{Proof of \ref{lat6c}}\label{applat}
\setcounter{equation}{0}
\renewcommand{\theequation}{\ref{applat}.\arabic{equation}}

In this Appendix we prove that in presence of Kekul\' e distortion the effective potential arising after the integration of the ultraviolet degrees of freedom has the form (\ref{lat6c}). With respect to the case $\D_0=0$, the only new check to perform is to show that the relevant terms, that is the {\it mass terms} for the fermion and photon field, have the form claimed in (\ref{lat15}), (\ref{lat6c}). The symmetry properties to which we shall refer here are the (1) -- (8) of Lemma \ref{lemlat}
\section{Fermionic mass terms}\label{applat1}
\setcounter{equation}{0}
\renewcommand{\theequation}{\ref{applat1}.\arabic{equation}}

By symmetry (8) it follows that 
\be
\widetilde W^{(j_0)}_{2,0,\r,\r,\ul{\o}}(\V0)=0\;.\label{applat1.0} 
\ee
Symmetry (4) implies that
\bea
&&\widetilde W^{(j_0)}_{2,0,\r,\r',\o,\o'}(\V0) = \label{applat1.1}\\&& = \widetilde W^{(j_0 + 1)}_{2,0,\r,\r',\o,\o'}(\V0)\exp\Big\{ \iu \vec p_{F}^{\o}(\vec\d_2 - \vec\d_1)(\r - 1) - \iu \vec p_{F}^{\o'}(\vec\d_2 - \vec\d_1)(\r' - 1) \Big\}\;,\nn
\eea
while symmetry (6.b) implies that
\be
\widetilde W^{(j_0)}_{2,0,\r,\r',\o,\o'}(\V0) = \widetilde W^{(j'_0)}_{2,0,\r,\r',-\o,-\o'}(\V0)\;;\label{applat1.1b}
\ee
therefore, from the combination (4) - (6.b) - (4) - (6.b) we find that:
\be
\widetilde W^{(1)}_{2,0,\r,\r',\o,\o}(\V0) = e^{2\iu \vec p_{F}^{\o}(\vec \d_{2} - \vec\d_1)(\r - \r')}\widetilde W^{(1)}_{2,0,\r,\r',\o,\o}(\V0)\;.\label{applat1.2}
\ee
Formula (\ref{applat1.2}) together with (\ref{applat1.1}) gives, if $\r\neq \r'$:
\be
\widetilde W^{(1)}_{2,0,\r,\r',\o,\o}(\V0)=0 = \widetilde W^{(j_0)}_{2,0,\r,\r',\o,\o}(\V0)\;.\label{applat1.2b}
\ee
Symmetry (4) implies in particular that 
\be
\widetilde W^{(j_0)}_{2,0,1,2,\o,-\o}(\V0) = \widetilde W^{(j_0 + 1)}_{2,0,1,2,\o,-\o}\eu^{-\iu \o \frac{2\pi}{3}}\;,\label{applat1.3}
\ee
while symmetry (5) implies that
\be
\widetilde W^{(j_0)}_{2,0,1,2,\o,-\o}(\V0) = \big[\widetilde W^{(j_0)}_{2,0,1,2,\o,-\o}(\V0)\big]^{*}\;;\label{applat1.4}
\ee
finally, symmetry (6.a) implies that
\be
\widetilde W^{(j_0)}_{2,0,\r,\r',\o,\o'}(\V0) = \widetilde W^{(j_0)}_{2,0,\r',\r,\o,\o'}(\V0)\;.\label{applat1.5}
\ee
Therefore, setting $\widetilde W^{(1)}_{2,0,\r,3-\r,\o,-\o}(\V0) =: \D^{(1)}_0$, formulas (\ref{applat1.3}), (\ref{applat1.5}) imply that:
\be
\D^{(j_0)}_{0,\o} := \widetilde W^{(j_0)}_{2,0,\r,3-\r,\o,-\o}(\V0) = \eu^{-\iu \vec p_{F}^{\o}(\vec\d_{j_0} - \vec\d_1)}\D^{(1)}_{0}\;;\label{applat1.6}
\ee 
moreover, formula (\ref{applat1.4}) implies that $\D^{(1)}_{0}\in \RRR$. This, together with (\ref{applat1.0}), (\ref{applat1.2b}), concludes the proof of (\ref{lat15}).

\section{Bosonic mass terms}\label{applat2}
\setcounter{equation}{0}
\renewcommand{\theequation}{\ref{applat2}.\arabic{equation}}

From (6.a) and (7) it follows that, if $\m\neq \n$:
\be
\widetilde W^{(j_{0})}_{0,2,\m,\n}(\V0) =0\;,\label{applat2.1}
\ee
while from (4) it follows that
\be
\widetilde W^{(1)}_{0,2,0,0}(\V0) = \widetilde W^{(2)}_{0,2,0,0}(\V0) = \widetilde W^{(3)}_{0,2,0,0}(\V0)\;.\label{applat2.1b}
\ee
Symmetry (6.b) implies
\be
\widetilde W^{(2)}_{0,2,i,i}(\V0) = \widetilde W^{(3)}_{0,2,i,i}(\V0)\;,\label{applat2.2} 
\ee
and from (4) we get
\be
\begin{pmatrix} \widetilde W^{(1)}_{0,2,1,1} & 0 \\ 0 & \widetilde W^{(j_0)}_{0,2,2,2} & 0 \end{pmatrix} = T\begin{pmatrix} \widetilde W^{(2)}_{0,2,1,1}(\V0) & 0 \\ 0 & \widetilde W^{(2)}_{0,2,2,2}(\V0) \end{pmatrix} T^{-1}\;;\label{applat2.3}
\ee
therefore, formulas (\ref{applat2.1b}) -- (\ref{applat2.3}) imply that:
\bea
\widetilde W^{(1)}_{0,2,\m,\m}(\V0) &=& \widetilde W^{(2)}_{0,2,\m,\m}(\V0) = \widetilde W^{(3)}_{0,2,\m,\m}\;,\nn\\
\widetilde W^{(j_0)}_{0,2,1,1}(\V0) &=& \widetilde W^{(j_0)}_{0,2,2,2}(\V0)\;.\label{applat2.4}
\eea
Formulas (\ref{applat2.1}), (\ref{applat2.4}) prove the structure of the bosonic mass terms claimed in (\ref{lat6c}); therefore, this together with what has been discussed in the previous Section concludes the proof of \ref{lat6c}.

\begin{acknowledgements}[Ringraziamenti]
\addcontentsline{toc}{chapter}{Acknowledgements}

Ringrazio il Prof. G. Gallavotti per avermi introdotto nel mondo della ricerca scientifica, per aver seguito con grande disponibilit\` a ed interesse il mio lavoro, e per essere stato fonte di preziosi consigli e di importanti stimoli in tutti questi anni.\\

Ringrazio il Prof. A. Giuliani per essere stato un costante punto di riferimento durante il mio dottorato, e per la profonda influenza che ha avuto nella mia formazione scientifica; il suo contributo a questo lavoro \` e stato essenziale.\\

Infine, un ringraziamento speciale va al Prof. V. Mastropietro, per avermi fornito gli strumenti indispensabili per affrontare il lavoro di Tesi, e per il ruolo di guida che ha svolto durante il mio dottorato. Senza le sue idee ed il suo entusiasmo niente di quello che \` e stato fatto in questa Tesi sarebbe stato possibile.

\end{acknowledgements}

\thispagestyle{plain}


\end{document}